\documentclass[PhD, colorlinks=true, linkcolor=black, citecolor=black, urlcolor=black]{dfaunictthesis}
\usepackage{lipsum}
\usepackage{fancyhdr}
\usepackage{tikz}
\DeclareMathAlphabet{\mathbbb}{U}{bbold}{m}{n}
\usepackage{bm} 
\usepackage{feynmf}
\usepackage{float}
\usepackage[Lenny]{fncychap}
\usepackage{amsmath} 
\usepackage{amssymb} 
\usepackage{bm,dsfont} 
\usepackage{slashed} 
\usepackage{tensor}
\usepackage{amsfonts}
\usepackage{physics}

\usepackage[numbers,sort&compress]{natbib}

\usepackage[absolute,overlay]{textpos}

\usepackage{verbatim}
\usepackage{amsmath}
\usepackage{graphicx}
\usepackage{caption}
\usepackage{subcaption}
\usepackage{bbold}
\usepackage{graphics}
\usepackage{geometry}
\newcommand{\cm}[2]{\left[ #1, #2 \right]}
\newcommand{\nn}{\nonumber \\}

\newcommand{\p}{\partial}
\newcommand{\dg}{\dagger}

\renewcommand{\Im}{ \mathrm{Im} }
\newcommand{\fm}{\mathrm{fm}}

\usepackage[sort,compress]{cleveref}
\crefname{section}{Sec.\!}{Secs.\!}
\crefname{equation}{Eq.\!}{Eqs.\!}
\crefname{figure}{Fig.\!}{Figs.\!}
\crefname{table}{Tab.\!}{Tabs.\!}
\crefname{appendix}{App.\!}{Apps.\!}
\crefname{chapter}{Chapter}{Chapters}

\newcommand\matthree[9]{%
  \begin{pmatrix}
  #1 & #2 & #3 \\ #4 & #5 & #6 \\ #7 & #8 & #9
  \end{pmatrix}%
}

\usepackage{upgreek}
\usepackage[acronym]{glossaries-extra}
\glssetcategoryattribute{acronym}{nohyperfirst}{true}
\setabbreviationstyle[acronym]{long-short}

\newacronym{rg}{RG}{Renormalization Group}
\newacronym{frg}{FRG}{Functional Renormalization Group}
\newacronym{uv}{UV}{ultraviolet}
\newacronym{ir}{IR}{infrared}
\newacronym{lpa}{LPA}{local potential approximation}
\newacronym{mc}{MC}{Monte Carlo}
\newacronym{cb}{CB}{Conformal Bootstrap}
\newacronym{qcd}{QCD}{Quantum Chromodynamics}
\newacronym{kt}{KT}{Kurganov-Tadmor}
\newacronym{pde}{PDE}{partial differential equation}
\newacronym{ode}{ODE}{ordinary differential equation}
\newacronym{QPR}{QPR}{quasi-particle regulator}
\newacronym{SR}{SR}{standard regulator}

\newacronym{SL}{SL}{smooth Litim-like}
\newacronym{E}{E}{exponential}
\newacronym{L}{L}{Litim}
\newacronym{SSB}{SSB}{spontaneous symmetry breaking}
\newacronym{QMM}{QMM}{quark-meson model}

\usepackage[normalem]{ulem}

\def\d{\,\mathrm{d}}

\definecolor{goethe-blau}{cmyk}{1.0,0.2,0.0,0.4}

\begin{document}
	
	\author{Gabriele Parisi}
	\title{Heavy Quarks in the initial stages of Proton-Ion Collisions}
 \aayear{2024/2025}
\begin{supervisors}
   \supervisor{Chiar.mo}{Prof.}{V. Greco}
   \supervisor{Chiar.mo}{Prof.}{M. Ruggieri}
\end{supervisors}
	\maketitlepage

	
	\thispagestyle{empty}
	\tableofcontents
	\newpage
	\chapter*{List of publications}

    This Thesis is largely based on the following papers:
\begin{description}
    
\item[-] \cite{Parisi:2025slf} Gabriele Parisi, Vincenzo Greco and Marco Ruggieri, ``Anisotropic fluctuations of momentum and angular momentum of heavy quarks in the pre-equilibrium stage of pA collisions at the LHC'', Phys. Rev. D \textbf{112} (2025) no.7, 074030, arXiv: 2505.08441 [hep-ph].

\item[-] \cite{Oliva:2024rex} Lucia Oliva, Gabriele Parisi, Vincenzo Greco and Marco Ruggieri, ``Melting of $c\bar{c}$ and $b\bar{b}$ pairs in the pre-equilibrium stage of proton-nucleus collisions at the Large Hadron Collider'', 
Phys. Rev. D \textbf{112} (2025) no.1, 014008, arXiv: 2412.07967 [hep-ph].

\item[-] \cite{Parisi:2026uhh} Gabriele Parisi, Fabrizio Murgana, Vincenzo Greco and Marco Ruggieri, ``Elliptic flow of charm quarks produced in the early stage of pA collisions'', arXiv: 2601.11123 [hep-ph].

\end{description}
Other works of the PhD candidate:
\begin{description}

\item[-] \cite{Parisi:2025gwq} Gabriele Parisi, Vincenzo Nugara, Salvatore Plumari and Vincenzo Greco, ‘‘Shear viscosity of a binary mixture for a relativistic fluid at high temperature'', Phys. Rev. D \textbf{113} (2026) no.1, 014001, arXiv: 2510.20704 [hep-ph].
    
\item[-] \cite{Sambataro:2024mkr} Maria Lucia Sambataro, Vincenzo Greco, Gabriele Parisi and Salvatore Plumari,
``Quasi particle model vs lattice QCD thermodynamics: extension to $N_f=2+1+1$ flavors and momentum dependent quark masses'',
Eur. Phys. J. C \textbf{84} (2024) no.9, 881, arXiv: 2404.17459 [hep-ph].

\item[-] \cite{Coci:2025drb} Gabriele Coci, Gabriele Parisi, Salvatore Plumari and Marco Ruggieri,
``Entropy from decoherence: a case study using glasma-based occupation numbers'', arXiv: 2507.04809 [hep-ph].

\end{description}

\chapter*{Abstract}
Collisions among heavy ions, like Pb or Au, are a great tool to study the theory of strong interactions, that is Quantum Chromodynamics (QCD). In particular, these experiments are able to give insights on all the complex phases of matter that the theory of QCD allows. In this PhD Thesis we have investigated the initial stages of proton-ion collisions: in particular, we will focus on the first $\sim 0.4$ fm/c ($\sim 10^{-24}$ s) after the collision, which are dominated by very intense gluon fields, in a state called glasma. We investigated the effect of such fields on the dynamics of heavy quarks (charm and beauty) which are created and evolve in this medium. The effect of the initial gluon fields on heavy quarks is quite substantial, in particular we observe that the glasma provokes a $50\%$ dissociation rate on quark-antiquark pairs. Moreover, glasma fields have a large momentum anisotropy, and transmit a large part of such anisotropy to the heavy quarks which evolve in this medium. Finally, we have generalized our study to a non-boost invariant medium, and shown that fluctuations in rapidity do not lead to significant isotropization within glasma timescales.

	\chapter{Introduction}\label{intro}

The core of the research in Physics is the comprehension of the underlying laws that govern the phenomena of the Universe, from the smallest subatomic particles to the largest cosmic structures. In this context, one of the most significant challenges in contemporary physics lies in understanding the behavior of strongly interacting matter. The theory that serves as the primary tool to elucidate its properties and interactions is the so-called {\sl Quantum ChromoDynamics} (QCD). Despite not being fully understood yet, this theory stands as a foundational element in modern particle physics, offering profound insights into the underlying structure of matter and the powerful force that binds it together. At its essence, QCD explores the dynamics of quarks and gluons, the basic building blocks of particles like protons, neutrons, and other hadrons.\\

In the 20$^\text{th}$ century, scientists embarked on a quest to unravel the fundamental structure of matter. This endeavor led to the current formulation, finalized around the end of the '70s, of the {\sl Standard Model} of particle physics, a theoretical framework that elegantly explains how subatomic particles interact and the nature of the such underlying forces. Within this model, QCD emerges as the theory governing the {\sl strong force}, one of the fundamental forces alongside gravity, electromagnetism, and the weak force. The strong force, mediated by particles known as {\sl gluons}, is responsible for binding {\sl quarks} together to form composite particles called {\sl hadrons}. Besides the degree of freedom of flavor, each characterized by a different mass and electric charge, quarks possess a fundamental property called ‘‘color charge'', in correspondence to the standard electric charge but with a broader scope. In fact, in the name QCD, “chromodynamics'' refers to the fact that the interaction between quarks and gluons happens through the exchange of color charge (the name {\sl color} is not related to any form of electromagnetic perception). While QCD has made considerable strides in uncovering the nature of strongly interacting matter, most importantly in understanding its behavior at high energies ({\sl asymptotic freedom} \cite{Gross:1973id,Politzer:1973fx,Coleman:1973sx}) and in providing explanations for hadron masses \cite{BMW:2008jgk}, several aspects remain elusive. For instance, the mechanism behind {\sl color confinement}, i.e. the fact that no color-charged object can be observed, is only poorly understood. It should be sufficient to mention that the behavior of QCD matter is so complex that a whole plethora of different phases can be obtained, by varying either the temperature or the matter density, and those form a non-trivial phase diagram, whose study is under active research. Even more importantly, we know only little about how the transition between one state of matter and another occurs, and to get some theoretical insights on that one must consider the use of effective models or numerical calculations.

In general a huge effort, both theoretically and experimentally, has been put on this research field. Being able to deepen our current knowledge on the behavior of strongly interacting matter would have a significant impact on several areas of Physics, ranging from our understanding of the very hot early stages of our universe, to the comprehension of the mechanisms that regulate dense and relatively cold neutron stars. \\

From an experimental point of view, the study of the properties of strongly interacting matter and the investigation of the QCD phase diagram are some of the main goals of modern experiments. Part of these experiments is performed via {\sl heavy-ion collisions}, like, for example, the ones that take place at the Large Hadron Collider (LHC for short, at the European Organization for Nuclear Research in Geneva) and at the Relativistic Heavy Ion Collider (RHIC for short, at the Brookhaven National Laboratory in Upton, New York). In the near future, more experiments are expected to be added to the previous ones, namely the ones performed at the Facility for Antiproton and Ion Research (FAIR, in Darmstadt), and at the Joint Institute for Nuclear Research  (JINR, in Dubna).

From a theoretical viewpoint, a great advancement in the direction of the understanding of the QCD phase diagram has been made with the introduction of {\sl lattice simulations}, which, as the name suggests, are able to perform non-perturbative calculations by discretizing the theory on a lattice. In particular, lattice studies obtained stunning results in the low chemical-potential region of the phase diagram, for instance confirming the hypothesis that in this regimes the phase transition between hadronic matter and a quark-gluon plasma is a smooth crossover. This result is just the tip of the iceberg, since the QCD phase diagram incorporates many other interesting features, but the reliability of lattice-QCD  results is limited to the case of vanishing or  small chemical potentials, due to the presence of the so-called {\sl sign problem} (see e.g. \cite{Pan:2022fgf,Allton:2002zi,Philipsen:2008zz,Anagnostopoulos:2001yb}). Moreover, even for small chemical potentials, lattice QCD is limited to the description of equilibrium properties, while the study of time-evolving systems is much more challenging. In all the cases that numerical studies cannot handle, {\sl effective models} can play a crucial role in capturing essential features of QCD, albeit with simplifications.  The study of effective field theories and models is crucial for our understanding of Physics, especially when we deal with fundamental interaction theories like QCD. In particular, these effective models are different from QCD in some aspects, but are built in such a way that they can capture some essentials features of the underlying theory, using a simplified subset of symmetries and degrees of freedom. The ultimate goal is the understanding of the critical features of these models, which within their applicability range are much more simple to deal with than QCD, but can give us an insight on the actual phenomena of QCD. \\

Heavy-ion collision experiments such as the ones at RHIC and at LHC provide fascinating insights into the properties of strongly interacting nuclear matter under extreme conditions. The main phenomenon of interest in these collisions is the creation and evolution of the {\sl Quark-Gluon Plasma} (QGP), an exotic state of matter in which the constituents of nuclei, neutrons and protons, come apart and break into their fundamental building blocks, namely quarks and gluons. The study of the QGP is of crucial importance for testing the ability of QCD to explain and predict experimental data. Moreover, the QGP created in relativistic nuclear collisions can be used as a model for the early universe immediately after the Big Bang, therefore its study also has implications for cosmology \cite{Heinz:2013th}.

Although the basic equations of QCD, which describe the interactions between quarks and gluons, are well established, the theoretical description of heavy-ion collisions in terms of the full theory is possible only partially. This is due to the phenomenon, which we have just mentioned, of confinement: the strong coupling constant becomes large at low momenta and consequently perturbative techniques are bound to fail. Since we are missing a comprehensive description of the phenomenon from first principles, the evolution of the ``fireball'' created in heavy-ion collisions is therefore split into various stages with different appropriate effective models used for each stage, each model approximating the underlying QCD processes \cite{Brambilla:2014jmp}. Roughly speaking, the three main stages are the {\sl pre-equilibrium}, the {\sl equilibrium} and the {\sl freeze-out}. The pre-equilibrium describes the earliest stage, starting directly after the collision until the evolving matter is in thermal equilibrium, which is a process known as thermalization. The fireball then evolves as a QGP in thermal equilibrium until the freeze-out, where quarks and gluons recombine and form a gas of hadrons. This gas stops interacting after some time and the free streaming hadrons travel towards the detectors. 

One of the most striking results to come out of heavy-ion collision experiments is that the QGP in the equilibrium stage almost behaves like an ideal fluid \cite{Adcox:2004mh, Arsene:2004fa, Adams:2005dq, Romatschke:2007mq}.
For this reason, the time evolution of the QGP can be described in large parts by relativistic viscous hydrodynamics \cite{Schenke:2010nt, Gale:2013da, Romatschke:2017ejr}, but this successful description only applies to the evolution of the QGP fireball itself and not to the earliest stages of the collision: as mentioned, different phases require different effective models. Relativistic hydrodynamical simulations therefore require initial conditions from the pre-equilibrium stage, for which there are different types of models. Popular choices are variants of the phenomenological MC-Glauber model \cite{Miller:2007ri}, but a more sophisticated approach to the initial state of nuclear collisions, developed over the past two and a half decades, is the {\sl Color Glass Condensate} (CGC) \cite{Gelis:2010nm, Iancu:2003xm, Iancu:2012xa}. It is a classical effective theory for high energy QCD and provides a first-principles description of the early stages of relativistic heavy-ion collisions. The main idea behind the CGC, as for every effective theory, is a separation of scales: the hard constituents of a relativistic nucleus, i.e. partons which carry most of total momentum such as valence quarks, are described as highly Lorentz-contracted, thin sheets of classical color charge. These fast color charges generate a highly occupied color field, which represents the soft partons of the nucleus, namely mostly gluons at low momenta.
The cutoff at which one performs the separation into soft and hard partons, is an entirely arbitrary longitudinal momentum scale: by requiring that observables do not depend on this artificial cutoff one can obtain a set of renormalization group equations known as the JIMWLK equations \cite{Iancu:2000hn,Ferreiro:2001qy}.  
The CGC is therefore an effective description of high energy nuclei in terms of color fields and color currents. They are approached as classical quantities, therefore their dynamics is governed by classical Yang-Mills field theory. In the CGC model, the result of a collision of two nuclei is a state called glasma (i.e. a gluon plasma), which is a precursor to the QGP \cite{Lappi:2006fp}. Because of the high energy which is produced right after the impact of the two nuclei, we can have the production of {\sl heavy quarks} (charm and beauty) in the first instants after the collisions: those then evolve within the glasma, with which they interact by exchanging energy, momentum, and color charge.\\

In this Thesis we will deal with various aspects of the glasma produced in heavy ion collisions and in proton-ion collisions, and with the dynamics of heavy quarks evolving in glasma. First of all, since most of the glasma studies has been performed assuming boost invariance in the initial stages, we will investigate the effect that initial state rapidity-dependent fluctuations have on the diffusion of heavy quarks, which for simplicity we assume to be infinitely massive. We will then move on to deal with realistic heavy quarks (i.e. charm and beauty) and study the effect that the color dense fields of the glasma have on heavy quark-antiquark pairs. In particular, we will investigate the effect of glasma on the color correlator of the pairs: we will see that the glasma fields are able to provoke the dissociation of heavy quark-antiquark pairs. Finally, we will study the anisotropies which are present in the initial stages of heavy ion collisions and of proton-ion collisions. More specifically, we will study the elliptic flow (the $v_2$, which encodes the anisotropy in momentum space) produced by the dense gluon fields, and also investigate how such anisotropy is then transmitted to heavy quarks in such medium.

This Thesis is organized as follows. In Chapter \ref{chap:The study of Quantum Chromodynamics via Heavy Ion Collisions} we will introduce the main features of QCD, i.e. its Lagrangian, the symmetries of such Lagrangian and the behavior of the coupling constant in this theory. Then, we will move on to describe heavy-ion collisions, our current understanding of these processes, and also argument on why they are useful in order to get insights into the underlying theory which governs them, i.e. QCD. In Chapter \ref{chap:The initial stages of heavy ion collisions} we will go deep into the initial stages of heavy-ion collisions, which is the main topic of this work. In particular, after recapping the current understanding we have on this phase, we will introduce the effective model which is able to describe it, i.e. the color glass condensate. At the end of the Chapter we will also mention some details of the numerical implementation of CGC and various results hereby obtained. In Chapter \ref{chap:Heavy quarks in anisotropically fluctuating glasma} we will study the effect that fluctuations in rapidity, which generalize the typical picture of a boost-invariant glasma, have on the momentum shifts of heavy quarks in the limit of infinite mass. In Chapter \ref{chap:Melting of heavy quark pairs in glasma} we will deal with the color decorrelating effect that the glasma fields have on charm--anticharm and beauty--antibeauty pairs produced early in heavy ion collisions. We also study the effect that this decorrelation has on the melting of the pairs themselves. Finally, in Chapter \ref{chap:Anisotropies of gluons and heavy quarks in glasma} we will study the amount of transverse momentum anisotropy that the gluon fields and the heavy quarks immersed in these fields carry in the early stages of heavy ion collisions. Such anisotropies are usually encoded in quantities like, but not limited to, elliptic flow ($v_2$) and higher order flows ($v_n$ for $n>2$). After studying the effect that the glasma fields have on the charm quark spectrum, we will move on to the elliptic flow $v_2$, which will be evaluated both for gluons and for charm quarks, to highlight the effect that the initial stages of ultrarelativistic collisions have on the experimental observables.

At the end of the Thesis we summarize various technical details, which have not been included in the main text for the sake of clarity, in various Appendices. In particular, in Appendix \ref{app:conv} we clarify on various conventions used in the text, namely on the system of natural units, the Gell-Mann matrices, and Milne coordinates. In Appendix \ref{appendix:spectrum_stuff} we get deeper into the various definitions of momentum spectrum of gluons which are present in the literature.

	\chapter[The study of QCD via Heavy Ion Collisions]{The study of Quantum Chromodynamics via Heavy Ion Collisions}
    \label{chap:The study of Quantum Chromodynamics via Heavy Ion Collisions}
    
	\section{Introduction to Quantum Chromodynamics}
	The theory that we currently use to describe  strongly interacting matter is called {\sl Quantum ChromoDynamics} (QCD). It is a Quantum Field Theory (QFT) and the fields used to represent the fundamental degrees of freedom of the theory are {\sl quarks}, which are spin-$1/2$ fermions, and {\sl gluons}, which are bosons with spin $1$ and mediate the strong force among themselves and among quarks. As a gauge theory, QCD is {\sl non-abelian}: this means that the local transformations under which the Lagrangian must remain invariant belong to a non-commutative gauge group, namely SU$(3)$. This represents a fundamental feature of QCD since, as we will see soon in detail, this implies the possibility for classical 3 and 4-gluon interactions. This feature is absent in abelian gauge theories, like for example Quantum ElectroDynamics (QED), whose gauge group is U$(1)$ which is commutative. 
     
    Quarks can interact via the electromagnetic force, mediated by the photon $\gamma$, since they carry an electric charge (which appears in fractions of either $\pm1/3$ or $\pm2/3$ of the electron elementary charge) and also via the weak force, mediated by the bosons $W^{\pm}$ and $Z^0$. However, the most relevant aspect for us is that they carry the so-called {\sl color charge}, which is the reason why quarks interact via the strong interaction through the exchange of gluons. In particular, from the formal point of view quarks belong to the fundamental (vectorial) 3-dimensional representation of the SU$(3)$, usually referred to as ``triplet'', meaning that they can have $N_c=3$ different colors, usually indicated as {\sl red, blue} and {\sl green} ($r,b,g$), and the corresponding anti-colors for antiquarks, called {\sl anti-red, anti-blue} and {\sl anti-green} ($\bar{r}, \bar{b}, \bar{g}$). Already from the names of the charges we can guess that some combinations of quarks may lead to color-less states, i.e. to a ``white'' color. This can occur if we combine all three colors (or anti-colors) together or if we have the combination of a color and its corresponding anti-color. Similarly, also gluons carry color charge, but they belong to the adjoint representation of SU$(3)$, which is a $N_c^2-1=3^2-1=8$-dimensional representation, therefore we have $8$ possible colors for gluons (hence the name ``octet'' to describe the adjoint representation of SU$(3)$). From the group theoretical point of view, the adjoint representation can be formally obtained by superposing a triplet and an anti-triplet, since
    \begin{equation}
        \textbf{3}\otimes \bar{\textbf{3}}=\textbf{8}\oplus \textbf{1}.
        \label{eq:3x3=8+1}
    \end{equation}
    Gluon colors are therefore usually indicated as combinations of quark colors and anti-colors. Naively, we can think of gluons as able to ‘‘change'' the color of quarks: for instance, a quark $g$ interacting with a $b\bar{g}$ gluon will have its color turned into a $b$.\\

    A fundamental feature of QCD is the so-called {\sl confinement}, which implies that only white (colorless) particles can be observed directly, while single colored quarks and gluons cannot. This is an experimental evidence, since it has never been proven from QCD first principles. The white particles that we observe are the so-called {\sl hadrons}, and we have already mentioned two possibilities for the formation of a colorless state. In particular, we can have a combination of a quark and an antiquark of opposite color, which form a {\sl meson} (e.g. $\pi, K, \eta$), or a combination of three quarks/antiquarks having all three colors, hence summing up to white and forming a {\sl baryon} (e.g. $p,n, \Lambda$). As it seems natural, there are more possibilities for color-neutral objects, like for example {\sl tetraquarks} (formed by two quark-antiquark pairs) or {\sl pentaquarks} (composed by a baryon and a meson), and so on. These combinations are called {\sl exotic} states, check \cite{Huang:2023jec,Lebed:2023vnd,Fini:2023ewy,Bicudo:2022cqi,Cowan:2018fki} and references therein for further details.\\
    Quarks also come in six different types, named {\sl flavors},  which have different masses \cite{ParticleDataGroup:2024cfk}:
    \begin{itemize}
        \item {\sl up}, mass $2.16$ MeV and electric charge $+2/3~e$.
        \item {\sl down}, mass $4.70$ MeV and electric charge $-1/3~e$.
        \item {\sl strange}, mass $93.5$ MeV and electric charge $-1/3~e$.
        \item {\sl charm}, mass $1.27$ GeV and electric charge $+2/3~e$.
        \item {\sl bottom} or {\sl beauty}, mass $4.18$ GeV and electric charge $-1/3~e$.
        \item {\sl top}, mass $172$ GeV and electric charge $+2/3~e$.
    \end{itemize}

    The quarks up and down, with a bit of stretch also the strange, are commonly referred to as {\sl light flavors}. The quarks charm, beauty and top are instead referred to as {\sl heavy flavors}.

    \subsection{The QCD Lagrangian}
	
	As for every gauge theory, the starting point of QCD is its Lagrangian, which reads:\footnote{Summation over repeated indices is always implied, unless otherwise stated.}
	\begin{equation}\label{eq:lagqcd}
		\mathcal{L}_{\text{QCD}}=\bar{\Psi}(x)\left(i\gamma^\mu D_\mu-M\right)\Psi(x)-\frac{1}{4}G^a_{\mu\nu} G_a^{\mu\nu}.
	\end{equation}
    Let us discuss on each term appearing in \eqref{eq:lagqcd}.\\
    
    Starting from the first term on the right hand side of \eqref{eq:lagqcd}, $\Psi(x)$ is a Dirac spinor which collects all the various possible flavors and colors for the quarks. As we discussed, quarks are in the fundamental representation of the SU$(3)$ group, meaning that they have 3 color degrees of freedom. This implies that, since they come in 6 possible flavors and each quark is described by a 4-component Dirac spinor, $\Psi(x)$ is a $4\times6\times3$ component spinor. From that, we define the Dirac adjoint of ${\Psi}(x)$ as
    \begin{equation}
        \bar{\Psi}(x)\equiv \Psi^\dagger(x)\gamma^0,
        \label{eq:bar_psi}
    \end{equation}
    where $\Psi^\dagger(x)$ denotes the hermitian adjoint of the spinor $\Psi(x)$, and $\gamma ^{0}$ is the time-like gamma matrix, which acts only on the four-dimensional Dirac structure of $\Psi(x)$. The quantity $M$, which in principle is a $(4\times6\times3)\times (4\times6\times3)$ matrix, contains the masses of the quarks and it is proportional to the identity both in Dirac and in color space.
    
    The term $D_\mu$ is called the {\sl covariant derivative}, and it is defined as:
    \begin{align}
        D^{ff', ij}_\mu=\delta^{ff'}D^{ij}_\mu,~~~~~D^{ ij}_\mu=\delta^{ij}\partial_\mu - ig_s A_\mu^{a}(x) T_a^{ij},
        \label{eq:covariant_derivative}
    \end{align}
    where $f,f'=1,\dots, 6$ are flavor indices and $i,j=1,2,3$ are color indices. Here $A^a(x)$ indicates the gluon fields, with $a=1,\dots, N_c^2-1 = 8$ labeling the eight kinds of gluon. The matrices $T_a$ are the eight $3 \times 3$ matrices standing as the generators of the SU$(3)$ color group in the fundamental representation: those are the so-called {\sl Gell-Mann matrices} divided by two. The quantity $g_s$ is the {\sl QCD coupling constant} and determines the strength of both quark-gluon and gluon-gluon interactions. 

    The term ‘‘covariant'' stems from the following property of $D_\mu$. One can show that, given $\theta(x)$ a generic function of coordinates, the first term in the right hand side of \eqref{eq:lagqcd} is invariant under local SU$(3)$ transformations of the form:
    \begin{equation}
        U(x)=\exp\left[-ig\,\theta^a(x)\, T_a\right].
        \label{eq:local_SU3_transformation}
    \end{equation}
    This means that, just by imposing the local SU$(3)$ invariance of the theory (gauge principle), the fermionic part of the QCD Lagrangian can be written as:
    \begin{equation}
		\mathcal{L}_{\text{Dirac}}=\bar{\Psi}(x)\left(i\gamma^\mu D_\mu-M\right)\Psi(x)=\bar{\Psi}(x)(i\gamma^\mu \partial_\mu-M)\Psi(x)+g_s \bar{\Psi}(x)\gamma^\mu A_\mu^{a}(x) T_a\Psi(x).
        \label{eq:fermionic_part_LagQCD}
	\end{equation}
    Here we separated the two contributions: the first term in the right hand side of \eqref{eq:fermionic_part_LagQCD} is the free Dirac Lagrangian, while the second one represents the interaction current mediated by the gluons. From the Feynman-rules perspective, this last term corresponds to a quark-antiquark-gluon vertex proportional to $g_s$.\\
	
	Let us now deal with the second part of the right hand side of \eqref{eq:lagqcd}. The term
	\begin{equation}\label{eq:gauge}
		\mathcal{L}_{\text{gauge}}=-\frac{1}{4}G^a_{\mu\nu} G_a^{\mu\nu} 
	\end{equation}
	represents the pure gluonic contribution to the Lagrangian. Here
	$G_{\mu\nu}$  corresponds to the gluon field-strength tensor, it being defined as
	\begin{equation}\label{eq:gluon}
		G_{\mu\nu}^a=\partial_\mu A^a_\nu-\partial_\nu A^a_\mu+g_s f^{abc} A_\mu^b A^c_\nu,
	\end{equation}
    where the $f^{abc}$ are the antisymmetric  structure constants of the SU$(3)$ group  appearing in the corresponding group algebra:
	\begin{equation}
		[T^a, T^b]=if^{abc}T^c.
        \label{eq:lie_algebra_su3}
	\end{equation}
	The presence of a quadratic term in the gluon fields in \eqref{eq:gluon} is a direct consequence of the theory being non-abelian, since for abelian theories the structure constants are trivially zero. From this it also follows that the insertion of \cref{eq:gluon} into \cref{eq:gauge} leads to
    \begin{equation}
    \mathcal{L}_{\text{gauge}}=-\frac{1}{4}(\partial_\mu A_\nu^a-\partial_\nu A_\mu^a)^2+g_sf^{abc}(\partial_\mu A_\nu^a)A^{b\mu}A^{c\nu}-\frac{g_s^2}{4}f^{abc}f^{ade}A_\mu^bA_\nu^cA^{\mu}_dA^{\nu}_e,
        \label{eq:LagQCD_full_3_4_gluon_terms}
    \end{equation}
    that is, to terms which are cubic and quartic in the gluon fields. This means that the theory allows already at tree level 3-gluon and 4-gluon self interactions, which are proportional, respectively,  to $g_s$ and  $g_s^2$. This feature is not present in abelian gauge theories, like QED, where photons do not interact with themselves at tree level (higher-order diagrams in QED, e.g. the fermion box diagram, actually lead to small photon-photon interactions). 

\subsection{The running of the coupling constant}
The {\sl deep inelastic scattering} experiments, performed via collisions of leptons over nuclei, pointed out the behavior of the strong force at high energy, actually giving credit to the QCD as the correct field theory describing these phenomena. In particular, these experiments pointed out that the coupling constant $g_s$ is not indeed ``constant'', but instead it has an energy dependence. This is not peculiar to QCD only: already in QED the creation (and the consequent annihilation) of lepton-antilepton pairs determines a process of vacuum polarization, from which it follows a dependence of the coupling constant on the energy scale. In particular, in QED, these corrections to the photon propagator result in a coupling constant which increases as a consequence of an increase in the energy scale.

Such a behavior appears also in the study of QCD, but with a crucial difference. In particular, the additional presence of the color charges and of the non-abelianity of the theory (which is manifest via the aforementioned 3-gluon and 4-gluon self interactions), implies that a gluon is able not only to interact with fermion-antifermion pairs, but it can also be involved in virtual loops with itself. The result is the combination of a screening over the color charge (simliarly as in QED) and of  an ``anti-screening'' caused by the gluons' self interactions. Since the latter phenomenon is found to be dominant, the final result is a decreasing behavior of the coupling constant with respect to an increase of the energy scale.

\begin{figure}[t]
    \centering
    \includegraphics[scale=0.35]{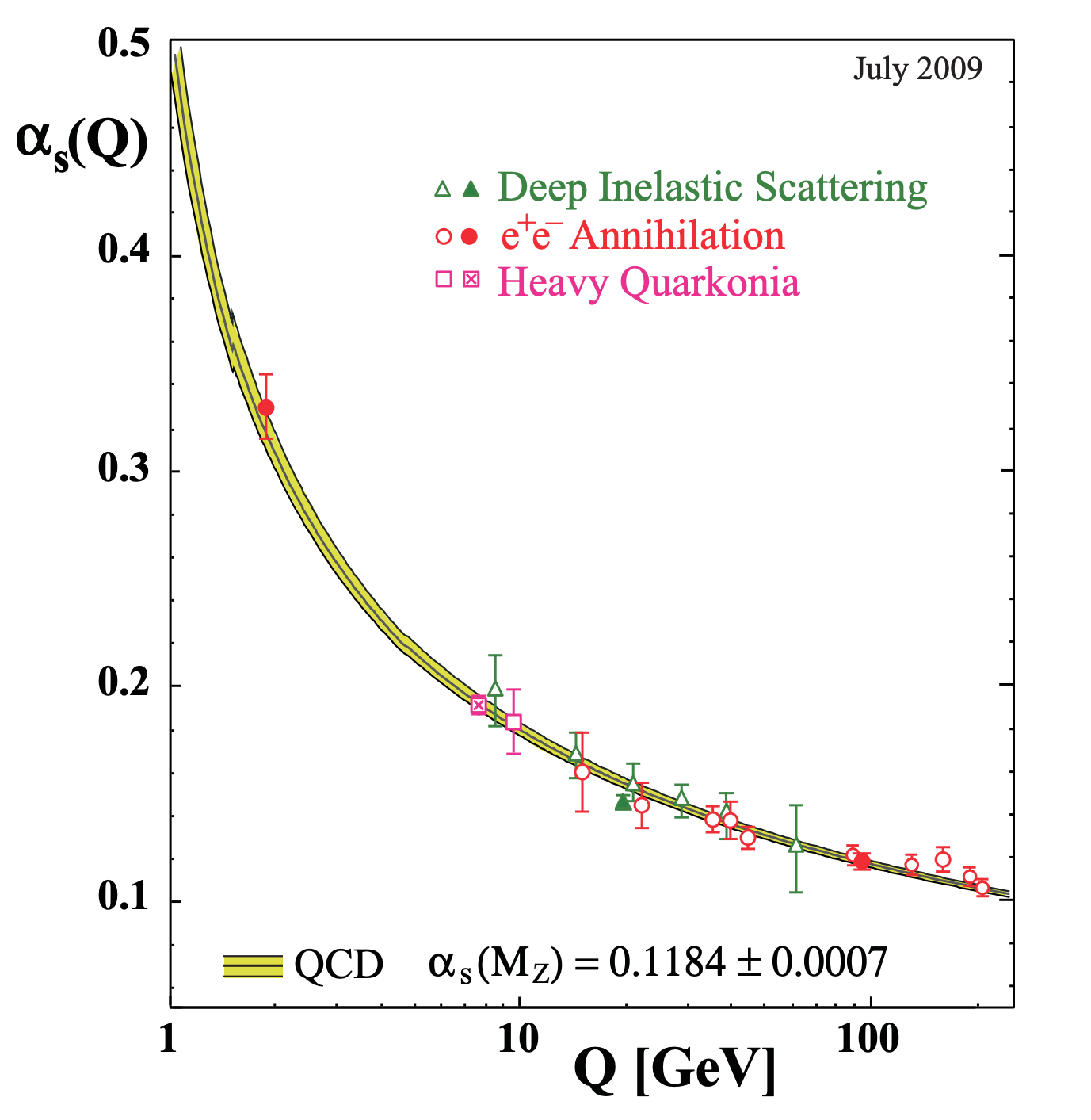}
    \caption{Summary of measurements of $\alpha_s$ as a function of the energy scale $Q$. The curves represent QCD predictions, overlaid with data from lattice QCD and more. Fig. from \cite{Bethke:2009jm}.}
    \label{fig1.3}
\end{figure}

Let us now be more quantitative, we know that in QCD, as for every quantum field theory, we have the appearance of ultraviolet divergences in the perturbative expansion, whose presence is seen in high order loops in the Feynman diagrams. Such divergences can be due to both fermionic and bosonic loops, but can be cured via the {\sl renormalization} procedure, whose result is a dependence of $\alpha_s\equiv g_s^2/4\pi$ on the exchanged four-momentum $Q$. Omitting the details, we find that the function $\alpha_s(Q^2)$ has to satisfy the {\sl renormalization group equation} \cite{Maggiore,Wilson:1973jj}:
\begin{equation}
    Q^2\frac{\partial\alpha_s}{\partial Q^2}=\beta(\alpha_s(Q^2)),
    \label{1.5}
\end{equation}
where the $\beta$-function in \eqref{1.5} is given by:
\begin{equation}
    \beta(\alpha_s)\equiv\mu^2\frac{\partial\alpha_s}{\partial \mu^2}.
    \label{eq:beta_function}
\end{equation}
In the above equation $\mu$ represents an arbitrary energy scale, called renormalization scale, a reference point for which we assume the corresponding value of $\alpha_s$ to be known. The behavior of $\alpha_s(Q^2)$ in QCD is then found to be: 
\begin{equation}
    \alpha_s(Q^2)=\frac{\alpha_s(\mu^2)}{1-[(2N_f-11N_c)/(12\pi)]\,\alpha_s(\mu^2)\log(Q^2/\mu^2)},
\label{1.6}
\end{equation}
being $N_f$ and $N_c$  the number of flavours and colors, respectively. Since we have $N_f=6$ and $N_c=3$, the result is a decrease in $\alpha_s(Q^2)$ with increasing $Q^2$: experimental data support this finding (Figure \ref{fig1.3}).
The previous expression \eqref{1.6} can be re-written in terms of the so-called {\sl QCD scale parameter} $\Lambda_{\text{QCD}}$, which is of the order of 200 MeV and defines the value of $Q^2$ for which $\alpha_s$ diverges. By doing so we get:
\begin{equation}
\alpha_s(Q^2)=\frac{4\pi}{(\frac{11}{3}N_c-\frac23 N_f)\log(Q^2/\Lambda_{\text{QCD}}^2)}.
\label{1.7}    
\end{equation}
We understand that $\Lambda_{\text{QCD}}$ gives us a reference scale for the validity of perturbative QCD (pQCD), and hence for the validity of \eqref{1.7}: if the energy scale $Q^2$ is much greater than $\Lambda_{\text{QCD}}^2$, then $\alpha_s(Q^2)$ drops, therefore we are in the {\sl asymptotic freedom} regime \cite{Gross:1973id,Politzer:1973fx,Coleman:1973sx}. Specifically, one has $\alpha_s(Q^2)\simeq 0.1$ when $Q=100$~GeV (the
typical scale for electroweak physics and also for hadronic jets at the
LHC).

On the other hand, when approaching lower momentum transfers \eqref{1.7} would predict a divergent behavior for $\alpha_s$ at the scale $\Lambda_{\text{QCD}}\sim200 $ MeV, i.e. for distances greater than 1 fm. It has to be said that this equation cannot be trusted for $Q\lesssim 1$~GeV, since it has been obtained
in perturbation theory, which is not valid in QCD at low energies. The fate of the QCD coupling for $Q\sim\Lambda_{\text{QCD}}$ is
still under debate, but various non--perturbative approaches suggest that
$\alpha_s(Q^2)$ should (roughly) saturate at a value close to one: for
all purposes, this is quite a strong coupling (it corresponds to
$g\simeq 3$). This means that below $\Lambda_{\text{QCD}}$ perturbative approaches are bound to fail and non-perturbative methods are required. For instance, in lattice QCD (lQCD) the action is discretized in a space-time lattice, whose vertices are occupied by quarks, while the interaction is encoded in gluonic fields lying in the ``links''. In this way we are able to solve QCD numerically, via simulations on the partition function expressed in the path integral formulation \cite{Borsanyi:2010cj}.
These numerical approaches are however applicable only for systems in equilibrium and with close-to-zero baryon chemical potentials.

\subsection{Symmetries of the QCD Lagrangian}
Let us now focus on the fermionic part of the QCD Lagrangian \eqref{eq:lagqcd}. In particular, let us consider only the two lighter quark flavours up ($u$) and down ($d$) and consider their masses to be both zero: $m_u= m_d= 0$. This is not exactly true in Nature, but since their masses are quite small we will see that this limit may be a good approximation. Our Lagrangian for the {\sl isospin doublet} $\Psi(x)=(u(x),d(x))^T$ will therefore be:
\begin{equation}\mathcal{L}_\text{Dirac}=\begin{pmatrix}
\Bar{u} &\Bar{d}
\end{pmatrix}\left[i\gamma^\mu\left(\partial_\mu-ig_sA_\mu^a(x)T_a\right)\right]\begin{pmatrix}
u \\d
\end{pmatrix}.
\label{L2}
\end{equation}
By construction, we know that $\mathcal{L}_\text{Dirac}$ is invariant with respect to local SU$(3)$ transformations. Let us now look for other global symmetries, i.e. transformations independent on the space-time point $x$, whose action over the isospin doublet leaves $\mathcal{L}_\text{Dirac}$ invariant. In order to find such symmetries let us introduce the chiral projectors:
\begin{equation}
    P_L=\frac{1-\gamma_5}{2},~~~~P_R=\frac{1+\gamma_5}{2}.
    \label{eq:chiral_projectors}
\end{equation}
Their action over the doublet components gives back the left-handed $\Psi_L\equiv P_L \Psi$ and the right-handed $\Psi_R\equiv P_R \Psi$ quark fields. These are eigenstates of the chirality operator $\gamma^5$, with eigenvalues $-1$ and $+1$, respectively. Moreover, in the massless limit these eigenstates also correspond to states with defined {\sl helicity} eigenvalues, with the helicity operator being defined as the projection of the spin vector over the momentum versor: $h=\Vec{\sigma}\cdot \hat{p}$. We can check that in this case the left-handed component $\Psi_L$ corresponds to the helicity eigenvalue -1, whereas the right handed component $\Psi_R$ corresponds to the eigenstate +1. For the antiquark fields the roles are inverted.\\

All that being said, we can now verify that the Lagrangian $\mathcal{L}_\text{Dirac}$ possesses a global invariance with respect to the following Lie group:
\begin{equation}
    \text{U}(1)_\text{V}\times \text{U}(1)_\text{A}\times \text{SU}(2)_\text{V}\times \text{SU}(2)_\text{A}.
    \label{eq:Lie_group_diracLag}
\end{equation}
The transformation laws and the related conserved Noether currents are the following:
\begin{align}
\text{U}(1)_\text{V}:\Psi(x)\to e^{-i\alpha}\Psi(x)&\implies
    J_{\text{U}(1)_\text{V}}^\mu(x)=\Bar{\Psi}(x)\gamma^\mu \Psi(x),\nonumber\\
\text{U}(1)_\text{A}:\Psi(x)\to e^{-i\beta \gamma^5}\Psi(x)&\implies J_{\text{U}(1)_\text{A}}^\mu(x)=\Bar{\Psi}(x)\gamma^\mu \gamma^5 \Psi(x),\nonumber \\
\text{SU}(2)_\text{V}:\Psi(x)\to e^{-i\alpha^i\sigma^i/2}\Psi(x)&\implies J_{\text{SU}(2)_\text{V}}^{i,\mu}(x)=\Bar{\Psi}(x)\gamma^\mu \frac{\sigma^i}{2}\Psi(x),\nonumber\\
\text{SU}(2)_\text{A}:\Psi(x)\to e^{-i\beta^i\gamma^5\sigma^i/2}\Psi(x)&\implies J_{\text{SU}(2)_\text{A}}^{i,\mu}(x)=\Bar{\Psi}(x)\gamma^\mu \gamma^5\frac{\sigma^i}{2}\Psi(x).\label{eq:conserved_currents_DiracLag}
\end{align}
We have used the labels V for {\sl vectorial} and A for {\sl axial}: the distinction between those two kinds of transformations can be understood by looking at the properties of the conserved currents. Indeed, the spatial part of $J_{\text{U}(1)_\text{V}}^\mu$ changes sign with respect to the action of a parity transformation, whereas the corresponding conserved charge $Q_{\text{U}(1)_\text{V}}$ does not: this means that $J_{\text{U}(1)_\text{V}}^\mu$ is a proper vector and $Q_{\text{U}(1)_\text{V}}$ is a proper scalar quantity (in particular $Q_{\text{U}(1)_\text{V}}$ is the {\sl baryonic number}, which we know to be conserved experimentally). On the other hand, the spatial components of $J_{\text{U}(1)_\text{A}}^\mu$ do not change sign for parity transformation, whereas $Q_{\text{U}(1)_\text{A}}$ does, therefore $J_{\text{U}(1)_\text{A}}^\mu$ is an axial vector and $Q_{\text{U}(1)_\text{A}}$ is a pseudo-scalar. The same analysis can be applied to $\text{SU}(2)_\text{V}$ and $\text{SU}(2)_\text{A}$, the only difference is that the resulting quantities are elements of the internal isospin space SU(2): the Pauli matrices are the generators of isospin rotations in the two flavour model.\\

After recapping the invariance properties of $\mathcal{L}_\text{Dirac}$, let us now focus on $\text{SU}(2)_\text{V}$ and $\text{SU}(2)_\text{A}$, which we have shown to be exact symmetries in the limit of massless $u$ and $d$ quarks. The corresponding quantized conserved charges $Q_{\text{V}}^i$ and $Q_{\text{A}}^i$ (which are vectors in the isospin space) can be shown to realize a closed algebra defined by the following rules:
\begin{align}
    \left[Q_\text{V}^i,Q_\text{V}^j\right]&=i\epsilon^{ijk}Q_\text{V}^k,\nonumber\\
    \left[Q_\text{A}^i,Q_\text{A}^j\right]&=i\epsilon^{ijk}Q_\text{V}^k,\nonumber\\
    \left[Q_\text{V}^i,Q_\text{A}^j\right]&=i\epsilon^{ijk}Q_\text{A}^k\label{1.8}.
\end{align}
What we see is that $\text{SU}(2)_\text{V}$ closes an angular momentum algebra, whereas $\text{SU}(2)_\text{A}$ does not. Moreover, the corresponding quantum number operators of the two groups do not commute with each other, which reflects the correlation between the associated symmetries. However, if we define the linear combinations:
\begin{equation}
    Q_\text{L}^i\equiv \frac{Q_\text{V}^i-Q_\text{A}^i}{2},~~~~Q_\text{R}^i\equiv \frac{Q_\text{V}^i+Q_\text{A}^i}{2},
    \label{eq:def_left_right_charges}
\end{equation}
we can see that the commutation relations \eqref{1.8} imply:
\begin{align}
    \left[Q_\text{L}^i,Q_\text{L}^j\right]&=i\epsilon^{ijk}Q_\text{L}^k\nonumber,\\
    \left[Q_\text{R}^i,Q_\text{R}^j\right]&=i\epsilon^{ijk}Q_\text{R}^k,\nonumber\\
    \left[Q_\text{L}^i,Q_\text{R}^j\right]&=0.\label{1.9}
\end{align}
This means that $Q_\text{L}^i$ and $Q_\text{R}^i$ close two independent angular momentum algebras. These charges correspond to the conserved quantum numbers associated to the global invariance of $\mathcal{L}_\text{Dirac}$ with respect to a new symmetry group. We call this group $\text{SU}(2)_\text{L}\times \text{SU}(2)_\text{R}$, and the corresponding symmetry is called {\sl chiral symmetry}.\\

\begin{figure}
    \centering
    \includegraphics[width=0.5\linewidth]{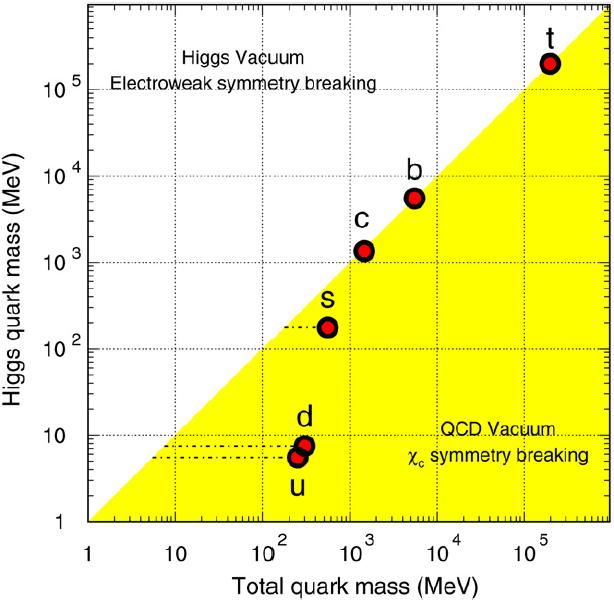}
    \caption{The contributions of chiral symmetry breaking and of the Higgs mechanism to the masses of the six flavors of quarks. The QCD interaction strongly affects the light quarks ($u$, $d$, $s$), whose total mass is significantly different than the Higgs mass, while the heavy quark masses ($c$, $b$, $t$) are mainly determined by the Higgs mechanism \cite{mass_generation_star}.}
    \label{fig:mass_generation_star}
\end{figure}

If we wanted to consider a finite mass in the Lagrangian \eqref{L2}, we would have to include an additional term in which the chirality eigenstates $\Psi_L$ and $\Psi_R$ are coupled: as we have already mentioned, chiral symmetry is exact only for vanishing quark masses. However, we can still look for evidence of an approximate chiral symmetry in the light quark sector ($u$, $d$, maybe $s$) at the  QCD ground state. For instance, such evidence may be provided by a vanishing value of the chiral condensate $\langle \Bar{\Psi}\Psi\rangle\equiv \langle0|\Bar{\Psi}\Psi|0\rangle$, which would be implied by the approximate current conservation law $\partial_\mu J^\mu_{\text{SU}(2)_\text{A}}\simeq0$. However, experimental observation on pion spectra showed strong deviations from the expected null value, even though the chiral symmetry is approximately present in the ground state. Nambu and Jona-Lasinio gave an explanation by showing that a classical QCD Lagrangian with a chiral invariance property can lead to an effective theory of mesons with a non-zero value of $\langle \Bar{\Psi}\Psi\rangle$ \cite{Nambu:1961tp,Nambu:1961fr}: this is an example of {\sl spontaneous symmetry breaking}, which enables the quarks to possess extra {\sl dynamical} QCD masses of about 300 MeV. Moreover, as a consequence of the spontaneous breaking of a continuous symmetry, by the virtue of the Goldstone theorem we expect the presence of massless bosons, which in the case of the chiral symmetry are predicted to be the pions. We know pions to have a non-zero mass, and this occurs since the chiral symmetry is only approximate \cite{Nambu:1960tm,Goldstone:1961eq}.\\
When the temperature crosses a critical value it is believed that quarks and gluons get free from the hadronic confinement and the chiral symmetry is restored at the ground state level: as a consequence of that, the production of quark-antiquark pairs is favored, since the QCD mass is missing and the threshold energy is lower. In this case, however, the absence of dynamical mass is partially filled by the acquiring of a {\sl thermal} mass.\\
It is worthwhile emphasizing that such a model allows for dynamical masses of the light quarks to be nonzero without the need of any Higgs mechanism for spontaneous electroweak symmetry breaking. On the other hand, for the heavy quarks ($c$, $b$, $t$) the dynamical mass is not the main contribution, and their mass comes instead almost uniquely from the Higgs mechanism, see Fig. \ref{fig:mass_generation_star}.

\section{The Physics of Heavy Ion Collisions}
We have mentioned that QCD can be treated analytically only in the perturbative regime, which means that a whole plethora of physical phenomena remains difficult to study, and theoretical studies are currently focused on either lattice simulations or effective models. Similarly, also from the experimental point of view the investigation on the properties of QCD matter is challenging because of confinement, in other words, because of our inability to directly study the color degree of freedom. 

In the past 50 years, as beams of ultrarelativistic protons and nuclei have become available, the collisions of protons with nuclei and of nuclei with nuclei at very high velocities (or total collision energies) have been studied in great detail. In this sense, studying ultrarelativistic {\sl Heavy Ion Collisions} (HIC) may give us a path to a more complete understanding
of how QCD matter particles behave. This
idea has caught scientists' attention for a long time: Heisenberg
and Heitler wrestled with it in the 1930s and 1940s \cite{Heisenberg:1949kqa,Hamilton:1986fj},
Fermi and Landau did so in the 1940s and 1950s \cite{Fermi:1950jd,Landau:1953gs,Florkowski:2010zz},
and Feynman tried his hand in the 1960s \cite{Feynman:1969ej}. We
can now gain new insights on old questions by studying high
energy collisions in a regime in which experimenters have many
knobs to dial, including, but not limited to: the size of each of the colliding nuclei,
(proxies for) the impact parameter, the energy of the collision. With the advent of the high--energy colliders RHIC (the Relativistic
Heavy Ion Collider operating at BNL since 2000) and LHC (the Large
Hadron Collider, which started operating at CERN in 2008), the physics of
ultrarelativistic HICs has entered a new era: the energies
available for the collisions are high enough -- up to 200 GeV per
interacting nucleon pair at RHIC and up to 5.5~TeV at the LHC  -- to ensure that
new forms of QCD matter are being explored in the collisions.
These challenges stimulated new ideas and the development of new
theoretical tools aiming at a fundamental understanding of QCD
matter under extreme conditions: high energy, high parton densities,
high temperature. The ongoing experimental programs at RHIC and LHC, and the forthcoming programs at the Electron Ion Collider (EIC, at BNL),
provide a unique and timely opportunity to test such new ideas, constrain
or reject models, and orient the theoretical developments. Over the last
decade, the experimental and theoretical efforts have gone hand in hand,
leading to a continuously improving physical picture of HICs, a picture which is by now
well rooted in QCD.

\subsection{Why do we study ultrarelativistic heavy ion collisions?\label{sec:Why-do-we}}

First of all, we will formulate variants of the ``why are we
doing this?'' question, to which we will provide a surprisingly wide spectrum of answers. In particular one may wonder:
are there insights, that we hope to gain from studying these collisions, that go beyond the understanding of the dynamics
of the collisions only? The affirmative answers to this question, which we
shall divide into three groups below, motivate much of the current experimental
and theoretical efforts \cite{Busza:2018rrf}.

\begin{figure}[t]
\centering
\includegraphics[width=\textwidth]{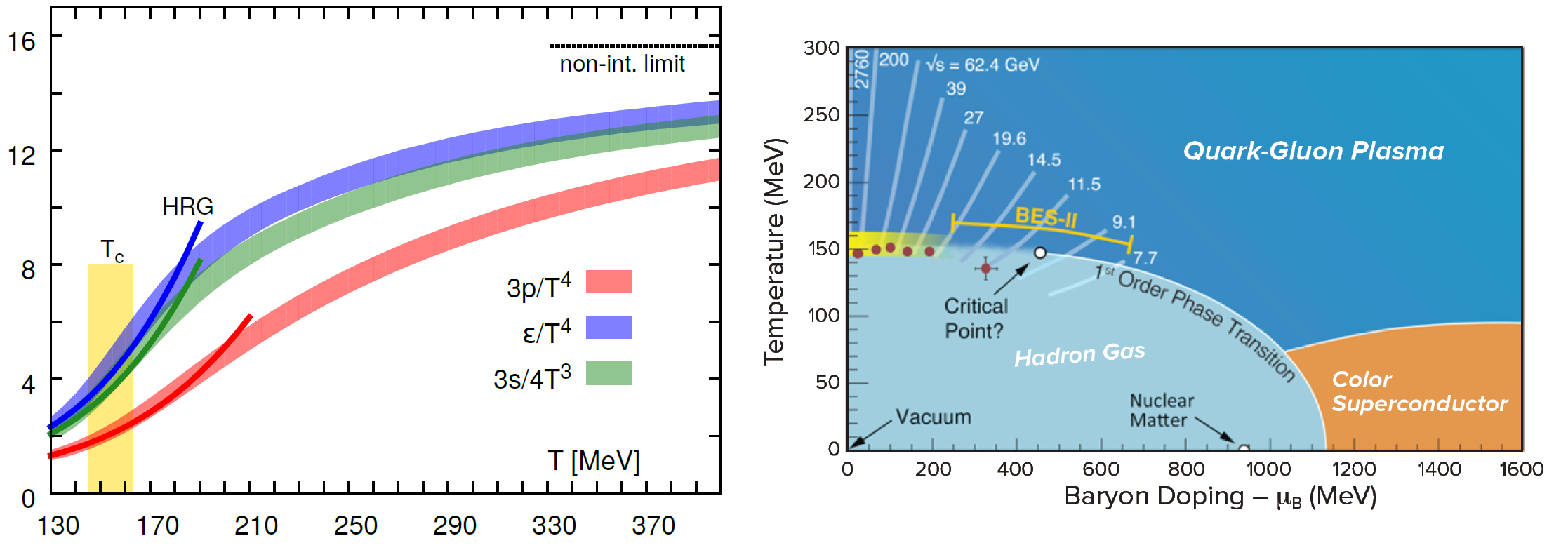}
\caption{\textbf{(left) }Lattice QCD calculations of the pressure $p$, energy
density $\varepsilon$ and entropy density $s$ of hot QCD matter
in thermal equilibrium at temperature $T$ \cite{Borsanyi:2013bia,Bazavov:2014pvz}
show a continuous crossover around $T\sim150$ MeV, from a hadron
resonance gas (HRG) at lower temperatures to QGP at higher temperatures.
Because QCD is asymptotically free, thermodynamic quantities will
reach the Stefan-Boltzmann limit (‘‘non-interacting limit'' in the figure) at extremely high
temperature. Within the range shown, however, they are around 20\% below
their Stefan-Boltzmann values. \textbf{(right)}
Our current understanding of the expected
features of the phase diagram of QCD as a function of temperature
and baryon chemical potential. The regions of the phase
diagram traversed by the expanding cooling droplets of QGP formed
in HICs with varying energies $\sqrt{s_{NN}}$ are sketched.
Figures from \cite{Bazavov:2014pvz,Aprahamian:2015qub}.\label{fig:lattice}}
\end{figure}

\subsubsection*{QCD in Cosmology}

Heavy ion collisions recreate droplets of the matter that filled the
universe around a microsecond after the Big Bang. Moreover, as it has been
understood since the mid 1970s \cite{Collins:1974ky,Linde:1978px},
when the universe was only a few microseconds old it was filled
with matter at temperatures above $\Lambda_{{\rm QCD}}$ and was too hot
for protons, neutrons, or any hadrons to have formed. This direct
and tangible connection to the earliest moments of the universe, together
with the insight that the primordial matter at such temperatures
had to be some new form of matter not made of hadrons, provides two
powerful motivations for studying ultrarelativistic HICs.
Historically, these were the motivations that provided much of the
initial impetus for the field.

In the 1980s, a fair amount of work was done on possible observable
consequences in cosmology of a first order phase transition between
the hot primordial matter and ordinary hadronic matter. These all
relied upon presuming a process of nucleation of widely separated ‘‘bubbles'' of the low temperature
hadronic phase. As the walls of these putative bubbles expanded through
the microseconds-old universe over distances as long as centimeters
or meters, they would have left behind matter that was inhomogeneous
over these long length scales \cite{Witten:1984rs,Applegate:1985qt}.
If this had happened, it would have modified the synthesis of light
nuclei that occurred right after, when the universe was minutes old. Starting in
the late 1990s, and culminating in the 2000s \cite{Karsch:2001vs,Aoki:2006we},
it became clear from first-principles lattice QCD calculations that the transition from primordial hot QCD matter to hadronic matter in the first few microseconds after
the Big Bang proceeded via a continuous crossover, not a first order
phase transition (assuming no imbalance between matter and antimatter). See Fig. \ref{fig:lattice}, left. This is in turn consistent with the modern understanding
of Big Bang nucleosynthesis, which is in agreement with cosmological
observations: these do not show any of the disruption that a strong first order
phase transition would have introduced \cite{Thomas:1993md}. In particular, a continuous crossover does not introduce any fluctuations on length scales much
longer than the natural length scales of QCD (i.e. around 1 fm=$10^{-15}$ m), meaning that it left no imprint in the microseconds-old universe that survived
so as to be visible in some way today. That is, we now understand
that we cannot use cosmological observations to ``trace back'' to the primordial hot QCD matter that filled the microseconds old universe, or to the crossover
transition at which ordinary protons and neutrons first formed.
A central goal of ultrarelativistic HICs, then, is
to recreate droplets of Big Bang matter in
controlled conditions, where we instead {\sl can} learn about its
properties as well as about its phase diagram, in ways that we will
never be able to do via observations made with telescopes or satellites.\\

What can we learn from such studies? What have we learned so far?
One of the most important discoveries made via studying ultrarelativistic
HIC is that matter that is a few trillions of degrees hot ($10^{12}$ K $\sim100$ MeV) is a {\sl liquid}. The early ideas that motivated the field
turned out to be half right: primordial matter at these temperatures
is not made of hadrons, as anticipated. However, at the temperatures
that have been achieved in HICs to date, this phase is not
a weakly coupled plasma of quarks and gluons as originally expected.
Instead, when the hadrons that make up ordinary nuclei are heated
to these extraordinary temperatures, the matter that results is better
thought of as a ‘‘soup'' of quarks and gluons, in which there are no hadrons
to be found but in which every quark and gluon is always strongly
coupled to its neighbors, with no quasiparticles that can travel long
distances between discrete scatterings. As we have already argued, such medium is usually referred to as the {\sl Quark-Gluon Plasma} (QGP).

The material property that quantifies the ‘‘liquidness'' of a fluid made
up of ultrarelativistic constituents is the ratio of its shear viscosity
$\eta$ to its entropy density $s$, called ``specific viscosity''. The {\sl shear viscosity} $\eta$, in particular, encodes how easily momentum can be exchanged between distant fluid cells and, consequently, how fast a gradient in fluid velocity (or a sound wave) dissipates into heat. The ratio $\eta/s$ plays a central role in the equations of relativistic hydrodynamics, since it governs the amount of entropy
produced within the fluid as a sound wave propagates through it, or
more generally, while the fluid itself flows in any nontrivial way. It is the natural
dimensionless measure of the effects of shear viscosity in a relativistic
fluid, and its temperature dependence can be constrained by a combination of experimental data and hydrodynamic calculations. Apart from technical details, here it is worth noting that $\eta/s$ for the liquid of quarks and gluons
produced in HICs is found close to the value $1/4\pi$ (cfr. \cite{Parisi:2025gwq} and references therein):
we have extraordinarily large values of both $\eta$ and $s$ relative
to those of any familiar fluid, but the specific viscosity $\eta/s$
turns out smaller than that of any other known fluid. The value $1/4\pi$, in particular, is the estimate of the ratio $\eta/s$ in the plasma of infinitely strongly coupled gauge particles: such value is calculated using the AdS/CFT conjecture, which exploits a duality between a weak coupling gravitational theory in a 4+1-dimensional Anti-de
Sitter space \cite{Policastro:2001yc,Casalderrey-Solana:2011dxg} and a strong coupling description of a corresponding Conformal Field Theory \cite{Hubeny:2011hd}. This implies that, to our best understanding, the medium formed in HICs is very strongly coupled. This connection, between the properties of the primordial matter recreated
in heavy ion collisions and a gravitational analogue in a higher dimensional space, provides strong motivation for pushing the determination of $\eta/s$ of quark-gluon plasma to higher accuracy. Unfortunately, lattice calculations are not of much help here, since those are build upon the path integral Euclidean formulation of equilibrium
thermodynamics. It is therefore quite challenging to use them to gain information
about transport coefficients, including shear viscosity, which describe the time-dependent processes via which infinitesimal perturbations away from equilibrium relax, producing entropy. Despite that, although many results have been obtained \cite{Nakamura:2004sy, Meyer:2007ic, Meyer:2009jp, Mages:2015rea, Borsanyi:2018srz}.
Lattice calculations of more dramatically time-dependent phenomena, including the quenching of jets in the liquid plasma or the initial formation of the plasma from a far-from-equilibrium collision, are still
beyond the horizon.

\subsubsection*{Phase diagram of QCD}

Among the most important reasons for studying ultrarelativistic collisions
is the expectation that doing so will teach us about the phase diagram
of hot QCD matter, as a function of both temperature
and baryon doping (see Fig. \ref{fig:lattice}, right). By baryon doping
(or net baryon number density) we mean the excess of quarks over antiquarks
in the hot matter, described by the baryon number chemical potential $\mu_{B}$.
To this point, we have assumed $\mu_{B}$ to zero, describing matter with
equal densities of quarks and antiquarks: this is a very good approximation
for the matter produced at mid-rapidity at the most energetic heavy
ion collisions at RHIC, an even better approximation at the LHC, and
an exceptionally good approximation in the early universe.
We have mentioned that when $\mu_{B}=0$ ordinary hadronic matter forms via a continuous crossover from the QGP, but this is only a very small part of the full QCD phase diagram. We would also like to study QCD when we have a significant excess of quarks over antiquarks, scanning different regions of the $\mu_B$--$T$ plane. This can be done by looking at heavy ion collisions with lower and lower collision energies, which imply higher $\mu_{B}$: this occurs since at lower energies the particle production processes are less likely, so most of the matter found right after the collision comes from the incident nuclei, so their quark-antiquark imbalance is not negligible. On the other hand, this does not occur at high energies, since the multiplicity of particles created in the collision is so large that any initial imbalance is basically irrelevant.

Low energy AA studies are underway at the RHIC Beam Energy Scan (BES) \cite{Luo:2015doi}, where tantalizing results are in hand. One of
the central questions that these experiments aim to answer is whether
the continuous crossover between liquid QGP and hadronic matter turns
into a first order phase transition above some critical
value of $\mu_{B}$, i.e. below some collision
energy. There are many effective models for QCD in which the phase diagram features a critical point. For instance, in QCD with
two massless quarks (i.e. in the chiral limit)
the crossover at $\mu_{B}=0$ becomes a sharp second order phase transition
at which chiral symmetry is restored, and we have a point with $\mu_{B}>0$ at which the transition becomes first order \cite{Rajagopal:2000wf,Stephanov:2004wx}. At the present we have no clear evidence for the existence of a critical point, but there are strong motivations for the experimental programs that aim to answer this question within the next few years.\\

Pushing to very high baryon doping while staying at low temperature, i.e. basically squeezing nuclei without heating them, takes us into another interesting region of the QCD phase diagram. Matter that is sufficiently dense cannot be made of well-separated nucleons, even at low temperatures: the nucleons are crushed into one another. In this regime we have a cold, dense state of matter, in which quarks fill momentum space up to some high Fermi momentum: what we obtain is a so-called {\sl color superconductor}, in which a condensate of correlated Cooper pairs of quarks creates a superfluid and yields the QCD-analogue of a Meissner effect. Extensive theoretical analyses of the phase diagram and the consequent properties of color superconducting quark matter have been performed \cite{Alford:2007xm}. These studies are well understood at asymptotic densities, but at densities of order 10 times or below that of nuclei, the analyses turn out to be sensitive to the ratio of the strange quark mass to the Fermi momentum, as well as to the strength of the Cooper pairing. This makes hard to quantitatively pin down observables, so experimental data is sorely needed.

Unfortunately, the only place in the universe where cold dense quark matter may be found is in the centers of neutron stars. In spite of their rarity, recent detection of the first gravitational waves \cite{TheLIGOScientific:2017qsa} emitted by collisions of neutron stars have provided a great opportunity to learn about the compactness and density profile of the incident neutron stars and, conceivably, to determine whether or not they feature dense quark matter cores. If they do, present constraints on heat transport in neutron stars, coming from X-ray observations of how they cool, will turn into constraints on the transport properties of cold and dense quark matter.

\subsubsection*{Emergence of complex quantum matter}
In the history of the universe, liquid quark-gluon plasma was the
earliest complex state of matter to form. At much earlier times, when
the temperature was a few orders of magnitude hotter than those of our
interest here, the matter that filled the universe actually \textit{was}
a weakly coupled plasma of quarks and gluons. We know this because
QCD is asymptotically free, hence quarks and gluons interact with each other only weakly when they scatter off each other with large momentum transfer.
Not only was liquid QGP the earliest complex matter to form, but from a certain  point of view, it is also the simplest form of complex matter that we know of. In particular, it is the complex matter that is ``closest'',
most directly connected to, the fundamental laws that govern all matter
in the universe, in this case the laws being the fundamental theory of QCD.

Again, because QCD is asymptotically free, we know that if we could hold
a droplet of the liquid QGP with temperature $T$ in place and study
its microscopic structure with a spatial resolution that is much finer
than $1/T$ (for example via scattering high energy electrons off
it in this thought experiment), what we would see is weakly coupled
quarks and gluons. This is the genesis of the strongest motivation
for developing experimental techniques for probing the structure of
the liquid QGP on different length scales. We know that at the shortest
length scales we must see weakly coupled quarks and gluons. We also
know that at length scales of order $1/T$ or longer we see a liquid
in which neighboring ``unit cells'' are tightly coupled to each
other, meaning that the liquid flows hydrodynamically with a small
$\eta/s$. If we can probe both these length scales and scales in
between, for example via studying how jets (which are intrinsically
multiscale probes, see Section \ref{sec:Probes and observables of HIC}), or heavy quarks with varying mass, or tightly
bound quarkonium mesons with varying sizes, ``see'' the plasma and
how the plasma responds to their passage through it, we have a chance
to maybe understand how the simplest form
of complex matter that we know emerges from weakly coupled, asymptotically
free, constituents at short length scales. The question of how the
almost infinite variety of complex forms of matter that we see in
the world around us emerges from laws of Nature that are so simple
that they can easily fit on a T-shirt is one of the great quests of
modern Physics. If we can answer this question for the case of liquid quark-gluon
plasma, which we have a chance to do by the virtue of this simple form
of complex matter being so close to its laws-of-Nature underpinnings,
maybe we have a chance of shedding light on the more general question, i.e. the behavior of QCD matter at all scales.

\subsection{Stages of a HIC}
    
\begin{figure}[t]
\begin{center}
\includegraphics[width=.9\textwidth]{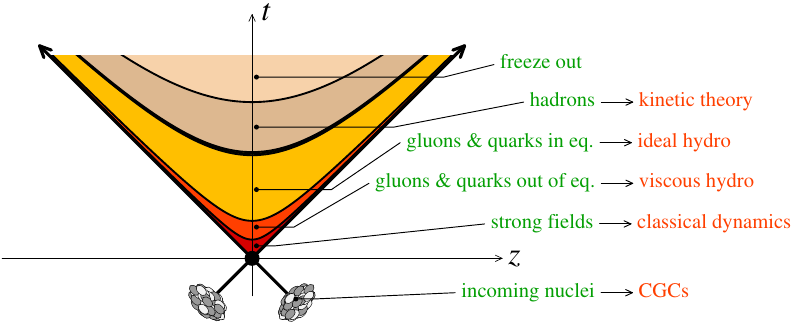}
\caption{Schematic representation of the various stages of a HIC as a function
of time $t$ and the longitudinal coordinate  $z$ (the collision axis).
The `time' variable which is used in the discussion in the text is the proper time $\tau\equiv\sqrt{t^2-z^2}$, which has a Lorentz--invariant
meaning and is constant along the hyperbolic curves separating
various stages in this picture. Figure from \cite{Iancu:2012xa}.} \label{fig:HIC}
\end{center}
\end{figure}

We now report a qualitative description, in the center of mass frame, of the sequence of events that
occur when two ultrarelativistic nuclei collide head on. This picture
follows from the observed phenomenology
special relativity, and our understanding of the workings of QCD. For a more comprehensive introduction into
heavy ion collisions, check \cite{Busza:2018rrf, Yagi:2005yb,Florkowski:2010zz,Wang:2016opj, Iancu:2012xa}. The theoretically motivated space--time picture of a heavy ion collision is depicted in Fig. \ref{fig:HIC}: this illustrates the various forms
of QCD matter intervening during the different phases of the collision, ordered in {\sl proper time} $\tau\equiv\sqrt{t^2-z^2}$.

\begin{enumerate}
\setcounter{enumi}{-1}
\item Prior to the collision ($\tau<0$), and in the center-of-mass frame (which
    at RHIC and the LHC is the same as the laboratory frame), the two
    incoming nuclei look as two Lorentz--contracted ‘‘pancakes'', with
    a longitudinal extent smaller than the transverse extent by a factor (the
    Lorentz boost factor) $\gamma\sim 100$ at RHIC and even $\gamma\sim 2500$ at LHC. These velocities correspond to beam rapidities $y=5.3$ and $8.5$, respectively.\footnote{Here we refer to the momentum rapidity $y$, such that $\cosh (y) \equiv \gamma$. Its definition is different than the space-time rapidity $\eta$, such that $\tanh\eta\equiv z/t$, with $z$ and $t$ space-time coordinates centered at the collision. However, in a boost invariant flow in which $v_z=z/t$, we have $\eta=y$.}
    Each disc includes many colored quarks and antiquarks, with three
    more quarks than antiquarks per nucleon in the incident nuclei and
    with $q\bar{q}$ pairs coming from quantum fluctuations in the initial
    state wave functions. These pairs are ``almost real'', as a consequence
    of time dilation. However, as we shall see in much more detail in this Thesis (cfr. Chapter \ref{chap:The initial stages of heavy ion collisions}), what the nuclei are mostly composed of at high speeds are gluons. These carry only tiny fractions $x\ll 1$ (Bjorken $x$) of the longitudinal momenta of their parent nucleons, but their density is rapidly increasing as $1/x$. By the uncertainty principle, the gluons which make up such a high--density system carry relatively large transverse momenta. A typical value for gluon momenta in a Pb or Au nucleus is $k_\perp\simeq 2$~GeV for Bjorken $x\sim10^{-4}$. By the asymptotic freedom property of QCD, the
    gauge coupling which governs the mutual interactions of these
    gluons is relatively weak. This gluonic form of matter,
    which is dense and weakly coupled, and dominates the wavefunction
    of any hadron (nucleon or nucleus) at sufficiently high energy,
    is {\sl universal}: its properties are the same for all
    hadrons. This phase of matter is known as the {\sl color glass condensate} (CGC), which we will describe thoroughly in Chapter \ref{chap:The initial stages of heavy ion collisions}. The area density of these virtual quarks, antiquarks and gluons increases with the velocity of the nuclei. It is not uniform across the area of the disc, and fluctuates from nucleus to nucleus. The spatial variation of the partons primarily reflects the instantaneous distribution of the nucleons inside the nuclei and of the valence partons inside the nucleons.
    
\item At time $\tau=0$, the two nuclei hit each other and the
    interactions start developing. When the two discs, each a tiny fraction of a Fermi thick, overlap and collide, most of the incident partons lose some energy but are not kicked by any large angle. Most of these interactions are ``soft'', meaning that they involve little transverse momentum transfer. These strong interactions can be described in terms of interacting fields or slabs of energy. In the language of fields and particles, as the
two discs of strongly interacting transverse color fields and associated
color charges collide, some color charge exchange occurs between the
discs, and longitudinal color fields are produced, which fill the
space between the two receding discs, reducing the energy in the discs
themselves. These fields then gradually decay into $q\bar{q}$ pairs and gluons.
A very small fraction of the incident partons suffer hard perturbative
interactions as the discs overlap initially. However, those interactions involving relatively large transferred momenta $Q\gtrsim 10$~GeV,
    are those which occur faster (within a time $\tau\sim 1/Q$, by
    the uncertainty principle) and also the ones responsible for the production of `hard particles', i.e. particles carrying transverse energies and
    momenta of the order of such $Q$. The high-energy
partons evolve, decay, radiate and finally produce a cone-shaped spray
or ``jet'' of hadrons and/or high-energy photons, leptons or heavy
$Q\bar{Q}$ pairs, all while traversing a region where QGP is in the
process of being produced and evolving. Such particles are generally the most striking ingredients of the final
state, since they are often used to characterize the topology of the
event: for instance one speaks about `a dijet event', cf.
Fig. \ref{fig:events} left, or `a photon--jet' event, cf. Fig. \ref{fig:events} right. They thus contain a wealth of information about the produced medium, and on how partons lose energy or disturb the medium
as they interact with it. 

In high energy heavy ion interactions, the maximum energy density
occurs just as the two highly Lorentz contracted nuclei collide. Clearly
this system is very far from equilibrium, and its very high energy
density is really just a consequence of Lorentz contraction. It is
much more interesting to ask what we can say about
the average energy density, say, 1 fm after the collision, by which
time the two discs are 2 fm apart in distance. The expanding high energy density
system produced around the midpoint between the two discs, where the
collision occurred, at this time has an energy density that is still
far greater the energy density inside
a typical hadron, which is around $500\text{ MeV/fm}^{3}$. In particular, a rough estimate can be obtained from the available
data for head-on LHC collisions with $\sqrt{s_{NN}}=2.76$ TeV (corresponding to $\gamma=1400$ and $y=8.0$). The
total transverse energy in particles with longitudinal velocity $-0.46<v<0.46$ is measured to be $1.65\pm0.1$ TeV \cite{Collaboration:2011rta}, meaning that the average energy
density 1 fm after the collision is greater than $1.65\text{ TeV}/[\pi\, (7\text{ fm})^{2}\,(0.92\text{ fm})]=12\text{ GeV/fm\ensuremath{^{3}}}$,
more than twenty times the energy density of a hadron. This estimate is further corroborated by lattice calculations for QCD thermodynamics: these show that matter in thermal equilibrium at a
temperature of 300 MeV, which is a good estimate for temperatures produced in HICs, has an energy density $\sim12\,T^{4}=12.7\,\text{GeV}/\text{fm}^{3}$.

The entropy produced in these collisions is also enormous. To get a sense of this, note that before the collision the entropy of the two incident nuclei is
essentially zero, whereas the final state after the collision can contain
as many as 30,000 particles, and hence it has a very large entropy. 

\item At a time $\tau\sim 0.2$~fm, the bulk of the partonic
    constituents of the colliding nuclei (meaning the gluons
    composing the respective CGCs) are liberated by the collision. The initial stage of the collision is dominated by intense chromoelectric and chromomagnetic fields, which are at first purely longitudinal, but then develop a transverse component.
    This is when most of the `multiplicity' in the final state is
    generated: in other words, most of the hadrons eventually seen in the
    detectors are produced via the fragmentation and the
    hadronisation of the initial--state gluons liberated at this
    stage. This non--equilibrium state of partonic matter, which
    besides its high density has also other distinguished features to
    be discussed later in Chapter \ref{chap:The initial stages of heavy ion collisions}, is known as the {\sl glasma}.
    Before ending up in the detectors, however, these partons
    undergo a complex evolution, which we will now discuss briefly.

\begin{figure}[ht]
\begin{center}\centerline{
\includegraphics[width=.4\textwidth, height=0.27 \textwidth]{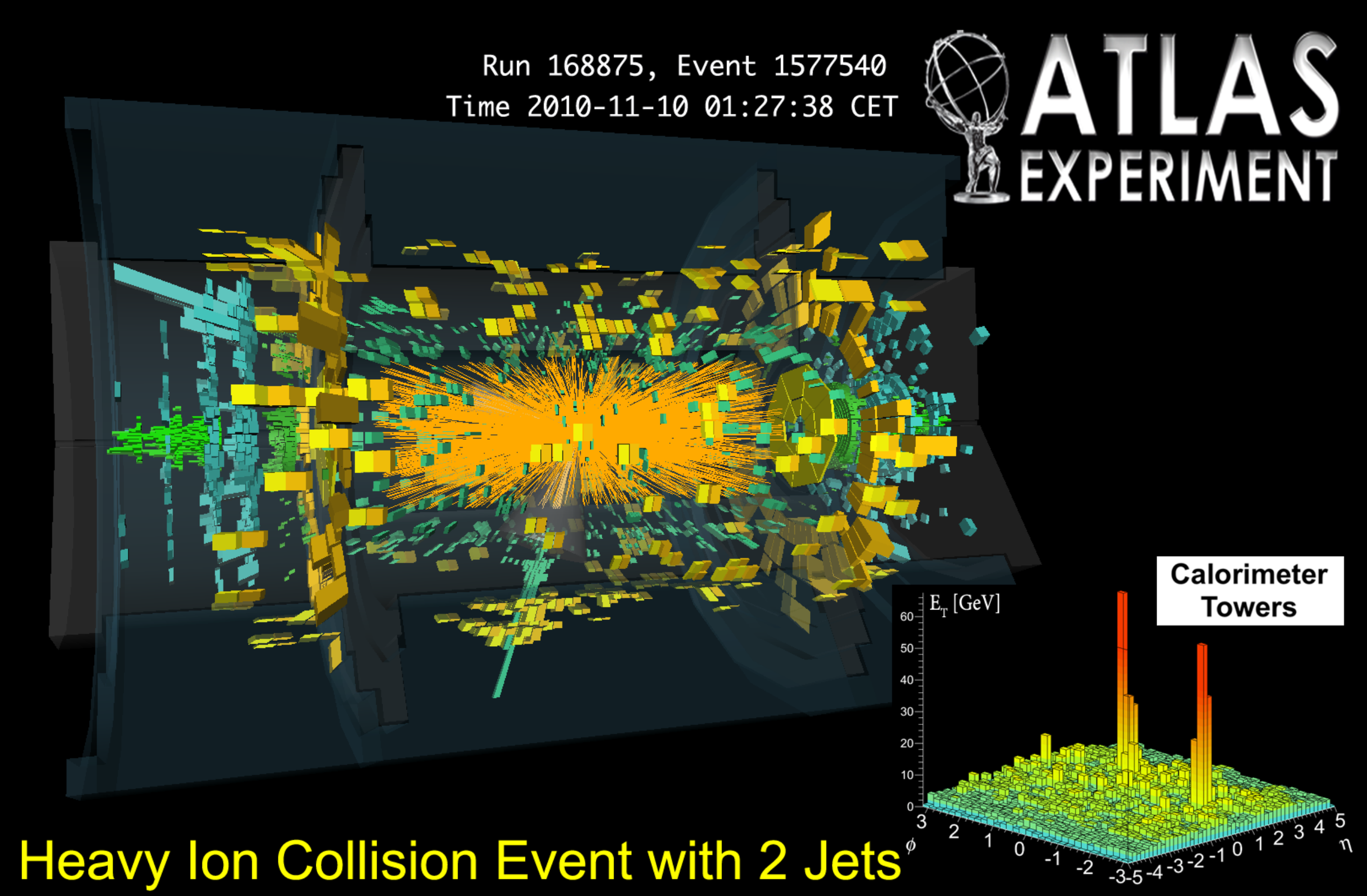}\qquad
\includegraphics[width=.4\textwidth, height=0.27 \textwidth]{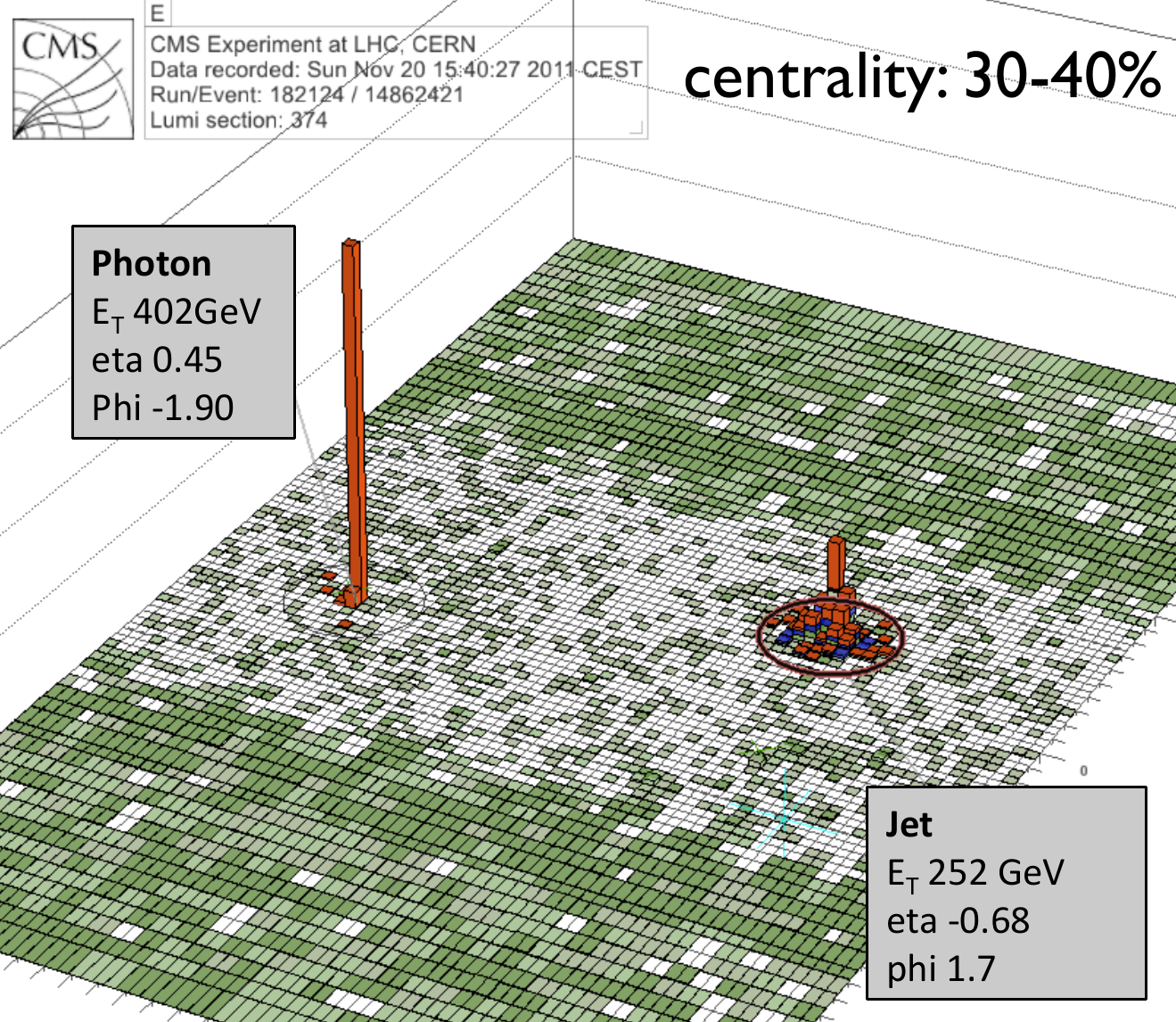}
}
\caption{A couple of di--jets events in Pb-Pb collisions at ATLAS
\textbf{(left)} and CMS \textbf{(right)}. Figures from \cite{Iancu:2012xa}.}
\label{fig:events}
\end{center}
\end{figure}

\item If the produced partons did not interact with each other, or if
    their interactions were negligible, then they would rapidly
    separate one another and independently evolve (via
    fragmentation and hadronization) towards the final--state
    hadrons. This is, roughly speaking, the situation in
    proton--proton collisions. But the data for heavy ion collisions
    at both RHIC and the LHC exhibit collective phenomena (like the
    `elliptic flow', to be discussed later in §\ref{sec:Probes and observables of HIC}) which clearly show that
    the partons liberated in the collision {\sl do} actually interact, and quite strongly. A striking consequence of
    these interactions is the fact that this partonic matter rapidly
    approaches towards {\sl thermal equilibrium}: the data are
    consistent with a relatively short thermalization time, of order
    $\tau\sim 1$~fm. This is striking since it requires rather
    strong interactions among the partons, since they have to outbalance the
    medium expansion: these interactions have to redistribute energy
    and momentum among the partons, in spite of the fact that the
    partons themselves separate quite fast away from each other. This evidence represents a main argument in favor of the existence of a strongly coupled partonic medium.

    Thus, the
quarks and gluons in this high energy density matter are far from
independent. They are so strongly coupled to each other that they
form a collective medium that expands and flows as a relativistic
hydrodynamic fluid with a remarkably low viscosity to entropy density
ratio $\eta/s\approx1/4\pi$ \cite{Heinz:2013th,Romatschke:2017ejr}, within a time that can be shorter
than or of order 1 fm in the rest frame of the fluid. Even
if the transverse velocity of the fluid is small initially, say 1
fm after the collision, the pressure-driven hydrodynamic expansion
rapidly builds up transverse velocities of order half the speed of
light. This very high pressure will drive fluid motion, expansion, and consequent
cooling throughout the whole the event. 

As the discs recede from each other and the QGP produced between
them is expanding and cooling, at the same time new QGP is continually
forming in the wake of each receding disc, see Fig. \ref{fig:movie} right. This happens because the quarks and gluons produced at high rapidity
are moving at almost the speed of light in one of the beam directions,
meaning that when enough time has passed in their frame for them to
form QGP, a long time has passed in the lab frame, around $330$ fm
for rapidity $y=6.5$, so QGP forms later at higher rapidities.
Throughout this QGP production process, each
disc gradually loses energy as partons with higher and higher rapidity
separate from it and form QGP.

The detailed study of quark--gluon plasma
is the \textit{Holy Grail} of the heavy ion programs at RHIC and the LHC.
The existence of this phase is well established via theoretical
calculations on the lattice, but its experimental analysis
within a HIC is very challenging, as one may have guessed. Indeed, the partonic matter keeps
expanding and cooling down (which implies that the
temperature is space and time dependent, i.e. thermal equilibrium
is reached only locally). Moreover, this matter eventually {\sl hadronizes}, i.e.
the `‘colored'' quark and gluons get trapped within colorless
hadrons, so we can reconstruct the dynamics of QGP only via the detected color-neutral ‘‘debris''.

\begin{figure}
\begin{centering}
\includegraphics[width=\textwidth]{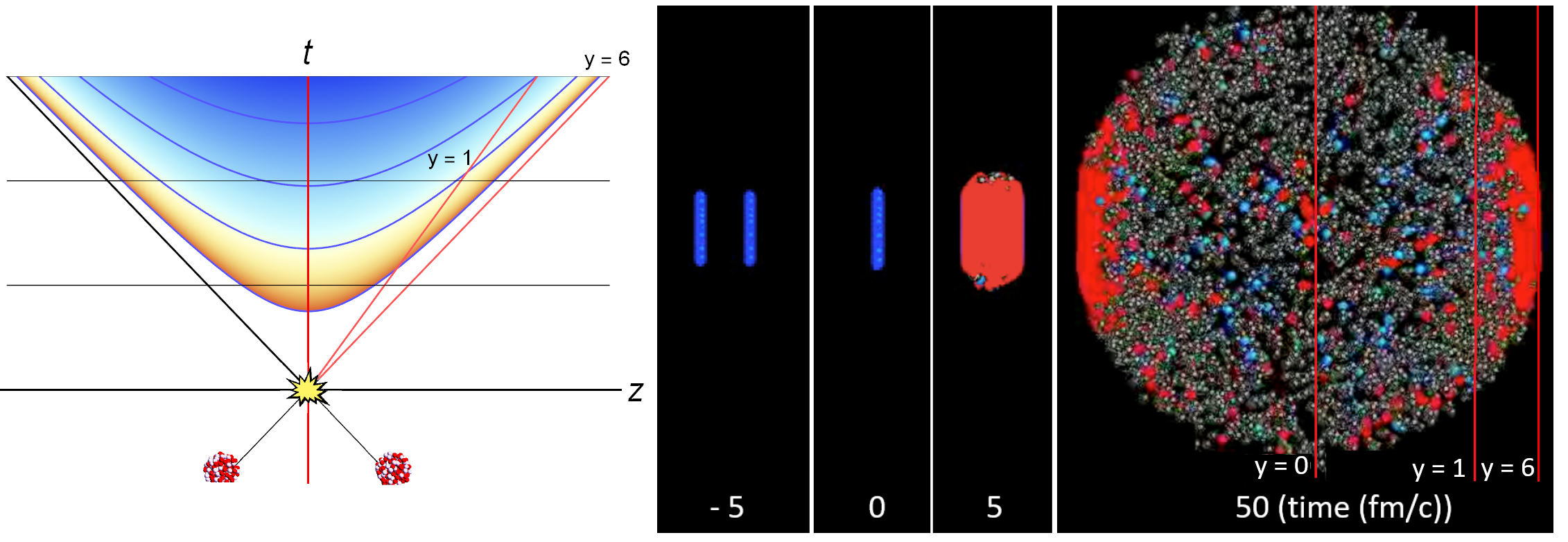}
\par\end{centering}
\caption{\textbf{(left) }Space-time picture of a heavy ion collision,
the color gives an indication of the temperature of the plasma formed.
Dynamics takes place as a function of proper time (blue curves), which
is why plasma forms later at higher rapidities. \textbf{(right) }Snapshots
of a central 2.76 TeV Pb-Pb collision at different times (different
horizontal slices of the space-time picture on the left) with hadrons
(blue and grey spheres) as well as QGP (red). In both figures, at
a given time the hottest regions can be found at high rapidity close
to the outgoing remnants of the nuclei and the red lines indicate
the approximate longitudinal location of particles with rapidity $y=0$,
$y=1$, and $y=6$. Figures adapted by \cite{Busza:2018rrf} from \cite{vanderSchee:2014qwa}.\label{fig:movie}}
\end{figure}

\item As the QGP cools down, the energy density within the fluid drops. At this point the fluid falls apart into a mist of hadrons that scatter off
each other: we have the transition of the QGP from a system of quarks and gluons to a state in which the degrees of freedom are color-neutral particles, namely hadrons. This process, called {\sl hadronization}, occurs when the (local) temperature
becomes of the order of the critical temperature $T_c$ for
deconfinement, known from lattice QCD studies to be $T_c\simeq
150- 180$~MeV. In Pb-Pb collisions at the LHC, this is
estimated to happen around a time $\tau\sim 10$~fm. At this time such hadronic system is still relatively dense, so it preserves local thermal equilibrium while expanding: one then speaks of a {\sl hot hadron gas}, whose temperature and density are decreasing with time. Moreover, since this process implies a sudden decrease in the entropy density with respect to a relatively small temperature change, the result is a great expansion of the system, since of course the total entropy cannot get smaller. As we have mentioned, in collisions at the highest energies of RHIC and at LHC the transition is a crossover, since the baryon chemical potential is very small. For smaller energies the transition may become of the first order.\footnote{In principle, the hadrons formed at highest rapidities are formed into an high baryon density region. Unfortunately, none of the LHC detectors are adequately instrumented around $y=6.5$, so we cannot study these products of the collision.}

\item Around a time $\tau\sim 20$~fm, the density becomes so low
    that the hadrons stop interacting with each other. That is, the
    collision rate becomes smaller than the expansion rate. This
    transition between a fluid state (where the hadrons undergo many
    collisions) and a system of free particles is referred to as the
    {\sl freeze--out}. From that moment on, the hadrons undergo free
    streaming until they reach the detector. One generally expects
    that the momentum distribution of the outgoing particles is
    essentially the same as their thermal distribution within the
    fluid, just before the freeze--out. This assumption appears to be confirmed by the data:
    the particle spectra as measured by the detectors can be well
    described as thermal (Maxwell--Boltzmann) distributions,
    using only few free parameters, like the fluid temperature and
    velocity at the time of freeze--out. This is generally seen as an
    additional argument in favor of thermalization, but one must be
    cautious on that, since the mechanism of hadronisation itself can
    lead to spectra which are apparently thermal. An indication of that is the freeze--out temperature extracted from the ratios of
    particle abundances at RHIC: it appears to be $T_f\simeq
    170$~MeV in both Au-Au and pp collisions, but in pp collisions we have no QGP formation, so no freeze-out altogether.

    \item When ions collide head-on the collision is called {\sl central}, whereas if the ions only partially overlap the collision is non-central or {\sl peripheral}. In this digression, we have actually considered only head-on collisions: how about non-central collisions? In the overlap region, the process is
the same as described above, except that the droplet of QGP is formed
with an initial approximately lenticular shape in the transverse plane.
In reality, because nuclei are made of individual nucleons, the energy
density of the QGP that forms is lumpy in the transverse plane, making
it neither perfectly circular in head-on collisions nor perfectly
lenticular in non-central collisions. Deviations from circular symmetry
in the initial shape of the QGP, whether due to off-center collisions
or to the lumpiness and fluctuations of the incident nuclei, result in
anisotropies in the pressure of the hydrodynamic fluid, which in turn
drive anisotropies in the expansion velocity and hence in the azimuthal
momentum distribution of the finally produced particles, see §\ref{sec:Probes and observables of HIC}.

In a non-central collision, the parts of the incident
nuclei that do not collide are referred to as ‘‘spectators''. At very
early times, they create a magnetic field in the collision zone, whose
possible effects are the subject of much discussion: see \cite{Kharzeev:2015znc} for further details. Later, the spectator partons fragment
into excited nuclei and hadrons, moving with almost the full rapidity
of the incident nuclei. The extreme limit of an off-center collision is one in which the nuclei themselves miss each other, but the Lorentz-contracted discs of electromagnetic fields around them do interact. These {\sl ultraperipheral} collisions give rise to copious $\gamma\gamma$ and $\gamma A$ interactions, which dominate the total nucleus-nucleus interaction cross-section in these kind of processes \cite{Baltz:2007kq}.

\end{enumerate}

Although extremely schematic, this simple enumeration of the various
stages of a HIC already illustrates the variety and complexity of the
forms of matter produced throughout a single event. In principle, all these forms of matter and their mutual transformations admit an unambiguous theoretical description in the framework of QCD, but although this theory has existed for about 40 years, it is still far from having delivered all the answers. Indeed, in spite of the apparent simplicity of its Lagrangian, which looks hardly more complicated than that of Quantum ElectroDynamics, the QCD dynamics is considerably richer and more complicated, which is the core reason why QCD can accommodate so many phases. What renders the theoretical study of HICs
so difficult is the extreme complexity of high density matter and of strong collective phenomena: for a theorist, the most efficient way to try and organize this complexity is to build effective theories.

\subsection{Effective theories for HIC}

As is evident from the above description, collisions of ultrarelativistic
nuclei are complex, consisting of several distinct stages, each probing
different aspects of QCD. What makes them interesting is that the
regimes of QCD we got experimental access to are not even close to be understood theoretically, because of the strength of the QCD interactions.
Heavy ion collisions are a laboratory that is rich with unique ways to probe fundamental aspects of QCD empirically, allowing some control over varying conditions. But at this point we have to ask ourselves: do we have, at least qualitatively, an understanding of all the stages of a heavy ion collision? Have any fundamentally new phenomena been seen or predicted, and what about any unexplained phenomena? And finally, what new insights have we obtained, or can we obtain, about the workings of QCD from analysis of heavy ion collision data and from the corresponding theoretical calculations? \\

As mentioned, effective theories may come to rescue. An {\sl effective theory} should not be confused with a ‘‘model'' (although we may regardless use this term as a synonym): its main purpose is not to provide a heuristic description of the data using some
physical guidance together with a set of free parameters. Rather, it aims
at a fundamental understanding and its construction is always guided by
the underlying fundamental theory, here QCD. Specifically, an
effective theory is a simplified version of the fundamental theory which
includes only the degrees of freedom which are relevant in the physical phenomena of our consideration. All the other degrees of freedom are ‘‘integrated out'' via some coarse--graining (or renormalization group)
procedure, which can be perturbative or non--perturbative.

Similarly to the previous Subsection, we now want to go through the phases of an heavy ion collision by describing the appropriate effective models of each regime: the different stages involve different forms of matter, with specific active degrees of freedom, hence their
theoretical description requires different effective theories. As before, most of the theoretical knowledge at our disposal comes from either QCD first principles and/or experimental data from jets, heavy quarks and high momentum particles.

\begin{itemize}

\item[\texttt{i.}]
During the early stages of the collision the parton density is very high, the typical transverse momenta are semi--hard (a few GeV), and the QCD
coupling is moderately weak, say $\alpha_s\sim 0.3$. In this case,
perturbation theory is (at least marginally) valid, but a correct study requires more than a
straightforward expansion in powers of $\alpha_s$. To construct the
correct effective theory for this phase, one needs to resum an infinite class of
Feynman graphs which are enhanced by high--energy and high gluon density
effects. This has been done in the recent years, and led to a formalism
-- the {\sl Color Glass Condensate} effective theory -- which offers a unified description from first principles for both the nuclear wavefunctions prior to the collision and for the very early stages of the collision itself. A key ingredient in
this construction is the proper recognition of the relevant degrees of freedom, which here are {\sl
quasi--classical color fields}. The concept of field is indeed
more useful, in this high--density environment, than that of particle, since the phase--space occupation numbers are large. This is because the would--be--particles overlap with each other and thus
form coherent states, which are more properly described as classical
field configurations. We will come back on this point extensively in Chapter \ref{chap:The initial stages of heavy ion collisions}.

\item[\texttt{ii.}] At later stages, the partonic matter expands, the
phase--space occupation numbers decrease, and the concept of {\sl
particle} becomes again meaningful: the classical fields break down into
particles. If these particles are weakly coupled (as one may expect by
continuity with the previous stages), then their subsequent evolution can
be described by {\sl kinetic theory}. This is an effective theory which
emerges under the assumption that the mean free path between two
successive collisions is much larger than any other microscopic scale
(e.g. the duration of a collision or the Compton wavelength
$\lambda=1/k_\perp$ of a particle). Over the last years, kinetic theory
has been extensively derived from QCD at weak coupling, but the results
appear to be deceiving. For instance, they cannot explain the rapid
thermalization suggested by the data at RHIC and at LHC: the
predicted thermalization times are much larger, $\tau\sim 10$ fm. Several alternative solutions have been proposed
so far, but the final outcome is still unclear. One of these proposals is
that the softer modes, which keep large occupation numbers and should be
better described as classical fields, become unstable due to the
anisotropy in the momentum distribution of the harder particles. This anisotropy should in turn follow from the disparity between longitudinal and transverse expansions. Anyway, numerical simulations of the coupled system ‘‘soft
fields--hard particles'' lead to thermalization times which are still too
large. Another suggestion is that the partonic matter is strongly coupled: the QCD coupling could indeed become larger,
because of the system being more dilute. In such a scenario, a candidate
for an effective theory is the {\sl AdS/CFT correspondence}, which we have already mentioned previously: it exploits a conjectured duality between a weak coupling gravitational description in a 4+1-dimensional Anti-de Sitter space and a strong coupling description of a corresponding Conformal Field Theory living in the boundary of this space.

\item[\texttt{iii.}] Assuming (local) thermal equilibrium, and hence the
formation of a quark--gluon plasma, the question is now whether
this plasma is weakly or strongly coupled. The maximal temperature of
this plasma, as estimated from the average energy density, should be
around $T\sim 500-600$~MeV. The thermodynamic
properties (like pressure or energy density) of a QGP at global
thermal equilibrium within this range of temperatures are, by now, well
known from numerical calculations and can serve as a
baseline of comparison for various effective theories. If the coupling is
weak, one can use the {\sl Hard Thermal Loop} (HTL) effective theory,
a version of kinetic theory which describes the long--range (or
`soft') excitations of the QGP. This effective theory lies at the basis
of a physical picture of the QGP as a gas of {\sl weakly--coupled
quasi--particles}, in particular as quarks or gluons with temperature--dependent
effective masses and couplings. Using this picture as a guideline for
a rearrangement of a perturbative expansion, the lattice data have been reproduced quite well. Thus, the thermodynamics appears
to be consistent with a weak--coupling picture for the QGP, although this
picture is considerably more complicated than that emerging from naive
perturbation theory (the strict expansion in powers in $g$). Yet, this is
not the end of the story, as we shall see.

\item[\texttt{iv.}] The QGP created in the intermediate stages of a HIC is
certainly not in global thermal equilibrium, but only in a {\sl
local} one, since it keeps expanding. Under very general assumptions, the
effective theory describing this flow is {\sl hydrodynamics}. The
corresponding equations of motion are simply the conservation laws for
energy, momentum, and other conserved quantities (like the electric
charge or the baryonic number), and as such they are universally valid. These equations also involve `parameters', like the {\sl viscosities}, which describe dissipative phenomena occurring during the flow and depend upon the specific microscopic dynamics, which are not specified in a hydrodynamic approach. The values
of these parameters are very different at weak vs. strong coupling. As mentioned, the
dimensionless viscosity--over--entropy-density ratio $\eta/s$ is one of such parameters, and data at RHIC and the LHC suggest a very small value for this
ratio, which is inconsistent with calculations at weak
coupling based on kinetic theory. On the other hand, such a small ratio
is naturally emerging at strong coupling (as shown by calculations within
the AdS/CFT correspondence), which is a strong argument in favor of a strongly coupled QGP.

This may look contradictory with the previous conclusions drawn from thermodynamics. However, one should remember that the QCD coupling
depends upon the relevant space--time scale, and hydrodynamics refers
to the long--range behavior of the fluid, as encoded in its softest
modes. By contrast, thermodynamics is rather controlled by the hardest
modes, those with typical energies and momenta of the order of the
(local) temperature. So, it is not inconceivable that the same system may look effectively weakly coupled for some phenomena and strongly coupled for some others.

\item[\texttt{v.}] Approaching the final stages of the collisions, we have the formation of color-less hadrons. This mechanism of particle production, via an intermediate epoch during which a hydrodynamic fluid forms and expands, is quite different from the current understanding of particle production in elementary collisions, in which only a few new particles are created. This is one of the reasons why the study of hadronization is very challenging and most its details are yet unknown, given the highly non perturbative nature of this phenomenon. The hadronization process, as well as the other phases of the collision, can be studied with different approaches. For instance, in the {\sl fragmentation} model \cite{Albino:2005me, Albino:2008fy} one takes into account the processes in which each parton produces a jet of hadrons, each carrying a fraction of the momentum of the initial parton. Instead, the {\sl coalescence} mechanism \cite{Greco:2003vf,Greco:2003xt, Greco:2003mm} allows for the recombination of two or three quarks into hadrons. Since the probability to find two or three partons in the same phase-space element decreases with momentum, coalescence becomes less important for high momenta ($p_\perp\gtrsim5$ GeV), for which instead fragmentation starts taking over. For further details, there are many reviews on hadronization in the literature, e.g. \cite{Fries:2025jfi}.

\end{itemize}

\subsection{Probes and observables of HIC}
\label{sec:Probes and observables of HIC}

In this Section we resume a list of the principal physical observables and probes that are able to give us any information about what happens during a HIC. Each may be affected, or can tell us some information about, different stages of
the evolution. For this reason it is also important to identify which observable can shed light on a particular phase without being influenced
by other stages, in order to have information about a particular
phase of the collision. We will describe the utility of both specific products of HICs (heavy quark hadrons and jet observables) and also global observables, describing the distribution of detected hadrons as a whole (nuclear modification factor and elliptic flow).

\subsubsection{Heavy Quark hadrons}
\label{sec:Heavy Quark hadrons}

In the context of HIC, heavy quarks and their bound states are considered very useful probes to study the evolution of the initial stage and also of the QGP deconfined state. More specifically, the quarks which serve our purpose are the {\sl charm}, whose bare mass is about 1.3 GeV, and the {\sl bottom}, which has a bare mass of 4.2 GeV \cite{ParticleDataGroup:2024cfk}. The top quark is usually not taken into account, since its huge mass (around 172 GeV) implies decay times (via the weak interaction) much smaller than 1 fm, therefore this particle is not able to hadronize since its lifetime is smaller than the QCD timescales. Check e.g. \cite{Apolinario:2017sob} for a study on top quarks in QGP. 

The main feature of Heavy Quarks is an high value of mass, much greater than the hadronization scale $\Lambda_{\text{QCD}}$. Since in such regimes the coupling constant is sufficiently small, we are entitled to use a perturbative approach in the calculation of the cross sections. The masses of the heavy quarks are not only greater than $\Lambda_{\text{QCD}}$, but also greater than the typical temperatures that characterize the system during the QGP formation. This implies that the number of heavy quark-antiquark pairs is determined in the first instants after the collision: the temperature is not high enough to determine a significant production of these pairs via thermal excitation of the vacuum. Moreover, the thermalization time of the heavy quarks is much greater than the one of the light quarks, and it is comparable with the lifetime of the QGP itself, i.e. around $\tau_{\text{therm}}\sim 10$ fm for typical Pb-Pb collisions at LHC energies \cite{Markert:2008jc}. This means that the charm quark, and even more the bottom quark, are not able to reach local thermal equilibrium with the QGP bulk. However, their interactions with the medium are able to modify the initial momentum spectra, so the final hadrons can give an indication of such interactions.

Finally, their thermal momentum is far greater than the momentum typically transferred by the medium, since we have
\begin{equation}
    p_{\text{thermal}}^2\sim 3M_QT\gg T^2
    \label{eq:thermal_momentum_HQ}
\end{equation}
in a non-relativistic approximation. This allows us to study the evolution of heavy quarks, for instance, in terms of a Brownian-like approach (e.g. via a Fokker-Planck equation, to derive the drag and diffusion coefficients of HQ in the medium \cite{Sambataro:2024mkr}). Several studies have however shown that the Brownian motion approximation is not strongly valid in the range of momenta of interest for heavy quarks $(p \approx 2-3 M_Q)$ and that, especially for the charm, the description of Brownian motion is in general a strong approximation, far from the dynamics of a realistic system. This is even more true if we consider the study of charm and bottom quarks in the initial stages of a HIC, as we will do in Chapters \ref{chap:Melting of heavy quark pairs in glasma} and \ref{chap:Anisotropies of gluons and heavy quarks in glasma}. Indeed, in the early stages the bulk is not made up of particles, but rather strong color fields, and their effect on heavy quarks has to be described by relativistic kinetic theory equations coupled to the gluon fields, which are called Wong equations \cite{Wong:1970fu}. Several works have investigated the impact of the early stages on heavy quarks~\cite{Mrowczynski:2017kso, Ruggieri:2018rzi, Sun:2019fud, Liu:2019lac, Liu:2020cpj, Boguslavski:2020tqz, Carrington:2020sww, Ipp:2020nfu, Khowal:2021zoo, Carrington:2021dvw, Pooja:2022ojj, Carrington:2022bnv, Das:2022lqh, Boguslavski:2023fdm, Avramescu:2023qvv, Pandey:2023dzz, Boguslavski:2023alu,Avramescu:2024poa,Avramescu:2024xts}:
these have shown that heavy quarks undergo significant diffusion in the 
early stage, suggesting an impact on experimental observables such as the nuclear modification factor and the elliptic flow, as analyzed in~\cite{Sun:2019fud}.

\subsubsection{Jet products}
Another strategy for studying the hadronic matter produced
in a HIC refers to the use of {\sl hard probes}. These are particles with large transverse energies (say, $E_\perp\gtrsim 20$~GeV at the LHC), which are produced in the very first instants of a collision and then cross the QCD matter liberated at later stages, along their way towards the detector. Some of these particles, like the (direct) photons and the dilepton pairs, do not interact with this matter and hence can be used as a baseline for comparisons. On the other hand, other particles like quarks, gluons, and
the {\sl jets} initiated by them (a jet is the narrow cone of hadrons and other particles produced via hadronization by a very energetic parton), {\sl do} interact. By measuring the
effects of these interactions -- say, in terms of energy loss, or the
suppression of multi--particle correlations -- one can infer
informations about the properties of the matter they crossed. 

The RHIC data have demonstrated that semi--hard partons can lose a substantial fraction of the transverse energy via interactions with the medium ({\sl jet quenching}), thus suggesting that these interactions can be quite
strong \cite{Gyulassy:2003mc}. These results have been confirmed at the LHC, where they found that even very hard jets ($E_\perp\gtrsim 100$~GeV) can be
strongly influenced by the medium, in the sense that they get strongly defocused: the energy distribution in the polar angle with respect to the jet axis becomes much wider after having crossed the medium. This is
visible for the photon--hadron di--jet event in the right panel of Fig. \ref{fig:events}: the photon and the parton which has initiated the hadronic jet have been created by a hard scattering, so they must have
been balanced in transverse momentum at the time of their creation. Yet, the central peak in the hadronic jet, which represents the final `jet' according to the conventional definition, carries much less energy than
the photon jet. This is interpreted as the result of energy transfer to large polar angles (i.e. angles outside the conventional `jet' definition) via
in--medium interactions.

In order to study such interactions, in particular high--density effects like multiple scattering and
coherence, it is again useful to build effective theories.  The problem is that, in this case, it is not so clear whether the physics is controlled by mostly
weak coupling, or mostly strong coupling: by itself, the jet is hard, but its coupling to the relatively soft constituents of the
medium may be still governed by a moderately strong coupling. In fact,
the unexpectedly strong jet quenching observed at RHIC is sometimes
interpreted as another evidence for strong coupling behavior. Within perturbative QCD, the effective theory we trust to study these processes is known as {\sl
medium--induced gluon radiation}, while at strong coupling one again relies on AdS/CFT techniques. Not surprisingly, {\sl
moderately} strong coupling turns out to be the most difficult situation to deal with.

\subsubsection{Nuclear Modification Factor $R_{\text{AA}}$}

\begin{figure}[ht]
    \centering
    \includegraphics[scale=0.4]{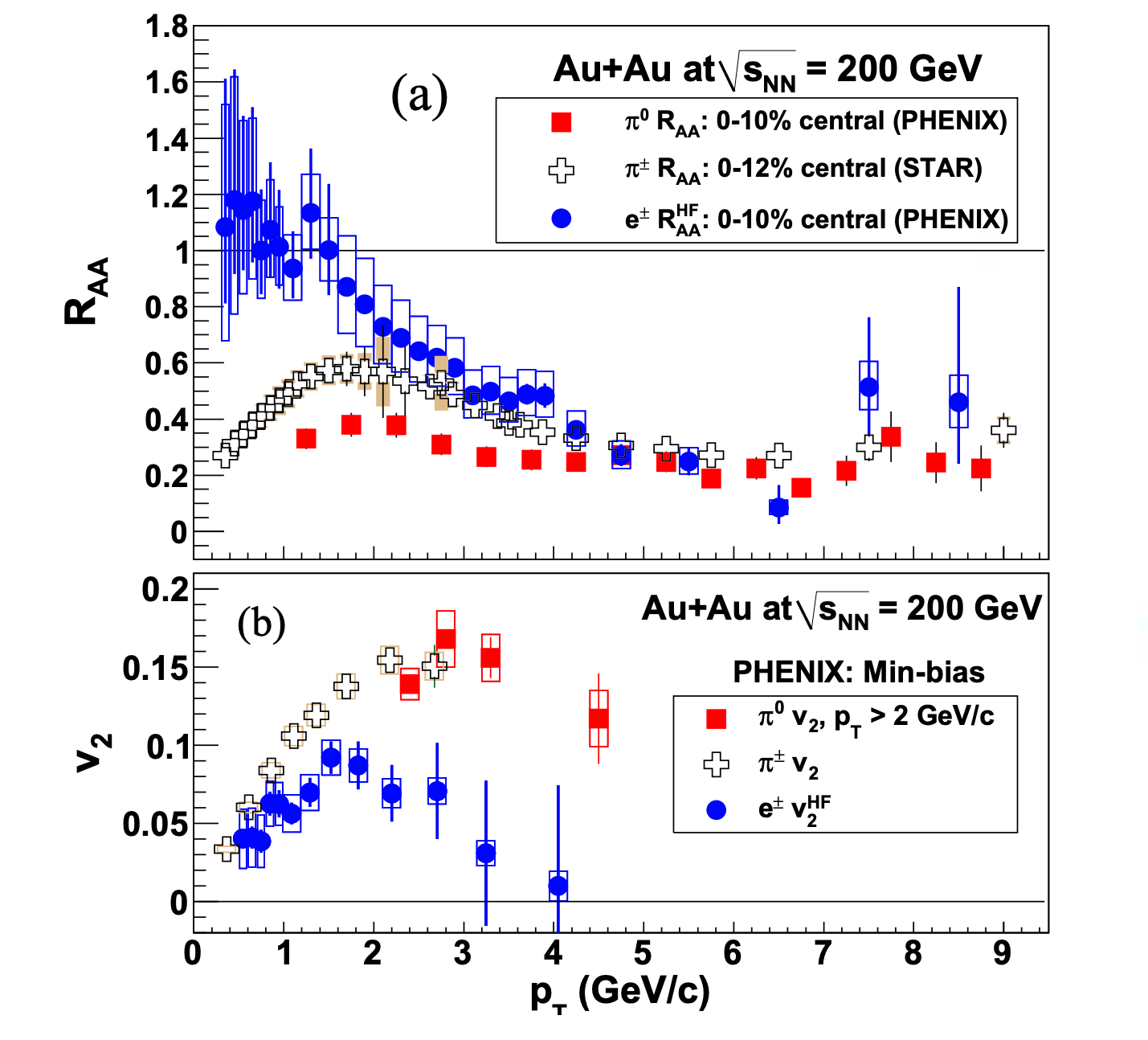}
    \caption{\textbf{(a)} Nuclear modification factor $R_{\text{AA}}^{\text{HF}}$ for heavy flavour electrons compared to the $R_{\text{AA}}$ of $\pi^0$ and $\pi^\pm$ in central Au-Au collisions at $\sqrt{s_{NN}}=200$ GeV. \textbf{(b)} Elliptic flow $v_2^{HF}$ of heavy flavour electrons compared with $v_2$ of $\pi^0$ and $\pi^\pm$ in minimum-bias Au-Au collisions at $\sqrt{s_{NN}}=200$ GeV \cite{Nouicer:2012pn,Nouicer:2009fy,Nouicer:2011zz,STAR:2007zea,PHENIX:2003qdw,PHENIX:2006ujp,PHENIX:2006mhb}.}
    \label{fig:RAAandv2}
\end{figure}

We have mentioned jet quenching, the process by which high $p_\perp$ particles, which are formed in the first stages after the collision, propagate inside the QGP and lose a significant amount of energy due to the interaction with the bulk constituents. Such energy, then, is soon dissipated in the plasma. However, in order to quantify the effect on the medium, we need a related physical observable which could give a quantitative estimate of the interaction among heavy quarks/jets and the partons of the bulk. In particular, such observable should compare the total observed momentum distribution of hadrons in a real HIC against the hypothetical momentum distribution of hadrons of a collision in which there is no medium interaction. The latter should act as a reference point, a baseline from which one can study the energy suppression. Such suppression is described by the {\sl nuclear modification factor} $R_{\text{AA}}$, whose definition is
\begin{equation}
    R_{\text{AA}}(p_\perp)\equiv\frac{1}{N_{\text{coll}}}\frac{\mathrm{d}^2N_{\text{AA}}/\mathrm{d}\eta\, \mathrm{d}p_\perp}{\mathrm{d}^2N_{\text{pp}}/\mathrm{d}\eta\, \mathrm{d}p_\perp}.
    \label{eq:nuclear_modification_factor}
\end{equation}
This observable is defined as the ratio between the invariant momentum distribution (also known as the {\sl yield}) of the particles produced in a nucleus-nucleus collision and the same yield of particles produced in elementary proton-proton collisions. The latter, of course, has to be normalized to the number of baryon-baryon binary collisions expected, by the Glauber model, at that center of mass energy and centrality: in other words, one basically counts the average number of colliding nucleons for a given impact parameter. Values of $R_{\text{AA}}$ close to one (see e.g. low $p_\perp$ blue points in Figure \ref{fig:RAAandv2} (a)) indicate that the medium created in a nucleus-nucleus collision has basically no effect on the final distribution of the particles, i.e. the medium is either weakly interacting or extremely dilute. On the other hand, large discrepancies with respect to the value 1 manifest the presence of a strongly interacting and dense medium, which makes high $p_\perp$ particles suffer many scattering processes (see high $p_\perp$ points in Figure \ref{fig:RAAandv2} (a)). These may eventually bring to thermalization with the medium itself.

\subsubsection{Elliptic flow $v_2$}
\label{subsection:elliptic_flow}
When the fireball is formed, pressure gradients generate a {\sl collective flow} of expanding matter which ``brings memory'' of the thermodynamic conditions of the early stages of the collision \cite{Heinz:2004qz}. In particular, collective flow is distinguished in a longitudinal component and in a transverse one: the former moves in the direction of the beam axis, which we can assume to be parallel to the $z$ axis, whereas the latter flows in the perpendicular plane. The initial properties of the transverse flow depend on the features of the collision: for central collisions (impact parameter $b=0$) between symmetric nuclei the transverse flow is isotropic in the whole $xy$ plane, i.e. it shows no dependence with respect to the azimuthal angle $\phi$. Such flow is called {\sl radial flow}. Instead, for non-central collisions ($b\neq 0$) the azimuthal isotropy is broken and therefore the radial flow fill acquire a dependence on $\phi$. Of course, the ability of the expanding system to hold a collective behavior depends on the ``stiffness'' of the Equation of State (EoS), which is not always the same. For instance, near the QCD phase transition the EoS becomes very soft (the speed of sound $c_s$, whose square is equal to $\partial P/\partial \epsilon$, is close to zero), therefore the system is hardly able to sustain a collective profile in the transverse plane. This observation allows us, in this case, to obtain information on the EoS close to the phase transition via measurements of the transverse flow.

\begin{figure}[t]
    \centering
    \includegraphics[scale=0.3]{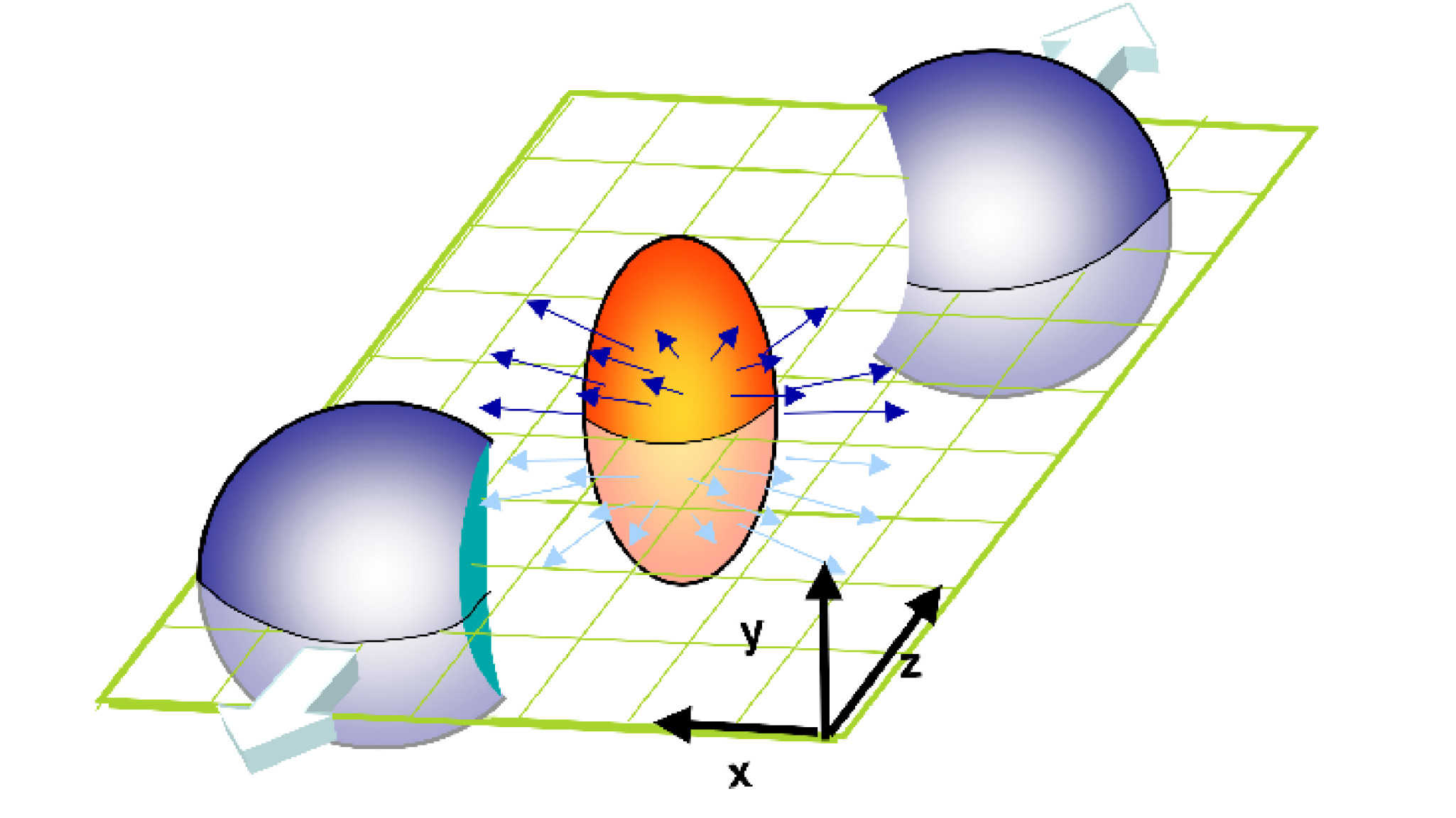}
    \caption{Sketch of deformed fireball and elliptic flow generated in non--central heavy--ion collisions \cite{Heinz:2008tv}.}
    \label{fig2.5}
\end{figure}

Spatial anisotropies can be investigated by studying collisions with fixed center of mass energy and different impact parameters $b$. In particular, in non-central collisions the overlapping nuclear region is pictured as an almond (Figure \ref{fig2.5}), in which the highest pressure is located in the center, whereas it is zero outside it. This implies that the pressure gradient is stronger along the ‘short side' of the almond rather than along the ‘long side', because the variation of pressure from the maximum value to zero occurs in a shorter length. This observation is crucial to the study of the final state, since it implies that the initial spatial anisotropies are transferred, by the action of particle interactions, to momentum space anisotropies, and this effect can be highlighted in the spectra of the detected hadrons.\\

By referring to the coordinate system in Figure \ref{fig2.5}, we can quantify the spatial eccentricity of the collision region as a function of the impact parameter $b$
\begin{equation}
    \epsilon_x(b)=\frac{\langle y^2-x^2\rangle}{\langle y^2+x^2\rangle}.
    \label{eq:spatial_eccentricity}
\end{equation}
In \eqref{eq:spatial_eccentricity} the averages are performed over the particle density distribution. As mentioned, once time passes such spatial eccentricity is translated into momentum space, and the invariant momentum distribution can be expressed by the means of a Fourier expansion in terms of the azimuthal angle $\phi$ as \cite{Voloshin:1994mz,Poskanzer:1998yz}
\begin{equation}
    E\frac{\dd^3N}{\dd\,\bm{p}^3}=\frac{1}{2\pi}\frac{\dd^2N}{p_\perp \dd p_\perp \dd y}\left\{1+2\sum_{n=1}^{\infty}v_n(p_\perp,b)\cos\left[n(\phi-\Psi_{\text{RP}})\right]\right\}.
    \label{2.1}
\end{equation}
Here the angle $\Psi_{\text{RP}}$ is the angle identified by the reaction plane and any axis orthogonal to the beam direction: for instance, in Figure \ref{fig2.5} we have $\Psi_{\text{RP}}=0$ with respect to the $x$ direction. In Eq. \eqref{2.1} we can observe the absence of the $\sin\left[n(\phi-\Psi_{\text{RP}})\right]$ terms, which vanish due to the reflection symmetry with respect to the reaction plane. Moreover, notice also the absence of the harmonic $v_0$, which gives the yield of particles per unit of rapidity and it is used to normalize all the other Fourier coefficients. The coefficients $v_n$ of the expansion \eqref{2.1} can be found by means of an average procedure:
\begin{equation}
    v_n(p_\perp, b)=\langle \cos\left[n(\phi-\Psi_{\text{RP}})\right]\rangle.
    \label{eq:vn_definition}
\end{equation}
It was believed that all the odd $v_n$ were zero for symmetry reasons, however the possibility to perform event-per-event analysis showed that the quantum oscillations allow also for non-zero odd harmonics.

Among all the Fourier coefficients, the $v_1$ coefficient gives the strength of the dipole flow, whereas $v_3$ gives the measure of the triangular flow. However, the {\sl elliptic flow} $v_2$ is the most interesting coefficient and gives us the highest indication of azimuthal anisotropy in momentum space. As well as all the other coefficients, it depends on the transverse momentum and on the modulus of the impact parameter. Its explicit expression is:
\begin{equation}
    v_2(p_\perp,b)=\langle \cos\left[2(\phi-\Psi_{\text{RP}})\right]\rangle=\left\langle\frac{p_x^2-p_y^2}{p_x^2+p_y^2}\right\rangle_{(p_\perp,b)}=\left\langle\frac{p_x^2-p_y^2}{p_\perp^2}\right\rangle_{(p_\perp,b)}.
    \label{eq:v2_definition}
\end{equation}

\begin{figure}[ht]
    \centering
    \includegraphics[scale=1.2]{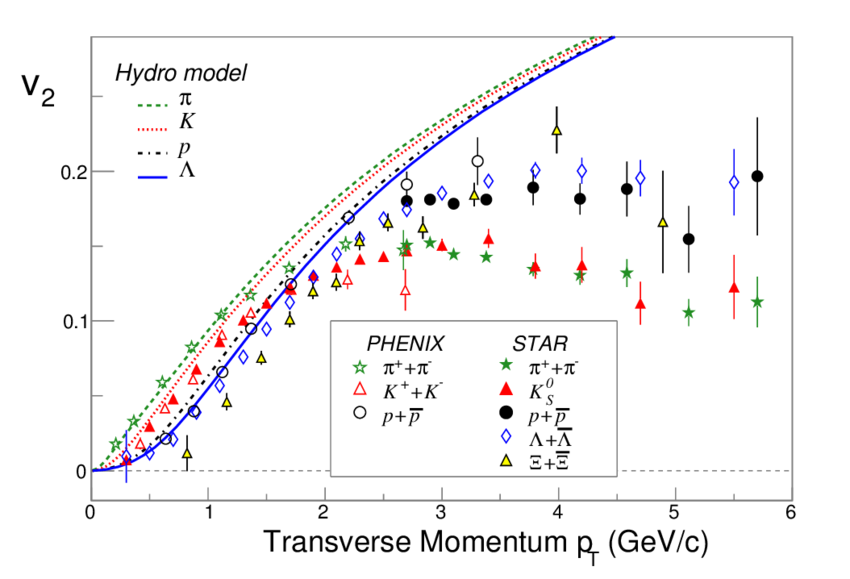}
    \caption{Measured $v_2$ for several hadron species for Au-Au collisions at $\sqrt{s_{NN}}=200$ GeV: experimental data refer to both PHENIX \cite{PHENIX:2003qra} and STAR \cite{STAR:2003wqp} collaborations at RHIC. The curves are instead hydrodynamical predictions from \cite{Huovinen:2001cy, Kolb:2003bf} for $v_2$ of various hadronic species. Picture from \cite{Fries:2008hs}.}
    \label{fig2.6}
\end{figure}

\begin{figure}[ht]
    \centering
    \includegraphics[scale=0.3]{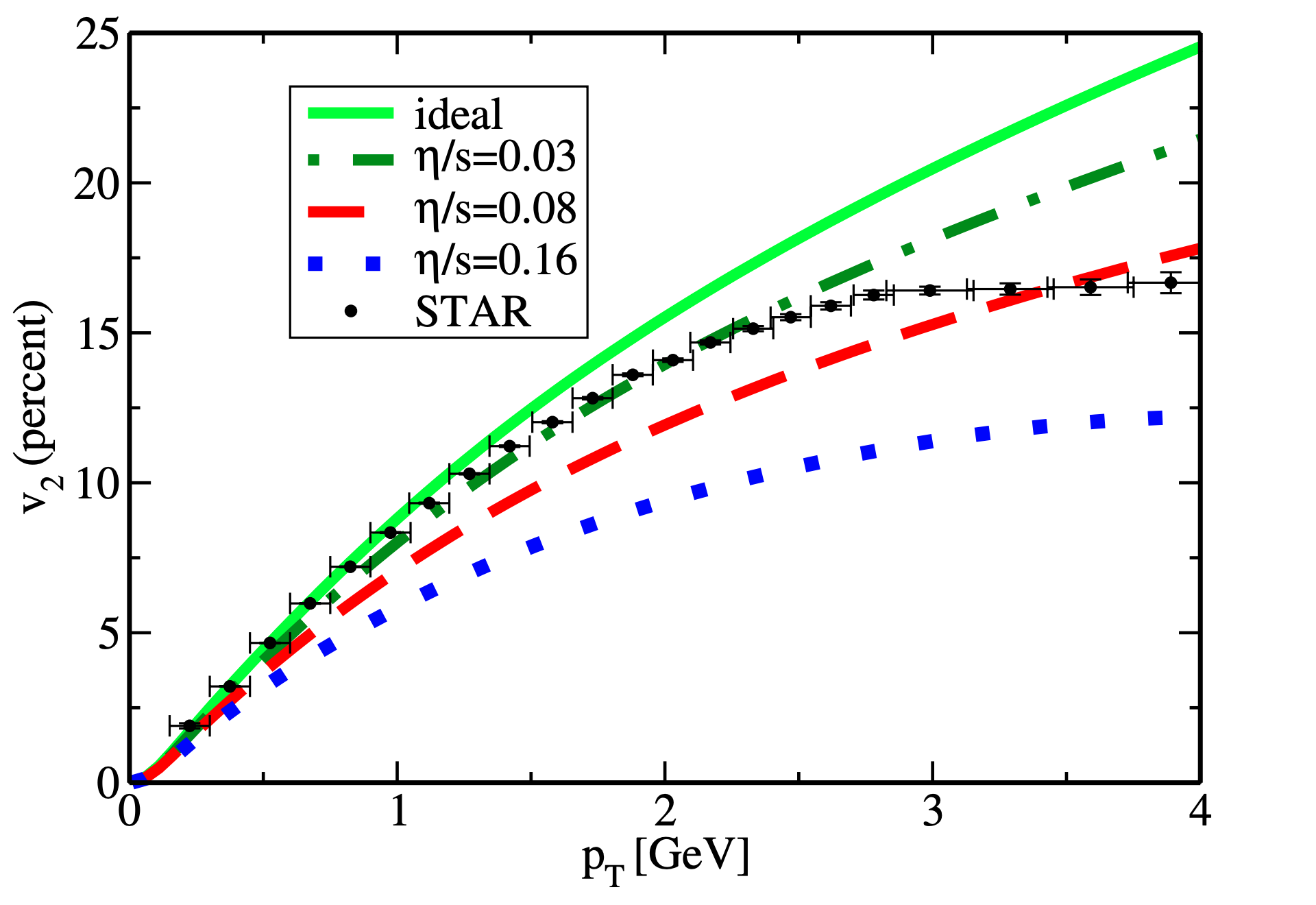}
    \caption{Charged hadron elliptic flow from STAR \cite{STAR:2008ftz} compared with both ideal and viscous hydrodynamic models at given values of the $\eta/s$ ratio \cite{Romatschke:2007mq,Luzum:2008cw}. The relevant value $\eta/s=1/4\pi\simeq 0.08$ is the lower bound provided by AdS/CFT \cite{Kovtun:2004de}.}
    \label{fig2.7}
\end{figure}

In Figure \ref{fig2.6} it is shown a comparison of the experimental results for $v_2$ at RHIC with the ideal hydrodynamical estimates, in which no dissipative effects (encoded by the shear viscosity) are taken into account. On the other hand, hydrodynamical estimates of $v_2$ at various $\eta/s$ ratios, are reported in Figure \ref{fig2.7}. Observation of quite large $v_2$ suggests that expanding matter is able to carry high collective flow, which is an indicator of a low-viscosity value. In the region $p_\perp< 2$ GeV the elliptic flow is in an excellent agreement with the hydrodynamic predictions in the hypothesis of a negligible $\eta/s$ ratio, whereas we have significant discrepancy for higher $p_\perp$ values.\\

Usually $R_{\text{AA}}$ and $v_2$ are inversely correlated: by referring to Figure \ref{fig:RAAandv2} (a) and (b), low $R_{\text{AA}}$ values correspond to high values of $v_2$, which is a sign of the presence of a dense plasma. One of the main goals related to the study of heavy quark dynamics is indeed the solution of the $R_{\text{AA}}$--$v_2$ {\sl puzzle}, the development of a physical model able to reproduce both the elliptic flow and the nuclear modification factor obtained experimentally. At the moment, neither pQCD nor the introduction of interactions among high $p_\perp$ particles and the bulk is able to accurately reproduce the experimental data.

\chapter[The initial stages of Heavy Ion Collisions]{The initial stages of Heavy Ion Collisions}
\label{chap:The initial stages of heavy ion collisions}

This Chapter is an introduction to the methods and key phenomena associated with the study of the initial stages of heavy ion collisions, described as a gluon plasma, i.e. a {\sl glasma}. At very high energies, or small values of Bjorken $x$, the density of partons per unit transverse area in hadronic wavefunctions becomes very large, leading to a saturation of partonic distributions. When the scale corresponding to the density per unit transverse area, called the saturation scale $Q_s$, becomes large ($Q_s \gg \Lambda_{\text{QCD}}$), the coupling constant becomes weak ($\alpha_s(Q_s)\ll 1$). This suggests that the high energy limit of QCD may be studied using weak coupling techniques, albeit always non perturbatively: the formalization of this simple idea is an eﬀective theory called the Color Glass Condensate (CGC), which describes the behavior of the small $x$ components of the hadronic wavefunction in QCD. This effective theory has a rich structure that has been explored using both analytical and numerical techniques. The CGC formalism can be applied to study a wide range of high energy scattering experiments, from deep inelastic scattering experiments at HERA and the oncoming Electron Ion Collider (EIC) to proton--nucleus and nucleus--nucleus experiments at the RHIC and LHC colliders \cite{Iancu:2003xm}.

In this Chapter, we will first of all discuss the physics of glasma, in particular how the BFKL evolution leads to a high saturation of gluons. Then we will move on to the Color Glass Condensate framework and its effectiveness to describe such phase. After that, we will deal with the numerical implementation of CGC on the lattice: this involves the description of the McLerran--Venugopalan (MV) model for the initial charge for both heavy ions and protons, as well as the time evolution which can be accomplished using real-time lattice gauge theory techniques. Using these methods it is possible to study quantities like the energy density and the pressure components of the glasma. To this purpose, we will show various result in the final pages of the Chapter.

\section{Nuclear matter at extreme conditions}

\subsection{Parton distribution functions at high energies}

The proton is a bound state of three ``valence'' quarks: two up quarks and one down quark. For this reason, it is not hard to believe that the simplest view of a proton reveals three quarks interacting via the exchanges of gluons, which ``glue'' the quarks together. However, experiments probing proton structure at the HERA
collider at Germany's DESY laboratory, and the increasing body of evidence from RHIC and LHC, suggest that this picture is far too
simple and strongly dependent on energy \cite{Accardi:2012qut}. Countless other gluons and a ``sea'' of quarks and antiquarks
pop in and out of existence within each hadron, and these fluctuations can be probed in high-energy scattering experiments. These fluctuations are, in principle, virtual. However, due to Lorentz time
dilation, the more we accelerate a proton, the closer it gets to the speed of light, therefore the longer are the lifetimes of the gluons that arise from the quantum fluctuations. For this reason an outside ``observer'', viewing a fast moving proton, would see the cascading of gluons last longer and longer, the larger the velocity of the proton.  So, in effect, by speeding the proton up, one can slow down the gluon fluctuations enough to ``take snapshots'' of them with a probe particle sent to interact with the high-energy proton.

\begin{figure}[t]
\begin{center}
\includegraphics[width=0.5\textwidth]{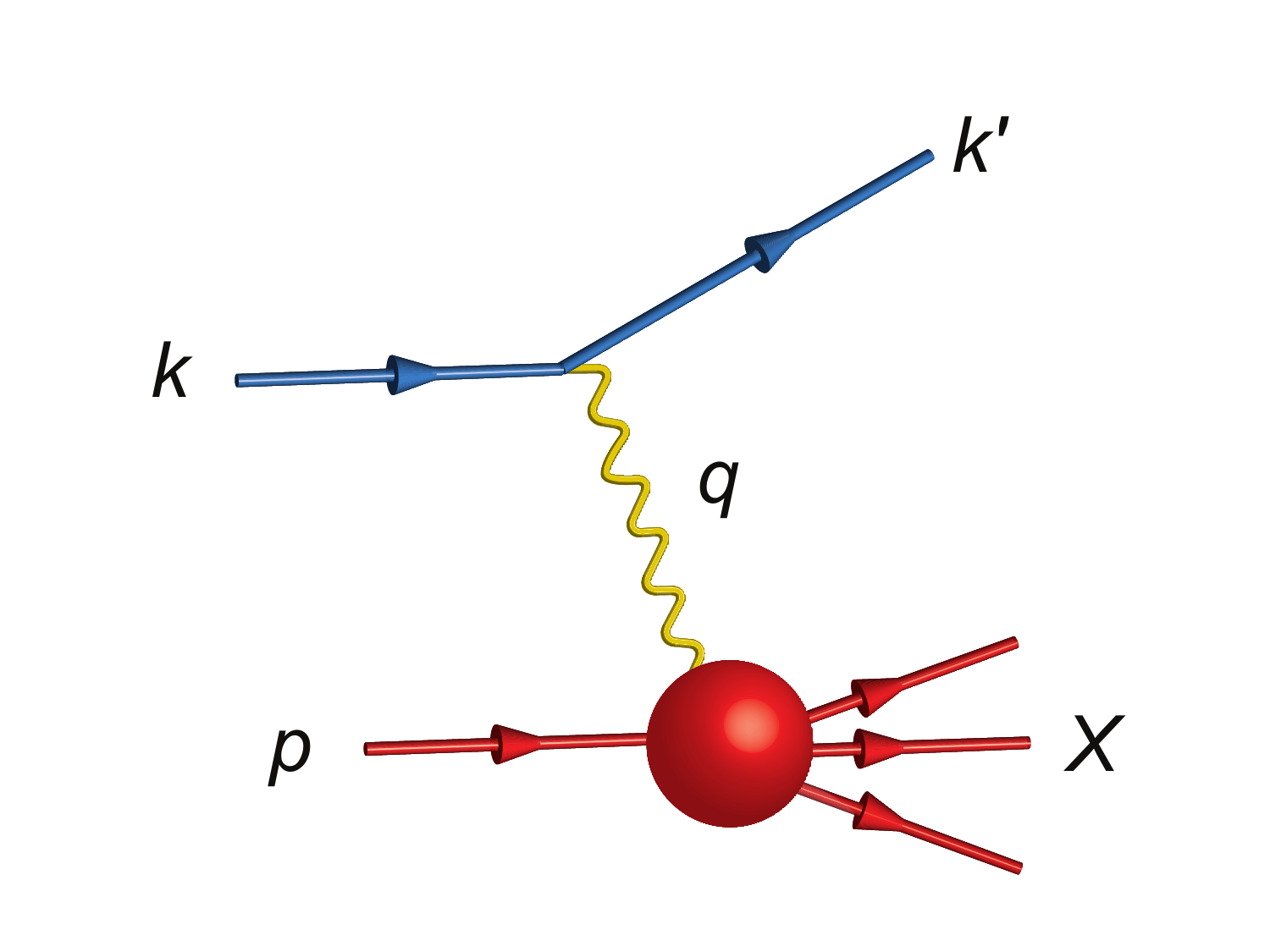}
\end{center}
\caption{A schematic diagram of the Deep
Inelastic Scattering (DIS) process. Figure from \cite{Accardi:2012qut}.}
\label{fig:sidebarDISGraph}
\end{figure}

This has been historically done in Deep Inelastic Scattering (DIS) experiments. In particular, in those experiments one probes the proton wave-function with a lepton, which interacts with the proton by exchanging a (virtual) photon with it (Figure \ref{fig:sidebarDISGraph}).  To explain each part of the terminology, {\sl scattering} refers to the deflection of leptons (mainly electrons or muons) off of hadrons. Measuring the angles of deflection gives information about the nature of the process. {\sl Inelastic} means that the target absorbs some kinetic energy. In fact, at the very high energies of leptons used, the target is shattered and emits many new particles: these particles are hadrons, which are produced via hadronization of the constituent quarks of the target. Finally, {\sl deep} refers to the high energy of the lepton, which gives it a very short De Broglie wavelength and hence the ability to probe distances that are small compared with the size of the target hadron. In this sense, we say that the lepton probes ‘‘deep inside'' the hadron. Also, note that in the perturbative approximation it is a high-energy virtual photon emitted from the lepton, and absorbed by the target hadron, which transfers energy to one of its constituent quarks, as shown in Fig. \ref{fig:sidebarDISGraph}. The {\sl virtuality} of the photon, i.e. the squared modulus $Q^2$ of its four-momentum, determines the size of the region (in the plane transverse to the beam axis) probed by the photon: in particular, by the uncertainty principle this region's width is of order $1/Q$. Another relevant variable is the {\sl Bjorken $x$}, which is the fraction of the proton momentum carried by the struck quark. At high energy $x \approx Q^2/W^2$ is small, where $W^2$ is the
center-of-mass energy squared of the photon-proton system. For this reason, small $x$ correspond to high-energy scattering.

\begin{figure}[ht]
\begin{center}
\includegraphics[width=0.48\textwidth]{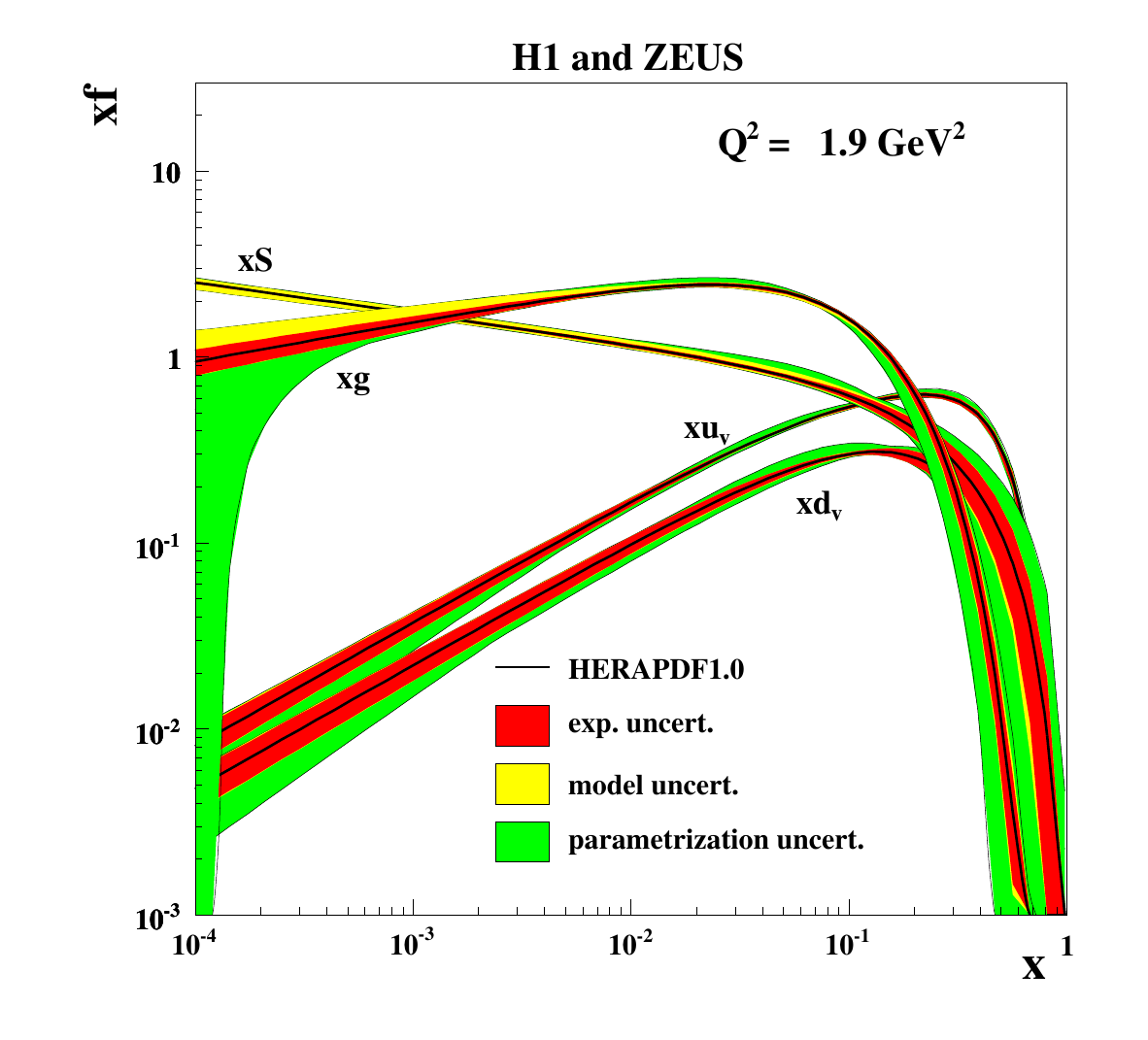}
\includegraphics[width=0.48\textwidth]{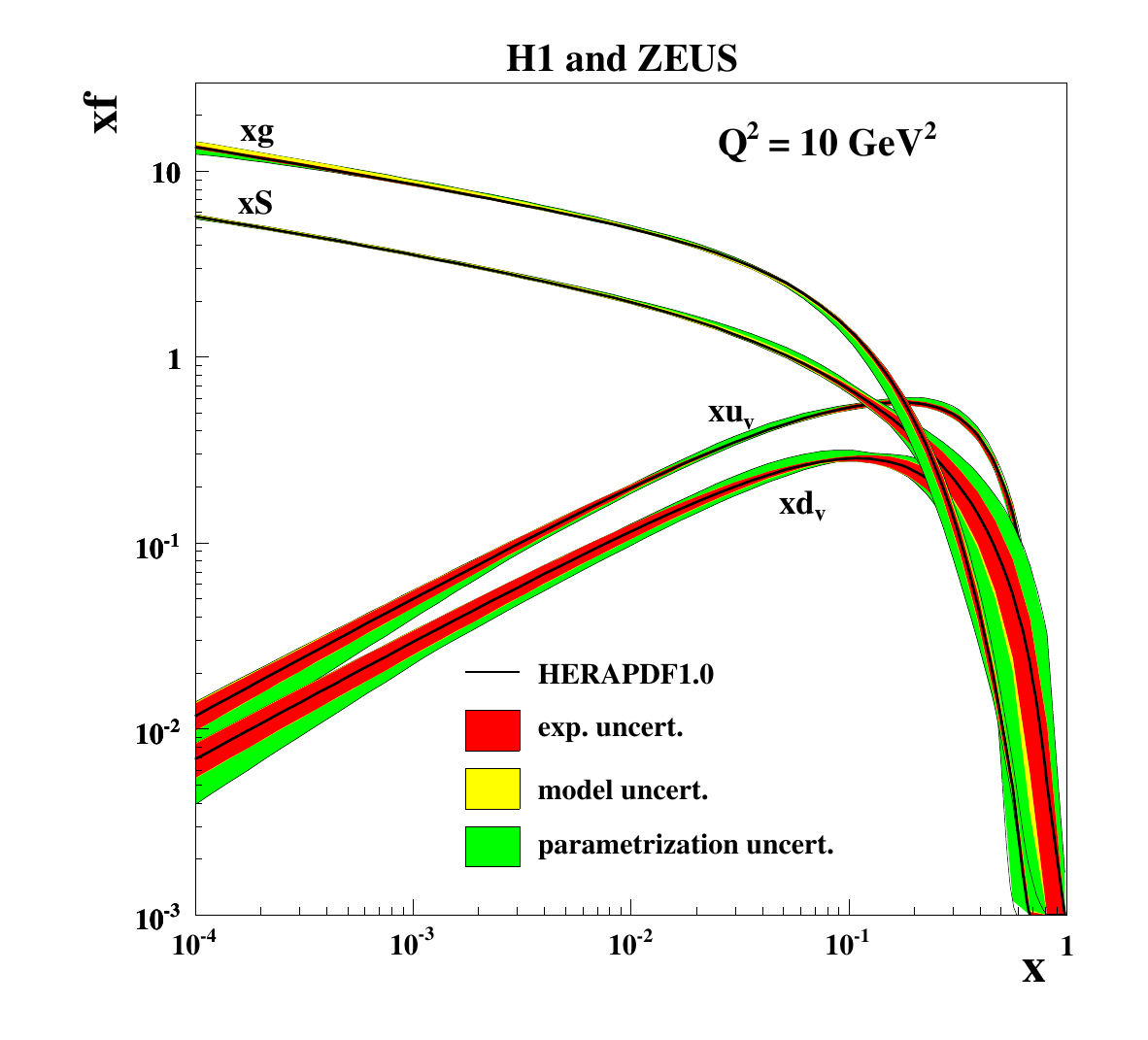}
\end{center}
\caption{The parton distribution functions from 
HERAPDF1.0 \cite{H1:2009pze}, $xu_\text{v},xd_\text{v},xS=2x(\bar{u}+\bar{d}),xg$, at
$Q^2 = 1.9$ GeV$^2$ \textbf{(left)} and $Q^2 = 10$ GeV$^2$ \textbf{(right)}.
The experimental, model and parametrisation uncertainties are shown separately.}
\label{fig:hera_1.9GeV2_and_10GeV2}
\end{figure}

As mentioned, DIS allow us to probe the deep structure of the wave-function of the proton, which depends on both the Bjorken $x$ of the struck parton and on the virtuality $Q^2$ of the incoming photon. An example of such a dependence is shown in Fig. \ref{fig:hera_1.9GeV2_and_10GeV2}, which is extracted from
the data measured at HERA for DIS on a proton \cite{H1:2009pze} for two different virtualities. There we show the
$x$-dependence of the {\sl Parton} (either quark or gluon) {\sl Distribution Functions} (PDFs) which are, at leading order, the probabilities to encounter
a parton carrying a fraction $x$ of the nucleon’s momentum. In Fig. \ref{fig:hera_1.9GeV2_and_10GeV2} one can see the PDFs of
the valence quarks in the proton, $xu_\text{v}$ and $xd_\text{v}$: these have a peak for values close to $2/3$ and $1/3$ respectively, as expected since a proton is made up of two valence up quarks and one valence down quark. Then, their PDFs decrease with decreasing $x$. Moreover, we see that this behavior is similar for both values of virtuality $Q^2 = 1.9$ GeV$^2$ and $Q^2 = 10$ GeV$^2$. On the other hand, what can we infer about the PDFs of the ``sea'' quarks and gluons, denoted by $xg$ and $xS$? In Fig. \ref{fig:hera_1.9GeV2_and_10GeV2} we see that they appear to grow very strongly towards the low $x$ (note the logarithmic scale of
the vertical axis), dominating the gluon distribution of the whole proton below $x=0.1$. Remembering that low--$x$ means high energy, we conclude that
the part of the proton wave-function responsible for the interactions
in high energy scattering consists mainly of gluons. Moreover, one can also observe that, for increasing $Q^2$, the PDFs of ``sea'' quarks and gluons at small $x$ increase quite significantly.

\begin{figure}[ht]
\begin{center}
\includegraphics[width=0.7\textwidth]{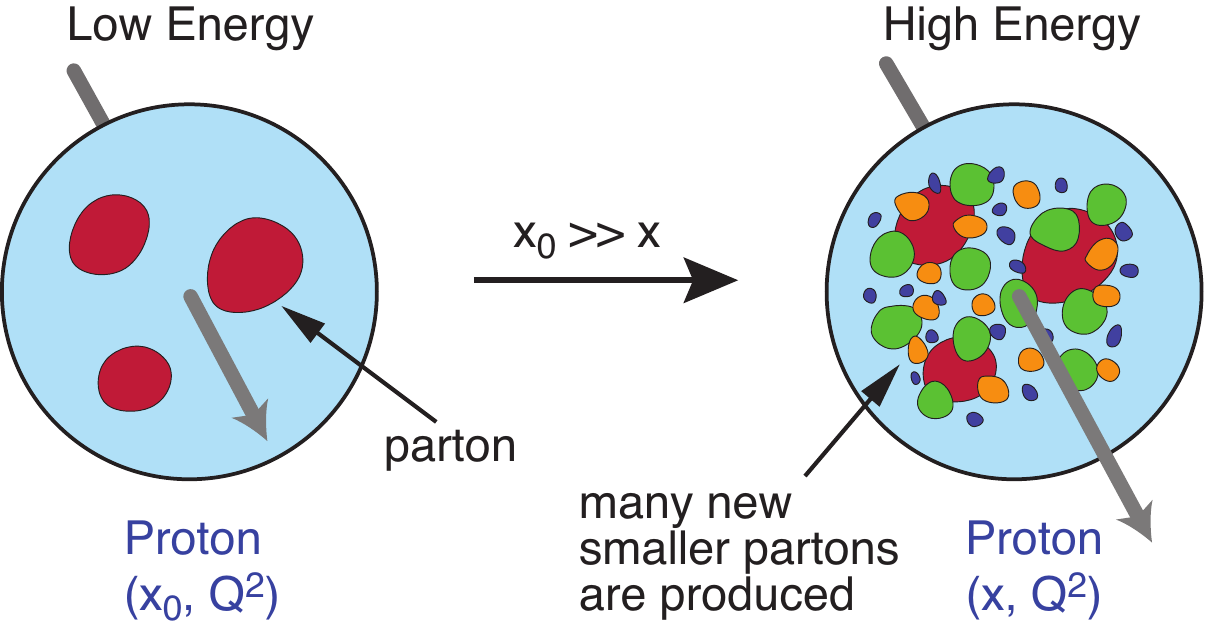}
\end{center}
\caption{The proton wave--function at small--$x$ (shown on the right)
  contains a large number of gluons (and virtual quarks) as compared to the
  same wave-function at a larger $x = x_0$ (shown on the left). The
  figure is a projection on the plane transverse to the beam axis (the
  latter is shown by arrows coming out of the page, with the
  length of the arrows reflecting the momentum of the proton). Figure adapted from \cite{Accardi:2012qut}.}
\label{fig:proton_wavefunction_smallx}
\end{figure}

We have seen that the small-$x$ proton wave-function is dominated by gluons, which populate the transverse area of the proton, creating a high density of gluons. This is pictorially shown in Fig. \ref{fig:proton_wavefunction_smallx}, which illustrates
how at lower $x$ (right panel), the partons (mainly gluons)
are much more numerous inside the proton than at larger-$x$ (left
panel), in agreement with Fig. \ref{fig:hera_1.9GeV2_and_10GeV2}. This dense small-$x$
wave-function of an ultrarelativistic proton or nucleus is referred to
as the {\sl Color Glass Condensate} (CGC), see next Sections.

\subsection{QCD evolution equations}

\begin{figure}[ht]
\begin{center}
\includegraphics[width=0.8\textwidth]{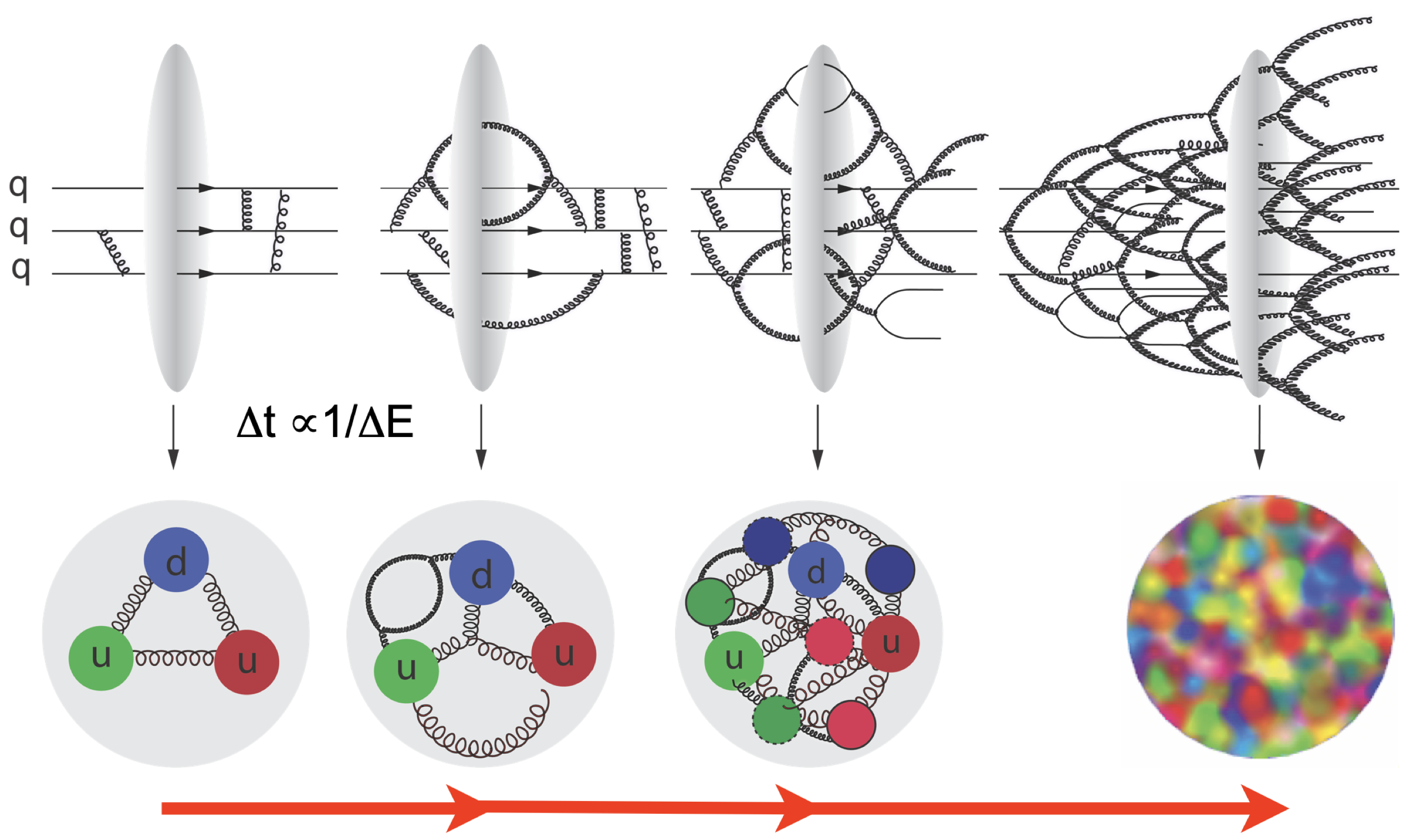}
\end{center}
\caption{Sketch of the BFKL evolution towards smaller $x$: at small $x$ a nucleon is mainly made up of soft gluons. Picture from \cite{picture_BFKL_Salazar}.}
\label{fig:BFKL_evolution}
\end{figure}

To understand the onset of the dense regime that DIS data highlight, one usually employs the so-called {\sl QCD evolution equations}. The main principle is as follows. While
the current state of the QCD theory does not allow for a
first-principles calculation of the quark and gluon distributions, the
evolution equations, loosely-speaking, allow one to determine these
distributions at some values of $(x, Q^2)$ assuming that they are known
at some other $(x_0, Q_0^2)$.\footnote{We will mostly refer to gluons from now on, because they are the partons we will be most interested with. Similar reasoning holds of course for virtual quarks.} The most widely used evolution equation
is the {\sl Dokshitzer-Gribov-Lipatov-Altarelli-Parisi} (DGLAP) equation
\cite{Gribov:1972ri,Altarelli:1977zs,Dokshitzer:1977sg}. If the PDFs
are specified at some initial virtuality $Q_0^2$, the DGLAP equation
allows one to find the parton distributions at $Q^2 > Q_0^2$, at all
$x$ where DGLAP evolution is applicable. In a similar fashion, there is also an evolution equation that allows one to construct the parton distributions at low-$x$, given their
value at some $x_0 > x$, for all $Q^2$. This is known as the {\sl Balitsky-Fadin-Kuraev-Lipatov} (BFKL) evolution equation \cite{Balitsky:1978ic,Kuraev:1977fs}, which can be written as
\begin{equation}\label{BFKL}
  \frac{\partial N (x, r_\perp)}{\partial \ln (1/x)} = \alpha_s 
  K_{\mathrm{BFKL}} \otimes  N (x, r_\perp),
\end{equation}
where $\otimes$ denotes a convolution with respect to transverse momenta. In \eqref{BFKL}, $K_{\mathrm{BFKL}}$ is an integral kernel known as the BFKL kernel, for which at the moment we have only perturbative expressions at first orders. Moreover, the dipole scattering amplitude $N(x, r_\perp)$ (whose Fourier transform is related to the parton transverse momentum distribution)\footnote{In
general, the dipole amplitude also depends on the impact parameter
$b_\perp$ of the dipole, but for simplicity
we suppress this dependence in $N (x, r_\perp)$.} probes the gluon distribution in the proton at the transverse distance $r_\perp \sim 1/Q$. Physically, Eq. \eqref{BFKL} can be read as follows: as we make one step of the evolution, by boosting the nucleus/proton to higher energy in order to probe its smaller-$x$ wave function, any of the partons can split into two partons, leading to an increase in the number of partons which is proportional to the number of partons $N$ at the previous step (see Fig. \ref{fig:BFKL_evolution}). The BFKL evolution
leads to a power-law growth of the parton distributions with
decreasing $x$, such that $N \sim (1/x)^\lambda$ with $\lambda$ a
positive number \cite{Balitsky:1978ic}. This behavior may account for
the increase of the gluon density at small-$x$ in the HERA data of
Fig. \ref{fig:hera_1.9GeV2_and_10GeV2}.\\

\begin{figure}[ht]
\begin{center}
\includegraphics[width=0.7\textwidth]{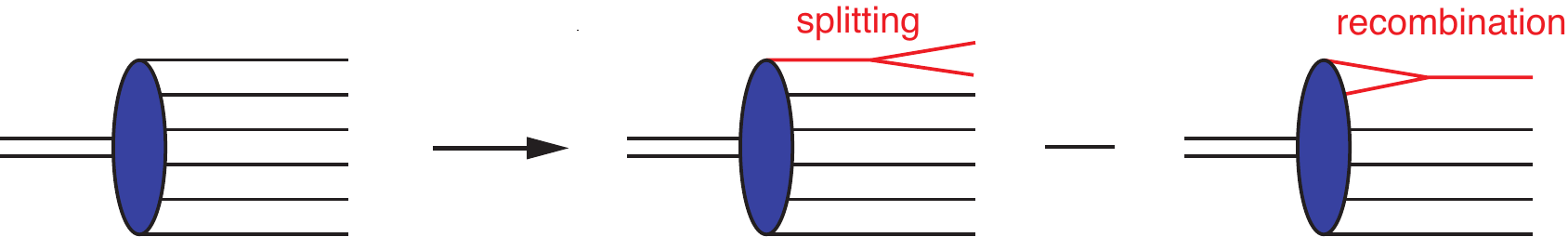}
\end{center}
\caption{The non-linear small-$x$ evolution of a hadronic or nuclear wave 
  functions. All partons (quarks and gluons) are denoted by straight
  solid lines for simplicity. Picture from \cite{Accardi:2012qut}.}
\label{fig:BKfig}
\end{figure}

The question arises whether the gluon and quark densities can grow
without limit at small--$x$. While there is no strict bound on the
number density of gluons in QCD, there is a bound on the scattering
cross-sections stemming from unitarity. This in turn translates to a limit on the density, since a proton (or nucleus)
with a lot of ``sea'' gluons is more likely to interact in high
energy scattering, i.e. it possesses a larger scattering cross-section.
The cross-section bound arises due to the black disk
limit known from quantum mechanics. In particular, the high-energy total scattering cross section of a particle on a sphere of radius $R$ is bounded by
\begin{equation}
  \sigma_{\mathrm{tot}} \leq  2 \pi R^2.
  \label{eq:black_disk}
\end{equation}
In QCD, the black disk limit translates into the {\sl Froissart--Martin
unitarity bound}, which states that the total hadronic cross-section
cannot grow faster than $\log^2 s$ at very high energies, being $s$ the
center-of-mass energy squared \cite{Froissart:1961ux}. This exposes the limits of the BFKL evolution, since the BFKL growth of the gluon density in the proton or nucleus wave-function grows as a power of energy,
$\sigma_{\mathrm{tot}} \sim s^{\lambda}$, and this clearly violates the Froissart--Martin bound at very high energy.

Given all of that, we see that something has to modify the BFKL evolution at high energy
to prevent it from becoming unphysically large. Referring to Fig. \ref{fig:BKfig}, while the first term on the right hand side represents the pure BFKL evolution, the modification we need to take into account in order to preserve unitarity is illustrated in the picture on the far right. In particular, at very high energies
(leading to high gluon densities), partons may start to recombine with
each other on top of the splitting. The recombination of two partons
into one is proportional to the number of pairs of partons, which in
turn scales as the dipole amplitude squared $N^2$. What we end up with is the following non-linear evolution equation:

\begin{align}
  \frac{\partial \, N (x, r_\perp)}{\partial \ln (1/x)} = \alpha_s 
  K_{\mathrm{BFKL}} \otimes  N (x, r_\perp) - \alpha_s [N (x, r_\perp)]^2. 
  \label{eq:BK}
\end{align}
This is the {\sl Balitsky-Kovchegov} (BK) evolution equation
\cite{Balitsky:1995ub,Kovchegov:1999yj,Kovchegov:1999ua},
which is valid for QCD in the limit of the large number of colors
$N_c$.\footnote{An equation of this type was originally suggested by
Gribov, Levin and Ryskin in \cite{Gribov:1983ivg} and by Mueller and
Qiu in \cite{Mueller:1985wy}. However, at the time it was believed that
the quadratic term was only the first non-linear correction, with
higher order terms expected to be present as well.  In
\cite{Balitsky:1995ub,Kovchegov:1999yj}, the exact form of the
equation was found, and it was shown that in the large--$N_c$ limit
Eq. \eqref{eq:BK} does not have any higher-order terms in $N$.} The physical impact of the quadratic term on the right hand side of \eqref{eq:BK} is
clear: this term slows down the small-$x$ evolution, and the effect of gluon merging processes becomes important when the quadratic term in the right hand side of Eq. \eqref{eq:BK} becomes comparable to the linear term. This eventually leads to {\sl parton saturation}, which occurs when the number density of partons stops growing with decreasing $x$. Most importantly, this gives rise to an energy scale (or equivalently, to a length scale) which is called the {\sl saturation scale} $Q_s$. This can be shown to grow as $Q_s^2 \sim (1/x)^\lambda$
with decreasing $x$ \cite{Gribov:1983ivg,Iancu:2002tr,Mueller:2002zm}: we will extensively come back to this point in Section \ref{sec:The saturation scale in HIC}.

As a final point, we have to mention that a generalization of Eq. \eqref{eq:BK} beyond the large-$N_c$ limit does exist, and it goes under the name of the
{\sl Jalilian-Marian--Iancu--McLerran--Weigert--Leonidov--Kovner} (JIMWLK) evolution equation \cite{Iancu:2001ad,Jalilian-Marian:1997jhx,Jalilian-Marian:1998tzv,Jalilian-Marian:1997ubg,
Iancu:2000hn}. This is a renormalization group equation, e.g. solvable by a Langevin--type step evolution \cite{Lappi:2012vw}.\\

\begin{figure}[t]
\begin{center}
\includegraphics[width=0.5\textwidth]{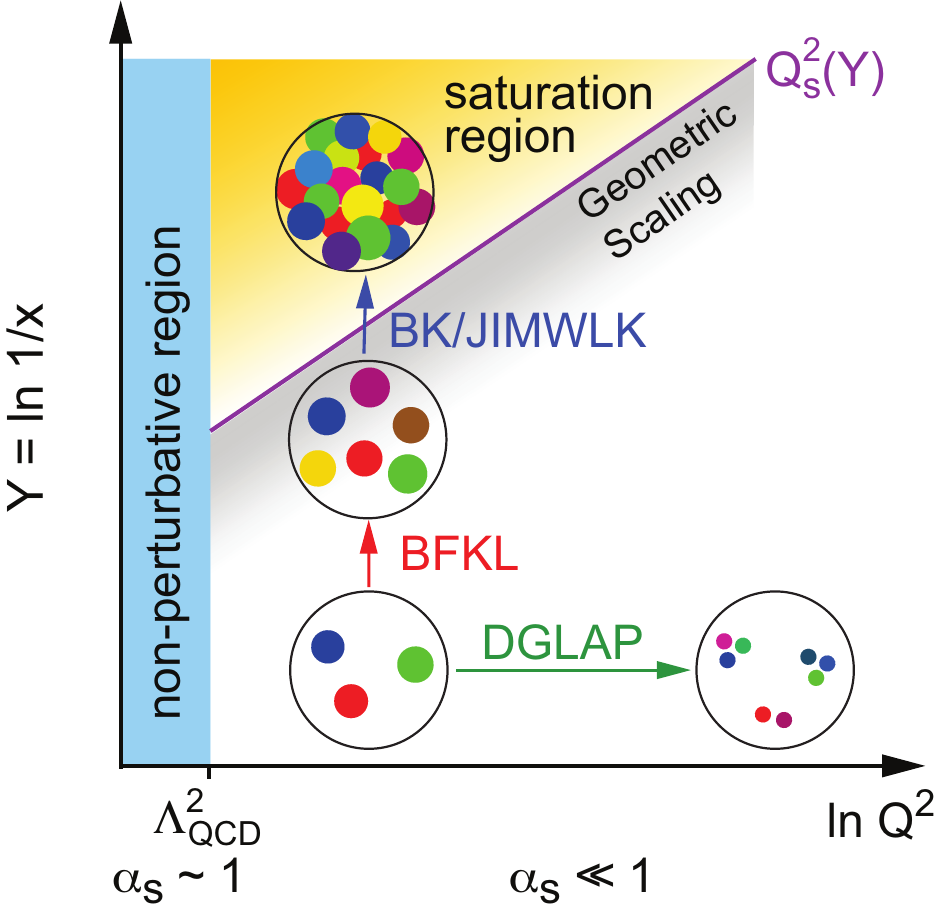}
\end{center}
\caption{The map of high energy QCD in the $(\ln Q^2,~\ln 1/x)$ plane. Figure from \cite{Accardi:2012qut}.}
\label{satbk}
\end{figure}

We summarize our theoretical knowledge of high energy QCD, as discussed
above, in Fig. \ref{satbk}, in which different regimes are plotted in the
$(\ln Q^2,~\ln 1/x)$ plane.  Let us start from the bottom left corner of Fig. \ref{satbk}, i.e. the region with $Q^2 \lesssim  \Lambda_{\text{QCD}}^2$ and large $x$: here the coupling constant is large, $\alpha_s \sim 1$, and small-coupling approaches do not work. Going to the right, towards the perturbative
region $Q^2 \gg \Lambda_{\text{QCD}}^2$ where the coupling is small, we see the standard DGLAP evolution. If we instead move upwards we see the linear BFKL evolution which evolves the gluon distribution
towards small-$x$, where the parton density becomes large and parton
saturation sets in. This transition to saturation happens at a value $Q_s^2 \gg \Lambda_{\text{QCD}}^2$, and it is described by the
non-linear BK and JIMWLK evolution equations.

\subsection{The saturation scale in HIC}
\label{sec:The saturation scale in HIC}
We have previously mentioned the phenomenon of parton saturation. This is a universal phenomenon, which is valid both for scattering on a proton or a nucleus. Here we demonstrate that nuclei provide an extra enhancement of the saturation
phenomenon, making it easier to observe and study experimentally.

\begin{figure*}[t]
\begin{center}
\includegraphics[width=0.65\textwidth]{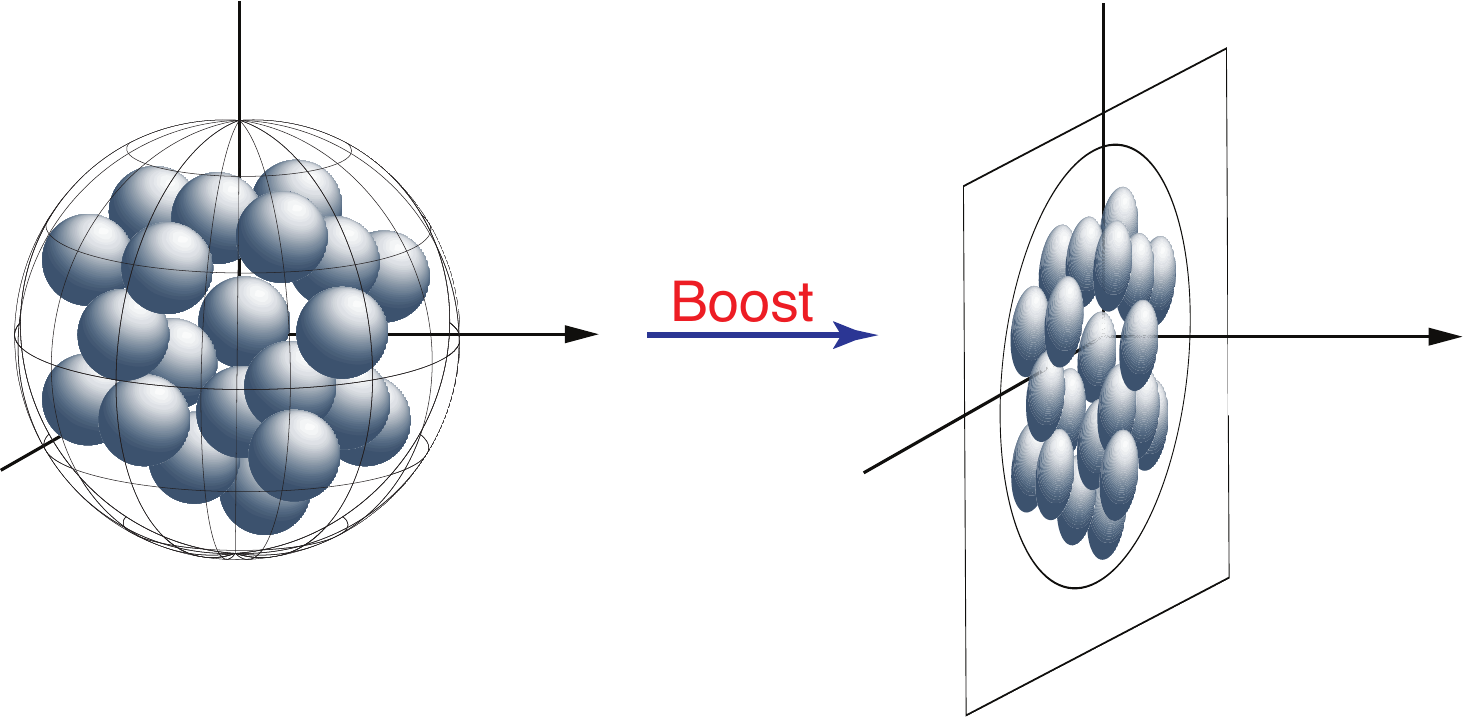}
\end{center}
\caption{A large nucleus before and after an ultrarelativistic boost. Figure from \cite{Accardi:2012qut}.}
\label{nucl_boost}
\end{figure*}

Imagine a large nucleus (a heavy ion), which was boosted to some
ultrarelativistic velocity, as shown in Fig. \ref{nucl_boost}. We are
interested in the dynamics of small-$x$ gluons in the wave-function of
this relativistic nucleus. One can show that, due to the Heisenberg
uncertainty principle, the small-$x$ gluons interact with the whole
nucleus coherently in the longitudinal (i.e. the beam) direction. Therefore,
only the transverse plane distribution of nucleons is important for
the small-$x$ wave-function. As one can see from Fig. \ref{nucl_boost},
after the boost the nucleons as ``seen'' by the small-$x$ gluons
with large longitudinal wavelength appear to overlap with each other
in the transverse plane, leading to high parton density. 

\begin{figure*}[t]
\begin{center}
\includegraphics[width=0.48\textwidth]{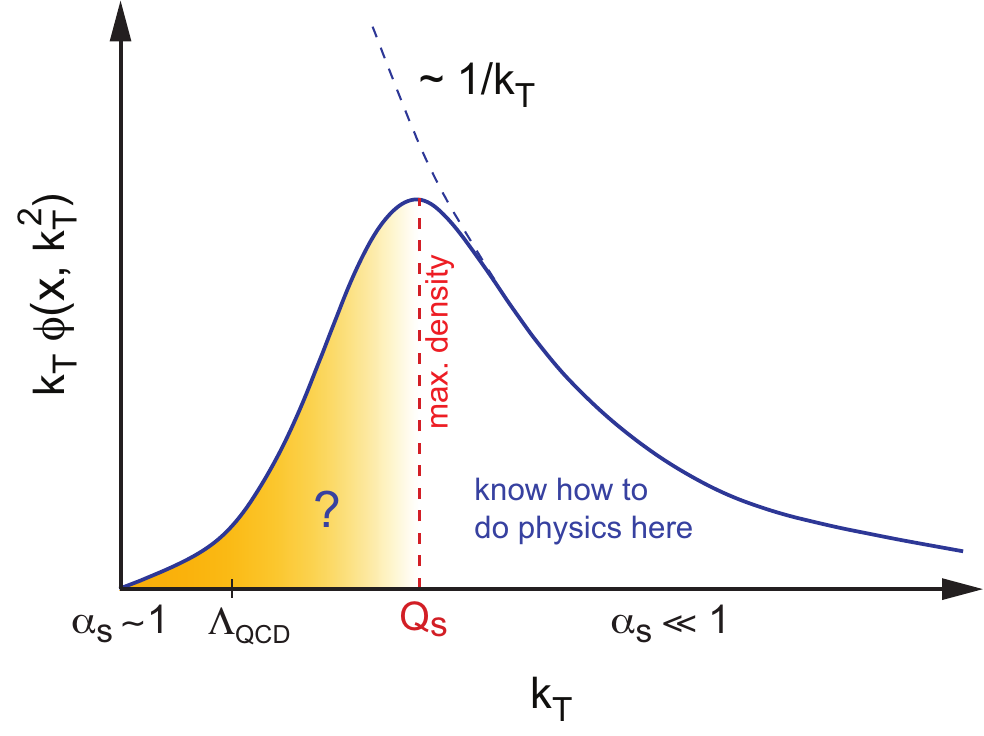}
\end{center}
\caption{The unintegrated gluon Transverse Momentum Distribution $\phi (x, k_\perp^2)$ (TMD, multiplied by the phase space factor $k_\perp$) of a large nucleus due to classical gluon fields (solid line). The dashed curve denotes the lowest-order perturbative result. Figure from \cite{Accardi:2012qut}.}
\label{mv2}
\end{figure*}

In the following Sections we will show that in this regime one can adopt the classical Yang-Mills equations \cite{Yang:1954ek}. These have been solved for a single nucleus exactly
\cite{Kovchegov:1996ty,Jalilian-Marian:1996mkd}, and their solution was
used to construct an unintegrated gluon Transverse Momentum Distribution (TMD) $\phi(x, k_\perp^2)$, shown in Fig. \ref{mv2} as a function of
$k_\perp$. This picture shows the emergence of the saturation scale $Q_s$. In particular, note that the majority of gluons in this classical distribution have transverse
momenta $k_\perp \approx Q_s$. Most importantly, the gluon distribution slows
down its growth with decreasing $k_\perp$ for $k_\perp < Q_s$ (from a
power-law of $k_\perp$ to a logarithm, as can be shown by explicit
calculations). The distribution {\sl saturates}, hence justifying the name for $Q_s$. \\

The gluon field arises from all the nucleons in the nucleus at a given
location in the transverse plane.  Away from the
edges, the nucleon density in the nucleus is approximately constant.
Therefore, the number of nucleons at a fixed transverse location is
simply proportional to the thickness of the nucleus in the
longitudinal (beam) direction. For a large nucleus that thickness, in turn, is
proportional to the nuclear radius $R\sim A^{1/3}$, being $A$ the nuclear mass number. The transverse momentum of the gluon can be thought
of as arising from many transverse momentum ``kicks'' acquired from
interactions with the partons in all the nucleons. Neglecting the correlations between nucleons, which is
justified for a large nucleus in the leading power of $A$
approximation, once can think of the ``kicks'' as being random.  Just
like in the random walk problem, after $A^{1/3}$ random kicks the
typical transverse momentum --- and hence the saturation scale ---
becomes $Q_s \sim \sqrt{A^{1/3}}$, i.e. $Q_s^2 \sim A^{1/3}$. We
see that the saturation scale for heavy ions, $Q_s^A$ is much larger
than the saturation scale of the proton $Q_s^p$ at the same $x$,
since $(Q_s^A)^2 \approx A^{1/3}  (Q_s^p)^2$
\cite{Gribov:1984tu,Mueller:1985wy,McLerran:1993ni,Mueller:1989st}. For instance, for a gold nucleus $A=197$ and the enhancement factor is $A^{1/3} \approx 6$.

As we argued above, the saturation scale grows not only with increasing mass number of a nucleus $A$, but also with decreasing $x$ (and, conversely, with
the increasing center-of-mass energy $\sqrt{s}$). This scaling can be combined as
\begin{align}\label{eq:Qs}
  Q_s^2 (x)  \sim  A^{1/3}  \left( \frac{1}{x} \right)^\lambda,
\end{align}
where the best current theoretical estimates of $\lambda$ give
$\lambda$ = 0.2 -- 0.3 \cite{Albacete:2007sm}, in agreement with the
experimental data collected at RHIC
\cite{Albacete:2007sm} and at
HERA \cite{Albacete:2009fh,Albacete:2010sy,Golec-Biernat:1998zce,Golec-Biernat:1999qor}. Therefore, for hadronic collisions at
high energy and/or for collisions of large ultrarelativistic nuclei,
the saturation scale becomes large, $Q_s^2 \gg \Lambda^2_{\text{QCD}}$. For
the total (and particle production) cross-sections, $Q_s$ is usually
the largest momentum scale in the problem.  We therefore expect it to
be the scale determining the value of the running QCD coupling constant, making it small: $\alpha_s (Q_s^2)  \ll  1$. This allows for first-principles calculations of total hadronic and
nuclear cross-sections, along with extending our ability to calculate
particle production and to describe diffraction in a small-coupling
framework. For detailed descriptions of the physics of parton
saturation and the CGC, we refer the reader to the review articles
\cite{JalilianMarian:2005jf,Weigert:2005us,Iancu:2003xm,Gelis:2010nm}
and to the book \cite{Kovchegov:2012mbw}.

\begin{figure}[t]
\begin{center}
\includegraphics[width=0.85\textwidth]{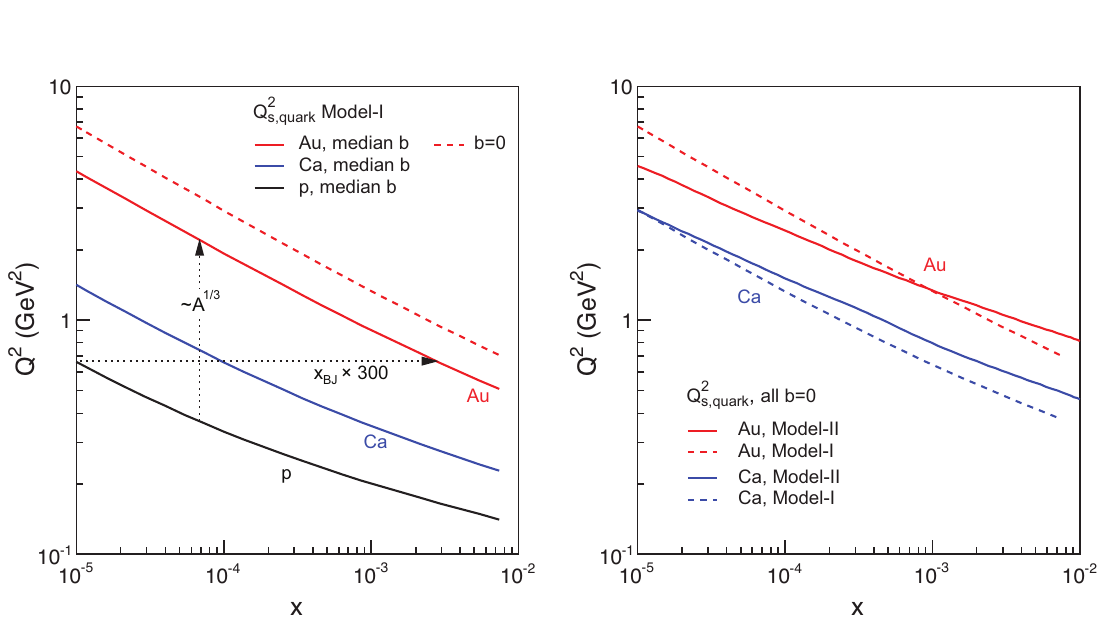}
\end{center}
\caption{Theoretical expectations for the saturation scale, as a
  function of Bjorken $x$, for the proton, along with Ca and Au
  nuclei. Figure from \cite{Accardi:2012qut}.}
\label{fig:QsModels}
\end{figure}

Let us note that \eqref{eq:Qs} can be written in a compact form. By putting $\lambda = 1/3$, which is close to the range of $\lambda$ quoted
above, one has
\begin{align}\label{oomph}
  Q_s^2 (x)  \sim  \left( \frac{A}{x} \right)^{1/3}.
\end{align}
From the pocket formula \eqref{oomph} we see that the saturation
scale of the gold nucleus ($A=197$) is as large as that of a proton
at a 197-times smaller value of $x$! This point is further illustrated in Fig. \ref{fig:QsModels}, which shows
our expectations for the saturation scale as a function of $x$ coming
from the saturation-inspired ‘‘Model-I'' \cite{Kowalski:2003hm} and from
the prediction of the BK evolution equation (with higher order
perturbative corrections included in its kernel), dubbed ‘‘Model-II''
\cite{Albacete:2009fh,Albacete:2010sy}. One can clearly see from the
left panel that the saturation scale for Au is larger than the
saturation scale for Ca, which, in turn, is much larger than the
saturation scale for the proton. The scaling as $A^{1/3}$, we see, is quite significant for heavy nuclei.

The saturation scale depends also on
the thickness of the nucleus at a given impact parameter
$b$. Consequently, it becomes larger for small $b \approx 0$ (for scattering through the
center of the nucleus) and smaller for large $b
\approx R$ (for scattering on the nuclear periphery). This can be seen in the left panel of
Fig. \ref{fig:QsModels}, where most values of $Q_s$ are plotted for ‘‘median
$b$'' by solid lines, while, for comparison, the $Q_s$ of gold is also plotted for $b=0$ (dashed line). One sees that the saturation
scale at $b=0$ is larger than at median $b$. On the other hand, the curves in the right
panel of Fig. \ref{fig:QsModels} are all plotted for $b=0$: this is why they
give higher values of $Q_s$ than the median-$b$ curves shown in the
left panel for the same nuclei.

\begin{figure*}[htb]
\begin{center}
\includegraphics[width=0.55\textwidth]{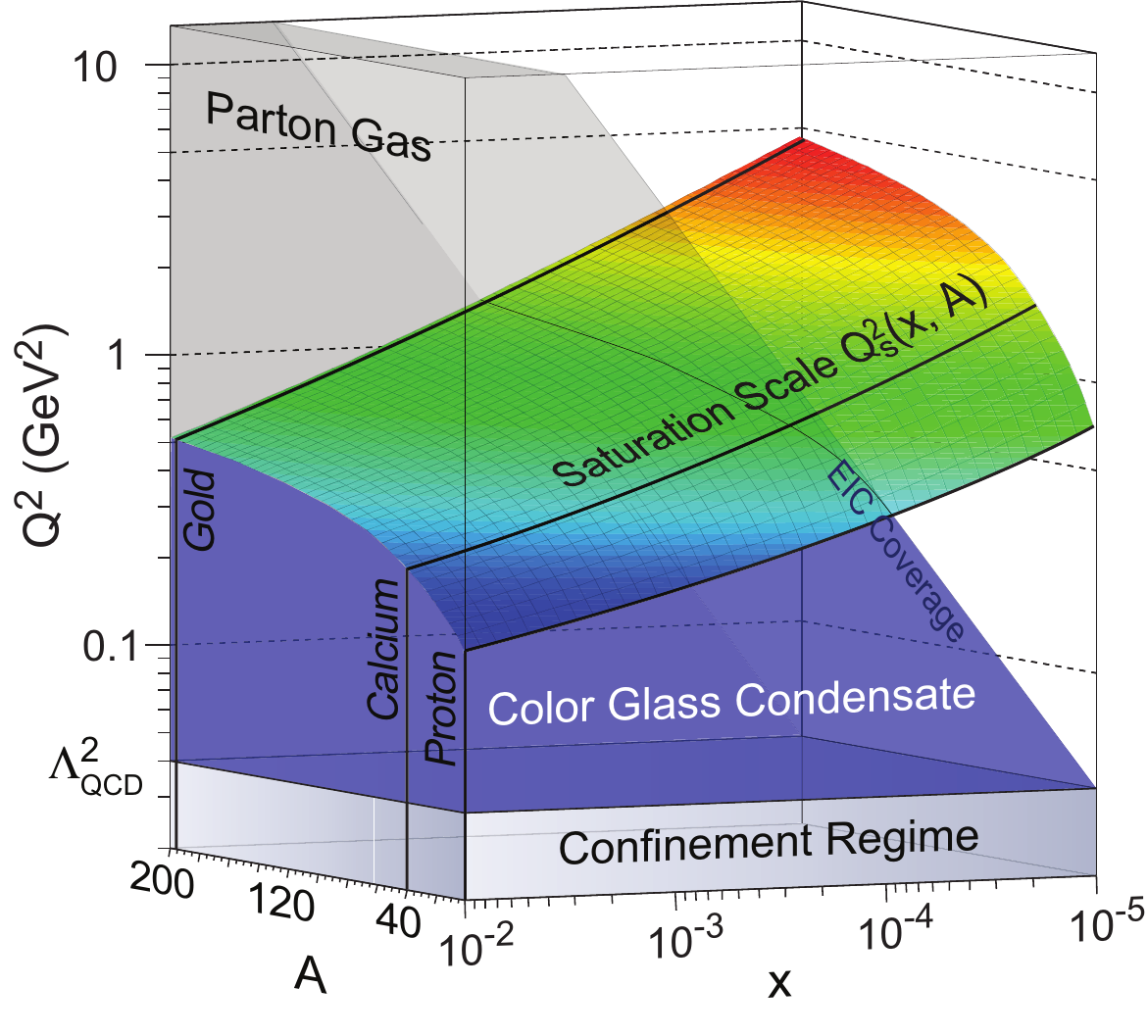}
\end{center}

\caption{The theoretical expectations for the saturation scale at medium impact parameter from ‘‘Model-I'' (see Fig. \ref{fig:QsModels}) as a function of Bjorken-$x$ and of the nuclear mass
  number $A$. Figure from \cite{Accardi:2012qut}.}
\label{fig:Qs3D}
\end{figure*}

Just to mention, this $A$-dependence of the saturation scale, including a realistic
impact parameter dependence, is the {\sl raison d'${\hat e}$tre} for
an electron-ion collider, as preferred to an electron-proton collider. Collisions of leptons with nuclei probe the same universal physics as seen with protons, but at values of $x$ at least two orders of magnitude lower (or equivalently, at an order of magnitude larger $\sqrt{s}$). Thus, the nucleus is an efficient {\sl amplifier} of the universal physics of high gluon densities, allowing us to study
the saturation regime in $e$+A at significantly lower energies than it would be possible in $e$+p collisions. For example, as can be seen from
Fig.~\ref{fig:QsModels}, $Q_s^2 \approx 0.7$~GeV$^2$ is reached at $x =
10^{-5}$ in $e$+p collisions, requiring a collision with a
center-of-mass energy of almost $\sqrt{s} \approx \sqrt{Q_s^2/x}
\approx 260$~GeV. On the other hand, in $e$+Au collisions, only $\sqrt{s} \approx
15$~GeV are required to achieve comparable gluon densities and the same
saturation scale. To illustrate the conclusion that $Q_s$ is an increasing function of both $A$ and $1/x$, we show a plot of its dependence on both variables
in Fig. \ref{fig:Qs3D}, using ‘‘Model-I'' of Fig. \ref{fig:QsModels} left. One can see
again, from Fig. \ref{fig:Qs3D}, that larger $Q_s$ can be obtained by
increasing the energy or by increasing mass number $A$.\\

Measurements extracting the $x$, $A$, and impact parameter $b$ dependence of the
saturation scale provide very useful information on the momentum
distribution and space-time structure of strong color fields in QCD at
high energies. The saturation scale defines the transverse momentum of
the majority of gluons in the small-$x$ wave-function, as shown in
Fig. \ref{mv2}, thus being instrumental to our understanding of the
momentum distribution of gluons. The impact parameter dependence of
the saturation scale tells us how the gluons are distributed in the
transverse coordinate plane, clarifying the spatial distribution of
the small-$x$ gluons in the proton or nucleus. Saturation/CGC physics therefore provides a new way of tackling the problem of
calculating hadronic and nuclear scattering cross-sections: it is
based on the theoretical observation that small-$x$ hadronic and
nuclear wave-functions -- and, therefore, the scattering
cross-sections -- are described by the internal momentum scale, the saturation scale $Q_s$ \cite{Gribov:1984tu}.

We close this Section by noting that a large
occupation number of color charges (partons) leads to a {\sl classical
gluon field} dominating the small-$x$ wave-function of the
nucleus. This happens since splitting and recombination compensate one another. The above is the essence of the so-called {\sl McLerran-Venugopalan} (MV) {\sl model}
\cite{McLerran:1993ni}, which we will extensively discuss in the following Sections. In particular, according to the MV model, the dominant gluon field is given by the solution of the classical Yang-Mills equations \cite{Yang:1954ek},
which are the QCD analogue of Maxwell equations of electrodynamics. These equations can be studied both analytically (e.g. by employing a small time expansion \cite{Carrington:2020ssh,Carrington:2021qvi}) or numerically \cite{Avramescu:2023qvv, Avramescu:2024poa}.

\section[Before the collision: the CGC framework]{Before the collision: the Color Glass Condensate framework}

As mentioned previously, nuclei at very high energies can be described within the Color Glass Condensate framework, which is a high energy effective theory for QCD and it enables an effectively classical description of nuclei in terms of classical color fields and color charges. In this Section we will apply the theoretical concepts which we have previously highlighted to the actual construction of the CGC model. In particular, here we will study the fields produced by each nucleus before an heavy-ion collision event at very high energies. On the other hand, in Section \ref{sec:glasma_initial} we will study what happens in the collision region once the collision has occurred.

\subsection{Classical gauge fields in a nucleus within CGC}

Even before the full development of the CGC framework as a genuine effective theory, McLerran and Venugopalan already proposed a classical model for high energy nuclei \cite{McLerran:1993ka, McLerran:1993ni}. They argued that the valence quarks, which carry most of the total momentum of the nucleus (
referred to as ``hard'' degrees of freedom), are to be treated as static, recoilless color charges whose dynamics is ``frozen'' due to time dilation. Furthermore, they proposed that the color charge density $\rho$ of the valence quarks provides an energy scale much larger than the QCD scale $\Lambda_{\text{QCD}}$. We now interpret this as the existence of the saturation scale $Q_s$ within the nuclei. This assumption implies that the Yang-Mills coupling constant $g$ can be considered to be weak, which justifies a classical (or eikonal) approximation. A priori the exact positions and charges of the valence quarks are unknown and therefore the classical color charges are considered to be fluctuating random variables. For this reason, the valence quarks are represented by a random classical color charge density $\rho^a(x)$ (or more generally a classical color current $J^a_\mu(x)$), whose probability distribution is specified by a probability functional $W[\rho]$.

On the other hand, the gluons, which carry only a fraction of the total momentum (the ``soft'' degrees of freedom), are considered to be dynamic. McLerran and Venugopalan realized that because of the large number of gluons in high energy nuclei, quantum mechanical effects should be, as a first approximation, neglected. More precisely, the gluon field is highly occupied and forms a coherent state, which is essentially a classical field state. The soft gluons are therefore represented by a classical color field $A^a_\mu(x)$. Since the classical Yang-Mills equations must hold, such color field $A^a_\mu$ is fixed by the color charge density $\rho^a$ of the valence quarks.\\

\begin{figure}[t]
	\centering
	\includegraphics{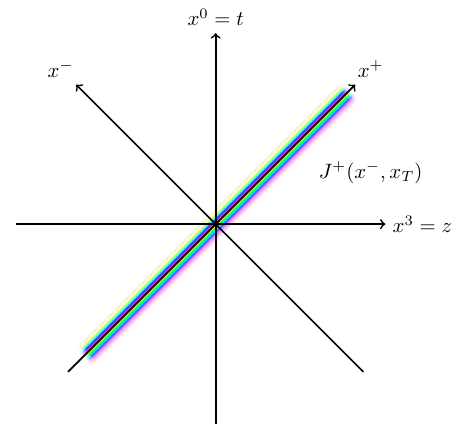}
	\caption{The schematic Minkowski diagram for a single nucleus moving at the speed of light. The transverse coordinate $\bm{x}_\perp$ is suppressed. The color structure of the nucleus depends on $x^-$, but is independent of $x^+$ since the current $J^+$ is assumed to be static. Figure from \cite{Muller:2019bwd}.}
		\label{fig:lightcone1}
\end{figure}

Using these assumptions we can start solving the classical problem. The equations of motion are most easily solved by employing {\sl Light Cone} (LC) {\sl coordinates}:
\begin{equation}
    x^{\pm} = ( x^0 \pm x^3) / \sqrt{2},
    \label{eq:light-cone coordinates}
\end{equation}
where $t = x^0$ and $z = x^3$ are the laboratory frame coordinates. When using light cone coordinates, Latin indices as in $x^i$ are reserved for transverse coordinates, i.e. $i\in \left\{ 1,2 \right\}$. We then consider (without loss of generality) the color current $J^\mu_a(x)$ of a nucleus moving in the positive $z$ direction (i.e. $x^3$) at the speed of light. In this hypothesis, the only relevant component is then $J^{+}_a(x)$, which we associate with the color charge density $\rho_a(x)$. Moreover, we can drop the $x^+$ dependency of the current because we assume the valence quarks to be static in $x^+$. We are then left with
\begin{equation} \label{eq:mv_color_current1}
J^\mu(x) = \delta^{\mu+} \rho_a(x^-,\bm{x}_\perp) t_a,
\end{equation}
where $\bm{x}_\perp$ are the coordinates in the transverse plane spanned by $x^1$ and $x^2$, and $t_a$ are the generators of the gauge group in the fundamental representation (see Appendix \ref{app:conv} for the conventions used in this Thesis). The fast moving nucleus that we are describing is highly Lorentz contracted in the $z$ direction, which tells us that the support along the $x^-$ direction must be very thin. In the ultrarelativistic limit the longitudinal support becomes infinitesimal and the color current is proportional to $\delta(x^-)$:
\begin{equation} \label{eq:mv_color_current2}
J^\mu(x) = \delta^{\mu+} \delta(x^-) \rho_a(\bm{x}_\perp) t_a. 
\end{equation}
For the present discussion, however, we will at the moment keep the more general form \cref{eq:mv_color_current1} with the color current being strongly peaked around $x^- = 0$. A schematic diagram is shown in \cref{fig:lightcone1}.
The next step is to solve the classical Yang-Mills equations, given by \cite{Yang:1954ek}
\begin{equation} \label{eq:ym_equations}
D_\mu F^{\mu\nu} = J^\nu,
\end{equation}
in the presence of the external color current $J^\nu$. In order to make progress we must choose a gauge condition and an appropriate ansatz for the color field. A useful starting point is the {\sl covariant} (or Lorenz) gauge condition
\begin{equation} \label{eq:cov_gauge}
\p_\mu A^\mu = 0,
\end{equation}
with the ansatz 
\begin{equation} \label{eq:mv_gf_ansatz}
A^\mu = \delta^{\mu +} A^+_a(x^-,\bm{x}_\perp) t_a.
\end{equation}
When using this ansatz, the only remaining field component is $A^+$. Moreover, the gauge condition forces us to drop the $x^+$ dependency of the color field, since it implies
\begin{equation}
\p_+ A^+ = 0.
\end{equation}
Notice that the ansatze \eqref{eq:mv_color_current2} and \eqref{eq:mv_gf_ansatz} are compatible with non-Abelian charge conservation, i.e.
\begin{equation} \label{eq:nonabelian_continuity_eq}
D_\mu J^\mu = \p_+ J^+ + i g \cm{A_+}{J^+} = 0,
\end{equation}
which holds since $A_+ = A^- = 0$. Inserting the ansatz \eqref{eq:mv_gf_ansatz} into the field strength tensor $F^{\mu\nu} = \p^\mu A^\nu - \p^\nu A^\mu + i g \cm{A^\mu}{A^\nu},$
we find that the only non-zero components are $F^{i+}$, given simply by
\begin{equation}
F^{i+} = \p^i A^+.
\end{equation}
Plugging this into the Yang-Mills equations \eqref{eq:ym_equations} yields
\begin{equation}
\p_i \p^i A^+ = J^+,
\end{equation}
i.e. a Poisson equation in the transverse plane
\begin{equation} \label{eq:mv_poisson}
- \Delta_\perp A^+_a(x^-,\bm{x}_\perp) = \rho_a(x^-,\bm{x}_\perp).
\end{equation}
This equation can be readily solved by inverting the two-dimensional Laplace operator ${\Delta_\perp = \sum_i \p^2_i}$, for instance using the Fourier transform. Defining the partial Fourier transform, with respect to the transverse coordinates only, as
\begin{align}
\tilde{\rho}_a(x^-,\bm{k}_\perp) &= \int d^2 \bm{x}_\perp \, \rho_a(x^-,\bm{x}_\perp)\, e^{- i \bm{k}_\perp \cdot \bm{x}_\perp},
\end{align}
and its inverse as
\begin{align}
\rho_a(x^-,\bm{x}_\perp) &= \int \frac{d^2 \bm{k}_\perp}{(2 \pi)^2} \tilde{\rho}_a(x^-,\bm{k}_\perp)\, e^{i \bm{k}_\perp \cdot \bm{x}_\perp},
\end{align}
we find the solution
\begin{equation} \label{eq:poisson_solution_momentum}
A^+_a(x^-, \bm{x}_\perp) = \int \frac{d^2 \bm{k}_\perp}{( 2 \pi )^2} \frac{\tilde{\rho}_a(x^-,\bm{k}_\perp)}{k_\perp^2} \, e^{i \bm{k}_\perp \cdot \bm{x}_\perp}. 
\end{equation}
Note that the color field $A^+$ inherits its longitudinal support and shape from the color charge density $\rho$.

It is also instructive to analyze the field strength of the color field: switching back to laboratory frame coordinates $t$ and $z$ we find 
\begin{equation}
A^0(x^-,\bm{x}_\perp) = A^3(x^-,\bm{x}_\perp) = \frac{1}{\sqrt{2}} A^+(x^-,\bm{x}_\perp),
\end{equation}
because $A^-=0$.
The only non-zero field strength components are then $F^{0i}$ and $F^{3i}$ with $i \in \{1,2\}$, which means that the nucleus has only transverse color-electric and color-magnetic fields
\begin{align}
E_i \equiv F_{0i} = - \p_i A_0, \hspace{3em} B_i \equiv - \frac{1}{2} \epsilon_{ijk} F^{jk} = + \epsilon_{ik} \p_k A_0,
\label{eq:electric_magnetic_fields_singlenucleus}
\end{align}
which are orthogonal to each other and have the same magnitude. In the above equation $\epsilon_{ijk}$ refers to the three-dimensional Levi-Civita symbol, while $\epsilon_{ij}$ is the two-dimensional Levi-Civita symbol in the transverse plane. It turns out that the solution of the Yang-Mills equations for a single propagating nucleus is completely analogous to the electromagnetic field of an electric charge moving at the speed of light, given by the ultrarelativistic Li\'{e}nard-Wiechert potentials. Due to the ansatz and a clever choice of gauge given by \cref{eq:cov_gauge}, all non-linear terms of the Yang-Mills equations can be ignored and the resulting solution ends up being remarkably simple.\\

The choice of the covariant gauge becomes inconvenient when discussing collisions of nuclei, where it turns out that the {\sl light cone} (LC) gauge $A^+ = 0$ is much better suited. Our goal therefore is to find the gauge transformation $V(x)$, acting on the gauge field via
\begin{equation}
A_\mu(x) \rightarrow V(x) \left( A_\mu(x) + \frac{1}{ig} \p_\mu \right) V^\dg(x),
\end{equation}
such that $A^+$ vanishes. This requirement leads us to
\begin{equation}
\p_- V^\dg(x^-,\bm{x}_\perp) = - ig A^+(x^-,\bm{x}_\perp) V^\dg(x^-,\bm{x}_\perp),
\label{eq:equation_for_wilson_line}
\end{equation}
and the solution of \eqref{eq:equation_for_wilson_line} is given by the path-ordered exponential
\begin{equation} \label{eq:lc_gauge_wilson_line}
V^\dg(x^-,\bm{x}_\perp) = \mathcal{P} \exp \left( - i g \intop_{-\infty}^{x^-} dz^- A^+(z^-,\bm{x}_\perp) \right).
\end{equation}
The symbol $\mathcal{P}$ denotes the path ordering operator, for which we use the ``left means later'' convention, i.e.  $\mathcal{P} \left[ A_\mu(x) A_\nu(y) \right] = A_\mu(x) A_\nu(y)$ if $x > y$, i.e. if $x$ comes after $y$ along the path. We identify the gauge transformation $V^\dg(x^-,\bm{x}_\perp)$ in \eqref{eq:lc_gauge_wilson_line} with the lightlike {\sl Wilson line} starting at $-\infty$ and ending at $x^-$. Equation \eqref{eq:lc_gauge_wilson_line} also implies that the Wilson line at the asymptotic boundary $x^- \rightarrow -\infty $ is a unit matrix.

Due to $\p_+ V^\dg = 0$, also here we have $A^- = 0$ like in the covariant gauge. On the other hand, now the transverse components of the gauge field are given by
\begin{equation}
A_i(x^-,\bm{x}_\perp) = \frac{1}{ig} V(x^-, \bm{x}_\perp) \p_i V^\dg(x^-, \bm{x}_\perp).
\end{equation}
The color current also has to be transformed accordingly, and it is now given by
\begin{equation}
J^+_{\mathrm{LC}}(x^-, \bm{x}_\perp) = \rho_{\mathrm{LC}}(x^-, \bm{x}_\perp) = V(x^-, \bm{x}_\perp) \rho(x^-, \bm{x}_\perp) V^\dg(x^-, \bm{x}_\perp), 
\end{equation}
where we inserted the subscript ``LC'' in order to differentiate this gauge current from the covariant gauge current $J^+$ (with no subscript). For ultrarelativistic nuclei we can derive a simple relationship between the transverse gauge fields $A_i$ and the LC gauge current $\rho_{\mathrm{LC}}$: a $\delta$-shaped charge density as in \cref{eq:mv_color_current2} implies that the transverse gauge field has the form of a Heaviside step function, in particular we have
\begin{align}
A^i(x^-, \bm{x}_\perp) & = \theta(x^-) \alpha^i(\bm{x}_\perp), \label{eq:urel_LC_gf}\\
\alpha^i(\bm{x}_\perp) & = \frac{1}{ig} V(\bm{x}_\perp) \p^i V^\dg(\bm{x}_\perp), \label{eq:urel_LC_gf2}
\end{align}
where $V^\dg(\bm{x}_\perp)$ is the asymptotic Wilson line
\begin{equation} \label{eq:asym_Wilson_line}
V^\dg(\bm{x}_\perp) = \lim_{x^- \rightarrow +\infty} V^\dg(x^-,\bm{x}_\perp) = \mathcal{P} \exp \left( - i g \intop_{-\infty}^{+\infty} dx^- A^+(x^-,\bm{x}_\perp) \right).
\end{equation}
We see that the space-time picture of the LC gauge solution \cref{eq:urel_LC_gf} is very different compared to the covariant gauge case: there the transverse gauge fields are everywhere zero, whereas in the LC gauge the transverse gauge fields are non-zero for $x^- > 0$ and extend to $x^-\rightarrow +\infty$. However, this is merely an artifact of the LC gauge condition, as the transverse gauge fields are pure gauge. In contrast, the actual color-electric and color-magnetic field strengths and the color current are still concentrated around $x^- = 0$.

Similarly as in \cref{eq:urel_LC_gf} and \cref{eq:urel_LC_gf2}, the ultrarelativistic LC current is given by
\begin{equation}
J^+_{\mathrm{LC}}(x^-, \bm{x}_\perp)  = \delta(x^-) V(\bm{x}_\perp) \rho(\bm{x}_\perp) V^\dg(\bm{x}_\perp)= \delta(x^-) \rho_{\mathrm{LC}}(\bm{x}_\perp).
\label{eq:lightcone_current}
\end{equation}  
Inserting the above current \eqref{eq:lightcone_current} and the field \eqref{eq:urel_LC_gf} into the Yang-Mills equations \eqref{eq:ym_equations} yields
\begin{equation}
    \p_i F^{i+} =\p_i \p_+ A^i(x^-,\bm{x}_\perp)=- \p_i \p_- A^i(x^-,\bm{x}_\perp)= - \delta(x^-) \p_i \alpha^i(\bm{x}_\perp)= \delta(x^-) \rho_{\mathrm{LC}}(\bm{x}_\perp),
\end{equation}
which finally results in the relation
\begin{equation}
\p_i \alpha^i(\bm{x}_\perp) = - \rho_{\mathrm{LC}}(\bm{x}_\perp).
\end{equation} 
The following Sections will be devoted to the actual solution of this equation, which involves a modeling of the initial color charge.

\subsection{The McLerran--Venugopalan model}
\label{sec:The McLerran--Venugopalan model}

Now that the relationship between the color charge density $\rho$ and the color field $A_\mu$ has been established, one has to specify what the color charge distribution of a large nucleus looks like. In their original formulation McLerran and Venugopalan (MV) assumed that the charge density is $\delta$-shaped as in \cref{eq:mv_color_current2}
and proposed a simple Gaussian probability distribution for the color charge density $\rho^a(\bm{x}_\perp)$. In particular, the distribution is defined by the one- and two-point functions
\begin{align}
\ev{\rho^a (\bm{x}_\perp)} & = 0, \label{eq:mv_onep} \\
\ev{\rho^a (\bm{x}_\perp) \rho^b (\bm{y}_\perp)} & = g^2 \mu^2 \delta^{ab} \delta^{(2)}(\bm{x}_\perp - \bm{y}_\perp), \label{eq:mv_twop}
\end{align}
where $\langle \cdot \rangle$ denotes ensemble average. The one-point function \cref{eq:mv_onep} guarantees that the nucleus is, on average, color neutral. The two-point function \cref{eq:mv_twop} instead fixes the average color charge fluctuation around zero. Here $\mu$ is a parameter of phenomenological nature, having units of energy, called the {\sl McLerran--Venugopalan parameter}. For a large nucleus with $A$ nucleons they estimated \cite{McLerran:1993ka}, from the average density of valence quarks, that
\begin{equation} \label{eq:mv_mu_phenomenological}
\mu^2 \approx 1.1 A^{1/3} \,\text{fm}^{-2}.
\end{equation}
For a gold nucleus with $A=197$ this estimate gives $\mu \approx 0.5\,\text{GeV}$. The MV model does not assume a finite transverse extent of the nucleus: instead, it approximates very large nuclei as infinitely thin, but transversely infinite, walls of color charge. Within the transverse plane, charges at different points are completely uncorrelated, due to the $\delta^{(2)}(\bm{x}_\perp-\bm{y}_\perp)$ term in \cref{eq:mv_twop}. On average, the MV model exhibits translational and rotational invariance in the transverse plane.

The MV model thus gives only a very crude approximation of a realistic nucleus, but due to its simplicity (it has only one dimensionful parameter $\mu$) and its high symmetry many otherwise complicated calculations can be performed analytically. Using the one- and two-point functions Eqs. \eqref{eq:mv_onep} and \eqref{eq:mv_twop}, and by exploiting the fact that these random color charges obey a Gaussian distribution (which is implicit in \eqref{eq:mv_twop}), one can define the probability functional $W[\rho]$ as
\begin{equation}
W[\rho] = Z^{-1} \exp{\left( - \int d^2 \bm{x}_\perp \frac{\rho_a(\bm{x}_\perp) \rho_a(\bm{x}_\perp)}{2 g^2 \mu^2} \right)}, 
\end{equation}
where $Z^{-1}$ is a normalization constant. The probability functional is used to define expectation values of arbitrary observables $\mathcal{O}[A_\mu]$ via a functional integral over all charge density configurations
\begin{equation}
\ev{\mathcal{O}[A_\mu]} \equiv \intop \mathcal{D} \rho\; \mathcal{O}[A_\mu] W[\rho],
\end{equation}
where it is implied that $A_\mu$ is the color field associated with the charge density $\rho$.\\

\begin{figure}[t]
	\centering
	\begin{subfigure}[b]{0.4\textwidth}
		\centering
		\includegraphics{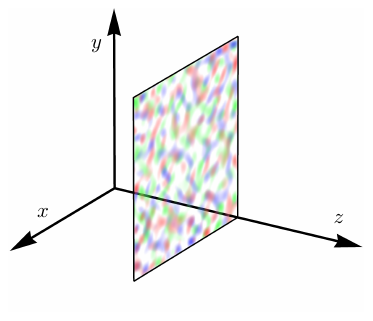}
		\caption{Single color sheet approximation}
	\end{subfigure}
	\qquad
	\begin{subfigure}[b]{0.4\textwidth}
		\centering
		\includegraphics{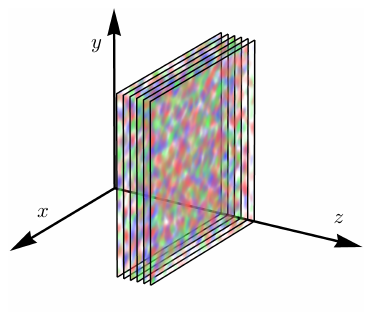}
		\caption{Generalized MV model}
	\end{subfigure}
	\caption{Illustration of the two versions of the MV model. \textbf{(a)} A nucleus described by the original MV model which can be thought of as an infinitesimal sheet of random color charge. \textbf{(b)} The generalized MV model sketched as a stack of uncorrelated sheets of color charge. In the ultrarelativistic limit the stack collapses to an infinitely thin sheet, but due to path ordering in the infinitesimal longitudinal extent, the generalized MV model does not reduce to its original formulation. Figure from \cite{Muller:2019bwd}.}
		\label{fig:color_sheets}
\end{figure}

It was later realized that for certain derivations the $\delta$-approximation of \cref{eq:mv_color_current2} can be problematic \cite{JalilianMarian:1996xn}. A more rigorous approach is to first regularize the $\delta$-peak using \cref{eq:mv_color_current1} for intermittent calculation steps and then performing the ultrarelativistic limit only in the final results. By doing so we get a generalization of the original MV model (Eqs. \eqref{eq:mv_onep} and \eqref{eq:mv_twop}) over a finite longitudinal support, which is given by \cite{JalilianMarian:1996xn}
\begin{align}
\ev{\rho^a (x^-, \bm{x}_\perp)} & = 0, \label{eq:mv2_onep} \\
\ev{\rho^a (x^-,\bm{x}_\perp) \rho^b (y^-,\bm{y}_\perp)} & = g^2 \mu^2(x^-) \delta^{ab} \delta(x^- - y^-) \delta^{(2)}(\bm{x}_\perp - \bm{y}_\perp), \label{eq:mv2_twop}
\end{align}
or equivalently
\begin{equation}
W[\rho] = Z^{-1} \exp{ - \int d x^- d^2 \bm{x}_\perp \frac{\rho_a(x^-, \bm{x}_\perp) \rho_a(x^-, \bm{x}_\perp)}{2 g^2 \mu^2(x^-)}}.
\end{equation}
Here the function $\mu^2(x^-)$ defines the average color charge fluctuation in the nucleus (as in the original MV model) and also the longitudinal shape along the $x^-$ direction. The physical picture of the generalized MV model is slightly different compared to the original MV model, see \cref{fig:color_sheets}: for a nucleus with thin but finite longitudinal support, one can think of the nucleus as a stack of infinitely thin and uncorrelated sheets of color charges (due to the additional term $\delta(x^- - y^-)$ in the charge density two-point function \cref{eq:mv2_twop}). On the other hand, the original MV model collapses this stack of color sheets into a single, infinitesimal sheet of color charge. Differentiating between the two formulations of the MV model is important, since in \cite{Fukushima:2007ki} it was found that the ultrarelativistic limit of the generalized MV model is actually not just simply obtained from the replacement $\mu^2(x^-) \rightarrow \mu^2 \delta(x^-)$, due to subtleties of the path ordering of color charges. Even in the case of infinitesimal width along $x^-$, the asymptotic Wilson line \cref{eq:asym_Wilson_line} ``remembers'' the ordering of the uncorrelated sheets of color charge.

In \cite{Fukushima:2007ki} a regularization of \cref{eq:asym_Wilson_line} is presented, which can be used for both numerical and analytical calculations. As described above, one can imagine the nucleus as a stack of $N_s$ separate, infinitesimally thin, uncorrelated sheets of color charge. The charge density two-point function can then be approximated as
\begin{equation} \label{eq:mv2_twop_reg}
\ev{\rho^a_n (\bm{x}_\perp) \rho^b_m (\bm{y}_\perp)} = g^2 \mu^2 \frac{1}{N_s} \delta_{nm} \delta^{ab} \delta^{(2)}(\bm{x}_\perp - \bm{y}_\perp),
\end{equation}
where the indices $n,m \in \{ 1, 2, \dots, N_s \}$ are introduced to denote the different sheets. This regularization effectively discretizes the term $\mu^2(x^-) \delta(x^- - y^-)$ in \cref{eq:mv2_twop}.
For each sheet one then solves the Poisson equation
\begin{equation} \label{eq:mv2_poisson_reg}
- \Delta_\perp A^+_n(\bm{x}_\perp) = \rho_n(\bm{x}_\perp),
\end{equation}
and the lightlike Wilson line \cref{eq:asym_Wilson_line} is given by multiplying over all the color sheets together:
\begin{equation} \label{eq:asym_Wilson_line_reg}
V^\dg(\bm{x}_\perp) = \prod_{n=1}^{N_s} \exp \left[- i g A^+_{n} (\bm{x}_\perp) \right].
\end{equation}
By taking the limit $N_s \rightarrow \infty$ in the above formulation one obtains the ultrarelativistic limit of the generalized MV model, while for $N_s = 1$ one trivially recovers the original formulation of the MV model. This limiting case is usually referred to as the {\sl single color sheet approximation}. Obviously, results generally depend on $N_s$, due to the nature of path ordering. This fact is relevant when one tries to relate the MV parameter $\mu$ to another central quantity, the saturation momentum $Q_s$.\\

\begin{figure}
    \centering
    \includegraphics[width=0.7\linewidth]{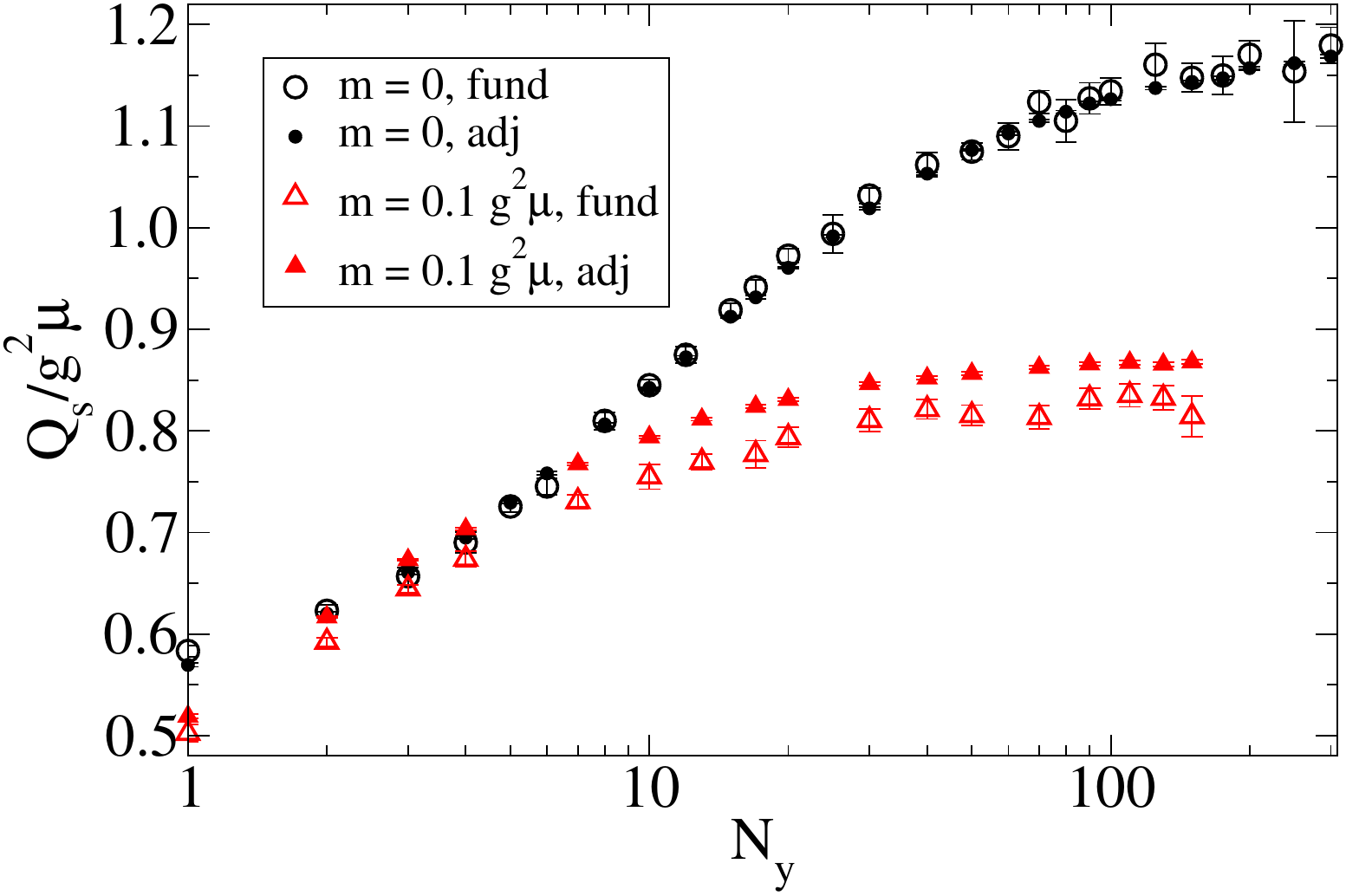}
    \caption{Dependence on the number of sheets (here denoted by $N_y$) of the fundamental and adjoint saturation scales $Q_s$ for $g^2\mu L_\perp = 100$ and $g^2\mu a_\perp = 0.5$, for $m=0$ and $m=0.1 g^2 \mu$. Notice how, for the regularized curve, the saturation scales converges for a number of color sheets of around $40$. Figure from \cite{Lappi:2007ku}.}
    \label{fig:Lappi_Qs}
\end{figure}

We have introduced the MV parameter as an external input having the dimension of energy. At this point one question arises naturally: how does it relate to the saturation scale $Q_s$ of the glasma? One way to define the saturation momentum $Q_s$ within the CGC framework is via the inverse correlation length of the Wilson line two-point function in the fundamental representation \cite{Lappi:2007ku}
\begin{equation} \label{eq:wilson_corr_fund}
C(|\bm{x}_\perp-\bm{y}_\perp|) \equiv \frac{1}{N_c} \ev{\text{Tr} \left( V^\dg(\bm{x}_\perp) V(\bm{y}_\perp) \right)},
\end{equation}
where the fundamental representation Wilson lines $V(\bm{x}_\perp)$ are given by \cref{eq:asym_Wilson_line}.
Translational and rotational invariance in the transverse plane imply that the two-point function only depends on the modulus $ |\bm{x}_\perp-\bm{y}_\perp|$.
An intuitive physical picture associated with \cref{eq:wilson_corr_fund} is that $C(|\bm{x}_\perp-\bm{y}_\perp|)$ represents the S-matrix for scattering of a quark-antiquark pair, ``probing'' the nucleus at different transverse coordinates $\bm{x}_\perp$ and $\bm{y}_\perp$, respectively, at the speed of light. Due to the quarks being recoilless (or eikonal), they pass through the nucleus without changing their lightlike trajectories along $x^-$ \cite{Kovchegov:2012mbw}. However, the quarks are not fully unaffected by the nucleus: even though their trajectories stay fixed, as they pass the color field their color charges rotate in accordance with non-Abelian charge conservation, i.e. the continuity equation $D_\mu J^\mu = 0$ has to hold. After passing through, one can compare the rotated color charges of the quarks: if the quarks are very close, meaning that the transverse separation $r \equiv |\bm{x}_\perp - \bm{y}_\perp|$ is much smaller than any transverse length scale of the nucleus (e.g.   $\mu^{-1}$), the quarks experience almost the same color rotation $V(\bm{x}_\perp) \simeq V(\bm{y}_\perp)$, thus we have $C(|\bm{x}_\perp-\bm{y}_\perp|) \simeq 1$. On the other hand, for a very large separation, the color fields and  Wilson lines $V(\bm{x}_\perp)$ and $V(\bm{y}_\perp)$ are uncorrelated, the color charges of the quarks are basically random and thus $C(|\bm{x}_\perp-\bm{y}_\perp|) \simeq 0$. The Wilson line two-point functions therefore define a characteristic length scale at which an initially correlated probing recoilless quark pair becomes de-correlated. The inverse of this length is associated with the characteristic transverse momentum scale $Q_s$. Specifically, one defines $Q_s$ as
\begin{equation} \label{eq:qs_definition}
C \left( r = \frac{\sqrt{2}}{Q_s}   \right) \equiv e^{-\frac{1}{2}}.
\end{equation}
There exist multiple, slightly different definitions of the saturation momentum (see \cite{Lappi:2007ku} for details), involving for instance the adjoint representation Wilson line two-point function instead of \cref{eq:wilson_corr_fund}, which is evaluated in the fundamental representation. Moreover, the above definition involves the two-point function in coordinate space, but it is also possible to define $Q_s$ using the momentum space representation of \cref{eq:wilson_corr_fund}. On dimensional and parametrical grounds, one finds as a rough estimate that
\begin{equation}
Q_s\simeq g^2 \mu.
    \label{eq:Qs_to_muMVparameter}
\end{equation}
One factor of $g$ is due to the two point charge correlator \cref{eq:mv2_twop}, while the second is the one appearing in the Wilson line \cref{eq:asym_Wilson_line}. In the color sheet regularization, the ratio $Q_s / \left( g^2 \mu \right)$ depends on $N_s$ and has to be determined numerically (see Fig. \ref{fig:Lappi_Qs}). 
The limiting case $N_s \rightarrow \infty$, however, can be performed analytically (see e.g.\ \cite{JalilianMarian:1996xn} and \cite{Iancu:2005jft} for a detailed derivation). Ignoring all of these complications, it will be sufficient for the purpose of this Thesis to assume that the ratio $Q_s / \left( g^2 \mu \right)$ for a heavy nucleus is roughly close to $1$.\\

Let us move on to another two-point function, which will serve as an excuse to tackle the problem of divergences in the MV model. We can study the correlator of the gauge field $A^+$, given by
\begin{equation}
\ev{A^{+}_a(x^-, \bm{x}_\perp) A^{+}_b(y^-, \bm{y}_\perp)}.
\label{eq:Aplus}
\end{equation}
In particular, \eqref{eq:Aplus} can be directly related to the charge density correlator \cref{eq:mv2_twop} via \cref{eq:poisson_solution_momentum}. One easily finds that:
\begin{equation}
\ev{A^{+}_a(x^-, \bm{x}_\perp) A^{+}_b(y^-, \bm{y}_\perp)} = g^2 \mu^2(x^-) \delta(x^- - y^-) \delta_{ab} \int \frac{d^2 \bm{k}_\perp}{\left( 2 \pi \right)^2} \frac{1}{k_\perp^4} e^{i \bm{k}_\perp \cdot \left( \bm{x}_\perp - \bm{y}_\perp \right)}.
\end{equation}
The momentum integral in the gauge field two-point function of the MV model turns out to be infrared divergent and has to be regularized. A popular way of curing this divergence is to introduce an infrared regulator $m$ by replacing \cref{eq:poisson_solution_momentum} with
\begin{equation} \label{eq:poisson_solution_momentum_regulated}
A^+_a(x^-, \bm{x}_\perp) = \int \frac{d^2 \bm{k}_\perp}{\left( 2 \pi \right)^2} \frac{\tilde{\rho}_a(x^-,\bm{k}_\perp)}{k_\perp^2 + m^2} e^{i \bm{k}_\perp \cdot \bm{x}_\perp},
\end{equation}
where $m$ has units of energy. Effectively, $m$ suppresses the long-range behavior of the gauge field $A^+$. This point can be illustrated in a simple case, in which we choose $\rho$ to be a point charge
\begin{equation}
\rho^a(x^-, \bm{x}_\perp) = q^a \delta(x^-) \delta^{(2)}(\bm{x}_\perp).
\end{equation}
Inserting this ansatz \eqref{eq:poisson_solution_momentum_regulated} yields the gauge field
\begin{equation}
A^+_a(x^-, \bm{x}_\perp) = \frac{1}{2 \pi} q_a K_0 \left( m |\bm{x}_\perp| \right) \delta(x^-),
\end{equation}
being $K_j(x)$ the modified Bessel functions of the second kind. At large distances $|\bm{x}_\perp| \gg m^{-1}$, the asymptotic behavior of the Bessel function shows that the gauge field falls off exponentially
\begin{equation}
A^+_a(x^-, \bm{x}_\perp) \simeq q_a \frac{e^{-m |\bm{x}_\perp|}}{\sqrt{8 \pi m |\bm{x}_\perp|}} \delta(x^-).
\end{equation}
The regulator $m$ thus plays the role of a screening mass. A possible physical interpretation one can assign to this infrared regulator is that its inverse $m^{-1}$ mimics the confinement radius (roughly $\Lambda_{\text{QCD}}^{-1} \approx 1\,\fm$), by imposing color neutrality at distances larger than the nucleon size. A more formally careful way of introducing this infrared regulation is the change of the color charge density via
\begin{equation} \label{eq:rho_reg_m}
\tilde{\rho}_a(x^-,\bm{k}_\perp) \rightarrow \frac{k_\perp^2}{k_\perp^2 + m^2} \tilde{\rho}_a(x^-,\bm{k}_\perp),
\end{equation}
without modifying \cref{eq:poisson_solution_momentum}. This circumvents the problem of adding a gauge-invariance breaking term in order to obtain \cref{eq:poisson_solution_momentum_regulated}. Moreover, it is now also clear that the color charge density is globally color neutral, because the zero mode of this newly-defined $\tilde{\rho}^a(x^-, \bm{k}_\perp)$ vanishes.
In any case, once we do so we can compute the (now finite) two-point function of the gauge field as
\begin{align} \label{eq:Ap_twopf}
\ev{A^{+}_a(x^-, \bm{x}_\perp) A^{+}_b(y^-, \bm{y}_\perp)} & = g^2 \mu^2(x^-) \delta(x^- - y^-) \delta_{ab} \int \frac{d^2 \bm{k}_\perp}{\left( 2 \pi \right)^2} \frac{1}{\left( k_\perp^2 + m^2 \right)^2} e^{i \bm{k}_\perp \cdot \left( \bm{x}_\perp - \bm{y}_\perp \right)} \nn
& = g^2 \mu^2(x^-) \delta(x^- - y^-) \delta_{ab} \frac{|\bm{x}_\perp - \bm{y}_\perp| K_1(m|\bm{x}_\perp - \bm{y}_\perp|)}{4 \pi m}.
\end{align}
The point we wanted to bring forward with this calculation is that the MV model contains an infrared divergence that has to be manually regularized. There are other ways to cure the divergence by imposing color neutrality: for instance, one can require only global color neutrality by eliminating the zero mode of $\rho$, which is equivalent to subtracting the monopole contribution of the charge density. However, by doing this way, results will then depend on the size of the system: since the MV model has no notion of finite size in the transverse plane, one is forced to manually choose some kind of system size. In practice, the system size is fixed by the transverse size of the lattice in numerical simulations. Optionally one can also choose to not only eliminate the monopole, but also subtract the dipole contribution of $\rho$, which again depends on system size and changes the infrared behavior of $\rho$ in a slightly different manner. For an extended discussion on how to impose color neutrality see \cite{Krasnitz:2002mn}. In this Thesis we use \cref{eq:rho_reg_m} not only for convenience, but also because this way of regularization has been established in more elaborate models of nuclei such as IP-glasma \cite{Schenke:2012wb, Schenke:2012fw}.

\subsection{Beyond naive MV model: pA collisions}
\label{sec:Beyond naive MV model: pA collisions}
The steps we have described in Section \ref{sec:The McLerran--Venugopalan model} to generate the initial color charge are shared for the study of either nucleus-nucleus collisions, proton-nucleus collisions, or even proton-proton collisions. Let us once again report the expressions which define the McLerran-Venugopalan model, i.e. Eqs. \eqref{eq:mv_onep} and \eqref{eq:mv_twop}
\begin{align}
\ev{\rho^a (\bm{x}_\perp)} & = 0, \label{eq:mv_onep_2} \\
\ev{\rho^a (\bm{x}_\perp) \rho^b (\bm{y}_\perp)} & = g^2 \mu^2 \delta^{ab} \delta^{(2)}(\bm{x}_\perp - \bm{y}_\perp). \label{eq:mv_twop_2}
\end{align}
The original MV model assumed a source charge density that is homogeneous in the transverse plane, which is feasible when dealing with large-A nuclei, of which we can neglect the details at the level of single nucleons. On the other hand, a reasonable description of protons involves a varying nuclear density in the transverse plane, in order to take into account for the underlying quark structure.

The main difference between those two cases lies in the choice of the MV parameter $\mu$. For a heavy nucleus (labeled by A), $\mu$ has been chosen as a constant, which translates into uniform fluctuations (i.e. basically white noise) of color charge throughout the lattice. The deal with the proton is instead more involved. The key node in the p-case is to consider a MV parameter $\mu$ which now depends on transverse position: $\mu=\mu(\bm{x}_\perp)$. If $\mu(\bm{x}_\perp)$ is able to take into account for the internal proton structure, also $\rho^a(\bm{x}_\perp)$ will.

\begin{figure}

\begin{center}
\includegraphics[scale=0.78]{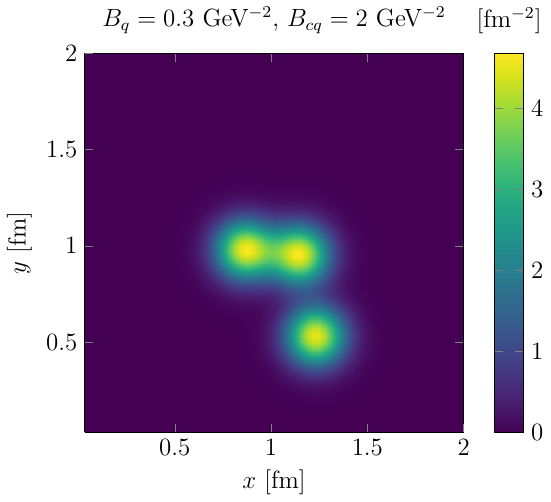} 
\hspace{1em}
\includegraphics[scale=0.78]{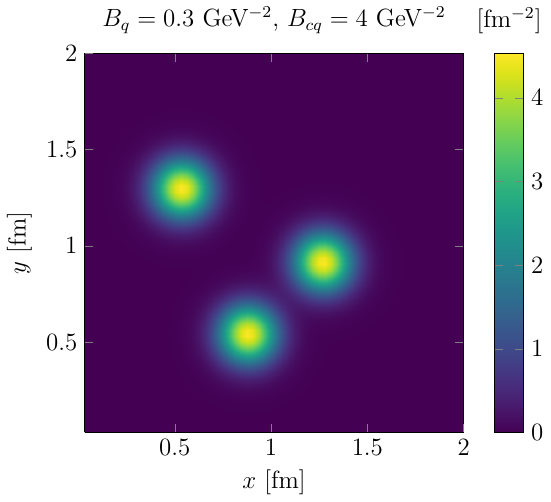} 
\end{center}
\begin{center}
\hspace{0.5em}
\includegraphics[scale=0.78]{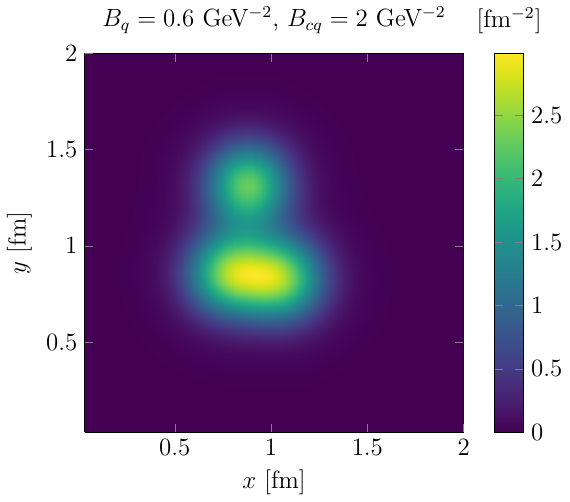} 
\hspace{.5em}
\includegraphics[scale=0.78]{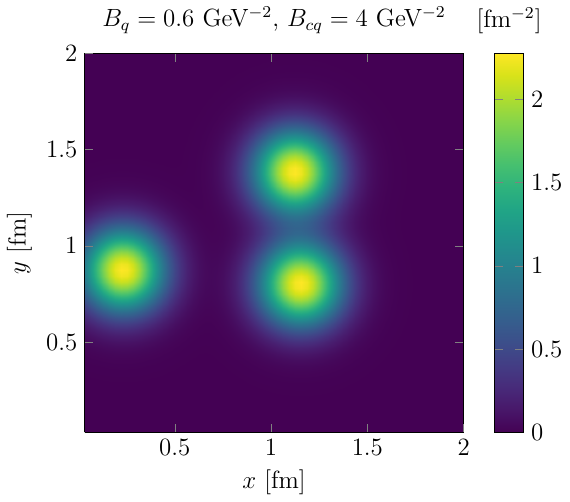} 
\end{center}
\caption{Profile of the thickness function of the proton $T_p$ for different values of the parameters $B_q$ (which describes the width of the color charge distribution around each quark) and $B_{cq}$ (which describes the width in the distribution of the centers of the quarks).}
\label{fig:proton_distribution_Bq_Bcq}
\end{figure}

Let us introduce the {\sl thickness function} of the proton, $T_p$, as \cite{Schenke:2020mbo}
\begin{equation}
T_p(\bm{x}_\perp) = 
\frac{1}{3} \sum_{i=1}^3 \frac{1}{2\pi B_q} \exp\left[-\frac{(\bm{x}_\perp-\bm{x}_\perp^i)^2}{2B_q}\right].
\label{eq:bd1}
\end{equation}
The expression in Eq. \eqref{eq:bd1} is clearly a superposition of three Gaussians, each of which centered at $\bm{x}_\perp^i$ for $i=1,2,3$ and having width equal to $B_q$, an external parameter. This is meant to be the distribution of the color charge within a proton, which is spread along the three constituent quarks. To this point, how are the positions $\bm{x}_\perp^i$ of the constituent quark fixed? These are extracted randomly from the following distribution
\begin{equation}
T_{cq}(\bm{x}_\perp^i) = \frac{1}{2\pi B_{cq}}
\exp\left(-\frac{\bm{x}_\perp^{i, 2}}{2B_{cq}}\right),
\label{eq:bd2}
\end{equation}
where $B_{cq}$ is another external parameter. In practice, we extract three pairs of random numbers from the distribution \eqref{eq:bd2}, which fix the positions of the three quarks in the transverse plane. These are then inserted in Eq. \eqref{eq:bd1} in order to obtain the charge profile of the proton. Those parameters are related to the geometric distribution of the charge in the proton. In particular, $B_q$ describes the width of the color charge distribution around each valence quark of the proton, while $B_{cq}$ describes the width in the distribution of the centers of the valence quarks themselves. For illustrative purposes, in Figure \ref{fig:proton_distribution_Bq_Bcq} we show the profile of the thickness function of the proton $T_p$ as in \eqref{eq:bd1}, with centers of the constituent quarks extracted from the distribution in \eqref{eq:bd2}, for various choices of the parameters $B_q$ and $B_{cq}$. An increase in $B_q$ induces a modelization of the proton whose charge is more distributed in the transverse plane, and whose peaks are lower. Instead, an increase in $B_{cq}$ makes the constituent quarks far apart from one another in the transverse plane.

The parameters $B_q$ and $B_{cq}$ we will use throughout this Thesis, for the modelization of the proton, have been fixed in \cite{Mantysaari:2016ykx}: they found that the only choice of these parameters which was able to fit the diffractive photoproduction cross sections of vector mesons, both coherent and incoherent, was
\begin{equation}
B_q=0.3 \text{ GeV$^{-2}$,}~~~~~B_{cq}=4 \text{ GeV$^{-2}$}.
    \label{eq:Bq_and_Bcq}
\end{equation}
The choice of parameters \eqref{eq:Bq_and_Bcq} (whose resulting distribution is depicted in the top right panel of Figure \ref{fig:proton_distribution_Bq_Bcq}) corresponds to a proton which is far from symmetric, in which the valence quarks are actually small and far apart from one other. Quantitatively, one sees that the widths of the two gaussians are around $\sqrt{B_q}=0.11$ fm and $\sqrt{B_{cq}}=0.39$ fm, i.e. they differ by a factor around 4. 

After doing so, for each point $\bm{x}_\perp$ and at fixed Bjorken $x$, we evaluate the saturation scale $Q_s$ in this model as \cite{Schenke:2020mbo}
\begin{equation}
Q_s^2(x,\bm{x}_\perp) = \frac{2\pi^2}{N_c}  \alpha_s\,
xg(x,Q_0^2)\, T_p(\bm{x}_\perp).
\label{eq:qs_2_77}
\end{equation}
from which $\mu(x,\bm{x}_\perp)$ of the proton is given by \cite{Schenke:2020mbo}
\begin{equation}
g^2\mu(x,\bm{x}_\perp) = c Q_s(x,\bm{x}_\perp),
\label{eq:gmu_ajk}
\end{equation}
with $c=1.25$. This choice of the constant $c$ has been determined numerically in \cite{Lappi:2007ku}: check for instance, the value reached by the red points at large number of sheets in Fig. \ref{fig:Lappi_Qs}, taken from \cite{Lappi:2007ku}. There we see the asymptotic value being close to $Q_s/g^2\mu\simeq 0.8=1/1.25$. Note that by virtue of Eq.~\eqref{eq:qs_2_77} not only $\mu$, but also $Q_s$ depends on the transverse plane coordinates. As far as the gluon distribution function $xg(x,Q_0^2)$ in \eqref{eq:qs_2_77} is concerned, that should in principle be computed at the scale $Q_s$, by means of the DGLAP equation with a proper initialization. This was done in \cite{Schenke:2020mbo,Rezaeian:2012ji}. Here instead, for simplicity, we assume that $xg$ is given by the initial condition at 
$Q_0^2=1.51$ GeV$^2$ as in \cite{Schenke:2020mbo,Rezaeian:2012ji}, namely
\begin{equation}
xg(x,Q_0^2)=A_g\, x^{-\lambda_g}\,(1-x)^{f_g},
\label{eq:anna23}
\end{equation}
with $A_g=2.308$, $\lambda_g=0.058$ and $f_g=5.6$. This simplification is partly justified by the fact that the average saturation scale for the proton is of the order of $Q_s\sim 1$--2 GeV, hence we do not expect the DGLAP evolution of the $xg(x,Q_0^2)$ in Eq.~\eqref{eq:anna23} to lead to significant changes. For pA collisions at the LHC energies, the relevant values of $x$ are in the range $[10^{-4},10^{-3}]$. These values of $x$ have been estimated in the literature, check for instance \cite{Schenke:2012hg} and \cite{Schenke:2020mbo}. The first paper estimates $x=Q_s/\sqrt{s}$, and if we assume $Q_s\sim 2$ GeV and $\sqrt{s}\sim 2-10$ TeV, we indeed get $x$ between $10^{-4}$ and $10^{-3}$ \cite{Schenke:2012hg}. Similarly in the second paper, in which a rougher estimate $x=\langle p_\perp\rangle/\sqrt{s}$ appears: there $\langle p_\perp\rangle\sim 1-2$ GeV is the mean transverse momentum of charged hadrons, and we are led back to the same estimate for $x$ \cite{Schenke:2020mbo}. All that said, can thus fix $x$ in the aforementioned range, then compute $xg$ by virtue of Eq.~\eqref{eq:anna23}. Throughout this work will use the corresponding value obtained for $x=10^{-4}$, that is $xg=3.94$.

\section{After the collision: the glasma phase} \label{sec:glasma_initial}

Equipped with the single nucleus solutions of the classical field equations from the last Section, we can start investigating ultrarelativistic collisions and the kinds of color fields therein produced, by interpolating the solutions of each of the two nuclei before and after the collision. Note that, as an obvious albeit necessary remark, all the results discussed in the previous Section apply to the MV model for nuclei moving in the positive $z$ direction (along $x^+$), but completely analogous calculations can be done along the opposite direction $x^-$.\\

\begin{figure}[t]
	\centering
	\includegraphics{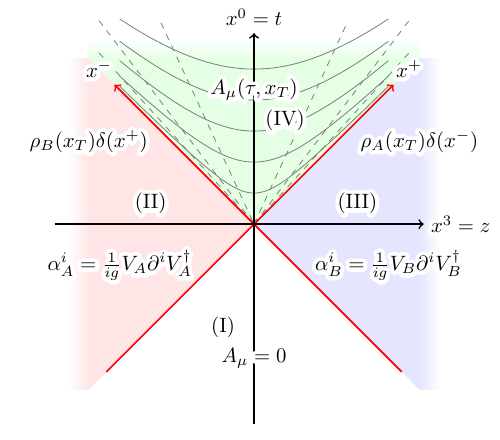}
	\caption{The schematic Minkowski diagram for an ultrarelativistic collision. The red diagonal lines along $x^+$ and $x^-$ are the infinitesimally thin color currents $J^+$ and $J^-$, respectively, which separate space-time into four distinct regions I - IV. The past light cone (region I) is completely unaffected by the collision and also by the single nuclei themselves, so the gauge field can be set to zero. In regions II and III we use the LC gauge solutions of the single nuclei $\alpha^i_A$ and $\alpha^i_B$. In the future light cone (region IV), the $(\tau, \eta)$ coordinate system is shown (as introduced in the main text). The glasma initial conditions are formulated at $\tau = 0^+$, which forms the boundary of the future light cone. The color field of the glasma $A_\mu(\tau, \bm{x}_\perp)$ for $\tau > 0$ is obtained by solving the boost invariant Yang-Mills equations. Figure from \cite{Muller:2019bwd}.}
		\label{fig:lightcone2}
\end{figure}

We denote the two nuclei by ``A'' and ``B''. The combined color current of both nuclei is given by
\begin{equation} \label{eq:combined_current}
J^\mu(x) = J_A^\mu(x) + J_B^\mu(x) =  \delta^{\mu+} \delta(x^-) \rho_A(\bm{x}_\perp) +  \delta^{\mu-} \delta(x^+) \rho_B(\bm{x}_\perp),
\end{equation}
where nucleus ``A'' moves along the $x^+$ axis and ``B'' along the $x^-$ axis. In the ultrarelativistic limit this color current defines the sharp boundary of the light cone in Fig. \ref{fig:lightcone2}, which separates the Minkowski diagram into four distinct regions I - IV. In this limit the two nuclei move almost at the speed of light, therefore the axes $x^-$ and $x^+$ basically coincide with the bisectors of the four quadrants. Due to causality, the regions I, II and III are unaffected by the collision and therefore we can use the single nuclei solutions from the last Section. However, in region IV, i.e. the future light cone, which in other words is the region enclosed by the two nuclei once they have crossed each other, non-linear interaction terms in the Yang-Mills equations result in a non-trivial color field. This color field is the {\sl glasma}, and our goal is to find this solution to the Yang-Mills equations in the presence of these moving color charges.

Formally, the solution in region IV is a functional of the two charge densities, i.e. we have 
$A_\mu [\rho_A, \rho_B]$.
Using this color field we can compute arbitrary observables via functional integration
\begin{equation}
\ev{\mathcal{O}[A_\mu[\rho_A, \rho_B]]} \equiv \intop \mathcal{D} \rho_B \mathcal{D} \rho_A\ \mathcal{O}[A_\mu[\rho_A, \rho_B]]\, W_A[\rho_A]\, W_B[\rho_B],
\end{equation}
where naturally one has to integrate over both color charge densities. An important assumption is that the color charges do not suffer any recoil during the collision, i.e. they do not lose any longitudinal momentum and stay on their fixed, lightlike trajectories. Consequently, both color currents must be separately conserved:
\begin{align}
D_\mu J^\mu_A &= \p_+ J^+_A + i g \cm{A^-}{J^+_A} = 0, \\
D_\mu J^\mu_B &= \p_- J^-_B + i g \cm{A^+}{J^-_B} = 0.
\end{align}
Even if the trajectories are unaffected, non-Abelian charge conservation still leaves the possibility that the color charges experience color rotation due to the presence of the field, leading to a non-trivial dependence of $J^+_A$ ($J^-_B$) on $x^+$ ($x^-$). In order to avoid this complication, let us once again change the choice of the gauge, and choose a gauge condition such that $A^-$ ($A^+$) vanishes along the $x^+$ ($x^-$) axis. A suitable choice is the {\sl Fock-Schwinger} gauge condition
\begin{equation}
x^+ A^- + x^- A^+ = 0,
\label{eq:Fock_schwinger_gauge_cond}
\end{equation}
which we will use in the future light cone $x^+, x^- \geq 0$. In regions I - III we will still use the light cone gauge condition $A^+ = A^- = 0$, which can be smoothly connected to the Fock-Schwinger gauge at the boundaries $\{x^+ = 0$, $x^- \geq 0 \}$ and $\{ x^- = 0$, $x^+ \geq 0 \}$. 

The combined color current \cref{eq:combined_current} features an important symmetry, namely invariance under longitudinal boosts, or {\sl boost invariance}. Specifically, this means that under the Lorentz transformation
\begin{align}
x^\pm &\rightarrow x'^\pm = e^{\pm \beta} x^\pm, \\
J^\pm(x) &\rightarrow J'^\pm(x') =  e^{\pm \beta} J^\pm(x),
\end{align}
the currents remain unchanged in their functional form:
\begin{align} 
J^+_A(x^-, \bm{x}_\perp) & \rightarrow e^{+\beta}\delta (e^{+\beta} x'^{\,-}) \rho_A(\bm{x}_\perp) = \delta (x'^{\,-}) \rho_A(\bm{x}_\perp), \label{eq:boosted_current_A} \\
J^-_B(x^+, \bm{x}_\perp) & \rightarrow e^{-\beta}\delta (e^{-\beta} x'^{\,+}) \rho_B(\bm{x}_\perp) = \delta (x'^{\,+}) \rho_B(\bm{x}_\perp). \label{eq:boosted_current_B}
\end{align}
Since the color current is the only input that we have for solving the Yang-Mills equations (apart from boundary conditions and the choice of gauge), the solutions and observables should also reflect this symmetry. It is therefore natural to introduce a coordinate system in which boost invariance becomes manifest, namely the {\sl Milne} coordinate set. Let us define
\begin{align}
x^+ \equiv \frac{\tau}{\sqrt{2}} e^{+\eta}, \qquad
x^- \equiv \frac{\tau}{\sqrt{2}} e^{-\eta},
\label{eq:Milne_coordinate_trans}
\end{align}
in which we introduced the {\sl proper time} $\tau \in (0, \infty)$ and the {\sl space-time rapidity} $\eta \in (-\infty, \infty)$. The boundary $\tau=0^+$ corresponds to the union of the axes $x^-$ and $x^+$ (red lines in Fig. \ref{fig:lightcone2}, but limited to the upper half-plane only). The lines of constant $\tau>0$ correspond to hyperbolas which are increasingly more distant from the $x^-$ and $x^+$ axes as $\tau$ increases. Inverting the above definitions yields
\begin{equation}
\tau = \sqrt{2\, x^+\, x^-},\qquad
\eta = \frac{1}{2} \ln \left( \frac{x^+}{x^-} \right).
\end{equation}
Check Appendix \ref{sec:Milne and Minkowski coordinates} for more details on the Milne coordinate set.
By applying the coordinate transformation \eqref{eq:Milne_coordinate_trans} to the gauge field we find
\begin{align}
A^\tau &= \frac{1}{\tau} \left( x^+ A^- + x^- A^+ \right) = A_\tau, \label{eq:a_tau_pm}\\
A^\eta &= - \frac{1}{\tau^2} \left( x^+ A^- - x^- A^+ \right) = - \frac{1}{\tau^2} A_\eta.
\end{align}
At this point, we can make a manifestly boost invariant ansatz for the color field by ``stiching'' together separate functions in the four regions of the Minkowski diagram \cite{Kovner:1995ja}, as
\begin{align}
A^i(x) &= \theta(x^+) \theta(x^-) \alpha^i(\tau, \bm{x}_\perp) + \theta(-x^+) \theta(x^-) \alpha^i_A(\bm{x}_\perp) + \theta(x^+) \theta(-x^-) \alpha^i_B(\bm{x}_\perp), \label{eq:sep_ansatz_1} \\
A^\eta(x) &= \theta(x^+) \theta(x^-) \alpha^\eta(\tau, \bm{x}_\perp), \label{eq:sep_ansatz_2}
\end{align}
where we suppressed all dependencies on $\eta$, making the ansatz explicitly boost invariant. In the above formula, the transverse gauge fields of the single nuclei fulfill $\p_i \alpha^i_{A,B}(\bm{x}_\perp) = - \rho_{A,B}(\bm{x}_\perp)$, and are pure gauge in region II (for nucleus A) and III (for nucleus B). This choice for \eqref{eq:sep_ansatz_1} and \eqref{eq:sep_ansatz_2} is driven by the sake of simplicity: in principle it would be possible to consider $\eta$--dependent gauge fields without breaking boost invariance in gauge invariant observables. However, here we require this symmetry also at the level of gauge fields.

Due to \cref{eq:a_tau_pm}, the Fock-Schwinger gauge condition \eqref{eq:Fock_schwinger_gauge_cond} renders the temporal component $A^\tau$ zero. Plugging the ansatze \eqref{eq:sep_ansatz_1} and \eqref{eq:sep_ansatz_2} into the Yang-Mills equations one obtains a few worrying terms such as $\delta^2$ and $\delta'$, due to partial derivatives acting on the Heaviside functions. Setting all coefficients of these problematic terms to zero, one finds the following matching conditions at the boundary $\tau \rightarrow 0$ \cite{Kovner:1995ja}:
\begin{align} \label{eq:glasma_initial_1}
\alpha^i(\tau \rightarrow 0, \bm{x}_\perp) &= \alpha^i_A(\bm{x}_\perp) + \alpha^i_B(\bm{x}_\perp), \\
\alpha^\eta(\tau \rightarrow 0, \bm{x}_\perp) &= \frac{ig}{2} \cm{\alpha^i_A(\bm{x}_\perp)}{\alpha^i_B(\bm{x}_\perp)}, \label{eq:glasma_initial_2}
\end{align}
together with $\p_\tau \alpha^i(\tau \rightarrow 0, \bm{x}_\perp) = \p_\tau \alpha^\eta(\tau \rightarrow 0, \bm{x}_\perp) = 0$. These matching conditions form the starting point for the time evolution of the color field in the future light cone, i.e. the glasma field.

At this point we may evaluate the glasma observables. The longitudinal electric and magnetic field in the laboratory frame are related to the field strength tensor in the co-moving frame via
\begin{align}
E_3 &= \frac{1}{\tau} F_{\tau \eta}, \\
B_3 &= - F_{12}.
\end{align}
which at $\tau \rightarrow 0$ yield
\begin{align}
E_3(\tau \rightarrow 0, \bm{x}_\perp) &= - i g\, \delta_{ij} \cm{\alpha_{A,i}(\bm{x}_\perp)}{\alpha_{B,j}(\bm{x}_\perp)}, \label{eq:bi_initial_EL_BL_1}\\
B_3(\tau \rightarrow 0, \bm{x}_\perp) &= - i g\, \epsilon_{ij} \cm{\alpha_{A,i}(\bm{x}_\perp)}{\alpha_{B,j}(\bm{x}_\perp)}, \label{eq:bi_initial_EL_BL_2}
\end{align}
where $\delta_{ij}$ and $\epsilon_{ij}$ are the Kronecker delta and the Levi-Civita symbol in two dimensions. What we see is that, while the non-zero components of the color-electric and color-magnetic fields of a single nucleus were the transverse ones (see Eq. \eqref{eq:electric_magnetic_fields_singlenucleus}), here we find the opposite. Right after the collision, in the region between the two nuclei the transverse electric and magnetic field components vanish, the collision of two nuclei therefore produces fields that are initially purely longitudinal. This leads to the following physical picture directly after the collision \cite{Lappi:2006fp}: as the purely transverse color fields of the single nuclei move away from the collision region, longitudinal electric and magnetic flux tubes span between the two nuclei. The size of these glasma flux tubes corresponds to the size of correlated domains in the color fields of the nuclei, which is roughly $Q^{-1}_s$ \cite{Dumitru:2014nka}. After the initial formation, the flux tubes quickly evolve, starting to expand in the transverse directions and dilute in intensity. In the next Section we will introduce methods to study the time evolution of this system numerically.

\subsection{Numerical implementation of CGC}
\label{sec:Numerical implementation of CGC}

We know get deep into the numerical details of the glasma simulations. Let us start form the implementation of the MV initial conditions, i.e. the one point function \eqref{eq:mv_onep} and the two point function \eqref{eq:mv_twop} (or rather, its generalization \eqref{eq:mv2_twop_reg}). By considering a generic number $N_s$ of color sheets, and by discretizing the Dirac delta onto the transverse lattice we get
\begin{align}
    \langle \rho^a_{n,\bm{x}}\rangle=&0,
    \label{eq:mv_numerical_implementation_1}\\
    \langle \rho^a_{n,\bm{x}}\rho^b_{m,\bm{y}}\rangle=&g^2 \mu^2 \frac{\delta_{n,m}}{N_s}\delta^{a,b} \frac{\delta_{\bm{x},\bm{y}}}{a_\perp^2}.
    \label{eq:mv_numerical_implementation_2}
\end{align}
Here $a,b=1,\dots, N_c^2-1$ are SU(3) color indexes, $n,m=1,\dots,N_s$ are the color sheet indexes and $\bm{x},\bm{y}$ represent the transverse coordinates, spanning each point of a $N_\perp\times N_\perp$ lattice. Such transverse lattice is a square with size $L_\perp$ and lattice spacing $a_\perp=L_\perp/N_\perp$. Operatively, the conditions \eqref{eq:mv_numerical_implementation_1} and \eqref{eq:mv_numerical_implementation_2} are satisfied by generating random gaussian numbers with mean zero and standard deviation equal to $\sqrt{(g^2\mu^2)/(N_s \,a_\perp^2)}$: for a gaussian distribution of mean zero and standard deviation $\sigma$, one has $\langle x\rangle=0$ and $\langle x^2 \rangle=\sigma^2-\langle x\rangle^2=\sigma^2$, moreover the appearance of the factor $a_\perp$ comes from the discretization of the Dirac delta. The coupling constant has been fixed as $g=2$ (corresponding to about $\alpha_s\simeq0.3$), and unless differently stated, we will use $N_s=50$ in Eq. \eqref{eq:mv_numerical_implementation_2} throughout this work.

This has of course to be done for each nucleus. Once the charge has been generated, we know that the hard and the soft sectors are coupled via the Yang-Mills equations $D_\mu F^{\mu\nu}=J^\nu$, in which $D_\mu=\partial_\mu-ig[A_\mu,\hspace{3pt} \cdot\hspace{3pt}]$ is the covariant derivative, $F^{\mu\nu}=\partial^\mu A^\nu-\partial^\nu A^\mu-ig[A^\mu,A^\nu]$ is the field strength tensor and $J^\mu$ the color current. After all the gauge choices we have already discussed in the previous Sections, the only component which is left is $A^+\equiv \alpha$. For each color sheet and each color component, we have to solve the Poisson equation
\begin{equation} \label{eq:mv2_poisson_reg_2}
- \Delta_\perp \alpha_{n, \bm{x}}^a = \rho_{n, \bm{x}}^a,
\end{equation}
where $\Delta_\perp=\partial_x^2+\partial_y^2$ is the Laplacian. This equation can in principle be solved by diagonalizing and inverting the discretized Laplacian on the lattice. However, since the Fourier transformations are easily performed using the 
\textsc{Fast Fourier Transform} (FFT) package in \texttt{Julia}, we rather Fourier-transform both sides of \eqref{eq:mv2_poisson_reg_2} in order to get
\begin{equation}
    \Tilde{k}^2_\perp \tilde{\alpha}_{n,\bm{k}}^a=\tilde{\rho}_{n,\bm{k}}^a.
    \label{9.1}
\end{equation}
From this we get, after a regularization of the IR modes
\begin{equation}
\tilde{\alpha}_{n,\bm{k}}^a=\frac{\tilde{\rho}_{n,\bm{k}}^a}{\Tilde{k}_\perp^2+m^2}.
    \label{9.4}
\end{equation}
The screening mass has been fixed as $m=0.2$ GeV $=1$ fm$^{-1}$. Naively, this discrete Fourier transform takes $\mathcal{O}(N_\perp^2)$ steps to be computed: however, if $N_\perp$ can be factored into small primes, the so-called {\sl Cooley–Tukey FFT algorithm} reduces the computational time to $\mathcal{O}(N_\perp\log N_\perp)$ \cite{Cooley:1965zz}. For this reason, in our work we have considered only values of $N_\perp$ which are powers of $2$.

In the above relations $\Tilde{k}_\perp^2$ is the lattice discretization of the squared transverse momentum. To obtain its expression, let us perform the discretized laplacian (as a 2--point formula) on the function
\begin{equation}
    \alpha_{n,\bm{x}}^a=\left(\frac{\Delta k}{2\pi}\right)^2\,\sum_{l=0}^{N_\perp-1} \sum_{m=0}^{N_\perp -1}  \tilde{\alpha}_{n,\bm{k}}^a \,
    e^{i x\, l\Delta k+i y\, m\Delta k},
    \label{eq:alpha_nx_discretized}
\end{equation}
where $\Delta k=2\pi/L_\perp$. By doing so we get
\begin{align}
&\Delta_\perp \alpha_{n, \bm{x}}^a=\frac{\alpha_{n, \bm{x}+ \bm{i}}^a-2\alpha_{n,\bm{x}}^a+\alpha_{n,\bm{x}- \bm{i}}^a}{a_\perp^2}+\frac{\alpha_{n, \bm{x}+ \bm{j}}^a-2\alpha_{n,\bm{x}}^a+\alpha_{n,\bm{x}- \bm{j}}^a}{a_\perp^2}\nonumber\\
&=\left(\frac{\Delta k}{2\pi}\right)^2\sum_{l=0}^{N_\perp-1} \sum_{m=0}^{N_\perp -1}\tilde{\alpha}_{n,\bm{k}}^a\, e^{i x\, l\Delta k+i y\, m\Delta k}\left[\frac{e^{ia_\perp l \Delta k}-2+e^{-ia_\perp l \Delta k}+e^{ia_\perp m \Delta k}-2+e^{-ia_\perp m \Delta k}}{a_\perp^2}\right]\nonumber\\
&=\left(\frac{\Delta k}{2\pi}\right)^2\sum_{l=0}^{N_\perp-1} \sum_{m=0}^{N_\perp -1}\tilde{\alpha}_{n,\bm{k}}^a\, e^{i x\, l\Delta k+i y\, m\Delta k}\left[\frac{(e^{ia_\perp l \Delta k/2}-e^{-ia_\perp l \Delta k/2})^2+(e^{ia_\perp m \Delta k/2}-e^{-ia_\perp m \Delta k/2})^2}{a_\perp^2}\right]\nonumber\\
&=-\left(\frac{\Delta k}{2\pi}\right)^2\sum_{l=0}^{N_\perp-1} \sum_{m=0}^{N_\perp -1}\tilde{\alpha}_{n,\bm{k}}^a\, e^{i x\, l\Delta k+i y\, m\Delta k}\left[\frac{4\sin^2\left(a_\perp l \Delta k/2\right)+4\sin^2\left(a_\perp m \Delta k/2\right)}{a_\perp^2}\right]\nonumber\\
&\equiv-\left(\frac{\Delta k}{2\pi}\right)^2\sum_{l=0}^{N_\perp-1} \sum_{m=0}^{N_\perp -1}\tilde{\rho}_{n,\bm{k}}^a\, e^{i x\, l\Delta k+i y\, m\Delta k}.
\label{eq:derivated_alpha_nx_discretized}
\end{align}
Above $\bm{i}$ and $\bm{j}$ are two vectors having magnitude $a_\perp$, oriented along the two transverse directions. By comparing the last two relations, we understand that 
\eqref{9.1} is satisfied if the discretized momentum $\Tilde{k}_\perp$ is given by
\begin{equation}
\Tilde{k}_\perp^2=\sum_{i=x,y}\left(\frac{2}{a_\perp}\right)^2\sin^2\left(\frac{k_ia_\perp}{2}\right).
    \label{9.2}
\end{equation}
Coming back to the Poisson equation, once $\tilde{\alpha}_{n,\bm{k}}^a$ has been obtained from \eqref{9.4}, by Fourier-transforming back as in \eqref{eq:alpha_nx_discretized} we get $\alpha_{n, \bm{x}}^a$ in coordinate space.\\

Starting from the solution \eqref{eq:alpha_nx_discretized}, our goal is to build up the glasma fields using a gauge covariant formalism. This is achieved by employing typical real time lattice gauge techniques. In particular, using the solution for $\alpha_{n, \bm{x}}^a$ we have just found, we will build the discretized version of the Wilson line in   \eqref{eq:lc_gauge_wilson_line} as
\begin{equation}
    V_{\bm{x}}=\prod_{n=1}^{N_s}\exp\{ig\,\alpha_{n,\bm{x}}^a t^a\}.
    \label{9.5}
\end{equation}
The integral over $x^-$ is therefore translated onto a ordered matrix multiplication over the color sheets $N_s$. From that, we define the {\sl gauge links} $U_{\bm{x},i}$, for each of the two nuclei A and B, as
\begin{equation}
U_{\bm{x},i}^{\text{A,B}}=V_{\bm{x}}^{\text{A,B}}\ V_{\bm{x}+\bm{i}}^{\dagger,\text{A,B}}.
\label{9.6}
\end{equation}
In the above expressions $t^a$ are the SU($N_c$) group generators (for $N_c=3$, the eight Gell-Mann matrices divided by 2). The index $\bm{x}$ refers to the position in the transverse plane, whereas $i=x,y$ labels the two different orientations along which \eqref{9.6} is evaluated. In \eqref{9.6} we first evaluate the Wilson line at a point $\bm{x}$, this is then multiplied by the Hermitian conjugate of the Wilson line at the point $\bm{x}+\bm{i}$, where $\bm{i}$ indicates a unit vector along $x$ or $y$. Basically, the gauge link is an object which, starting from $\bm{x}$ in the transverse plane, ‘‘points'' towards one of the two transverse directions. If the point $\bm{x}$ belongs to the edge of the simulation lattice, periodic boundary conditions have been implemented.\\

The gauge links in \eqref{9.6} have to be calculated for each nucleus, and they are ‘‘pure gauge'' links. Then, one would like to derive the gauge link $U$ for the combined system of the two nuclei immediately after the collision. We know that the gauge field at $\tau=0^+$ will be the sum of each nucleus' contribution, but since QCD is a non-Abelian gauge theory, the resulting gauge link will not be the product of the gauge links of each nucleus:
\begin{equation}
    U \neq U^{\text{A}}\,U^{\text{B}}.
    \label{eq:tot_gauge_link_notequal}
\end{equation}
It can be shown \cite{Krasnitz:1998ns} that the total gauge link $U_{\bm x,i}$ is determined by solving a set of $N_c^2-1$ equations, which are
\begin{equation}
    \text{Tr}[t_a(U_{\bm{x},i}^{\text{A}}+U_{\bm{x},i}^\text{B})(\mathbb{1}+U_{\bm{x},i}^\dagger)-\text{h.c.}]=0,
    \label{10}
\end{equation}
with $a=1,\dots,N_c^2 - 1$, and $U_{\bm{x},i}^{\text{A,B}}$ the gauge links determined in \eqref{9.6}. In other words, we have to require that $(U_{\bm{x},i}^\text{A}+U_{\bm{x},i}^\text{B})(\mathbb{1}+U_{\bm{x},i}^\dagger)$ has no anti-Hermitian traceless part. The system of equations \eqref{10} is non linear: its non-linearity arises from the requirement that the link matrix $U$ is an element of the SU($N_c$) group. It is easy to show that this equation has the correct formal continuum limit \eqref{eq:glasma_initial_1}: if we write $U^{\text{A,B}}$ as $\exp(i g a_\perp \alpha^\text{A,B})$ and $U$ as $\exp(i g a_\perp \alpha)$, for small $a_\perp$ we get \cite{Krasnitz:1998ns}
\begin{equation}
\alpha=\alpha^{\text{A}}+\alpha^{\text{B}},
    \label{eq:alpha_after_collision}
\end{equation}
as required.

In the case of $N_c=2$ there is an exact solution of \eqref{10} for $U_{\bm{x},i}$ \cite{Krasnitz:1998ns}. First of all, let us prove the simple following lemma: the sum of any two matrices belonging to SU(2) is a unitary matrix itself, times a real scalar (which can be zero). Let $U,V\in $ SU(2), and define $X\equiv U^\dagger\, V$. Then $X$ is unitary itself with $\det X=1$, so its eigenvalues are $e^{i\theta}$ and $e^{-i\theta}$, for some real $\theta$. This means that $X=P\, \text{diag}(e^{i\theta},e^{-i\theta})P^\dagger$ for some unitary $P$, hence
\begin{equation}
X+X^\dagger=P\,\text{diag}(e^{i\theta}+e^{-i\theta},e^{i\theta}+e^{-i\theta})P^\dagger=2\cos \theta\, P\,\mathbb{1}\,P^\dagger=2\cos\theta\,\mathbb{1}.
    \label{eq:X+Xdagger}
\end{equation}
This means that
\begin{equation}
(U+V)^\dagger(U+V)=\mathbb{1}+X+X^\dagger+\mathbb{1}=(2+2\cos\theta)\mathbb{1}.
    \label{eq:U+V_times_U+V}
\end{equation}
If $c^2\equiv 2+2\cos\theta=0$ then we have that $(U+V)^\dagger(U+V)=0$, that is $U+V=0$. Otherwise, the matrix defined as $W\equiv (U+V)/c$ satisfies $W^\dagger W=\mathbb{1}$, so indeed $U+V=cW$ where $W$ is unitary.
Once we proved this lemma, we can show that the following matrix satisfies Eq. \eqref{10}:
\begin{equation}
U_{\bm{x},i}=(U_{\bm{x},i}^{\text{A}}+U_{\bm{x},i}^{\text{B}})\,(U_{\bm{x},i}^{\text{A},\dagger}+U_{\bm{x},i}^{\text{B},\dagger})^{-1}.
\label{eq:total_gauge_link_SU2}
\end{equation}
Indeed, it is well defined, it is unitary by the virtue of the previous lemma and we can prove that it solves Eq. \eqref{10} by direct substitution:
\begin{align}
&\text{Tr}\left[t^a(U_{\bm{x},i}^{\text{A}}+U_{\bm{x},i}^\text{B})[\mathbb{1}+(U_{\bm{x},i}^{\text{A}}+U_{\bm{x},i}^{\text{B}})^{-1}\,(U_{\bm{x},i}^{\text{A},\dagger}+U_{\bm{x},i}^{\text{B},\dagger})]-\text{h.c.}\right]\nonumber\\
=&\text{Tr}\left[t^a(U_{\bm{x},i}^{\text{A}}+U_{\bm{x},i}^\text{B})+t^a
(U_{\bm{x},i}^{\text{A},\dagger}+U_{\bm{x},i}^{\text{B},\dagger})-(U_{\bm{x},i}^{\text{A},\dagger}+U_{\bm{x},i}^{\text{B},\dagger})t^a-(U_{\bm{x},i}^{\text{A}}+U_{\bm{x},i}^\text{B})t^a\right]\nonumber\\
=&\text{Tr}\left[t^a(U_{\bm{x},i}^{\text{A}}+U_{\bm{x},i}^\text{B})-(U_{\bm{x},i}^{\text{A}}+U_{\bm{x},i}^\text{B})t^a\right]+\text{Tr}\left[t^a
(U_{\bm{x},i}^{\text{A},\dagger}+U_{\bm{x},i}^{\text{B},\dagger})-(U_{\bm{x},i}^{\text{A},\dagger}+U_{\bm{x},i}^{\text{B},\dagger})t^a\right]\nonumber\\
=&0+0=0, \label{eq:proof_total_gauge_link_SU2}
\end{align}
by the virtue of the cyclic property of the trace.

On the other hand, for SU(3) there is no exact solution: the matrix in Eq. \eqref{eq:total_gauge_link_SU2} may not be even defined since $(U_{\bm{x},i}^{\text{A},\dagger}+U_{\bm{x},i}^{\text{B},\dagger})$ is not guaranteed to be invertible, and even if this was the case, the $U_{\bm{x},i}$ hereby obtained may not be unitary. For this reason, we have to solve \eqref{10} via an iterative method. This method has been developed in \cite{Cautun_thesis}: we will apply it to SU(3), but it can in principle be applied to SU($N_c$) for any number of colors $N_c$. In the following, for simplicity of notation, we drop the subscripts $\bm{x}$ and $i$, meaning that the following procedure has to be followed for every point $\bm{x}$ in the transverse plane and for both indices $i=x,y$. The first step of the algorithm involves taking a good initial guess for the link matrix $U$. We take as initial guess the naive estimate
\begin{equation}
    U = U^{\text{A}}\,U^{\text{B}}, \label{eq:initial_guess}
\end{equation}
which is the solution of \eqref{10} for the abelian case.  For each step, we insert our previous guess of $U$ in Eq. \eqref{10}, resulting in:
\begin{equation}
    \Im \,\text{Tr} \left[ t^a (U^{\text{A}} + U^{\text{B}})(1 + U^{\dagger}_{\text{old}}) \right] = f^a \label{eq:numerical_ev_initialgauge_1}
\end{equation}
where $f^a$ is a set of real numbers. If the $U_{\text{old}}$ at the current step was the exact solution of \eqref{10} then we would have $f^a=0$ for each $a$. On the other hand, this is often not the case, and we want to minimize the set of $f^a$ with respect to some measure. To do so, we want to improve our current guess for $U_{\text{old}}$. Since we want to remain in SU(3), the solution of Eq. \eqref{10} may be approached by considering
\begin{equation}
    U_{\text{new}} = e^{i x^a t^a} U_{\text{old}} \simeq (1+ix^a t^a+\dots)\,U_{\text{old}},\label{eq:numerical_ev_initialgauge_2}
\end{equation}
for a certain choice of the vector $x^a$. Using only the first two terms in the Taylor expansion of the exponential in Eq. \eqref{eq:numerical_ev_initialgauge_2}
and imposing the condition that $U_{\text{new}}$ satisfies Eq. \eqref{10}, i.e. 
\begin{equation}
    \Im \,\text{Tr} \left[ t^a (U^{\text{A}} + U^{\text{B}})(1 + U^{\dagger}_{\text{new}}) \right] = 0,
    \label{eq:numerical_ev_initialgauge_2.1}
\end{equation}
we have
\begin{align}
    &\Im \,\text{Tr} \left[ t^a (U^{\text{A}} + U^{\text{B}})(1 + U^{\dagger}_{\text{new}}) \right]\nonumber\\
    =&\Im \,\text{Tr} \left\{ t^a \left[U^{\text{A}} + U^{\text{B}}\right]\left[1 + U^{\dagger}_{\text{old}}(1-i x^b t^b)\right] \right\} \nonumber\\
    =&\Im \,\text{Tr} \left[ t^a (U^{\text{A}} + U^{\text{B}})(1 + U^{\dagger}_{\text{old}}) \right]-x^b \, \Im \, \text{Tr} \left[i\,t^b t^a (U^{\text{A}} + U^{\text{B}}) U^{\dagger}_{\text{old}} \right]\nonumber\\
    =&f^a-x^b \, \Re \, \text{Tr} \left[t^b t^a (U^{\text{A}} + U^{\text{B}}) U^{\dagger}_{\text{old}} \right]=0,
    \label{eq:numerical_ev_initialgauge_2.2}
\end{align}
so the components $x^b$ must satisfy
\begin{equation}
    x^b \, \Re  \text{Tr} \left[ t^b t^a (U^{\text{A}} + U^{\text{B}}) U^{\dagger}_{\text{old}} \right] = f^a.
    \label{eq:numerical_ev_initialgauge_3}
\end{equation}
This linear system of equations can be easily solved for $x^b$, which means that we can calculate $U_{\text{new}}$ using \eqref{eq:numerical_ev_initialgauge_2}. Notice that, since $U_{\text{new}}$ must still be an element of SU(3), to compute it from $U_{\text{old}}$ we must use Eq. \eqref{eq:numerical_ev_initialgauge_2} without any approximations for the exponential. Using the newly derived guess for $U$, the procedure starts again from \eqref{eq:numerical_ev_initialgauge_1}, to get a new set of $f^a$ and so on. The procedure stops once a desired precision is reached.

How fast do we expect the algorithm to converge? After one step, we move from a set of $f^a$ to a new set $f'^a$ given by (see Eq. \eqref{eq:numerical_ev_initialgauge_1}):
\begin{equation}
    \Im  \, \text{Tr} \left[ t^a (U^{\text{A}} + U^{\text{B}})(1 + U^{\dagger}_{\text{new}}) \right] = f'^a.
    \label{eq:numerical_ev_initialgauge_4}
\end{equation}
We can relate those two vectors by using the first two terms of the Taylor expansion for the exponential in Eq. \eqref{eq:numerical_ev_initialgauge_2}. The first non-zero term reads
\begin{equation}
    f'^a = \frac{1}{2} x^b x^c \, \text{Re} \, \text{Tr} \left[ t^b t^c t^a (U^{\text{A}} + U^{\text{B}}) U^{\dagger}_{\text{old}} \right].\label{eq:numerical_ev_initialgauge_5}
\end{equation}
This equation can give us a rough estimate on the speed of the algorithm. Close to the continuum limit, the Wilson lines can be approximated by unit matrices. This means that \eqref{eq:numerical_ev_initialgauge_3} reduces approximately to 
$x^a = f^a$, hence Eq. \eqref{eq:numerical_ev_initialgauge_5} can be simplified as
\begin{equation}
    f'^a = \frac{1}{4} f^b f^c d^{abc},
    \label{eq:numerical_ev_initialgauge_6}
\end{equation}
where $d^{abc}$ are the symmetric structure constants (see Eq. \eqref{eq:Tr_TaTbTc} in Appendix \ref{app:conv}). If the initial guess was a good one, then $f^a$ should be small, in which case the convergence of $f'^a$ is exponential in the number of steps, see \eqref{eq:numerical_ev_initialgauge_6}. By iterating the above process, one can get a converging algorithm for finding the solution of Eq. \eqref{10}. When we are far from the continuum limit, one can still expect $f'^a$ to be roughly proportional to $f^b f^c$ also in the general case, but the exact dependence will be more complicated. In particular, when the initial guess given by Eq. \eqref{eq:initial_guess} is not a very good one, the general method which we have described may not be particularly efficient.
This occurs because, as it is, this method usually overshoots the correct solution: by solving \eqref{eq:numerical_ev_initialgauge_3} we get the right direction
in which to move in order to get $f_a$ closer to zero, but the size of the move is often not the correct one. We will therefore modify the algorithm via the insertion of a factor $\alpha$ in the exponent of \eqref{eq:numerical_ev_initialgauge_2}, as follows
\begin{equation}
    U_{\text{new}} = e^{i \alpha x^a t^a} U_{\text{old}}. \label{eq:numerical_ev_initialgauge_improvement}
\end{equation}
This allows to overcome the initial stalemate, which occurs especially for lattices which are not fine enough and therefore \eqref{eq:initial_guess} is not very accurate.

Going into the details, we fix $\alpha=0.9$ in all our calculations. As far as the desired precision is concerned, keep in mind that for each point $\bm{x}$ in the transverse plane we will have two gauge links, so for each $\bm{x}$ and at each step of the procedure we will have two sets $f_x^a$ and $f_y^a$ of vectors, where $a=1,\dots N_c^2-1$. We have fixed a threshold precision as
\begin{equation}
    \sqrt{\sum_{a=1}^{N_c^2-1} f^a_x f^a_x}+\sqrt{\sum_{a=1}^{N_c^2-1} f^a_y f^a_y}<10^{-10},
    \label{eq:threshold_precision}
\end{equation}
which means that as soon as the sum of the norms of the two vectors gets smaller than $10^{-10}$ the procedure stops and we accept the gauge link that we obtain from this procedure. In case this does not occur fast enough, we stop the procedure anyway after a maximum number of iterations equal to 50. Both those values have been fixed in order to achieve a good compromise between speed and numerical precision, especially when we have a small number of lattice points, for which this procedure takes a significant percentage of the total computational time. On the other hand, when the lattice is fine this procedure is resolved in very few steps. Indeed, for fine lattices the naive abelian estimate \eqref{eq:initial_guess} is already very good: the exponential of the total gauge field $\alpha$ is basically the product of the exponential of the single nuclei gauge fields $\alpha^{\text{A,B}}$, since the higher order terms in the non-abelian Baker-Campbell-Hausdorff formula are proportional to $a_\perp$ and therefore small.

Using the above procedure, for each point $\bm{x}$ in the transverse plane we initialize
$U_{\bm{x},i}$ for both transverse directions $i=x$ and $i=y$. As far as the longitudinal component is concerned, at initial time $\tau=0^+$ the $\eta$ component of the gauge field is initialized as $U_{\bm{x},\eta}=\mathbb{1}$. Along with the gauge fields (which, we will see, will be needed to derive the color-magnetic field), we also need an initial expression for the color-electric fields, which are the canonical momenta of the gauge links. The initial transverse electric fields, as we have already mentioned, vanish:
\begin{equation}
E_x=E_y=0.
\label{eq:transverse_electric_fields_initial_time}
\end{equation}
On the other hand, the longitudinal component is of course non zero at initial time.
Its expression on the lattice is implemented as \cite{Krasnitz:1998ns,Fukushima:2011nq}
\begin{align}
    E^\eta=-\frac{i}{4ga_\perp^2}\sum_{i=x,y}\big[&(U_i(\bm{x})-\mathbb{1})(U_i^{\text{B},\dagger}(\bm{x})-U_i^{\text{A},\dagger}(\bm{x}))+\nonumber\\
   &(U_i^\dagger(\bm{x}-\bm{i})-\mathbb{1})(U_i^{\text{B}}(\bm{x}-\bm{i})-U_i^{\text{A}}(\bm{x}-\bm{i}))-\text{h.c.}\big].
\label{eq:longitudinal_electric_fields_initial_time}
\end{align}
It is easily seen that the above equation has the correct formal continuum limit. Indeed, if we write again $U^{\text{A,B}}_i$ as $\exp \normalsize(i g a_\perp \alpha^\text{A,B}_i\normalsize)$ and $U_i$ as $\exp(i g a_\perp \alpha_i)$, with $\alpha_i=\alpha^A_i+\alpha^B_i$, for small $a_\perp$ we get  
\begin{equation}
(U_i(\bm{x})-\mathbb{1})(U_i^{\text{B},\dagger}(\bm{x})-U_i^{\text{A},\dagger}(\bm{x}))-\text{h.c.}\simeq -2g^2a_\perp^2[\alpha^\text{A}_i,\alpha^\text{B}_i],
\label{eq:electric_field_continuum_limit_initialtime}
\end{equation}
where no summation over $i$ is intended here. Hence, in the limit of smooth fields, \eqref{eq:longitudinal_electric_fields_initial_time} reduces to \eqref{eq:bi_initial_EL_BL_1}.\\

We now discuss the time evolution of the fields whose initialization has just been illustrated. The electric fields are expressed in terms of link variables as \cite{Fukushima:2011nq}
\begin{align}
    \partial_\tau U_i(\bm{x})&=\frac{-iga_\perp}{\tau}E^i(\bm{x})U_i(\bm{x}),\label{eq:parisihatoltoilginger}\\
    \partial_\tau U_\eta(\bm{x})&=-iga_\eta\tau E^\eta(\bm{x}) U_\eta(\bm{x}).
    \label{11.3}
\end{align}
The term $a_\eta$ denotes the discretization step in the $\eta$-direction.

In order to reduce the discretization error in time, we drive the evolution through a leapfrog algorithm, i.e. by letting the gauge links and the electric fields evolve in different time steps alternatively. Beyond being a second-order accurate method in the time step $\Delta \tau$, the leapfrog scheme also possesses a crucial structural property: it is {\sl symplectic}. Symplectic integrators are designed to preserve the geometric structure of Hamiltonian systems, particularly the conservation of phase space volumes (Liouville's theorem). In practical terms, this implies that while the leapfrog method may introduce small, bounded oscillations in conserved quantities such as the energy, it avoids systematic drift. This is in stark contrast with standard Runge-Kutta methods (including RK4), which—despite their higher local accuracy—tend to violate energy conservation over long integration times.
Leapfrog's time-reversal invariance and symplectic nature make it particularly well suited for classical Yang-Mills dynamics, especially in the early stages of heavy-ion collisions where field interactions are highly nonperturbative and nonlinear, and symmetry preservation is critical. A non-symplectic integrator can lead to artificial energy growth or damping, ultimately corrupting the physical evolution of the system. 
For more on symplectic methods and their applications to Hamiltonian dynamics, we refer the reader to \cite{Hairer:Geometric,Leimkuhler:Reich,Young:Leapfrog}, where their properties are explored in depth.

\begin{figure}
    \centering
    \includegraphics[width=0.4\linewidth]{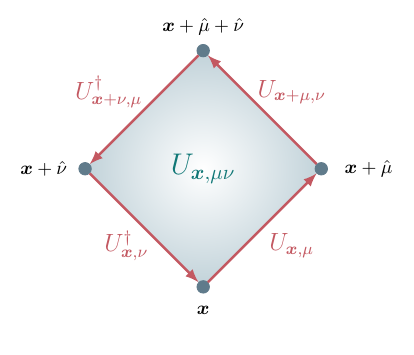}
    \caption{Representation of a plaquette variable $U_{\mu\nu}(\bm{x})$. The gauge links, starting from a point $\bm{x}$, make a closed loop around the neighbor points.}
    \label{fig:plaquette}
\end{figure}

The discretized time evolution for $U_i$ and $U_\eta$, at each position, is given by \cite{Fukushima:2011nq}
\begin{align}
U_i(\tau'')&=\exp\left[-iga_\perp\, \Delta \tau \, E^i(\tau')/\tau'\right]U_i(\tau),\label{11.4bisAAA}\\
U_\eta(\tau'')&=\exp\left[-ig a_\eta\, \Delta \tau\, E^\eta(\tau')\, \tau'\right]U_\eta(\tau),
\label{11.4}
\end{align}
where $\tau'=\tau+\Delta \tau/2$ and $\tau''=\tau+\Delta \tau$. Notice that the exponentiation of the electric field is important in order to keep the up-to-date gauge links as unitary matrices. In the same fashion, the equations of motion for the electric field at each position are discretized as:

\begin{align}
E^i(\tau')=E^i(\tau-\Delta \tau/2)+&\Delta \tau \frac{i}{2ga_\eta^2 a_\perp\tau}[U_{\eta i}(\tau)+U_{-\eta i}(\tau)-\text{h.c.}]\nonumber\\
+&\Delta \tau \frac{i\tau}{2g a_\perp^3}\sum_{j\neq i}[U_{ji}(\tau)+U_{-ji}(\tau)-\text{h.c.}],\nonumber\\
E^\eta(\tau')=E^\eta(\tau-\Delta \tau/2)+&\Delta \tau \frac{i}{2g a_\eta a_\perp^2 \tau} \sum_{j=x,y}[U_{j\eta}(\tau)+U_{-j\eta}(\tau)-\text{h.c.}],\label{12}
\end{align}
where
\begin{equation}
    U_{\mu\nu}(\bm{x})\equiv U_\mu(\bm{x})U_\nu (\bm{x}+\bm{\mu})U_\mu^\dagger(\bm{x}+\bm{\nu})U_\nu^\dagger(\bm{x})
    \label{13}
\end{equation}
are the {\sl plaquette} variables (see Fig. \ref{fig:plaquette}). 
Here $\bm{\mu}$ and $\bm{\nu}$ denote unit vectors along the respective directions. Notice that in this way the gauge links are initialized at time $\tau=0$, and evolve with integer steps: $\tau=\Delta \tau, \, 2\Delta \tau, \dots$ Instead, the electric field is initialized at $\tau=-\Delta \tau/2$ and is let evolve along half-integer time steps $\tau=\Delta \tau/2,\, 3\Delta \tau/2,\, 5 \Delta \tau/2,\dots$ At each iteration, the electric fields are updated using the gauge links at integer time, and subsequently the gauge links are updated using the electric fields at half-integer time. It is important to note that the initial condition for the electric fields is assumed to be given at time $\tau=-\Delta \tau/2$, even if for practical purposes the available configuration is defined at $\tau=0$. This shift of half a time step is consistent with the leapfrog scheme and introduces only an $\mathcal{O}(\Delta \tau^2)$ discretization error, which is within the overall accuracy of the algorithm. This convention allows one to start the evolution without the need to explicitly compute a backward half-step.

Along with the electric field, using the gauge links we can derive for each point in space the components of the color-magnetic field as \cite{Fukushima:2011nq}
\begin{align}
    B_x^a(\tau)&=\frac{2}{iga_\eta \tau a_\perp}\text{Tr}[t^a(\mathbb{1}-U_{\eta y}(\tau))]\label{eq:magnetic_field_x},\\
    B_y^a(\tau)&=\frac{2}{iga_\eta \tau a_\perp}\text{Tr}[t^a(\mathbb{1}-U_{\eta x}(\tau))]\label{eq:magnetic_field_y},\\
    B_z^a(\tau)&=\frac{2}{iga_\perp^2}\text{Tr}[t^a(\mathbb{1}-U_{xy}(\tau))].\label{eq:magnetic_field_z}
\end{align}

\section{Results and plots}

Once we have outlined the basics and the numerical details of the formalism, we now show our numerical results for various quantities of fundamental interest in the study of glasma. These are the energy density, the pressure components, the color electric and magnetic fields in the early stages of ultrarelativistic heavy-ion collisions.

The electric fields and gauge fields which we have dealt with so far fit into the definition of the longitudinal/transverse electric and magnetic components of the energy density. In particular, at time $\tau$ we have the electric components of the energy density given by
\begin{align}
E_L^2(\bm{x},\tau)&\equiv E^{\eta}(\bm{x},\tau)\, E^{\eta}(\bm{x},\tau),\nonumber\\
E_T^2(\bm{x},\tau)&\equiv \frac{1}{\tau^2}[E^{x}(\bm{x},\tau)\, E^{x}(\bm{x},\tau)+E^{y}(\bm{x},\tau)\, E^{y}(\bm{x},\tau)].\label{13.1}
\end{align}
The magnetic components of the energy density are given by\footnote{Note that these can be also obtained from Eqs. \eqref{eq:magnetic_field_x}, \eqref{eq:magnetic_field_y} and \eqref{eq:magnetic_field_z} \cite{Fukushima:2011nq}. In particular, $B_z=B_\eta$ since they are both equal to $F_{xy}$, and the change of coordinates from $x^{\mu'}=(\tau,\eta)$ to $x^\mu=(t,z)$ in Eqs. \eqref{eq:milne_coordinate_transformation} and \eqref{eq:milne_inverse_coordinate_transformation} does not affect the transverse components.}
\begin{align}
    B_L^2(\bm{x},\tau)&=\frac{2}{g^2a_\perp^4}\left[\mathbb{1}-U_{xy}(\bm{x},\tau)\right],\nonumber\\
    B_T^2(\bm{x},\tau)&=\frac{2}{(ga_\eta a_\perp\tau)^2}\sum_{i=x,y}\left[\mathbb{1}-U_{\eta i}(\bm{x},\tau)\right].\label{13.2}
\end{align}
The above quantities fit into the definition of $\varepsilon$ \cite{Fukushima:2011nq}
\begin{equation}
    \varepsilon(\bm{x},\tau)= \text{Tr}[E_L^2(\bm{x},\tau)+B_L^2(\bm{x},\tau)+E_T^2(\bm{x},\tau)+B_T^2(\bm{x},\tau)].
    \label{13.3}
\end{equation}

\begin{figure}[t]
\includegraphics[width=\textwidth]{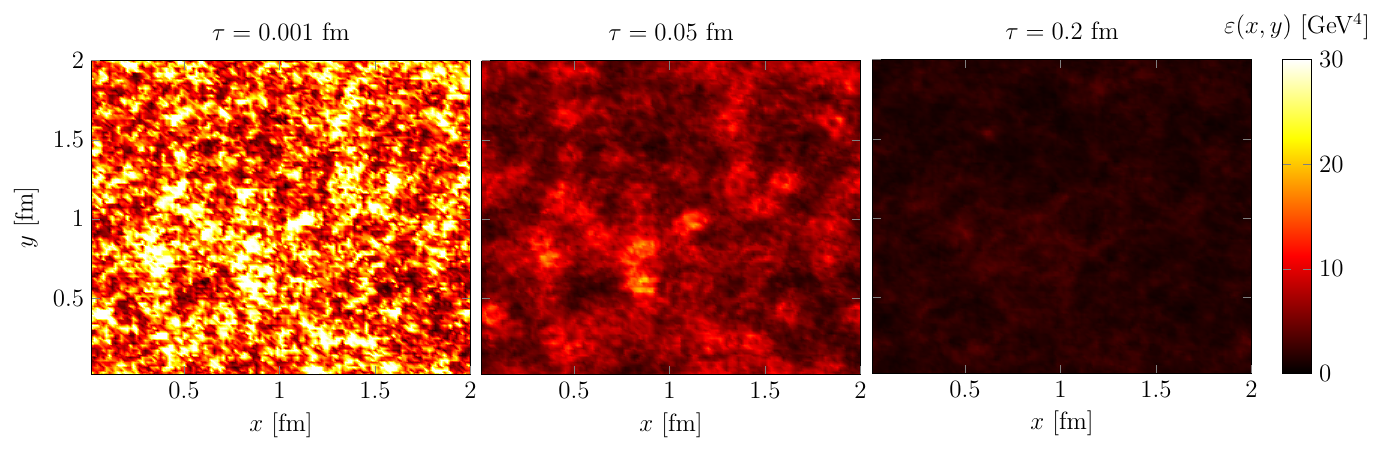}
\caption{Energy density (legend on the right side) produced in a AA collision
in one event, represented in the transverse plane for three values of proper time $\tau$. The choice of the parameters is $L_\perp=2$ fm and with $N_\perp=128$ points.}
        \label{Fig:energy_dens_AA}
\end{figure}

\begin{figure}[t]
\includegraphics[width=\textwidth]{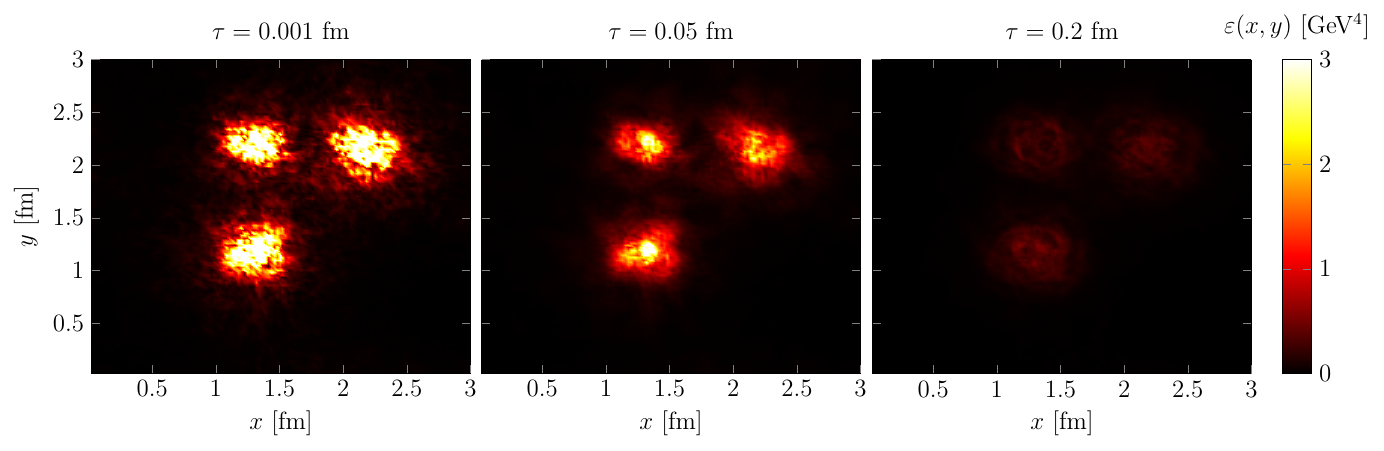}
\caption{Energy density (legend on the right side) produced in a pA collision
in one event, represented in the transverse plane for three values of proper time $\tau$. The choice of the parameters is $L_\perp=2$ fm and with $N_\perp=128$ points.}
        \label{Fig:1}
\end{figure}

In Figures \ref{Fig:energy_dens_AA} and \ref{Fig:1} we plot the profile in the transverse plane of the energy density, for a nucleus-nucleus and a proton-nucleus collision, respectively. These simulations have been performed in a lattice with transverse size $L_\perp=2$ fm and with $N_\perp=128$ points. In the left-most panel we show the energy density right after the collision ($\tau=0.001$ fm), which closely resembles the charge profile as initialized from the McLerran-Venugopalan model. In particular, the energy density in the AA case has a white noise profile, whereas in the pA case we notice the energy density concentrated over the three lumps, standing for the three constituent quarks in the proton. As time passes (in our Figures going from left to right), the energy density gets diluted, due to the longitudinal expansion of the system. The transverse expansion, instead, as one can see for instance in the pA case of Fig. \ref{Fig:1}, is negligible.

\begin{figure}[ht]
    \centering
\includegraphics[width=0.5\linewidth]{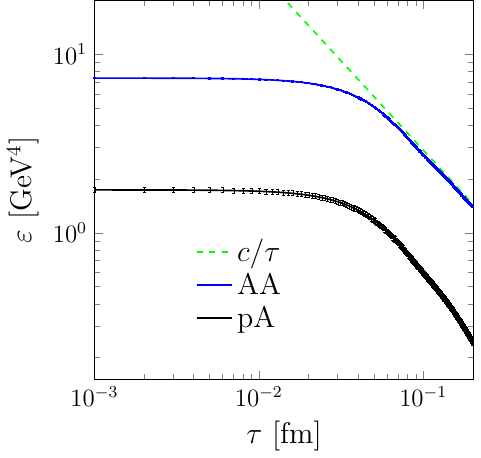}
    \caption{Energy density $\varepsilon$, versus 
    proper time $\tau$, averaged over numerical events and over the transverse plane (see main text for details on the transverse-plane average). Both AA and pA results are shown, along with their error bars (for the AA case the error is smaller than the line width). We also highlight a $1/\tau$ behavior at late times, typical of a free streaming regime. The choice of the parameters is $N_{\text{events}}=100$, $L_\perp=2$ fm and $N_\perp=32$.}
    \label{Fig:2}
\end{figure}

In Fig. \ref{Fig:2} we plot $\varepsilon$ versus $\tau$ 
for both AA and pA collisions,
averaged over $100$ events and over the transverse plane, for $L_\perp=2$ fm and $N_\perp=32$. The average over the transverse plane, in the AA case, is straightforwardly the mean over the points of the lattice, with equal weights. On ther other hand, in order to perform the transverse plane average consistently in the pA case, we have used the thickness function of the proton $T_p(\bm{x}_\perp)$ (already introduced in Fig. \eqref{eq:bd1}) as a weight. In other words, in Fig. \ref{Fig:2} we plot $\langle \langle \text{Tr}[E_L^2+B_L^2+E_T^2+B_T^2] \rangle \rangle$, where $\langle \langle A \rangle \rangle$ for any (transverse--space dependent) quantity $A$ is defined as 
\begin{equation}
\langle\langle
 A
\rangle\rangle =
\int d^2\bm{x}_\perp\langle A(\bm{x}_\perp)\;T_p(\bm{x}_\perp)\rangle,
\label{eq:rai3_22_02}
\end{equation} 
and the symbol $\langle \cdot \rangle$ indicates the average over the numerical events.
What we see, first of all, is that at initial time the mean energy densities in the AA and in the pA case are substantially different. This can be qualitatively seen also by comparing Figs. \ref{Fig:energy_dens_AA} and \ref{Fig:1}. The actual values, in particular, are $\varepsilon=7.4$ GeV$^4$ for nucleus-nucleus collisions and $\varepsilon=1.8$ GeV$^4$ for proton-nucleus collisions. In terms of energy over volume, those translate into around 960 GeV/fm$^3$ for AA and 240 GeV/fm$^3$ for pA. Moreover, the log-log scale let us notice that for $\tau\approx 0.1$ fm (i.e. around $1/Q_s$, being $Q_s=g^2\mu=2$ GeV) we have the onset of a free streaming regime, which is highlighted by a power-law behavior of the energy density as $1/\tau$ (green dashed line in Fig. \ref{Fig:2}). Along with the event averages, here we also plot the statistical uncertainty as an error band, obtained as the standard deviation of the mean value. More specifically, all the error bars in this Thesis have been obtained, from the $\sigma$ of the distribution of the numerical events, as
\begin{equation}
    \sigma_{\text{avg}}=\sigma/\sqrt{N_{\text{events}}}.
    \label{eq:sigma_over_the_mean}
\end{equation}
We see that in Fig. \ref{Fig:2} the uncertainties are quite small. Remarkably, for nucleus-nucleus collisions the error bar falls within the thickness of the blue line itself.

\begin{figure}[ht]
    \centering
\includegraphics[scale=0.77]{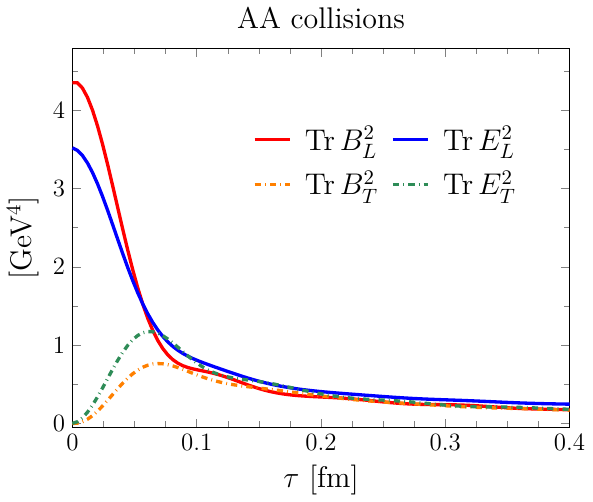}
\includegraphics[scale=0.77]{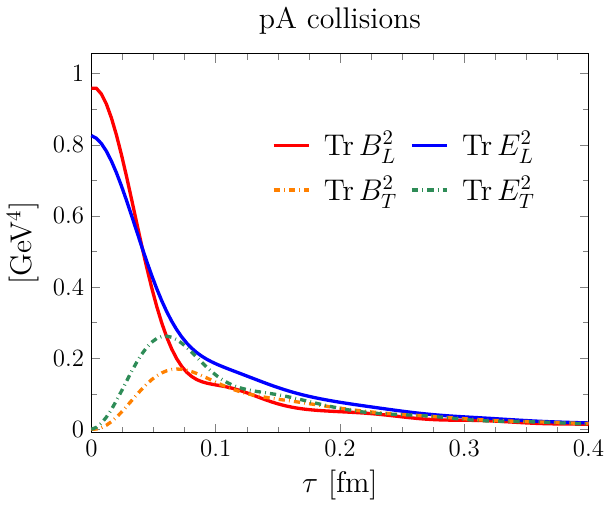}
    \caption{Electric and magnetic components of the energy density versus 
    proper time $\tau$, averaged over numerical events and over the transverse plane. We show results for both AA \textbf{(left)} and pA \textbf{(right)} collisions: see main text for details on the transverse-plane average for pA collisions. The choice of the parameters is $N_{\text{events}}=50$, $L_\perp=2$ fm and $N_\perp=32$.}
    \label{fig:electric_magnetic_fields_vst}
\end{figure}

In Fig. \ref{fig:electric_magnetic_fields_vst} we show, for both AA and pA collisions, the time behavior of the color-electric and color-magnetic components of the energy density, defined in \eqref{13.1} and \eqref{13.2}. These simulations have been performed in a lattice with transverse size $L_\perp=2$ fm, with $N_\perp=32$ points and $N_{\text{events}}=50$. First of all, let us note that the only components which are non-zero at initial time are the longitudinal ones, as already expected theoretically (cfr. Section \ref{sec:glasma_initial}). From their maximum values, these contributions to the energy density start to decrease. As far as the transverse components are concerned, those start from zero and rapidly increase, reaching a maximum at around $\tau\simeq 0.06$ fm. Around this value of $\tau$, all the four contributions to the energy density become equal, and from $\tau\simeq 0.1$ fm onwards the four curves follow the same behavior. These features are shared among AA and pA collisions.\\

\begin{figure}[ht]
    \centering
\includegraphics[scale=0.76]{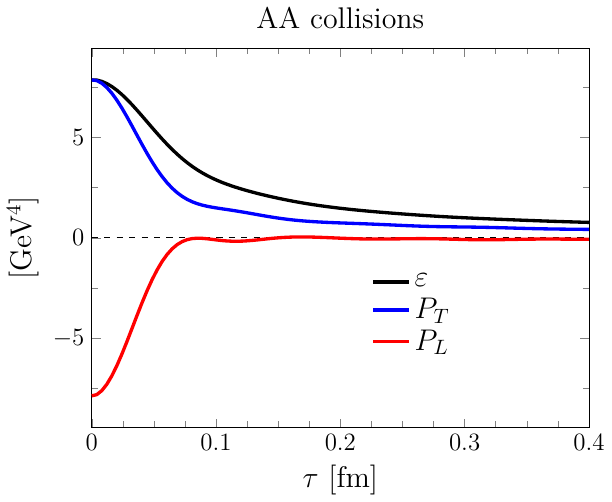}
\includegraphics[scale=0.76]{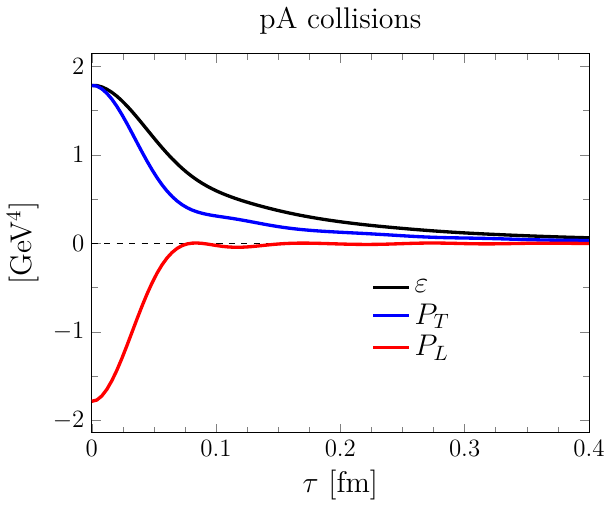}
   \caption{Transverse and longitudinal pressures $P_T,P_L$, and energy density $\varepsilon$ versus 
    proper time $\tau$, averaged over numerical events and over the transverse plane. We show results for both AA \textbf{(left)} and pA \textbf{(right)} collisions: see main text for details on the transverse-plane average for pA collisions. The choice of the parameters is $N_{\text{events}}=50$, $L_\perp=2$ fm and $N_\perp=32$.}
    \label{fig:PL_PT_eps_vst}
\end{figure}

Once we have introduced the time behavior of the color electric and magnetic fields, in Fig. \ref{fig:PL_PT_eps_vst} we show the time behavior of the average transverse and longitudinal pressures $P_T$ and $P_L$, them being defined as:
\begin{align}
P_T=&\langle \langle \text{Tr}[E_L^2+B_L^2] \rangle \rangle\nonumber,\\
P_L=&\langle \langle \text{Tr}[E_T^2+B_T^2-E_L^2-B_L^2] \rangle \rangle.
    \label{eq:PL_PT_eps_vst}
\end{align}
The results have been obtained in a lattice with transverse size $L_\perp=2$ fm, with $N_\perp=32$ points and $N_{\text{events}}=50$. Starting from the transverse component, we immediately see that it starts at a non-zero value: this coincides with the initial value of the energy density (also shown in Fig. \ref{fig:PL_PT_eps_vst} for reference), since the transverse components of the fields are zero, check Eqs. \eqref{13.3} and \eqref{eq:PL_PT_eps_vst}. As time passes, we see a decrease of $P_T$, which is seen to be faster than the decrease of $\varepsilon$: this occurs because at $\tau>0$ we have the emergence of the transverse fields, which appear in the expression for $\varepsilon$, but they do not appear in $P_T$. Moving on to the longitudinal component of the pressure $P_L$, a striking feature is that its initial value is negative, and equal to $P_L(\tau=0^+)=-P_T(\tau=0^+)=-\varepsilon(\tau=0^+)$. A negative pressure is a sign of a contracting system: indeed, the longitudinal color flux tubes work against the expansion of the two nuclei. Once we have the onset of also the transverse fields, the negative contribution gets smaller and smaller in absolute value. Once again, at times of the order of $1/Q_s$, all the contributions get dilute and close to zero.\\

\begin{figure}[ht]
    \centering
\includegraphics[scale=0.77]{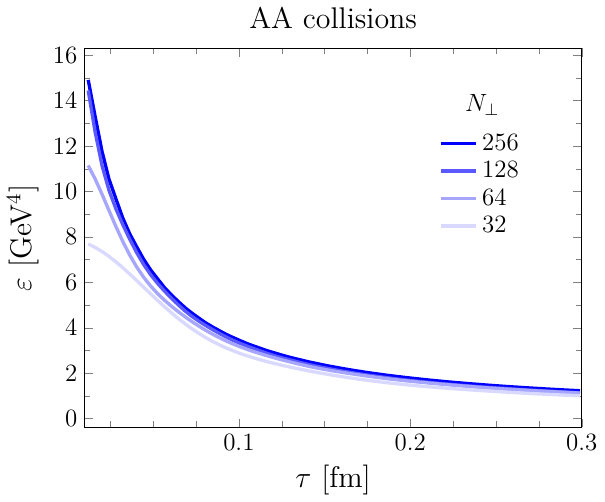}
\includegraphics[scale=0.77]{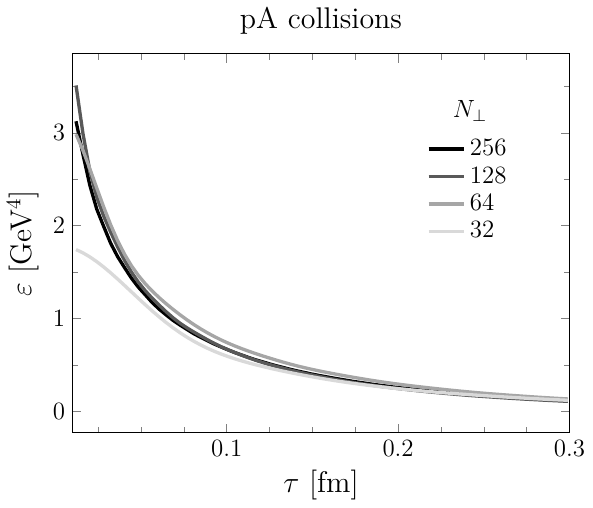}
    \caption{Energy density $\varepsilon$ versus 
    proper time $\tau$, averaged over numerical events and over the transverse plane, for different choices of the number of points in the transverse lattice $N_\perp$. We show results for both AA \textbf{(left)} and pA \textbf{(right)} collisions: see main text for details on the transverse-plane average for pA collisions. The choice of the parameters is $N_{\text{events}}=50$ and $L_\perp=2$ fm.}
    \label{fig:eps_vs_Nt}
\end{figure}

We now move on to discuss the dependence of our quantities on the numerical parameters of our simulation. In Fig. \ref{fig:eps_vs_Nt} we plot $\varepsilon$ versus $\tau$ for different numbers of lattice points in the transverse plane. We see that, at very small $\tau$, the energy density has a non-physical dependence on $N_\perp$, which is well known in the literature \cite{Fukushima:2011nq,Lappi:2003bi}. From $\tau\sim 0.05$ fm onwards, in both AA and pA collisions, we see that the discrepancy in $N_\perp$ vanishes, and all the curves follow the same behavior.

\begin{figure}[ht]
    \centering
\includegraphics[scale=0.77]{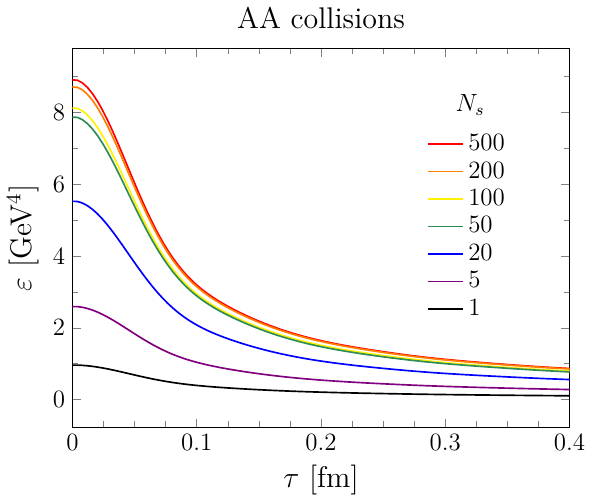}
\includegraphics[scale=0.77]{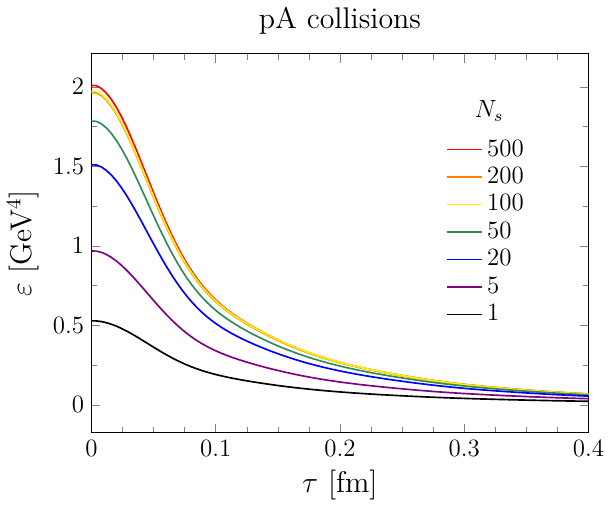}
    \caption{Energy density $\varepsilon$ versus 
    proper time $\tau$, averaged over numerical events and over the transverse plane, for different choices of the number of color sheets $N_s$. We show results for both AA \textbf{(left)} and pA \textbf{(right)} collisions: see main text for details on the transverse-plane average for pA collisions. The choice of the parameters is $N_{\text{events}}=50$, $L_\perp=2$ fm and $N_\perp=32$.}
    \label{fig:eps_vs_N_sheets}
\end{figure}

Instead of varying $N_\perp$, we can also study the dependence on the number of color sheets $N_s$. The results are shown in Fig. \ref{fig:eps_vs_N_sheets}. For small $N_s$, we see that an increase in the number of sheets brings a significant contribution to the energy density, both at early and at late times. However, at around $N_s\sim 50$ we see that the results start to saturate, until the energy density does not present any dependence on $N_s$ anymore: we see that the three curves for $N_s=100,200,500$ in Fig. \ref{fig:eps_vs_N_sheets} are basically indistinguishable. Once again, the same considerations hold for both nucleus-nucleus and proton-nucleus collisions.

\chapter[HQs in anisotropically fluctuating glasma]{Heavy quarks in anisotropically fluctuating glasma}
\label{chap:Heavy quarks in anisotropically fluctuating glasma}

This Chapter is largely based on the work contained in \cite{Parisi:2025slf}:
\begin{center}
    Gabriele Parisi, Vincenzo Greco and Marco Ruggieri, ``Anisotropic fluctuations of momentum and angular momentum of heavy quarks in the pre-equilibrium stage of pA collisions at the LHC'', Phys. Rev. D \textbf{112} (2025) no.7, 074030, arXiv: 2505.08441 [hep-ph]. \\
\end{center}

So far, the success of the glasma simulations has been largely limited by the boost invariance assumption, which shows good agreement with experimental data only around the midrapidity region. Recently, much attention has been paid to the 3+1--dimensional (3+1D) glasma simulations,
which aim to go beyond the boost invariance assumption, as necessary to understand observables across a broader region of rapidity \cite{Pang:2012uw,Pang:2014pxa}. In fact, on top of the glasma it is possible to add quantum
fluctuations \cite{Fukushima:2013dma, Romatschke:2005pm, Romatschke:2006nk, Fukushima:2011nq, Iida:2014wea, Epelbaum:2013ekf, Epelbaum:2013waa, Ryblewski:2013eja, Ruggieri:2015yea,Tanji:2011di, Berges:2012cj, Berges:2013fga, Berges:2013lsa, Berges:2013eia}, which are known to trigger plasma instabilities \cite{Bazak:2023kol} and make the classical solutions of the Yang-Mills equations exponentially sensitive to their initial conditions \cite{Romatschke:2005pm,Romatschke:2006nk}. Moreover, fluctuations are helpful in the production of entropy during
the early stages of high energy nuclear collisions \cite{Iida:2014wea, Tsukiji:2016krj,Tsukiji:2017pjx,Matsuda:2022hok}. Quantum fluctuations appear when one considers the finite
coupling corrections to the glasma solution (which is obtained in the small coupling limit): the spectrum of these
fluctuations has been computed within a perturbative
calculation in \cite{Epelbaum:2013ekf} and it has been shown that they affect
both the gauge potential and the color electric field.

In this Chapter we simulate the real-time evolution of the SU(3)--glasma generated in the early stages of high-energy proton–nucleus collisions, employing the classical lattice gauge theory techniques we have discussed in §\ref{sec:Numerical implementation of CGC}. More specifically, 
our generalization incorporates not only a realistic modeling of the proton’s internal structure (via the modified MV initial condition, see §\ref{sec:Beyond naive MV model: pA collisions}), but also longitudinal fluctuations in the initial state, which enable the study of genuinely non-boost-invariant collision dynamics via full 3+1 D Yang-Mills simulations. What we will focus on, in particular, are the effect of such fluctuations on heavy quarks in the infinite mass limit. Indeed, in this particular limit case it can be shown that the accumulated momentum of hard probes in the glasma can be evaluated from glasma lattice field correlators \cite{Boguslavski:2020tqz,Avramescu:2023qvv}, without explicitly solving the particle equations of motion (i.e. the Wong equations). We will study the momentum and angular momentum anisotropies of heavy quarks in this limit, to look for effects of such anisotropy across a range of fluctuation amplitudes.

\section{Quarks of infinite mass in glasma}
\label{Quarks of infinite mass in glasma}

Let us deal with the diffusion of HQs in the evolving glasma fields. This problem has attracted a lot of interest recently,
see for example \cite{Ruggieri:2018ies,Ruggieri:2018rzi,Sun:2019fud,Liu:2019lac,Jamal:2020fxo,Liu:2020cpj,Sun:2021req,Khowal:2021zoo, Pooja:2022ojj,Pooja:2024rnn, Pooja:2024rsw, Avramescu:2024poa, Avramescu:2024xts} as well as references
therein.
So far, this problem has been solved using a classical approximation 
in which
the equations of motion for the HQs are the Wong equations~ \cite{Wong:1970fu, Heinz:1984yq}.
In this work, we limit ourselves to analyze the diffusion of heavy quarks
in the very-large mass limit, similarly to
what has been done in \cite{Boguslavski:2020tqz}. 

\subsection{Momentum broadening}
\label{sec:Momentum broadening}
One of the quantities we want to study 
is the {\sl momentum broadening}, $\delta p_i^2$, 
of the HQs. It has already been studied, for instance, in~\cite{Ruggieri:2018ies,Ruggieri:2018rzi,Sun:2019fud,Liu:2019lac,Jamal:2020fxo,Liu:2020cpj,Sun:2021req,Khowal:2021zoo, Pooja:2022ojj,Pooja:2024rnn, Pooja:2024rsw, Avramescu:2024poa, Avramescu:2024xts}. We analyze it in the present study, bringing the novelty of introducing longitudinal fluctuations. As a benchmark, we firstly show the results obtained without fluctuations.

The momentum broadening is defined as
\begin{equation}
\delta p_i^2(\tau)\equiv p_i^2(\tau)-p_i^2(\tau_{\text{form}}),~~~
i=x,y,z,
\label{eq14.1}
\end{equation}
where $\tau_{\text{form}}$ is the formation time of the HQ and $p^i(\tau_{\text{form}})$ the $i$-th component of the momentum at the formation time. 
In order to compute the momentum broadening, 
we need to specify the initial distribution of the HQs in the
configuration space. To this end, we assume that
HQs are produced near the hotspots of the energy density. The idea behind this claim is that heavy quarks, being quite massive, can only be produced by hard QCD scatterings in the regions with highest energy density and color charges (i.e. around the hotspots of the proton, see Fig.~\ref{Fig:1}). Hence, it is reasonable to distribute the
HQs according to the $T_p(\bm{x}_\perp)$ in Eq.~\eqref{eq:bd1}.
Following the same arguments presented
in~\cite{Avramescu:2023qvv}, from the Wong equations we can write the momentum broadening of each heavy quark,
in the $M\rightarrow\infty$ limit, as
\begin{align}
\delta p_L^2(\tau) &=
g^2\int_{0}^\tau d\tau' \int_{0}^\tau d\tau''\, \text{Tr}[E_z(\tau')E_z(\tau'')],
\label{eq:filmnoiososuamazonprime}\\
\delta p_T^2(\tau) &=
g^2\int_{0}^\tau d\tau' \int_{0}^\tau d\tau'' 
\,\frac{1}{\tau^\prime \tau^{\prime\prime}}
\text{Tr}[E_x(\tau')E_x(\tau'')+E_y(\tau')E_y(\tau'')],
\label{eq:italia1_21_34}
\end{align}
where we took the formation time of the infinitely massive HQs
equal to zero, since $\tau_{\text{form}}=\mathcal{O}(1/M)$ and $M\to \infty$. The integrands in the right-hand side of 
Eqs.~\eqref{eq:filmnoiososuamazonprime}
and~\eqref{eq:italia1_21_34} depend also on the transverse plane coordinates. Integrating over the whole transverse plane and ensemble-averaging 
Eqs.~\eqref{eq:filmnoiososuamazonprime} and~\eqref{eq:italia1_21_34},
results in
\begin{align}
\langle\langle\delta p_L^2(\tau)\rangle\rangle=&
g^2\int_{0}^\tau d\tau' \int_{0}^\tau d\tau'' \int d^2\bm{x}_\perp
\langle
\text{Tr}[E_z(\tau')E_z(\tau'')] T_p(\bm{x}_\perp)\rangle,
\label{eq:oraguardiquesto}\\
\langle\langle\delta p_T^2(\tau)\rangle\rangle =&
g^2\int_{0}^\tau d\tau' \int_{0}^\tau d\tau'' \int d^2\bm{x}_\perp
\frac{1}{\tau^\prime \tau^{\prime\prime}}
\langle \text{Tr}[E_x(\tau')E_x(\tau'')+E_y(\tau')E_y(\tau'')]T_p(\bm{x}_\perp)\rangle,
\label{eq15}
\end{align}
i.e. the definition of $\langle \langle \cdot \rangle \rangle$ already introduced in \eqref{eq:rai3_22_02}. In writing Eqs.~\eqref{eq:oraguardiquesto} and~\eqref{eq15}
we took advantage of the fact that the HQs are infinitely massive, therefore their distribution in the configuration space does not change in time. Let us notice that in the infinite mass limit Eqs.~\eqref{eq:oraguardiquesto} and~\eqref{eq15} are exact,
in the sense that the color-magnetic field gives no contribution
in this limit, as its effect is proportional to the
velocity of the quarks, which vanishes for $M\to \infty$.
For $i=x,y$ the resulting equation \eqref{eq15} for the momenta shift can be evaluated directly from the $E_x$ and $E_y$ fields. On the other hand, 
in principle would we need to 
explicitly compute the $E_z$ component of the color-electric field from the $(\tau,\eta)$ components. 
This can be easily achieved as under the change of coordinates 
from $x^{\mu'}=(\tau,\eta)$ to $x^\mu=(t,z)$ in Eq. \eqref{eq:milne_inverse_coordinate_transformation} we have
\begin{equation}
    E_z=F_{0z}=\frac{\partial x^{\mu '}}{\partial t}\frac{\partial x^{\nu'}}{\partial z}F_{\mu' \nu'}=F_{\tau \eta}=E_\eta.
    \label{15.1}
\end{equation}
Consequently,
the knowledge of the $\eta$--component of the field, which we 
easily extract from the
Yang-Mills equations, is enough to compute the momentum spreading
along the $z$ direction.

The equations~\eqref{eq:oraguardiquesto}
and~\eqref{eq15} are formally similar to what we would obtain
in the case of a standard Langevin equation
for a purely diffusive motion: $dp^i/dt=\xi^i$,
where $\xi^i$ denotes the random force. In fact, in this case
we would obtain~\cite{Liu:2020cpj}
\begin{equation}
\langle \delta p_i^2\rangle = \int_0^t dt_1\int_0^t dt_2\,
\langle
\xi_i(t_1)\xi_i(t_2)
\rangle.
\label{eq:tuttivoglionoFrattesi}
\end{equation}
We can indeed notice the similarity among \eqref{eq:tuttivoglionoFrattesi}
and Eqs.~\eqref{eq:oraguardiquesto} and~\eqref{eq15} after realizing that
the role of the random force in the context of HQs
is taken by the force exerted by the color-electric field. Within a Langevin dynamics, one can also include ‘‘memory effects'', which here are not considered, in order to account for coherent field-induced dynamics \cite{Ruggieri:2022kxv}. Moreover, one can in principle also include radiation effects, but these have been shown to be basically irrelevant within the dynamics of heavy quarks in glasma \cite{Liu:2020cpj}.

\begin{figure}[t!]
    \centering
    \includegraphics[width=0.6\linewidth]{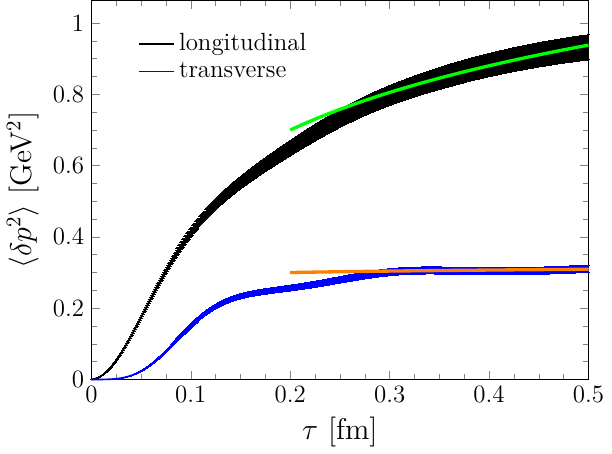}
    \caption{Longitudinal momentum broadening $\langle \delta p_L^2(\tau)\rangle$ (black), and transverse momentum broadening
$\langle \delta p_T^2(\tau)\rangle$ (blue) for a boost invariant pA collision, along with the corresponding error bars. The green line
corresponds to the function $a+b\log(\tau/\bar{\tau})$ where we put
$\bar{\tau}=0.4$ fm, $a=0.88$ GeV$^2$ and $b=0.26$ GeV$^2$.
Similarly, the orange solid line represents the function
$c + d \log(\tau/\tilde{\tau})$ where we put
$\tilde{\tau}=0.2$ fm, $c=0.30$ GeV$^2$ and 
$d=0.01$ GeV$^2$.}
    \label{Fig:3}
\end{figure}

In Fig.~\ref{Fig:3} we show $\langle \delta p_L^2(\tau)\rangle$ and
$\langle \delta p_T^2(\tau)\rangle$ versus proper time for a boost invariant collision, calculations correspond to an average over $100$ events, for a transverse plane with $32\times 32$ points and transverse size $L_\perp=2$ fm. The MV parameter has been fixed as $\mu=0.5$ GeV. The evolution is shown up to $\tau=0.5$ fm. We notice that the transverse momentum shift is substantially lower than the longitudinal one: this is a consequence of the predominance of the longitudinal color-electric fields in the first instants after the collision. The anisotropy of the evolving glasma,
which reflects onto the difference between the longitudinal and the
transverse field correlators, is transmitted to the momentum broadenings. We find that, as a result of the diffusion of the HQs in the evolving glasma fields, both $\langle \delta p_L^2(\tau)\rangle$ and $\langle \delta p_T^2(\tau)\rangle$ grow up with proper time. After a short transient lasting up to $\tau\approx 0.3$ fm,
the growth of $\langle \delta p_T^2(\tau)\rangle$ drastically slows down,
while that of $\langle \delta p_L^2(\tau)\rangle$ remains substantial
for the whole time range considered here. The values obtained can be compared, for example, with those found in \cite{Avramescu:2023qvv} for AA collisions: our momenta shifts are smaller since we are dealing with pA collisions, in which the charge density (and hence the energy density) is smaller.

The late-time behavior of $\langle\delta p_L^2(\tau)\rangle$
and $\langle\delta p_T^2(\tau)\rangle$
can be interpreted by assuming an idealized $\delta-$like correlator
of the color-electric fields. For instance, let us assume that for $E_z$ we have
\begin{equation}
\langle E_z(\tau^\prime)E_z(\tau^{\prime \prime})\rangle = 
\frac{  \langle E_z^2(\tau^\prime)\rangle}{Q_s}
\delta(\tau^\prime - \tau^{\prime\prime}),
\label{eq:sonobravisullavoro}
\end{equation}
where the overall $1/Q_s$ on the right hand side of the above equation
is added in order to balance the units of the 
$\delta$ function. The particular choice of $Q_s$
in the equation is due to the fact that $Q_s$ is the only
energy scale in the MV model.
Using the ansatz~\eqref{eq:sonobravisullavoro} in Eq.~\eqref{eq:oraguardiquesto}
we easily get
\begin{equation}
\langle \delta p_L^2(\tau)\rangle_{M\to \infty} \approx g^2 Q_s^{-1}
\int_{\bar{\tau}}^\tau d\tau^\prime \langle E_z^2(\tau^\prime)\rangle,
\label{eq:messicaniinfuga}
\end{equation}
where we cut the time-integral down to $\bar{\tau}$, which represents a value of the proper time of the
order of $1/Q_s$. We introduce it because 
we are interested in the late-time behavior only, since it is
large enough that it allows us to use
\begin{equation}
    \langle E_z^2(\tau^\prime)\rangle  \propto Q_s^4/(Q_s\tau^\prime),
    \label{eq:asymptotic_Ez2_chapter4}
\end{equation}
in agreement with the late-time behavior of the energy
density, see Fig.~\ref{Fig:2}.
We have also inserted
the needed powers of $Q_s$ in order to get the correct dimension of energy to the fourth power.\footnote{For this argument we do not need to be precise and fix
the overall coefficient, as we are only interested in
extracting the late-time behavior of the momentum
broadening.}
We thus get
\begin{equation}
\langle \delta p_L^2(\tau)\rangle_{M\to \infty} 
\propto
g^2 Q_s^2
\log(\tau/\bar{\tau}),~~~\text{for }\tau/\bar{\tau} >1.
\label{eq:conoscebeneilmestiere}
\end{equation}
The late time behavior~\eqref{eq:conoscebeneilmestiere} of 
$\langle \delta p_L^2(\tau)\rangle$ 
(and similar calculations hold for $\langle \delta p_T^2(\tau)\rangle$)
is in fair agreement with the results 
of the full numerical simulations,
see orange and green lines in Fig.~\ref{Fig:3}. Here, by ‘‘late time'' we refer to timescales which are at the upper limit of the validity of the glasma framework, i.e. around $\tau\sim 0.4$ fm. As a final comment on the results shown in
Fig.~\ref{Fig:3}, we notice that we measure some tiny fluctuations
of both $\langle \delta p_L^2(\tau)\rangle$ and 
$\langle \delta p_T^2(\tau)\rangle$, similarly to what observed  in~\cite{Avramescu:2023qvv}. The amplitude of these fluctuations
is however very small in comparison with
the bulk value of these quantities, thus they are practically irrelevant.

\subsection{Angular momentum broadening}
The momentum anisotropy we have highlighted in the previous Section is also transmitted to other
observables of the heavy quarks.
In particular,
one can look at the {\sl angular momentum anisotropy parameter} 
$\Delta_2$, defined as~\cite{Pooja:2022ojj}
\begin{equation}
    \Delta_2\equiv \frac{\langle L_x^2-L_z^2\rangle}{\langle L_x^2+L_z^2\rangle},
    \label{22}
\end{equation}
where $L_i$, with $i=x,y,z$, denotes the $i^{th}$ component of the
orbital angular momentum of the HQ. The $\Delta_2$ therefore measures the anisotropy
of the fluctuations of the orbital angular momentum of the HQs: the fluctuations of the HQ spin are suppressed by a power
of $M$~\cite{Pooja:2022ojj}. It is also worth noting that $\Delta_2$ receives contribution from both the anisotropy of the color
fields and from the geometry of the system~\cite{Pooja:2022ojj}. 

Such quantity can be rewritten 
in terms of the averages of the squared components of the momentum of the HQs. This can be done following the same steps given in~\cite{Pooja:2022ojj} and assuming that momenta and positions of the HQs are not correlated with each other: this holds since we are working in the static limit for HQs. We can write
\begin{align}
\langle L_x^2\rangle&=\langle (yp_z-zp_y)^2\rangle\simeq \langle y^2\rangle\langle p_z^2\rangle+\langle z^2\rangle\langle p_y^2\rangle,\nonumber\\
\langle L_z^2\rangle&=\langle (xp_y-yp_x)^2\rangle\simeq \langle x^2\rangle\langle p_y^2\rangle+\langle y^2\rangle \langle p_x^2\rangle.
\label{38}
\end{align}
We can insert the above expressions \eqref{38} into Eq. \eqref{22} to get
\begin{equation}
    \Delta_2=\frac{[\langle z^2\rangle-\langle x^2\rangle-\langle y^2\rangle]\langle p_T^2\rangle/2+\langle y^2\rangle \langle p_z^2\rangle}{[\langle z^2\rangle+\langle x^2\rangle+\langle y^2\rangle]\langle p_T^2\rangle/2+\langle y^2\rangle \langle p_z^2\rangle}.
    \label{39}
\end{equation}
This expression can be further specialized
to the case of a pA collision as follows.
Firstly, 
we estimate the mean squared value of the longitudinal $\langle z^2\rangle$ and of the transverse coordinates $\langle x^2\rangle$ and $\langle y^2\rangle$. 
As far as the extension along the beam axis $z$ is concerned, we will assume that the HQs are generated uniformly along the whole $\eta$ interval. This means that $\langle z^2\rangle$ will be given by:
\begin{equation}
    \langle z^2\rangle =\frac{\displaystyle{\int_{-L_\eta/2}^{L_\eta/2}}d\eta\, (\tau \sinh \eta)^2}{\displaystyle{\int_{-L_\eta/2}^{L_\eta/2}}d\eta}=\frac12 \tau^2 \left(\frac{\sinh L_\eta}{L_\eta}-1\right).
    \label{40}
\end{equation}
We fix the $\eta$ extension as $L_\eta=2$, corresponding to the rapidity range $\eta \in[-1,1]$. For what concerns the transverse coordinates, firstly
we assume that the geometrical distribution of the HQs in the transverse plane follows the same profile of the initial energy density,
which in turn resembles that of the color charges of the colliding proton. We have already discussed on the validity of this approximation in §\ref{sec:Momentum broadening}, and within these assumptions we can estimate 
$\langle\bm{x}_\perp^2 \rangle =\langle  x^2 + y^2 \rangle$. Firstly, we fix the location of the
three constituent quarks, $\bm{x}_i$, and we average over the transverse coordinates using the thickness function $T_p( \bm{x}_\perp)$ in \eqref{eq:bd1}, getting
\begin{equation}
    \langle \bm{x}_\perp^2\rangle_{\bm{x}_i}=\frac{\displaystyle{\int} d^2\bm{x}_\perp\, \bm{x}_\perp^2\, T_p(\bm{x}_\perp)}{\displaystyle{\int} d^2\bm{x}_\perp\, T_p(\bm{x}_\perp)}=2B_q+\frac13 \sum_{i=1}^3 \bm{x}_i^2.
    \label{41}
\end{equation}
Then, an average over the locations of the constituent quarks
$\bm{x}_i$ is needed: we perform this
using the distribution $T_{cq}(\bm{x}_i)$ in~\eqref{eq:bd2}, resulting in 
\begin{equation}
    \langle \bm{x}_\perp^2\rangle=\frac{\displaystyle{\int} d^2\bm{x}_i\, \langle \bm{x}_\perp^2\rangle_{\bm{x}_i} T_{cq}(\bm{x}_i)}{\displaystyle{\int} d^2\bm{x}_i\, T_{cq}(\bm{x}_i)} =2B_q+2B_{cq}.
    \label{42}
\end{equation}
We obtain,
\begin{equation}
    \langle x^2\rangle= \langle y^2\rangle=\frac12 \langle \bm{x}_\perp^2\rangle=B_q+B_{cq}.
    \label{43}
\end{equation}
Using the results~\eqref{40} and~\eqref{43} 
in Eq.~\eqref{39}, along with the momenta shifts we previously got from \eqref{eq:oraguardiquesto} and \eqref{eq15},
we can compute $\Delta_2$ for a pA collision.\\

We have already mentioned that a nonzero $\Delta_2$ may not only be due to the anisotropy of the fields (which is then reflected on to a anisotropic momentum distribution), 
but entails a geometric contribution as well. The latter is related to the shape of the fireball and can be nonzero also when momenta are isotropic. 
In order to 
extract the geometric contribution to $\Delta_2$, we start with Eq.~\eqref{39} in which we evaluate $\langle p_z^2\rangle$ as follows. Since we have
\begin{equation}
p_z=p_T\sinh y\simeq p_T \sinh \eta,
    \label{eq:pz_to_pTsineta}
\end{equation}
by assuming once again that the HQs are uniformly spread in $\eta$ we have:
\begin{equation}
    \langle p_z^2\rangle =\frac{\displaystyle{\int_{-L_\eta/2}^{L_\eta/2}}d\eta\, \langle p_T^2\rangle (\sinh \eta)^2}{\displaystyle{\int_{-L_\eta/2}^{L_\eta/2}}d\eta}=\frac12 \langle p_T^2\rangle \left(\frac{\sinh L_\eta}{L_\eta}-1\right).
    \label{eq:avgd_pz_to_pTsineta}
\end{equation}
This allows us to define
the geometric component of $\Delta_2$, which we call $\Delta_2^{\text{geom}}$, as follows
\begin{align}
&\Delta_2^{\text{geom}}=\frac{\langle z^2\rangle+\langle x^2\rangle-\langle y^2\rangle+\langle y^2\rangle\left(\sinh L_\eta/L_\eta-1\right)}{\langle z^2\rangle+\langle x^2\rangle+\langle y^2\rangle+\langle y^2\rangle\left(\sinh L_\eta/L_\eta-1\right)}=\nonumber\\
&=\frac{[\frac12 \tau^2+(B_q+B_{cq})]\left(\sinh L_\eta/L_\eta-1\right)}{[\frac12 \tau^2+(B_q+B_{cq})]\left(\sinh L_\eta/L_\eta-1\right)+2(B_q+B_{cq})}.
    \label{44}
\end{align}
From Eq.~\eqref{44} we get the initial value of $\Delta_2^{\text{geom}}$, which is
\begin{equation}
    \Delta_2^{\text{geom}}(\tau=0)=
(\sinh L_\eta/L_\eta-1)/(\sinh L_\eta/L_\eta+1)\simeq 0.289,
\label{eq:initial_value_Delta2geom}
\end{equation}
therefore we have a geometric correction already at initial time. Notice also that $\Delta_2^{\text{geom}}(\tau=0)$ is independent with respect to the widths of the proton charge, namely $B_q$ and $B_{cq}$. 

\begin{figure}
    \centering
    \includegraphics[width=0.6\linewidth]{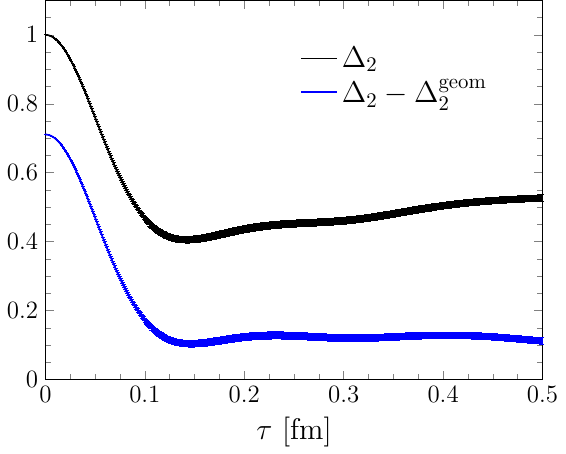}
    \caption{$\Delta_2$ (black) and  $\Delta_2-\Delta_2^{\text{geom}}$ (blue) versus proper time for pA collisions, along with the corresponding error bars.}
    \label{Fig:4}
\end{figure}

In Figure~\ref{Fig:4} we show our results for both $\Delta_2$ and $\Delta_2-\Delta_2^{\text{geom}}$, along with their error estimates. 
The calculations correspond to a pA collision for a $32\times 32$ lattice
and an average over $100$ events. For times up to $\tau \sim 0.15 $ fm the two curves in Figure~\ref{Fig:4} differ basically by a constant: this occurs since $\Delta_2^{\text{geom}}$ varies with time but quite slowly (see the $\tau^2$ term in \eqref{44}), hence $\Delta_2^{\text{geom}}(\tau)\simeq \Delta_2^{\text{geom}}(\tau=0)$ for small $\tau$. During these initial times we observe most of the damping for both curves. For larger times we see that $\Delta_2$ (black curve) shows an increasing behavior, but a look at $\Delta_2-\Delta_2^{\text{geom}}$ in the same time scales shows that the increase in angular momentum anisotropy is likely due to an expansion of the system, mostly along the longitudinal direction. Indeed, by subtracting the geometric component we see that the pure anisotropy of the glasma fields (blue curve) remains constant and small over time.

\section{Inclusion of field fluctuations}
\label{Initial state fluctuations}

As already explained in the Introduction of the Chapter,
in relativistic nuclear collisions 
one clearly does not have exact boost-invariance. 
In particular, 
the glasma is the solution of the Yang-Mills equations
only in the case of weak coupling $g$: 
for realistic values of $g$ it is likely that quantum fluctuations develop, with a strength that 
increases with $g$. For instance,
in \cite{Epelbaum:2013ekf} it has been shown that increasing $g$ leads to a larger contribution
of fluctuations to the longitudinal pressures. 
Since a finite value of $g$ is relevant for realistic collisions,
it is meaningful to repeat the analysis presented in the previous Section to the case in which we add fluctuations on top of the glasma.

We construct the fluctuations by adding terms to the boost-invariant electric fields at initial time. These additions
must always satisfy the Gauss' law, i.e. the total fields must be such that
\begin{equation}
    D_i E_i+D_\eta E_\eta=0,
    \label{45}
\end{equation}
where $D$ denotes the covariant derivative. Firstly, we evaluate random configurations $\xi_i(\bm{x}_\perp)$, for $i=x,y$, such that \cite{Fukushima:2011nq,Romatschke:2006nk}
\begin{equation}
    \langle \xi_i(\bm{x}_\perp) \xi_j(\bm{y}_\perp)\rangle=\delta_{ij} \delta^{(2)}(\bm{x}_\perp-\bm{y}_\perp),
    \label{46}
\end{equation}
which are operatively generated by sampling random Gaussian numbers with zero average and standard deviation $1/a_\perp$. 
After that, the additional terms to the 
glasma color-electric fields are evaluated as \cite{Fukushima:2011nq,Romatschke:2006nk}
\begin{align}
        \delta& E^i(\bm{x}_\perp,\eta)=-\partial_\eta F(\eta)\,\xi_i(\bm{x}_\perp),\label{eq:ottimathuram}\\
        \delta& E^\eta(\bm{x}_\perp,\eta)=F(\eta)\sum_{i=x,y}D_i \xi_i(\bm{x}_\perp).\label{47}
    \end{align}
The information about the $\eta-$distribution of the
fluctuations is encoded in the function $F(\eta)$, which will be specified later.
By construction, these fluctuations \eqref{eq:ottimathuram} and \eqref{47} satisfy the Gauss law
constraint~\eqref{45}. 
The above relations are discretized on the lattice as~\cite{Fukushima:2011nq}
\begin{align}
\delta& E^i(\bm{x}_\perp,\eta)=a_\eta^{-1}[F(\eta-a_\eta)-F(\eta)]\xi_i(\bm{x}_\perp),\\
\delta& E^\eta(\bm{x}_\perp,\eta)=
        -a_\perp^{-1}F(\eta)\sum_{i=x,y}[
        \Xi^i(\bm{x}_\perp)
        -\xi^i(\bm{x}_\perp)],\label{47bis}
    \end{align}
where
\begin{equation}
\Xi^i(\bm{x}_\perp) = 
U_i^\dagger(\bm{x}_\perp-\bm{i})\xi^i(\bm{x}_\perp-\bm{i})U_i(\bm{x}_\perp-\bm{i}).
\end{equation}
In the above equations, $a_\eta=L_\eta/N_\eta$ is the lattice spacing along the $\eta$ direction, $\bm{i}$ is a unit vector pointing towards the $i=x,y$ transverse direction, while the term  $\Xi^i(\bm{x}_\perp)-\xi^i(\bm{x}_\perp)$
is the covariant derivative of $\xi^i$, discretized in terms of the gauge links. Moreover, also in the longitudinal direction as well as we did in the transverse directions, periodic boundary conditions have been implemented for points at the edge of the $\eta$ domain.

As far as the choice of the rapidity-dependent term $F(\eta)$ is concerned, one may consider different approaches. For instance, in~\cite{Romatschke:2006nk} 
the authors
implemented a model in which $F(\eta)$ 
is a Gaussian random number with zero mean and 
standard deviation equal to one.
In this work, we instead follow a simpler model of fluctuations, similar to the one introduced in~\cite{Fukushima:2011nq}, in which $F(\eta)$ takes the form
\begin{equation}
F(\eta)=\frac{\Delta}{N_\perp}\sum_{n\in \mathcal{I}}
\frac{1}{|\mathcal{I}|}
 \cos\left(\frac{2\pi\, n\, \eta}{L_\eta}\right),
    \label{48_bis}
\end{equation}
where $\mathcal{I} \in \mathbb{N}$ denotes a set of positive
integers with cardinality $|\mathcal{I}|$. The higher the order of the 
harmonics that contribute to $F(\eta)$ in~\eqref{48_bis}
the higher the energy carried by the 
transverse fields of the
fluctuations, see the $\eta$-derivative in Eq.~\eqref{eq:ottimathuram}. Notice the presence of a free 
parameter, $\Delta$, which encodes the strength of the
fluctuating fields.
 
\subsection{Momentum broadening fluctuations\label{sec:prescel}}

Here we present our results on the impact of the initial fluctuations on the momentum broadening of the HQs in the pre-equilibrium stage of pA collisions.
The fluctuations are added 
on top of the boost-invariant glasma at 
$\tau\equiv\tau_0=0.05$ fm:
we checked that the results are not significantly
affected by changing $\tau_0$ in the range
$[0.01,0.05]$ fm. The plots which follow have been obtained in a $32\times 32 \times 32$ lattice, as an average over $30$ events, for different values of $\Delta$.

\begin{figure}[t]
\centering
\includegraphics[width=0.6\linewidth]{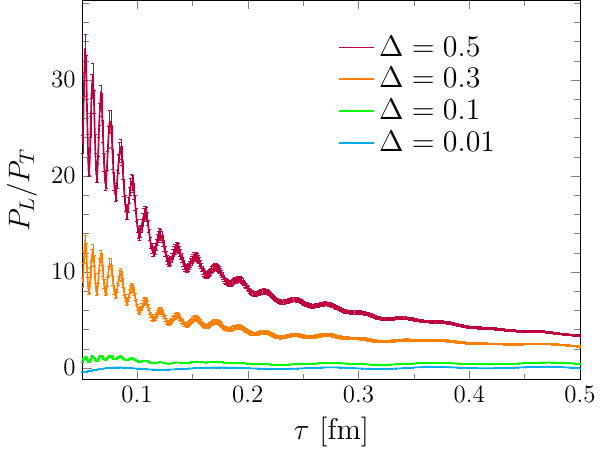}
\caption{The ratio of the longitudinal pressure over the transverse pressure $P_L/P_T$, versus proper time for different values of $\Delta$. Error bars are also shown.}
\label{Fig:PL_over_PT}
\end{figure}

In Fig.~\ref{Fig:PL_over_PT} we show, for the set of harmonics $\mathcal{I}=\{1,\dots,10\}$ and several
values of $\Delta$, the ratio of longitudinal pressure over transverse pressure, $P_L/P_T$, versus proper time: the expressions for these quantities has been provided in Eqs. \eqref{eq:PL_PT_eps_vst}. The main purpose of the results shown in Fig.~\ref{Fig:PL_over_PT}
is to highlight the net effect of the fluctuations
on the bulk gluon fields. 
Indeed, for the highest values of $\Delta$, we find a non-trivial behavior of $P_L/P_T$, which 
is quite different from the one we find for smaller values of $\Delta$ as well as for $\Delta=0$ (the result for $\Delta=0$ is not shown, since we checked that it is indistinguishable from the $\Delta=0.01$ result). In particular, for large $\Delta$, $P_L/P_T$ stands significantly higher than the close-to-zero values obtained for small $\Delta$. We remark that the largest value of $\Delta$ used in
Fig.~\ref{Fig:PL_over_PT}, that is $\Delta=0.5$, is probably
too large as it corresponds to a system whose energy is carried
half by glasma and half by fluctuations. Nevertheless,
it is shown to illustrate that fluctuations are actually introduced
in the bulk and substantially modify it.

\begin{figure}[t]
\centering
\includegraphics[width=0.49\linewidth]{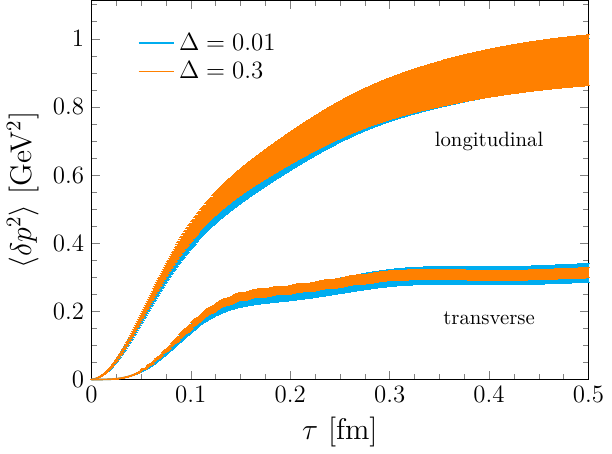}
\includegraphics[width=0.49\linewidth]{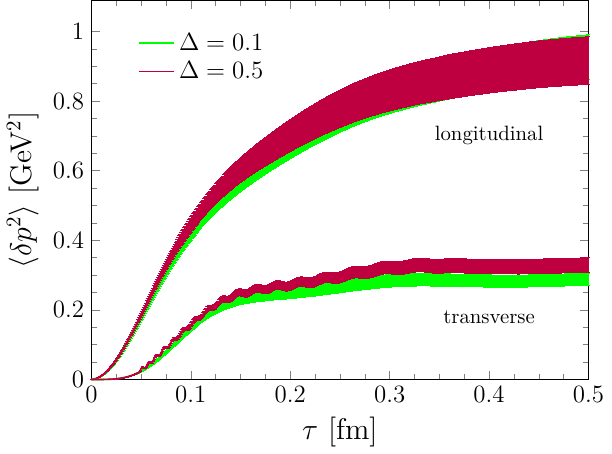}
\caption{The longitudinal and transverse momentum shifts, $\langle \delta p_L^2(\tau)\rangle$ and
$\langle \delta p_T^2(\tau)\rangle$,
versus proper time for different values of $\Delta$. We also show the corresponding error bars. We used $\mathcal{I}=\{1,\dots,10\}$ in Eq.~\eqref{48_bis}. The four curves have been split into two plots for the sake of clarity.}
\label{Fig:shifts_pt3}
\end{figure}

Next, we turn to
the momentum broadening. 
In Fig.~\ref{Fig:shifts_pt3} we plot 
$\langle \langle\delta p_L^2(\tau)\rangle\rangle$ and
$\langle \langle\delta p_T^2(\tau)\rangle\rangle$
versus proper time. We show results for the same set of harmonics $\mathcal{I}=\{1,\dots,10\}$ and for the same parameters $\Delta$ we have already considered in Fig. \ref{Fig:PL_over_PT}: we have split the datasets into two different plots, just for the sake of clarity. In addition to the averaging over the transverse plane (as in \eqref{eq:rai3_22_02}),
we average on the space-time rapidity of the HQs,
restricting ourselves to the rapidity range $\eta\in[-1,1]$. We find no net effect of the rapidity fluctuations on the momenta shifts, since each result overlaps with the error bars of all the others, including the highest value $\Delta=0.5$. Our conclusion is that the fluctuations do not affect 
substantially
the HQ momenta shifts in the static limit.
Since momentum broadening is related, in the static limit, to the time-correlator of the color-electric fields, see Eqs.~\eqref{eq:oraguardiquesto}
and~\eqref{eq15}, we see that these correlators are not
very much affected by the longitudinal 
fluctuations. Our interpretation is that, since these fluctuations add a non-trivial dependence of the
fields on $\eta$, correlations along the $\eta$-direction are the most affected here \cite{Ruggieri:2017ioa}, but these are not relevant for the momentum shifts.

    \chapter[Color melting of HQ pairs in glasma]{Color melting of heavy quark pairs in glasma}
    \label{chap:Melting of heavy quark pairs in glasma}

This Chapter is largely based on the work contained in \cite{Oliva:2024rex}:
\begin{center}
    Lucia Oliva, Gabriele Parisi, Vincenzo Greco and Marco Ruggieri, ``Melting of $c\bar{c}$ and $b\bar{b}$ pairs in the pre-equilibrium stage of proton-nucleus collisions at the Large Hadron Collider'', 
Phys. Rev. D \textbf{112} (2025) no.1, 014008, arXiv: 2412.07967 [hep-ph].\\
\end{center}

A natural extension of the works concerning the transport properties and the diffusion of heavy quarks in the initial stages, is the study of the impact of the gluon-dominated initial state on the evolution and dissociation of {\sl heavy quarkonia}. These are bound states of heavy quarks and antiquarks of the same flavour, namely {\sl charmonium} $c\bar{c}$ and {\sl bottomonium} $b\bar{b}$.
The production and suppression of heavy quarkonium states in hadronic collisions, in particular in QGP, has been widely studied for almost four decades~\cite{Matsui:1986dk, Datta:2003ww, Burnier:2009yu, Brambilla:2011sg, Srivastava:2012pd, Casalderrey-Solana:2012yfo, Brambilla:2013dpa, Singh:2015eta, Andronic:2015wma, Brambilla:2016wgg, Song:2017phm, Akamatsu:2020ypb, Yao:2021lus, He:2021zej, Miura:2022arv, Villar:2022sbv, Brambilla:2022ynh, Song:2023zma, Song:2023ywt, Brambilla:2024tqg, Bai:2024xmm,Delorme:2024rdo, Du:2015wha, Du:2017qkv, Du:2018wsj, Yan:2006ve, Zhou:2014kka}. In particular, the ‘‘melting'' emerges from the interplay of several mechanisms coming from the different phases of the collision, stemming from initial HQ pair production, through the QGP stage to final hadronization, see Ref.~\cite{Rothkopf:2019ipj, Andronic:2024oxz} for a review.
As in the QGP, $c\bar{c}$ and $b\bar{b}$ pairs can dissociate also in the evolving glasma~\cite{Pooja:2024rnn}, with important implications for their evolution in the subsequent QGP medium and for the determination of the heavy-flavoured final states produced at hadronization (both open heavy-flavour hadrons and regenerated quarkonia).
In the evolving glasma, it is the HQ interaction with the classical color fields that leads to a modification of their spatial and momentum coordinates, as well as of their color charge, thus affecting the probability that they survive to the pre-equilibrium 
stage and enter into the possibly-formed QGP phase as a pair.

In this Chapter, we investigate the melting of 
$c\bar c$ and $b\bar b$
pairs during the pre-equilibrium stage of high-energy pA collisions, the latter being modeled within the glasma framework.
The primary objective of this work is to determine the fraction of pairs that dissociate during the propagation in the evolving glasma, aiming also at disentangling the role of color-charge decorrelation of the HQ pair in the melting,
which was only partly addressed before.
Within our model the HQs in each pair, formed at
$\tau=\tau_\mathrm{form}$, interact with the strong gluon fields
via the Wong equations~\cite{Wong:1970fu,Heinz:1984yq}, which are relativistic kinetic theory equations coupled to the glasma fields. With respect to the approach presented in~\cite{Pooja:2024rnn}, in which a similar study has been performed,
we implement a different criterion for the calculation of the number of melting pairs. This criterion is based on a survival pair-by-pair probability, $\mathcal{P}_\mathrm{survival}$,
that takes into account the decorrelation of the color charges
induced by the interactions of $Q$ and $\bar Q$ with the gluon fields
in the early stage. Such probability $\mathcal{P}_\mathrm{survival}$
is built similarly to the hadronization probability introduced in coalescence models~\cite{Greco:2003xt, Fries:2003vb, Greco:2003mm, Fries:2003kq, Greco:2003vf, Plumari:2017ntm}.
Differently from~\cite{Pooja:2024rnn}, we consider $N_c=3$ and
an initialization for pA collisions
that takes into account the event-by-event fluctuations of the color charges
in the proton and of the initial distribution of the 
heavy quarks in the transverse plane. Moreover,
within our model we allow for the expansion both along the longitudinal
direction and in the transverse plane, while in~\cite{Pooja:2024rnn}
only a static geometry was considered.

\section{Heavy quark dynamics in the pre-equilibrium stage}
\label{sec:devrijTOP}

In this Section, we discuss how we model the diffusion of the 
heavy quark pairs in the pre-equilibrium stage.
Following~\cite{Pooja:2024rnn},
we model 
the dynamics of the quarks in the $c\bar c$ and $b\bar b$ pairs
using semi-classical equations of motion,
namely the Wong equations. 
They describe the time evolution of the coordinate $\bm{x}$, 
momentum $\bm{p}$, and color charge $Q_a$ (with
$a=1,\dots,N_c^2-1$) of the quarks~\cite{Wong:1970fu, Heinz:1984yq}.
In the laboratory frame, the Wong equations read
\begin{align}
    \dfrac{\mathrm{d}x^i}{\mathrm{d}t} &=\frac{p^i}{E},
    \label{eq:wong1}\\
    \dfrac{\mathrm{d}p^i}{\mathrm{d}t} & =gQ^aF^{i\mu,a}\frac{p_\mu}{E}-\frac{\partial V}{\partial x_i},
    \label{eq:wong2}\\
    \dfrac{\mathrm{d}Q_a}{\mathrm{d}t} & =-gf^{abc} A_\mu^bQ_c\frac{p^\mu}{E},
    \label{eq:wong3}
\end{align}
where $i=x,y,z$ and $\mu=0,\dots,3$. 
Moreover, $f^{abc}$ are the structure constants of the gauge group SU(3) and $E=\sqrt{ \bm p^2+M^2}$ is the kinetic energy of the quark with mass $M$. 
The Wong equations have been recently used to study the dynamics of the heavy quarks in the pre-equilibrium stage of high-energy nuclear collisions; see \cite{Ruggieri:2018ies,Ruggieri:2018rzi,Sun:2019fud,Liu:2019lac,Jamal:2020fxo,Liu:2020cpj,Sun:2021req,Khowal:2021zoo, Pooja:2022ojj,Pooja:2024rnn, Avramescu:2024poa, Avramescu:2024xts} and references therein. Here we introduced a potential $V$, which produces an internal force $F^i=-\partial V/\partial x_i$ within each pair, in the transverse plane. We will come back to this point later.

Let us give more details on the numerical implementation of the Wong equations \eqref{eq:wong1}, \eqref{eq:wong2} and \eqref{eq:wong3}. All three equations have been solved using the Euler method. The equation \eqref{eq:wong1} on the position is easily solved, by performing the transformation to milne coordinates we get:
\begin{equation}
        x^\mu(\tau+\Delta \tau)=x^\mu(\tau)+\Delta \tau\cdot \frac{p^\mu}{p^\tau},
    \label{eq:wong1_discretized}
\end{equation}
where $x^\mu \in \{x,y,\eta\}$ and $\left(p^\tau\right)^2=\left(p^x\right)^2+\left(p^y\right)^2+\tau^2\left(p^\eta\right)^2+M^2.$ Moving on to the Wong equations \eqref{eq:wong2} for the momenta, by expressing the field strength tensor through the color-electric and color-magnetic fields we can write
\begin{align}
    \tau\dfrac{\mathrm{d}p^\eta}{\mathrm{d}\tau}+2 p^\eta &=\frac{g}{T_R}\left(\text{Tr}\,[Q E_\eta] -  \text{Tr}\,[QB_x]\frac{p^y}{p^\tau}  +  \text{Tr}\,[QB_y] \frac{p^x}{p^\tau}\right),\label{eq:wonginmilne_1}\\
    \frac{dp^x}{d\tau}&=\frac{g}{T_R}\left(\frac{1}{\tau}\text{Tr}\,[QE_x]  +  \text{Tr}\,[QB_\eta]\frac{p^y}{p^\tau} -  \text{Tr}\,[QB_y] \frac{\tau\, p^\eta}{p^\tau}\right)+F^x,\label{eq:wonginmilne_2}\\
    \frac{dp^y}{d\tau}&=\frac{g}{T_R}\left(\frac{1}{\tau}\text{Tr}\,[QE_y] - \text{Tr}\,[QB_\eta] \frac{p^x}{p^\tau}  +  \text{Tr}\,[QB_x]\frac{\tau\, p^\eta}{p^\tau}\right)+F^y,\label{eq:wonginmilne_3}
\end{align}
using the expressions \eqref{12} and \eqref{eq:magnetic_field_x}, \eqref{eq:magnetic_field_y}, \eqref{eq:magnetic_field_z}. The Euler method can be straightforwardly applied to the last two equations \eqref{eq:wonginmilne_2} and \eqref{eq:wonginmilne_3}, while for the equation in $p^\eta$ this is not trivial. However, it is sufficient to write
\begin{equation}
    \tau\dfrac{\mathrm{d}p^\eta}{\mathrm{d}\tau}+2 p^\eta=\frac{1}{\tau}\frac{\mathrm{d}(\tau^2p^\eta)}{\mathrm{d}\tau},
    \label{eq:peta_transformation_toEuler}
\end{equation}
to write also the first equation \eqref{eq:wonginmilne_1} in a Euler-solvable form. What we obtain is the following discretized set of time evolution equations
\begin{align}
    (\tau+\Delta \tau)^2p^\eta(\tau+\Delta \tau)&=\tau^2p^\eta(\tau)+\Delta\tau\left[\tau\,\frac{g}{T_R}\left(\text{Tr}\,[Q E_\eta] -  \text{Tr}\,[QB_x]\frac{p^y}{p^\tau}  +  \text{Tr}\,[QB_y] \frac{p^x}{p^\tau}\right)\right],\nonumber\\
    p^x(\tau+\Delta \tau)&=p^x(\tau)+\Delta\tau\left[\frac{g}{T_R}\left(\frac{1}{\tau}\text{Tr}\,[QE_x]  +  \text{Tr}\,[QB_\eta]\frac{p^y}{p^\tau} -  \text{Tr}\,[QB_y] \frac{\tau\, p^\eta}{p^\tau}\right)+F_x\right],\nonumber\\
    p^y(\tau+\Delta \tau)&=p^y(\tau)+\Delta\tau\left[\frac{g}{T_R}\left(\frac{1}{\tau}\text{Tr}\,[QE_y] - \text{Tr}\,[QB_\eta] \frac{p^x}{p^\tau}  +  \text{Tr}\,[QB_x]\frac{\tau\, p^\eta}{p^\tau}\right)+F_y\right].\label{eq:wong2_discretized}
\end{align}
These should in principle accompanied by a dynamic equation for the temporal component $p^\tau$ \cite{Avramescu:2023qvv}, but at each time we can simply enforce the normalization $\left(p^\tau\right)^2=\left(p^x\right)^2+\left(p^y\right)^2+\tau^2\left(p^\eta\right)^2+M^2$ to update also $p^\tau$. It has been shown, e.g. in Fig. 11 from \cite{Avramescu:2023qvv}, that the difference between this extraction of $p^\tau$ and the actual solution of its equation of motion produces results which differ by less than $1\%$. Finally, moving on to the equation \eqref{eq:wong3} for the color charge $Q=Q^a T^a$, we can express it in terms of commutators as
\begin{equation}
Q(\tau+\Delta \tau)=Q(\tau)+\Delta \tau\left[ ig\left([Q,A_x]\frac{p^x}{p^\tau}+[Q,A_y]\frac{p^y}{p^\tau}+[Q,A_\eta]\frac{p^\eta}{p^\tau}\right)\right],
    \label{eq:wong3_discretized}
\end{equation}
where the gauge field $A_\mu$ is determined by\footnote{The $A_\tau$ component is zero because of the gauge choice.}
\begin{align}
    A_i(\bm{x})&=-\left(\frac{i}{g a_\perp}\right)\log U_i(\bm{x}) \qquad \text{for }i=\{x,y\}, \label{eq:gauge_field_xy}\\
    A_\eta(\bm{x})&=-\left(\frac{i}{g a_\eta}\right)\log U_\eta(\bm{x}).\label{eq:gauge_field_eta}
\end{align}
For $N_c=3$
the evolution of the color charges governed by Eq.~\eqref{eq:wong3} conserves the Casimir invariants
\begin{equation}
q_2 = Q_a Q_a,~~~q_3=d_{abc}Q_a Q_b Q_c,
\label{eq:mostrasumarte}
\end{equation} 
with
\begin{equation}
q_2=\frac{N_c^2-1}{2} ~~\text{and}~~q_3=\frac{(N_c^2-4)(N_c^2-1)}{4N_c}, \label{eq:casimir_values}
\end{equation}
and $d_{abc}$ are the symmetric structure constants \cite{Avramescu:2023qvv}. Numerically, the discretization \eqref{eq:wong3_discretized} may present a violation of those invariants, since it is not written in a explicitly Casimir preserving form. To overcome this problem one could write the time evolution from \eqref{eq:wong3} in a explicitly Casimir conserving form using Wilson lines, in order to maintain the charge $Q$ always within the color algebra:
\begin{equation}
    Q(\tau+\Delta \tau)=\mathcal{U}^\dagger(\tau)Q(\tau)\,\mathcal{U}(\tau)
    \label{eq:exponentiation_charge_evolution}
\end{equation}
where
\begin{equation}
    \mathcal{U}(\tau)=\exp\left[i(A_xv_x+A_yv_y)\Delta \tau\right]
    \label{eq:gauge_exponential}
\end{equation}
and $v_i=p_i/E$ is the $i^\text{th}$ component of the velocity of the HQ here considered. This approach has been adopted in \cite{Avramescu:2023qvv}, one can check that the results are exactly the same, provided the time step in the evolution is chosen as sufficeintly small.

Equations \eqref{eq:wong1}, \eqref{eq:wong2} and \eqref{eq:wong3} are those that govern the motion of HQ in the evolving glasma fields. Within our model, 
we follow previous works and neglect the back-reaction of the quarks 
in the Yang-Mills equations: it has been shown~\cite{Liu:2020cpj} that 
such back-reaction  does not substantially affect the diffusion of heavy quarks
within the lifetime of the pre-equilibrium stage.\\

In equation~\eqref{eq:wong2} we have introduced $V$, denoting the interaction potential between 
a quark and its respective antiquark in the pair. 
In QCD,
the tree-level static potential between a quark and an 
antiquark can be written as
\begin{equation}
V_{\alpha\beta\gamma\delta}=
\frac{\alpha_s}{r_\text{rel}}T^a_{\alpha\beta}
\bar T^a_{\gamma\delta}.
\label{HQ_potentialsing_buffet}
\end{equation}
Equation \eqref{HQ_potentialsing_buffet} is related to the static limit of the amplitude of the 
one-gluon-exchange
interaction between two quarks
in the color states $\beta$ and $\delta$ that 
go out respectively in the color states
$\alpha$ and $\gamma$.
The $\bar T^a$ correspond to the 
generators in the $\bar{\bm 3}$
representation, while
$r_\mathrm{rel}=|\bm{x}_{Q}-\bm{x}_{\bar{Q}}|$,
is the relative distance between the quark and
the antiquark. This potential depends on the
matrix element of the operator $T^a\otimes\,\bar T^a$,
where the first color generator acts on the Hilbert space of the
quark, and the second on that of the antiquark.
The potential~\eqref{HQ_potentialsing_buffet} can be
projected onto irreducible representations of 
SU(3). More specifically, if the quark-antiquark pair belongs to the 
irreducible representation $R$
we find
\begin{equation}
V_{R}=\lambda_R
\frac{\alpha_s}{r_\text{rel}},
\label{HQ_potentialsing_buffet_lambdaR}
\end{equation}
where
\begin{equation}
\lambda_R = \frac{1}{2}
\left[
C_2(R) - C_2(3) - C_2(\bar 3)\right].
\label{eq:giuliettamivuoisposare}
\end{equation}
Here, $C_2(R)$ denotes the eigenvalue of the quadratic Casimir
in the representation $R$, while
$C_2(3) = C_2(\bar 3) = 4/3$ are the Casimir for the quark
and the antiquark in the $\bm 3$ and $\bar{\bm 3}$ 
representations respectively.
If the quark-antiquark pair is in the
color-singlet representation then $C_2(1)=0$ and
\begin{equation}
V_1=-\frac{4}{3}\frac{\alpha_s}{r_\text{rel}}.
\label{HQ_potentialsing}
\end{equation}
Similarly, for the octet $C_2(8)=3$ and 
\begin{equation}
V_8=
+\frac{1}{6}\frac{\alpha_s}{r_\text{rel}}.
\label{HQ_potentialsingoct}
\end{equation}
Within our semi-classical approach, 
the most natural choice 
for $V$ in Eq.~\eqref{eq:wong2}
consists in replacing
$T^a\otimes T^a$ in Eq.~\eqref{HQ_potentialsing_buffet}
with $Q_a\bar Q_a/N_c$.
Hence, in Eq.~\eqref{eq:wong2} we consider
\begin{equation}
V=
\frac{ Q_a  \bar{Q}_a}{N_c}
\frac{\alpha_s}{r_\text{rel}}.
\label{HQ_potential}
\end{equation}
The overall $1/N_c$ in the right hand side of~\eqref{HQ_potential}
is included in order to reproduce $V=V_1$ in Eq. \eqref{HQ_potentialsing}
when we initialize the pairs in a color-singlet
state (see below Eq.~\eqref{eq:giornicontati}).
For each pair,
$Q_a  \bar{Q}_a$ evolves with time,
therefore $V$ is modified 
following the evolution of the color charges, as a result
of the interaction with the background gluon fields.
For a justification of Eq.~\eqref{HQ_potential}
starting from the expectation value of the QCD potential
see Section~\ref{sec:coequi}.

Before going ahead, we notice that 
in~\cite{Pooja:2024rnn} a similar problem was studied,
although for the simpler geometry of a static box
and for the SU(2) gauge group.
The improvement that the potential~\eqref{HQ_potential}
brings with respect to the one used in~\cite{Pooja:2024rnn},
which is a pure singlet potential,
is that our $V$ dynamically evolves with the color charges
of the pair. Our potential therefore takes into account that even if the pairs are
initialized in color-singlet states, they can get
an octet component due to the interaction with the
background gluon fields. 

\section{Heavy quarks initialization}
\label{sec:HQ_init}

Both $c\bar c$ and $b\bar b$  quark pairs 
are produced by hard QCD scatterings among the nucleons of the colliding nuclei at the very early stage of high-energy nuclear collisions, within 0.1 fm from the overlap of the two initial nuclei.
In this work, we assume
the pairs to form at proper time 
$\tau_\mathrm{form} = 1/(2 M)$, with $M$ being the heavy-quark mass: $M=1.3$ GeV for $c$ and $M=4.2$ GeV for $b$. Therefore $\tau_\mathrm{form} \sim 0.08$ fm 
for charm pairs, and $\tau_\mathrm{form} \sim 0.02$ fm
for bottom pairs.

\subsection{Initialization in the coordinate space and momentum space}
\begin{figure}[t!]
\centering
\includegraphics[width=0.47\columnwidth]{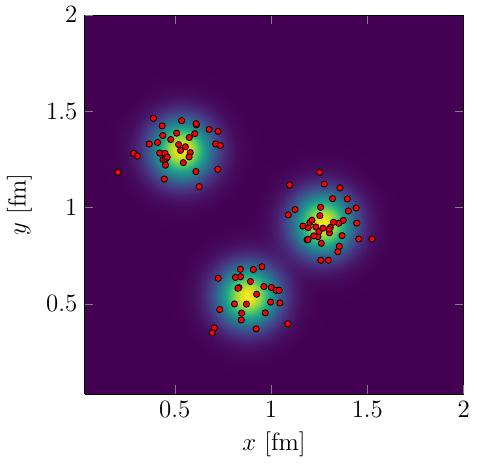}
\includegraphics[width=0.5155\columnwidth]{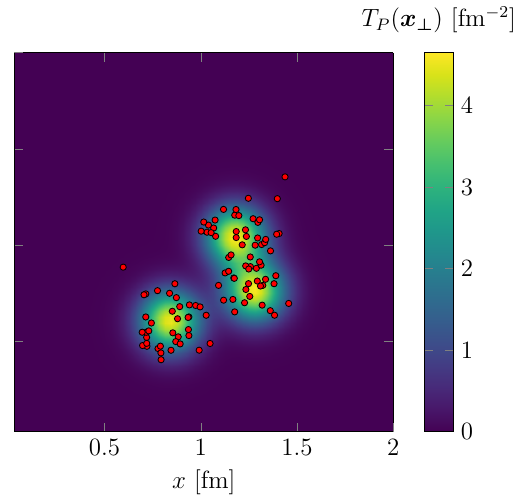}
\caption{Centers of mass of the production points of heavy-quark pairs (red dots) on the transverse plane for proton-nucleus collisions. 
The background image corresponds to the contour plot of the profile $T_P(\bm{x}_\perp)$ in Eq. \eqref{eq:bd1}, characterized by three hotspots that develop around the valence quarks in that event. The two panels correspond to two different events.}
\label{fig:init_HQ_xT}
\end{figure}

In this work we consider the pairs produced at mid-rapidity, so the longitudinal coordinate of the heavy quarks is kept $z=0$, if we indicate with positive $z$ and negative $z$ respectively the forward-going and backward-going directions of the colliding nuclei.
In the transverse plane, 
the initial positions of the heavy quarks
are extracted on an event-by-event basis.
For each event, after extracting the position of the three valence
quarks via the Gaussian profile~\eqref{eq:bd2}, we extract the position of the center of mass of each quark-antiquark pair via the profile~\eqref{eq:bd1}. In this way, the heavy quarks are distributed around the hotspots of the
initial energy density of the color fields. For illustrative purposes, we show in Fig.~\ref{fig:init_HQ_xT} the centers of mass of 100 heavy-quark pairs in two events. 
We remark that this number of pairs is chosen only for a good visualization
and does not correspond to the real number of quarks produced in the considered collision system. 
In both panels of the Figure, the three hotspots 
correspond to the profile~\eqref{eq:bd1} 
that develops around
the three valence quarks, whose location changes event-by-event;
these are the same locations where most of the energy density
gets deposited at the initial time.
The red dots represent the centers of mass of the production points of the heavy quark and antiquark in the pairs.

Once we have chosen the position of the center of mass, we extract the relative distance between each quark and antiquark in the pair (in its rest frame) according to the distribution:
\begin{equation}
P_\mathrm{HQ}(r_\text{rel}) = r_\text{rel}\,\exp(-r_\text{rel}^2/r_\text{0}^2),
\label{eq:HQ_extraction}
\end{equation}
where $r_\text{0}=0.4$ fm for a $c\bar{c}$ pair and $r_\text{0}=0.2$ fm for a $b\bar{b}$ \cite{Zhao:2024vqp}. in particular, using the $r_\text{rel}$ we extract for each pair, each quark is placed at a distance $r_\text{rel}/2$ from the center of mass of the pair, while the polar angle is extracted uniformly in the range $(0,2\pi)$. The corresponding antiquark is generated opposite to the quark, with respect to the center of mass. 

Moving on to the initialization in momentum space, we extract the modulus of the relative momentum in transverse plane of each quark and antiquark in the pair (in its rest frame) according to the distribution:
\begin{equation}
P_\mathrm{HQ}(p_\text{rel}) = p_\text{rel}\,\exp(-p_\text{rel}^2 r_\text{0}^2).
\label{eq:HQ_extraction_momentum}
\end{equation}
Once $p_\text{rel}$ is extracted for each pair, we set the modulus of the quark momentum as $p_\text{rel}/2$, while the polar angle is extracted uniformly in the range $(0,2\pi)$. From that, the components $p_x$ and $p_y$ of the heavy-quark momentum are determined: those of the companion antiquark are fixed opposite to them. 
We assume $p_z=0$ for quarks and their companion antiquarks,
as we simulate the mid-rapidity region and assume that the space-time rapidity of the produced particles $\eta$
is equal to the energy-momentum rapidity $y$.
With this initialization of momenta the quark and antiquark in a pair have momenta, ${\bm p}_q$ and ${\bm p}_{\bar q}$ respectively, such that the total momentum is zero:
\begin{equation}
{\bm p}_q + {\bm p}_{\bar q}=0
    \label{eq:sum_pq_pQbar_zero}
\end{equation}
Note also that the width of the gaussian \eqref{eq:HQ_extraction_momentum} is set as $1/r_\text{0}$, being $r_\text{0}$ the same parameter used in \eqref{eq:HQ_extraction} for the initialization in coordinate space. The HQs in a $b\bar{b}$ pair will therefore be initialized with an average momentum which is larger than the HQs in a $c\bar{c}$ pair.

\subsection{Initialization in the color space\label{sec:colorspace}}
Let us deal with the initialization of the color charges appearing in the Wong equations~\eqref{eq:wong2} and~\eqref{eq:wong3}. 
The initial value of the color charges 
needs to satisfy the constraints~\eqref{eq:mostrasumarte}.
In order to get these, we firstly 
fix the charge of one heavy quark 
in the ensemble
to be~\cite{Avramescu:2023qvv}
\begin{equation}
Q_{0a}=-1.69469\,\delta_{a5}-1.06209\,\delta_{a8}
    \label{eq:HQcharge0}
\end{equation}
which explicitly satisfies 
the conditions~\eqref{eq:mostrasumarte}. Then, we generate the charges of all the other heavy quarks as
\begin{equation}
Q_a=Q_{0b} U^{ab}, ~~~U^{ab}=2\text{Tr}[T^a U T^b U^\dagger],
    \label{eq:HQ_charge_all}
\end{equation}
where $U$ is a SU(3) matrix which is extracted for each heavy quark according to the Haar measure, which ensures uniform coverage of the manifold. By construction, it can be shown that all the HQs hereby obtained satisfy \eqref{eq:mostrasumarte}, and this condition will be satisfied throughout the time evolution, as already stated.

As anticipated in Section~\ref{sec:devrijTOP},
we initialize the $c\bar c$  and $b\bar b$
pairs as an ensemble of color-singlet states. 
In a quantum-mechanical framework, this would amount to initialize the color part of the wave function of
each pair as
\begin{equation}
|S\rangle = \frac{1}{\sqrt{3}}(|r\bar r\rangle + |g\bar g\rangle + |b\bar b\rangle ).
\label{eq:parisi_1113}
\end{equation}
For the state~\eqref{eq:parisi_1113} 
we have
\begin{equation}
\langle S |\sum_a(T^a + \bar T^a)^2 |S\rangle = 0,
\label{eq:visonobruttenotizie}
\end{equation}
where $T^a$ and $\bar T_a$ denote the SU(3) color generators
acting on the Hilbert spaces of the quark and the antiquark
respectively. In the classical limit, we replace
operators with their expectation values.
Denoting by $Q_a$ and $\bar{Q}_a$ the color charges of, respectively, the quark and the antiquark in a pair, the classical version of Eq.~\eqref{eq:visonobruttenotizie} reads
\begin{equation}
\sum_a (Q_a + \bar Q_a)^2=0,
\label{eq:giocateallaguerra}
\end{equation}
which immediately implies
\begin{equation}
Q_a = - \bar Q_a,\qquad \text{for } a=1,\dots,N_c^2-1.
\label{eq:dottoresenzapantaloni}
\end{equation}
Hence, we initialize the color charge of each antiquark by imposing the condition~\eqref{eq:dottoresenzapantaloni}.
This also implies that, by taking into account the normalization
of the classical charges~\eqref{eq:mostrasumarte}, 
at initial time we have:
\begin{equation}
\sum_a Q_a \bar Q_a = -q_2.
\label{eq:giornicontati}
\end{equation}

\section{Results}
\label{sec:results}

In this Section we show our results on the heavy-quark evolution and heavy quark-antiquark pair dissociation in proton-nucleus collisions at LHC energy.
The results shown in this section have been obtained for a square grid of transverse size $L_\perp=3$ fm, with $N_\perp=64$. The MV parameter has been fixed as $\mu=0.5$ GeV. In realistic collisions,
the pre-equilibrium stage lasts for few fractions of fm,
but for illustrative purposes
we performed simulations up to 1 fm with 1000 time steps
(we checked that this number of time steps is enough to have
numerical convergence and to preserve the Casimir invariants \eqref{eq:mostrasumarte}).

In each simulation we considered 1000 heavy quark pairs, for a total of 2000 particles.
We remark that 
the pairs used in our simulation serve as test particles and their number does not correspond
to the number of actual pairs produced in the collisions. Since we will only focus on
percentages and fractions over the total number of pairs, such value is chosen purely for statistical reasons. Moreover, each quark pair interacts only within itself, since we consider all the pairs to be independent: this is motivated by the net average number
of formed HQ pairs in a pA collision being very small.
This can be estimated as follows.
For Pb-Pb collisions in the $0-10\%$ centrality class, 
it has been estimated that
$\langle N_{c\bar c~\mathrm{ pairs}}\rangle\approx 15$ (see for example \cite{Minissale:2023dct}).
From this, we can extrapolate the average number of pairs 
in p-Pb collisions
by scaling $\langle N_{c\bar c~\mathrm{ pairs}}\rangle$ with 
the number of binary collisions, $N_\mathrm{coll}$.
The latter can be computed
by the average $\mu^2$ of the proton and the nucleus,
being $\mu^2 \propto T$ where $T$ denotes the thickness
function. 
In particular, 
within our implementation, for the nucleus
we have $\langle \mu^2\rangle_{A}=(0.5$ GeV)$^2$, it being trivially constant in the interaction region. 
For the proton, we use Eqs.~\eqref{eq:qs_2_77} and~\eqref{eq:gmu_ajk} to get $\mu(\bm{x}_\perp)$, then we integrate over all constituent quark configurations and over the transverse plane, and finally divide by
the area of the proton. We get
\begin{equation}
\langle \mu^2 \rangle_{p}=\frac{\displaystyle{\int \mathrm{d}^2 \bm{x}_i \int \mathrm{d}^2 \bm{x}_\perp\, \mu^2(\bm{x}_\perp)\, T_{cq}(\bm{x}_i)}}{\pi R_p^2}=\frac{c^2 \, xg}{6 g^2 R_p^2},
    \label{average_mu_pA}
\end{equation}
where we consider $R_p=1$ fm for the radius of the proton, and $T_{cq}(\bm{x}_\perp)$ is given by \eqref{eq:bd2}. 
Consequently, 
\begin{equation}
\frac{N_\mathrm{coll}^{\text{pA}}}{N_\mathrm{coll}^{\text{AA}}}
=
\frac{\langle \mu^2 \rangle_{\text{p}}\langle \mu^2 \rangle_{\text{A}}}
{\langle \mu^2 \rangle_{\text{A}}^2}\approx \frac{1}{25}.
    \label{ratio_mu_pA_AA}
\end{equation}
This
gives $\langle N_{c\bar c~\mathrm{ pairs}}\rangle\approx 0.6$ in p-Pb. For $b\bar b$ the numbers are even smaller: one can estimate a cross section for $b\bar{b}$ production which is 25 smaller with respect to $c \bar{c}$ production \cite{Cacciari:2012ny, ALICE:2014aev}.
Therefore the system of $c\bar{c}$ and $b\bar{b}$ pairs is very dilute and interaction among different pairs is negligible, which justifies our assumption of independent pairs.

In our numerical implementation,
following~\cite{Pooja:2024rnn}
we cure the numerical divergence for $r_\text{rel}=0$ in~\eqref{HQ_potential} 
by using  in Eq.~\eqref{eq:wong2}
the regularized potential
\begin{equation}
V=\frac{Q_a  \bar{Q}_a}{N_c}
\frac{\alpha_s}{r_\text{rel}}(1-e^{-A\, r_\text{rel}}),
    \label{HQ_potential_regularized}
\end{equation}
instead of~\eqref{HQ_potential}.
Here we present results obtained with
$A=(0.017\text{ fm})^{-1}$: 
we checked that taking $A$ larger than this value makes the screening of the divergence at $r_\text{rel}=0$ ineffective, not leading to the convergence of the numerical algorithm.

\subsection{Spreading and color decorrelation of the pairs\label{subse:rideremmo}}

\begin{figure}[t!]
\centering
\includegraphics[width=.49
\linewidth]{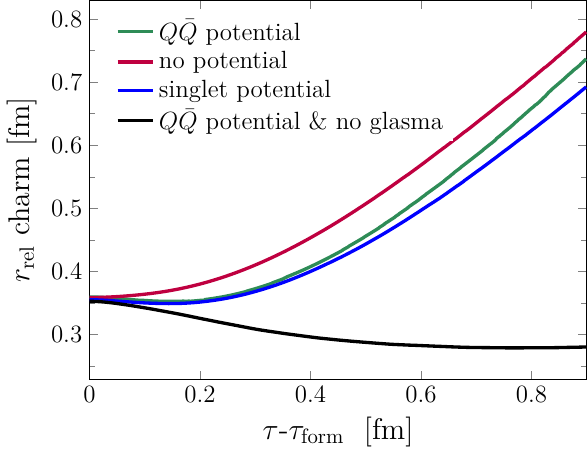}
\includegraphics[width=.49
\linewidth]{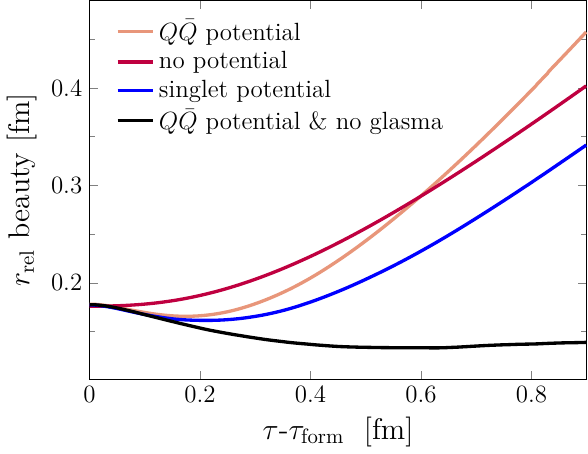}
\caption{Averaged relative distance for $c\bar c$ \textbf{(left)} and $b\bar b$ \textbf{(right)} pairs,
computed in the pair rest frame.
The ‘‘no potential'' data (brown curves) correspond to calculations without the
potential~\eqref{HQ_potential} in the Wong equation~\eqref{eq:wong2}. The
‘‘$Q\bar Q$ potential'' data (green curve on the left, salmon curve on the right) denote the results obtained considering 
the potential as well as the evolving glasma fields. The blue line ‘‘singlet potential'' stands
for the results obtained assuming a pure color-singlet potential \eqref{HQ_potentialsing} in
Eq.~\eqref{eq:wong2}. Finally, ‘‘$Q\bar Q$ potential \& no glasma'' (black curve) denotes the results obtained by solving the equations of motion
switching off the interaction of the quarks with the fields.}
\label{Fig:rreldifferentpotentials}
\end{figure}

Let us focus on heavy quark-antiquark pairs and investigate the effect of the evolving glasma in dissociating the initially produced pairs.
The relative position of the quark and antiquark in the pairs is modified during their propagation in the fireball, both in the pre-equilibrium and in the thermalized QGP phases.
In particular, the interaction with the gluon fields in the initial stages
leads to a diffusion in coordinate (as well as momentum) space,
resulting on average 
in the increase of the relative distance of the two particles in a pair.

In Fig.~\ref{Fig:rreldifferentpotentials} we plot the relative 
distance $r_{\mathrm{rel}}$ of quark and antiquark in 
$c\bar{c}$ (left) and $b\bar{b}$ (right) pairs, in their rest frame, versus 
the proper time $\tau$ (measured with respect to the formation time
$\tau_\mathrm{form}$). The results of each pair have been averaged over the HQ pairs as well as over the gauge field configurations.  We show our results in four different cases: along with the result obtained with the dynamic potential~\eqref{HQ_potential} (solid green or salmon, for charm or beauty respectively), we plot $r_{\mathrm{rel}}$ obtained by considering $V=0$ (solid red),  by using $V=V_1$ 
defined in Eq.~\eqref{HQ_potentialsing} (solid blue), 
as well as
by turning off the gluon fields (solid black).
The latter is shown in order to illustrate the quantitative effect of the HQ potential and of the gluon fields
on $r_{\mathrm{rel}}$. 

In the scenario in which only the potential is considered (black curves), 
$r_{\mathrm{rel}}$ decreases over time due to the attractive interaction between the quark and the antiquark in the pair. Indeed, in this case the dynamical potential coincides with the singlet potential, since the color charges do not evolve under the action of the glasma.
On the other hand (green and salmon curves), we see that
the effect of the glasma is to increase the separation of the pair, 
due to the intense bombardment of gluons on both the quark and antiquark
of the pair. Notably, during the time interval where the glasma framework 
is phenomenologically relevant, i.e., up to approximately $0.3{-}0.4$ fm, 
$r_{\mathrm{rel}}$ remains nearly constant. This behavior results from the 
competition between the attractive potential and the interaction with the 
color fields of the glasma. For completeness, in Fig.~\ref{Fig:rreldifferentpotentials} we also display the case where the potential 
is absent (brown curves). 
By comparing this with the full calculation, we observe that, 
as expected, the presence of an attractive potential leads to an initial 
decrease in $r_{\mathrm{rel}}$. Finally, when the potential is modeled as a pure singlet interaction (blue curves),
$r_{\mathrm{rel}}$ remains very close to the result of the full calculation at early times, within the time range where the glasma is phenomenologically relevant. 
This suggests that, during this period, the 
interaction between the quarks in the pair is not significantly different from that of a pure singlet. On the other hand, moving on at later times, $r_{\mathrm{rel}}$ in the pure singlet 
scenario is slightly lower if compared to the full calculation. 
This is a natural consequence of the fact that the singlet potential remains stronger than the full potential at larger times, leading to a tighter binding of the quark-antiquark pairs.

\begin{figure}[t!]
\centering
\includegraphics[width=.49
\linewidth]{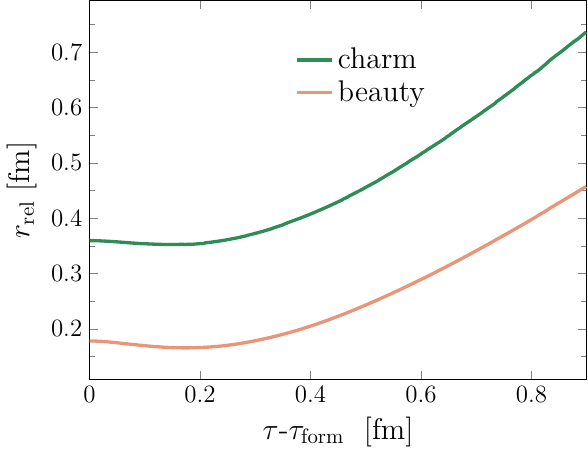}
\includegraphics[width=.49
\linewidth]{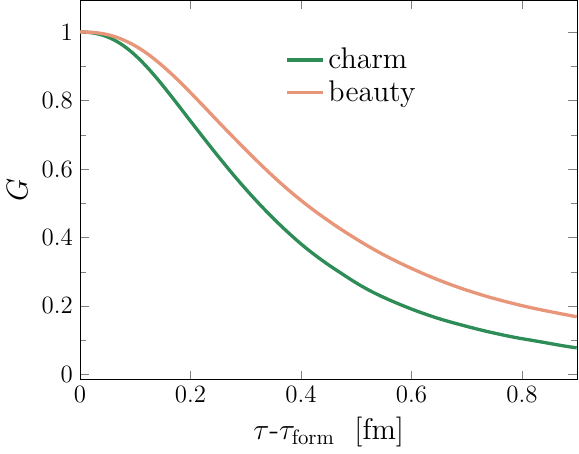}
\caption{Averaged relative coordinate in the pair rest frame \textbf{(left)} and color-charge correlator \textbf{(right)} of the heavy quark-antiquark pairs versus time $\tau-\tau_\mathrm{form}$, with $\tau_\mathrm{form}$ being the heavy-quark formation time. Green and salmon curves correspond to charm and beauty heavy-quark pairs, respectively.}
\label{fig:r_rel_c_cor_charm_vs_beauty}
\end{figure}

By now limiting ourselves to the physical case of the potential in Eq.~\eqref{HQ_potential}, in the left panel of Fig.~\ref{fig:r_rel_c_cor_charm_vs_beauty} we show $r_{\mathrm{rel}}$ in the pair rest frame versus $\tau-\tau_\mathrm{form}$ for a $c\bar{c}$ pair (same green curve of Fig.~\ref{Fig:rreldifferentpotentials} left) and for a $b\bar{b}$ pair (same salmon curve of Fig.~\ref{Fig:rreldifferentpotentials} right). We once again see that, within the glasma timescales, the relative distance remains basically constant. Moreover, we see that even when we take into account the different initialization in position for charm and beauty, we interestingly note that the $c\bar c$ system spreads slightly more quickly with respect to the $b\bar b$ one. This is particularly remarkable if one thinks that beauty quarks are produced earlier, therefore they experience a stronger diffusive effect of glasma fields. Nonetheless, their mean relative distance grows slower than for the charm quarks, since the higher masses of $b$ make them more static in the gluon fields.\\

Not only coordinates and momenta of heavy quarks are affected by
the background gluon fields, but also their color charges are, see Eq.~\eqref{eq:wong3}.
The quark-antiquark pairs can undergo transitions from singlet to octet (and viceversa) both in glasma and in QGP. In the latter stage, such transitions have been studied especially within the framework of open quantum systems \cite{Casalderrey-Solana:2012yfo, Brambilla:2016wgg, Miura:2022arv}. 
These singlet-octet transitions may also occur in the glasma stage, due to the interaction of the quark and antiquark in the pair with the initial color fields. For this reason we study the decorrelation of the color charges of the quark and the antiquark in a pair, which serves as an estimate of the probability of singlet-to-octet transitions, and hence gives indications on the survival probability
of the pair in the gluon fields.

As explained in Section~\ref{sec:colorspace},
we initialize the quark-antiquark pairs
in color-singlet states, then the interactions with the gluonic environment changes their color charges. 
Denoting by $Q_a$ and $\bar Q_a$ the color charges of a quark
and its antiquark respectively, let us consider for each pair the time-dependent quantity
\begin{equation}
\mathcal{G}(\tau) = -\frac{1}{q_2}\mathrm{Tr}\left[Q_a(\tau) \bar Q_b(\tau)\right]
=-\frac{1}{q_2}\sum_a Q_a(\tau) \bar Q_a(\tau),
\label{eq:intascailtelefonino}
\end{equation}  
where $q_2$ is the quadratic Casimir invariant of
SU(3) introduced in \eqref{eq:casimir_values}, and the trace is understood over the $(a,b)$ indices. 
For the color-singlet initialization we have $Q_a   = - \bar Q_a $ at
$\tau=\tau_\mathrm{form}$, therefore for each pair
\begin{equation}
\mathcal{G}(\tau_\mathrm{form})=1,
\label{eq:mascheradoro}
\end{equation}
indicating that with our initialization the color charges of the two particles at $\tau=\tau_\mathrm{form}$ are maximally anti-correlated. We expect that on average the
magnitude of $\mathcal{G}$ decreases with time, as a result of the 
decorrelation of $Q_a$ and $\bar Q_b$ due to the random interactions with the gluon fields of the evolving glasma. 
This decrease can be interpreted as
the raising of the weight of the color-octet component in the pair,
which eventually leads to the melting of the pair itself. 
We will use this idea to define a melting probability of the
$c\bar c$ and $b\bar b$ pairs, see Section~\ref{subsec:melting_of_the_pairs}.

From the above definition \eqref{eq:intascailtelefonino} of $\mathcal{G}$, we define the gauge-invariant color-charge correlator as~\cite{Pooja:2024rnn}
\begin{equation}
    G(\tau) \equiv   \langle   \mathcal{G}(\tau)
    \rangle.
    \label{eq:c_cor}
\end{equation}
The brackets in~\eqref{eq:c_cor} 
indicate the average 
over the pairs as well as over the gauge field configurations (i.e. numerical events). 
In the right panel of Fig.~\ref{fig:r_rel_c_cor_charm_vs_beauty} we plot
$G(\tau)$ versus time $\tau-\tau_\mathrm{form}$,  
for charm (green line) and beauty (salmon line) pairs.
We notice that $G$ decreases with time, 
as expected,
indicating a decorrelation of the color charges of the quark and of the companion antiquark in the pair.

In order to be more quantitative
we define a decorrelation time,
$\tau_\mathrm{dec}$,
via the requirement that $G(\tau_\mathrm{dec})=1/2$.
We notice that
$\tau_\mathrm{dec}-\tau_\mathrm{form}$ stays in the range
$(0.3,0.4)$ fm for both charm and beauty.
We also find that $\tau_\mathrm{dec}-\tau_\mathrm{form}$ 
for $b$ quarks is larger than the corresponding quantity 
computed for the $c$ quarks: this is most likely due to the fact that
$b\bar b$ pairs are tighter, hence they spend more time within 
a single correlation domain
of the gluonic background, and hence get decorrelated more slowly. We could interpret $G(\tau)$ as the probability for pairs to stay
in the singlet channel during propagation in the 
pre-equilibrium stage, or equivalently,
$1-G(\tau)$ is the probability to have singlet-to-octet
fluctuations.
In fact, during the evolution,
the decrease of the correlator can be interpreted as the fluctuations
of the color charges becoming increasingly more important, making transitions to a color-octet state more probable.

One of the interesting aspects shown in 
Figs.~\ref{Fig:rreldifferentpotentials} and~\ref{fig:r_rel_c_cor_charm_vs_beauty} 
is that, while the interquark relative distance within the pairs remains nearly 
constant 
up to $\tau \approx 0.3$~fm for both $c$ and $b$ quark pairs, color decorrelation occurs rather quickly. 
In this time range, it is therefore color decorrelation that primarily 
drives 
the melting of the pair. We will keep this in mind in Section~\ref{subsec:melting_of_the_pairs}, when we will define a melting probability for HQ pairs.

\subsection{Color equilibration\label{sec:coequi}}

The results shown in the right panel of Fig.~\ref{fig:r_rel_c_cor_charm_vs_beauty} can be rephrased in terms of the expectation values of the color-projector operators,
that allow us to measure the probability of pairs 
to be in the color-singlet and the color-octet states.
To this end, we recall that in QCD one can introduce
the projectors onto the singlet, $\mathcal P_S$, and the octet, 
$\mathcal P_O$, spaces as
\begin{align}
\mathcal P_S &= -\frac{2}{3}T^a \otimes \bar T^a   + \frac{ 1}{9}\mathbb{1},
\label{eq:proiettori_singoletto}\\
\mathcal P_O &= \frac{2}{3}T^a\otimes \bar T^a + \frac{8 }{9}\mathbb{1},
\label{eq:proiettori_ottetto}
\end{align}
which can be easily obtained from the eigenvalues
of the operator $T^a \otimes \bar T^a$ in the color-singlet,
$\lambda_S$, and 
color-octet, $\lambda_O$, representations, namely
\begin{equation}
\lambda_S = -\frac{4}{3},~~~\lambda_O = + \frac{1}{6}.
\label{eq:autovalori}
\end{equation}
It is easy to verify that these projectors actually satisfy the following properties
\begin{equation}
\mathcal P_S^2 = \mathcal P_S,\quad \mathcal P_O^2 = \mathcal P_O, \quad \mathcal P_O\mathcal P_S = \mathcal P_S \mathcal P_O = 0,\quad \mathcal P_O + \mathcal P_S=\mathbb{1}.
    \label{eq:properties_color_projectors}
\end{equation}
The classical counterpart of
the projectors~\eqref{eq:proiettori_singoletto}
and~\eqref{eq:proiettori_ottetto} can be written as
\begin{align}
P_S &= -\frac{2}{3}\frac{Q_a \bar Q_a}{N_c}
+ \frac{1}{9},\label{eq:pro_sin_cla}\\
P_O &= \frac{2}{3}\frac{Q_a \bar Q_a}{N_c} 
+ \frac{8}{9},\label{eq:pro_oct_cla}
\end{align}
which can be obtained from
Eqs.~\eqref{eq:proiettori_singoletto}
and~\eqref{eq:proiettori_ottetto}
via the formal replacement
$T^a \otimes\bar T^a \rightarrow Q_a \bar Q_a/N_c$, 
similarly to what we have done in writing the classical 
counterpart of the potential~\eqref{HQ_potentialsing_buffet}
in Eq.~\eqref{HQ_potential}.
It is easy to check that
the singlet condition~\eqref{eq:giornicontati}
implies $P_S=1$ and $P_O=0$.

We now want to relate $P_S$ to the probability
that the quark-antiquark pair is in the
color-singlet state. In fact, 
in the quantum theory, the probability 
that the state $|\psi\rangle$ is a
color-singlet is 
\begin{equation}
p_S = 
|\langle\psi| \mathcal P_S|\psi \rangle |^2.
\label{eq:cb_ooo}
\end{equation}
If $|\psi\rangle$ represents a semi-classical
state, we can neglect
the quantum fluctuations and we can write
\begin{equation}
p_S=\langle\psi| \mathcal P_S^2|\psi \rangle,
\label{eq:cb_pbs}
\end{equation}
and taking into account that $\mathcal P_S^2 =
\mathcal P_S$, we get
\begin{equation}
p_S=\langle\psi| \mathcal P_S|\psi \rangle=P_S,
\label{eq:cb_f.it}
\end{equation}
where the last equality stands in the classical limit. 
Hence, we can interpret $P_S$ as the probability
that the semiclassical quark-antiquark
state is in the color-singlet state. Similarly, $P_O=1-P_S$ gives the probability that the pair is in the color-octet state.

The identification of $P_S$ and $P_O$ as the probabilities for the state to be in the singlet and octet representations, respectively, allows us to justify {\sl a posteriori} the potential~\eqref{HQ_potential}, as it coincides with the classical limit of the QCD potential.
 In fact, 
let us consider
a linear combination of 
singlet, $|S\rangle$, and octet 
$|O_i\rangle$ with $i=1,\dots,8$,
states, namely
\begin{equation}
|\psi\rangle = c_S |S\rangle + \sum_{i=1}^8c_{i}|O_i\rangle,
\label{eq:psiso1}
\end{equation}
with $|c_S|^2 + \sum |c_i|^2=1$.
The QCD potential is
\begin{equation}
V=V_1\,\mathbb{1}_S + V_8\,\mathbb{1}_O.
    \label{eq:V_v1_v8}
\end{equation}
where $\mathbb{1}_S$ and $\mathbb{1}_O$ are identities in the singlet
and the octet representations, and
the potentials $V_1$ and $V_8$ are given by Eqs.~\eqref{HQ_potentialsing} and~\eqref{HQ_potentialsingoct}. All that said, the projection of this $V$ on the state $|\psi\rangle$ is
\begin{equation}
\langle \psi | V | \psi \rangle =
|c_S|^2 V_1 + (1-|c_S|^2) V_8.
\label{eq:projection1641}
\end{equation}
Within our semi-classical approach, we replace 
$|c_S|^2$  with $P_S$ 
in the
right hand side of 
Eq.~\eqref{eq:projection1641}, 
see Eq.~\eqref{eq:cb_f.it},
and 
the state $|\psi\rangle$ with a classical state
representing our quark-antiquark pairs. 
We then obtain
\begin{equation}
  V   = P_S V_1 + (1-P_S) V_8.
\label{eq:projection1643}
\end{equation}
Taking into account Eq.~\eqref{HQ_potentialsing}, \eqref{HQ_potentialsingoct} and~\eqref{eq:pro_sin_cla}, we 
finally get
Eq.~\eqref{HQ_potential}.
Therefore, we can identify the potential~\eqref{HQ_potential}
with the classical limit of the QCD potential~\eqref{eq:projection1641}.

\begin{figure}[t!]
\centering
\includegraphics[width=.6
\linewidth]{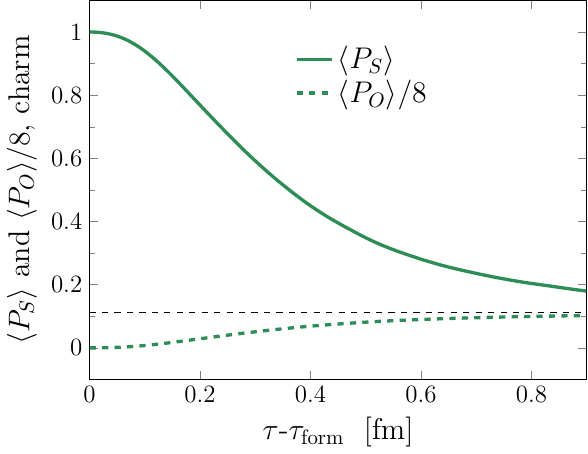}
\caption{Expectation values of the singlet and the (normalized)
octet projectors versus proper time for $c\bar c$ pairs.
The dashed horizontal black line stands for $1/9$.
The results show the almost-perfect equipartition
of probability 
among color-singlet and color-octet states at large times.}
\label{fig:Ps_Po}
\end{figure}

In Fig.~\ref{fig:Ps_Po} we plot $\langle P_S\rangle$ and 
$\langle P_O\rangle/8$
versus proper time, computed for $c\bar c$ pairs
(the results for $b\bar b$ pairs are similar).
The dashed horizontal black line stands for the value $1/9$.
In particular, we divided $\langle P_O\rangle $ 
by eight in order to
count the probability of being in one
of the eight states
of the octet.
The results shown in the Figure have been obtained by
taking the ensemble average of Eqs.~\eqref{eq:pro_sin_cla}
and~\eqref{eq:pro_oct_cla}, similarly to the procedure we 
followed to produce the results for $G(\tau)$ shown in Fig.~\ref{fig:r_rel_c_cor_charm_vs_beauty}.

We notice in Fig.~\ref{fig:Ps_Po}
that at the formation time the pairs 
are in 
the singlet state 
with probability $1$,
due to our initialization of the color
charges. As the pairs evolve and interact with the
gluon fields, there are transitions from the singlet
to the octet states as a result of color decorrelation.
At large times, when the colors of the quark and the antiquark
in the pair are uncorrelated, the averaged projectors
tend to the same value,
\begin{equation}
\langle P_S \rangle \simeq  \langle P_O \rangle/8
\simeq 1/9.
\label{eq:equipartition_1656}
\end{equation}
This shows that asymptotically there is an equipartition
of probability for the pairs to be in the singlet
and in one of the eight octet states.

It is very interesting to notice that, from the
qualitative point of view, the color equilibration 
shown in Fig.~\ref{fig:Ps_Po} is in agreement
with the results of Ref.~\cite{Delorme:2024rdo}, in which 
the singlet-to-octet 
transitions have been studied for quarkonia states
in a thermalized QGP
within the framework of the quantum Brownian motion.
In that context, color equilibration takes place
on a longer timescale, likely because of
the different energy scales which are relevant in the
two problems. Nevertheless,
the interaction with the external environment (the QGP 
in~\cite{Delorme:2024rdo} and the background gluon fields
in the present work) in both cases leads to the equipartition
between singlet and octet states.

\subsection{Melting of the pairs}
\label{subsec:melting_of_the_pairs}

We make use of the results discussed in Section~\ref{subse:rideremmo}
to define a survival probability of the pairs in the pre-equilibrium stage,
$\mathcal{P}_\mathrm{survival}$, and from this a melting probability:
\begin{equation}
\mathcal{P}_\mathrm{melting}=1-\mathcal{P}_\mathrm{survival}.
\label{eq:presepi_a_napoli}
\end{equation}
In order to define $\mathcal{P}_\mathrm{survival}$, 
we notice in Fig.~\ref{fig:r_rel_c_cor_charm_vs_beauty}
that the relative distance is almost constant during the early evolution
of the system, while the color of the quarks in the pairs decorrelates
quickly. Therefore, it is reasonable to assume that color rotation is the leading mechanism to dissolve the pairs, hence $\mathcal{P}_\mathrm{survival}$ 
can be connected to the color-charge correlator $\mathcal{G}$ defined in
Eq.~\eqref{eq:intascailtelefonino}. 
We assume a Gaussian shape of $\mathcal{P}_\mathrm{survival}$ in color space, namely
\begin{equation}  \mathcal{P}_\mathrm{survival}=\exp\left[-\kappa(\mathcal{G}-1)^2\right],\label{eq:conte_non_contento}
\end{equation}
where $\kappa$ is a parameter that measures
the width of the fluctuations of the color charges
that are necessary to break the pair.
The form~\eqref{eq:conte_non_contento} of $\mathcal{P}_\mathrm{survival}$ is inspired by the Wigner function of coalescence models to form a vector meson from a quark and an antiquark of the same flavor \cite{Greco:2003vf}. 
In our model, at each time, the status of melted for each quark-antiquark pair is assigned with the probability given by Eq.~\eqref{eq:presepi_a_napoli}. We have investigated the effect of 
the pre-equilibrium stage
on pair dissociation for several values of $\kappa$.
Here, for both charm and beauty
we show results for $\kappa=4$, 
which corresponds
to require that 
we have $\mathcal{P}_\mathrm{survival}=1/e$ when $\mathcal{G}=1/2$.

\begin{figure}[t!]
\centering
\includegraphics[width=.6
\linewidth]{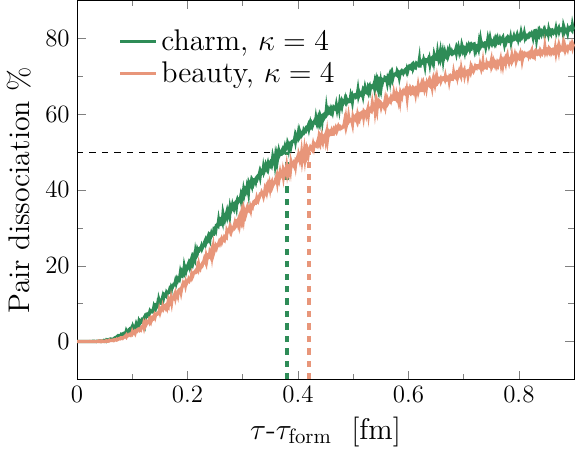}
\caption{Percentage of dissociated charm and bottom quark-antiquark pairs versus time $\tau-\tau_\mathrm{form}$, with $\tau_\mathrm{form}$ being the heavy quark formation time. We highlight, using dashed lines, the time at which we have a dissociation percentage of $50\%$ for charm and beauty pairs.}
\label{fig:dissoc}
\end{figure}

We quantify the impact of the early stage on the melting of the pairs by calculating the number of dissociated pairs with respect to their initial value. We show this quantity, in percentage, as a function of time $\tau-\tau_\mathrm{form}$ in Fig.~\ref{fig:dissoc} for both $c\bar{c}$ and $b\bar{b}$ pairs. 
By introducing the time to break half of the initial pairs, $\tau_\mathrm{break}$,
we find $\tau_\mathrm{break}-\tau_\mathrm{form}=0.37$ fm
for $c\bar c$ pairs, 
and  $\tau_\mathrm{break}-\tau_\mathrm{form}=0.42$ fm for $b\bar b$ pairs. This estimate is not strongly dependent on $\kappa$: for instance, by taking $\kappa=9/4$, which amounts to require that 
$\mathcal{P}_\mathrm{survival}=1/e$ for $\mathcal{G}=1/3$, we find
$\tau_\mathrm{break}-\tau_\mathrm{form}=0.43$ fm
for $c\bar c$ pairs, 
and  $\tau_\mathrm{break}-\tau_\mathrm{form}=0.54$ fm for $b\bar b$ pairs. We conclude that the effect of the interaction of the heavy quarks
with the gluon fields in the pre-equilibrium stage is to gradually
melt the pairs due to the color rotation of the quark and its companion
antiquark,
and the time needed to break half of the initial pairs
can be as small
as $0.4-0.5$ fm from the formation time. 
We finally notice 
that our results on the dissociation percentage and timescales are in the same 
ballpark as those shown in~\cite{Pooja:2024rnn},
in which the melting of pairs in the early stage was already studied, albeit within a simpler model.

\chapter[Anisotropies of gluons and HQs in glasma]{Anisotropies of gluons and heavy quarks in glasma}
\label{chap:Anisotropies of gluons and heavy quarks in glasma}

This Chapter is largely based on the work contained in \cite{Parisi:2026uhh}:
\begin{center}
    Gabriele Parisi, Fabrizio Murgana, Vincenzo Greco and Marco Ruggieri, ``Elliptic flow of charm quarks produced in the early stage of pA collisions'', arXiv: 2601.11123 [hep-ph].\\
\end{center}

In realistic heavy ion collisions, immediately after the impact, an initial distribution with a huge energy density is left in the central region. However, the shape of such a medium is far from being isotropic: in the naive picture of two disks colliding with a certain impact parameter $b$, this region can be described by a typical almond shape, which can be characterized by the geometrical ellipticity. However, in a more refined picture which includes event-by-event fluctuations, the initial shape of the fireball has a fully irregular shape, which is expected to reduce to an almond only if an average over the events, namely in the same centrality class, is performed. The event-by-event fluctuations do have a measurable impact on the final observables and therefore one has to take into account the irregularity of the initial state to reproduce experimental data. This irregular shape can be described via an (infinite) series of coefficients, called {\sl eccentricities}, which we have briefly introduced in §\ref{subsection:elliptic_flow}:
namely, the ellipticity $\epsilon_2\ne 0$ indicates an almond shape, $\epsilon_3\ne 0$ a triangular shape,  $\epsilon_4\ne 0$ a quadrangular shape and so on. Obviously, in a realistic initial condition only a few of all coefficients $\epsilon_n$ have a significantly large value. As already argued in §\ref{subsection:elliptic_flow}, we know that the initial eccentricities determine late-time collective observables, such as the anisotropic flows $v_n$. The interaction happening within the medium, indeed, converts this spatial anisotropy into a corresponding momentum anisotropy. 

The previous considerations do not hold only for the quark-gluon plasma phase, but also for the initial gluon-dominated glasma phase. In particular, the fact that the early stage already possesses a finite momentum anisotropy is well known, see for example \cite{Schenke:2015aqa}. However, the results on the collective flows of the evolving glasma depend on the gluon spectrum used in the actual
calculations, that is, on the conversion from fields to on-shell particles. On the other hand, this ambiguity does not hold for the momentum distribution of heavy quarks, which is uniquely determined: consequently, we can use HQs to quantify the amount of anisotropy carried by the system produced in the early stages of heavy ion collisions. 

In this Chapter we move on to deal with momentum anisotropies in the glasma and in heavy quarks moving within such glasma. Most works in the literature have focused on the elliptic flow $v_2$ produced in the QGP, using various approaches, e.g., relativistic transport theory and hydrodynamics \cite{Plumari:2015cfa, Plumari:2019gwq, Romatschke:2007mq, Song:2008si, Schenke:2010rr, Ferini:2008he, Xu:2008av, Greco:2008fs, Plumari:2011re, Konchakovski:2012yg} which aim to reproduce the experimental data. On the other hand, one of the main points of this Chapter is to show that despite the short lifetime of the early, out-of-equilibrium,  stage in AA and pA collisions, a significant amount of anisotropy from the gluon bulk is transmitted to HQs, in particular to charm quarks. Since we want to study the anisotropies generated by the event-by-event fluctuations, we will improve our modelization of an heavy nucleus to take into account for the positions of the single nucleons within each nucleus. This will resemble what we already explained in §\ref{sec:Beyond naive MV model: pA collisions} for the structure of a proton. In this Chapter we will first study the momentum anisotropy of the glasma medium itself, and then move on to the charm quarks evolving within such medium. In particular, we will highlight the effect of glasma on the momentum spectrum and on the elliptic flow, showing that a significant part of the momentum anisotropy observed in experimental runs of proton-lead collisions may be traced back to the initial stages. Moreover, we will see that these considerations do not depend on the specific method used for the calculation of $v_2$, either event-plane or 2-particle correlator, since the results obtained with the two methods are in excellent agreement with each other.

\section{Generalization of the nucleus structure}
As mentioned, in this Chapter we extend the definition of color charge for an heavy nucleus, in order to study event-by-event fluctuations. The physical rationale is simple. The MV 'wall model' we have introduced in Eqs.
\eqref{eq:mv_numerical_implementation_1} and \eqref{eq:mv_numerical_implementation_2}, which we have employed to describe the initial color charge without the need to model single nucleons (see Fig. \ref{Fig:energy_dens_AA}), is reasonable only when dealing with heavy nuclei (e.g. Pb at LHC and Au at RHIC). However, for smaller systems one has to take into account the sub-nucleonic structure, especially when studying anisotropies. For this reason, what we want to do is to generalize the modelization of the proton given in §\ref{sec:Beyond naive MV model: pA collisions}, by extending Eqs. \eqref{eq:bd1} and \eqref{eq:bd2} to obtain the thickness function for a general number of participant nucleons $N_\text{part}$:

\begin{equation}
T_p(\bm{x}_\perp) = 
\frac{1}{3} \sum_{i=1}^{3N_\text{part}} \frac{1}{2\pi B_q} \exp\left[-\frac{(\bm{x}_\perp-\bm{x}_\perp^i)^2}{2B_q}\right].
\label{eq:bd1_generalized}
\end{equation}
\begin{equation}
T_{cq}(\bm{x}_\perp^i) = \frac{1}{2\pi B_{cq}}
\exp\left(-\frac{\bm{x}_\perp^{i, 2}}{2B_{cq}}\right),
\label{eq:bd2_generalized}
\end{equation}
As for the proton, what one does in practice is to extract $3N_\text{part}$ pairs of random numbers from the distribution \eqref{eq:bd2_generalized}, which fix the positions of the three quarks of each nucleon. Those are then inserted in Eq. \eqref{eq:bd1_generalized} in order to obtain the charge profile of the nucleus. 
The parameters $B_q$ and $B_{cq}$ have been fixed once again as $B_q=0.3$ GeV$^{-2}$ and
$B_{cq}=4$ GeV$^{-2}$ \cite{Mantysaari:2016ykx}. 

\begin{figure}[t!]
    \centering
\includegraphics[width=0.45\linewidth]{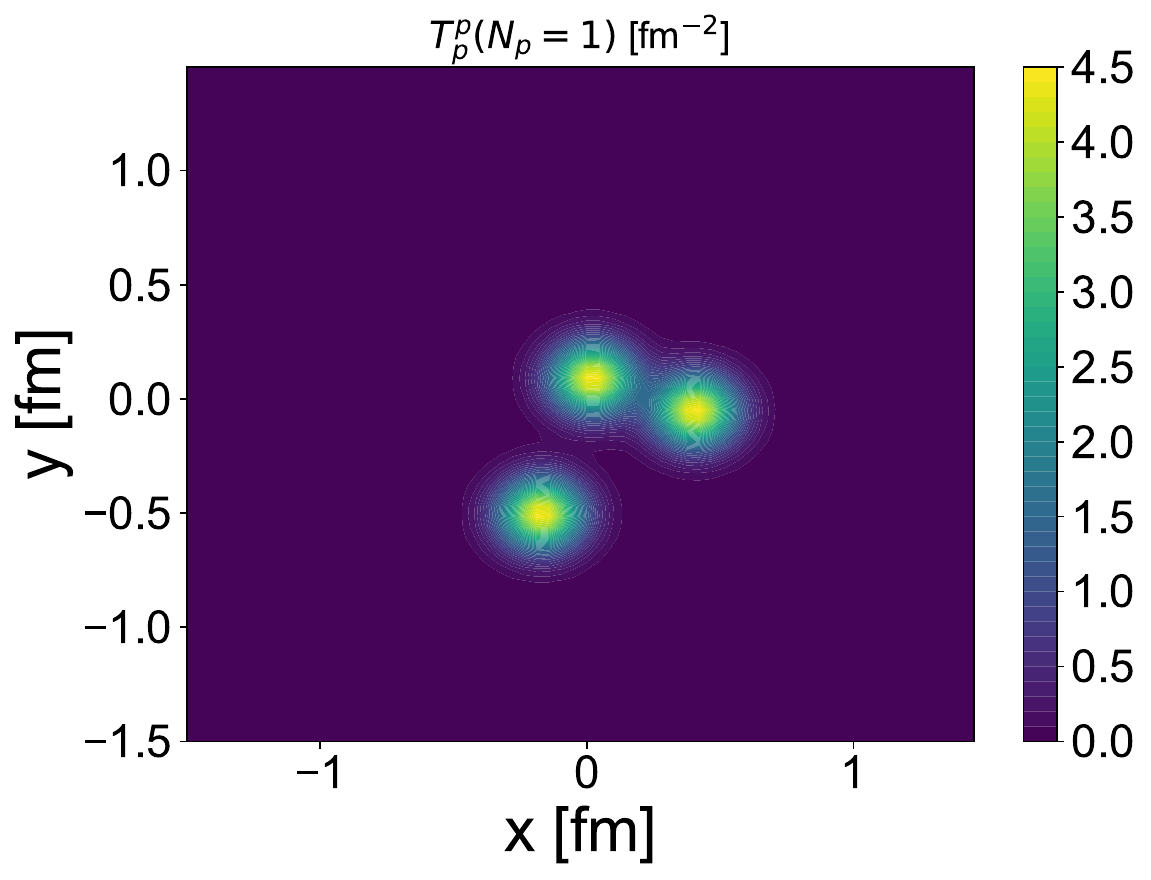}~~~
\includegraphics[width=0.37\linewidth]{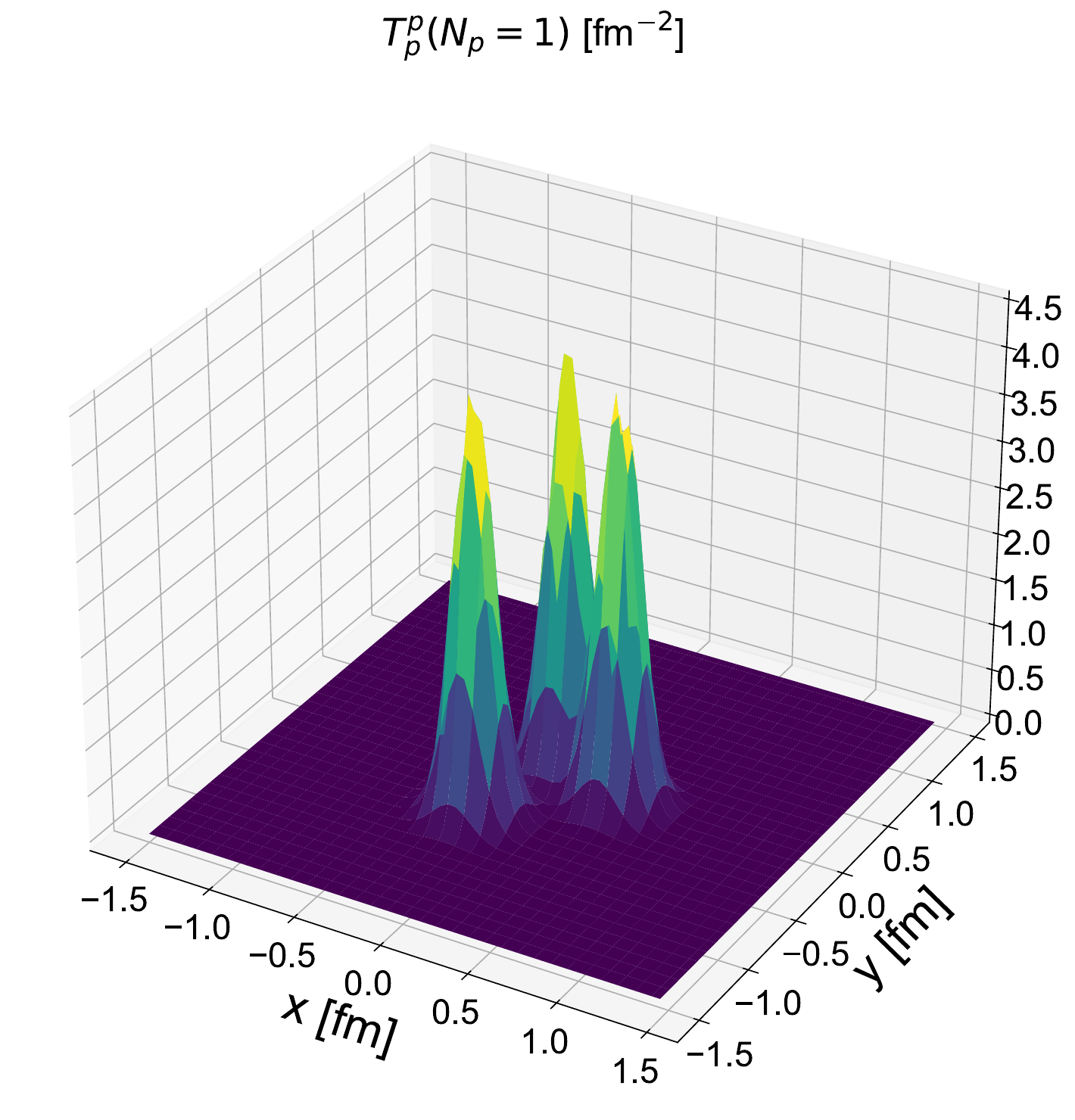}\\
\includegraphics[width=0.45\linewidth]{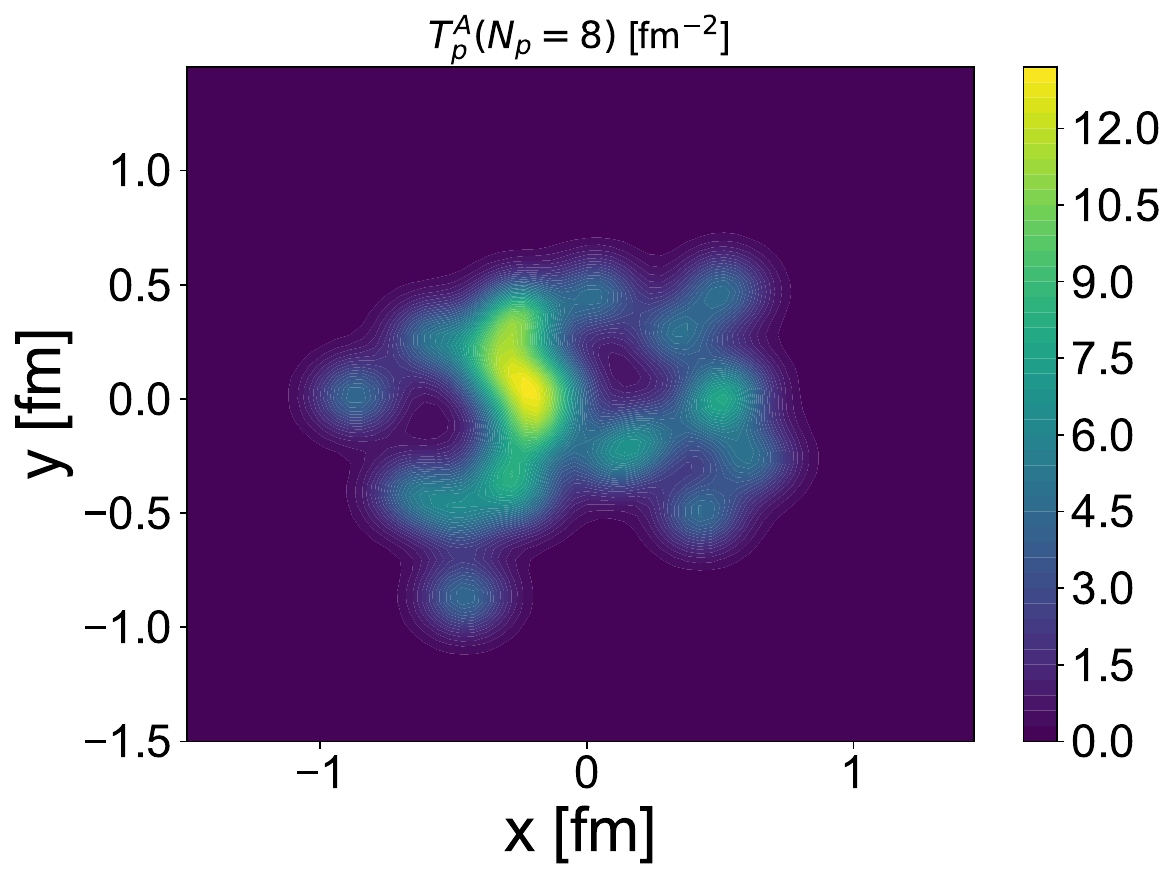}~~~
\includegraphics[width=0.37\linewidth]{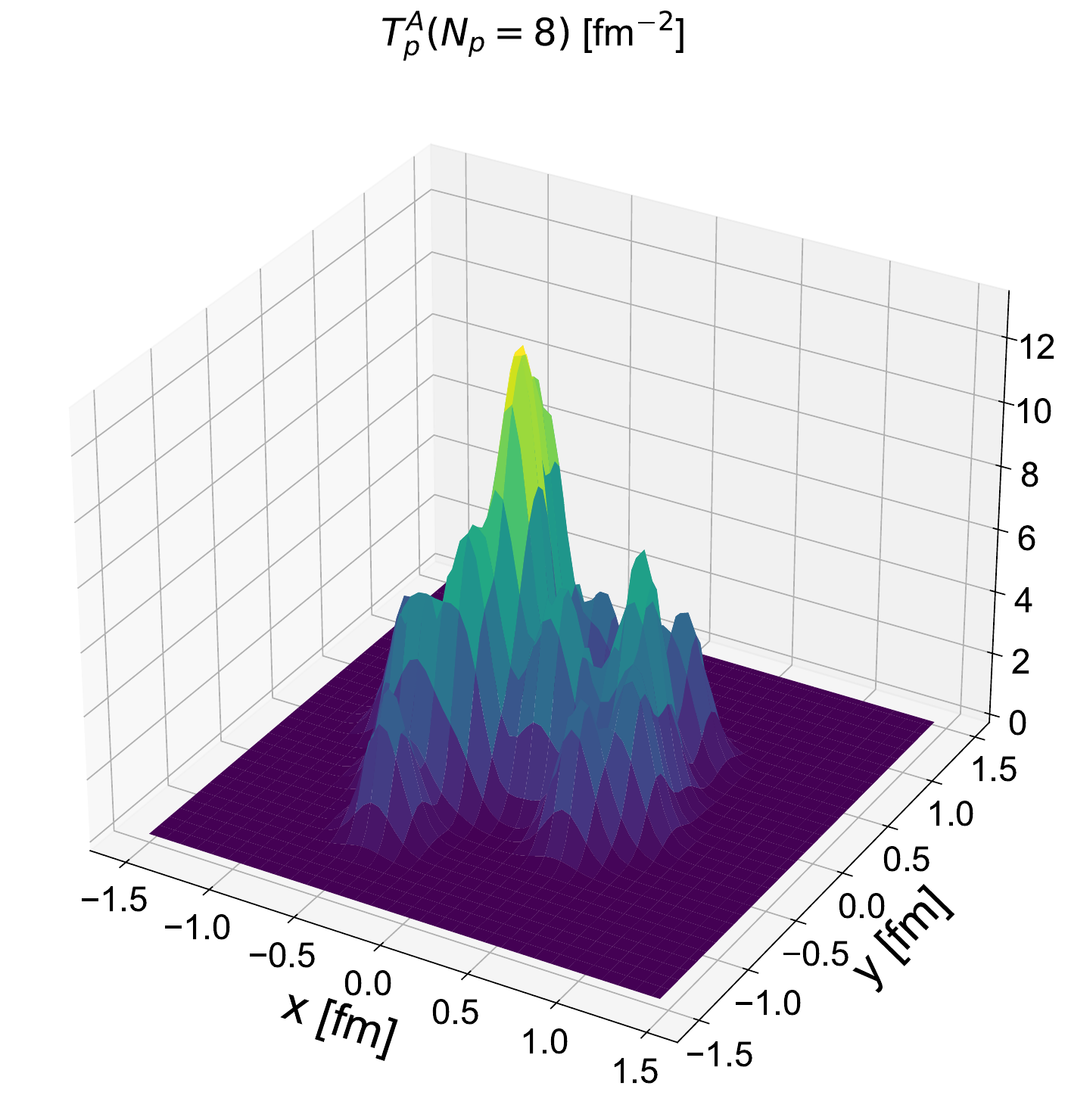}\\
\includegraphics[width=0.45\linewidth]{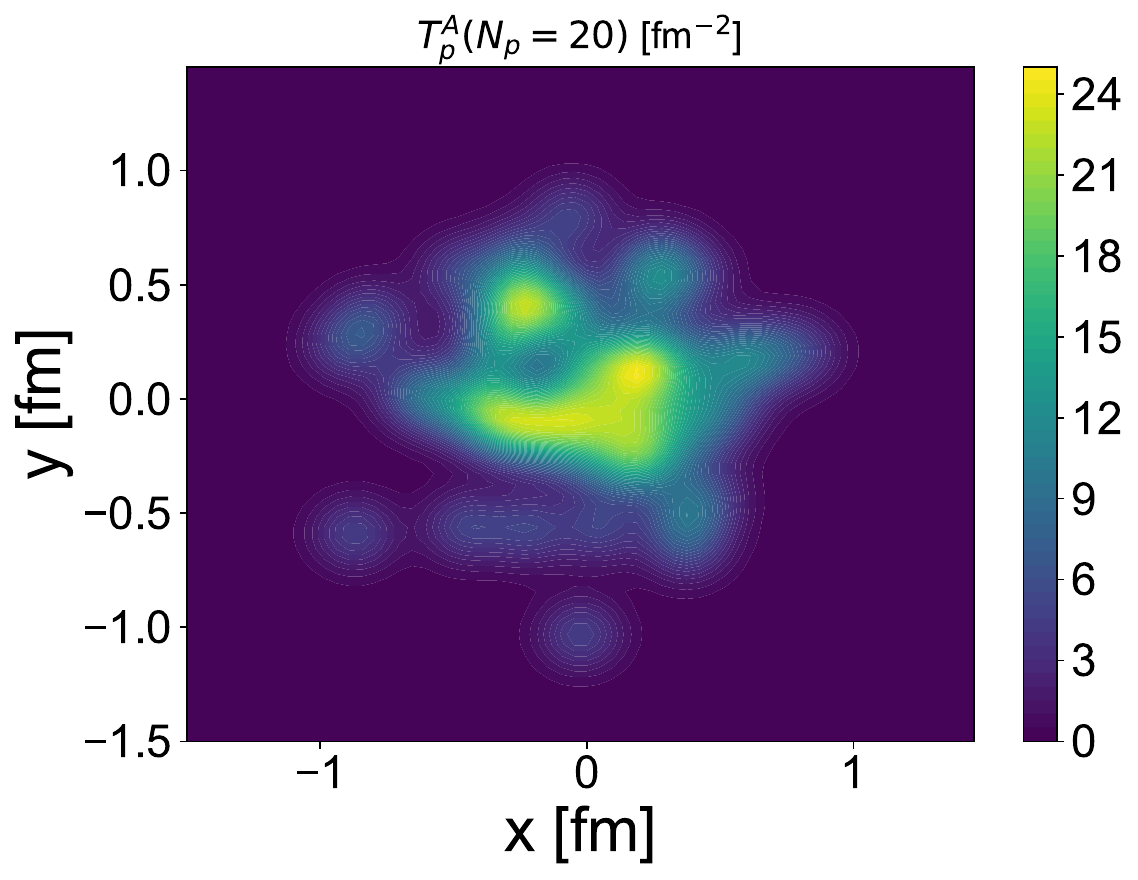}~~~
\includegraphics[width=0.37\linewidth]{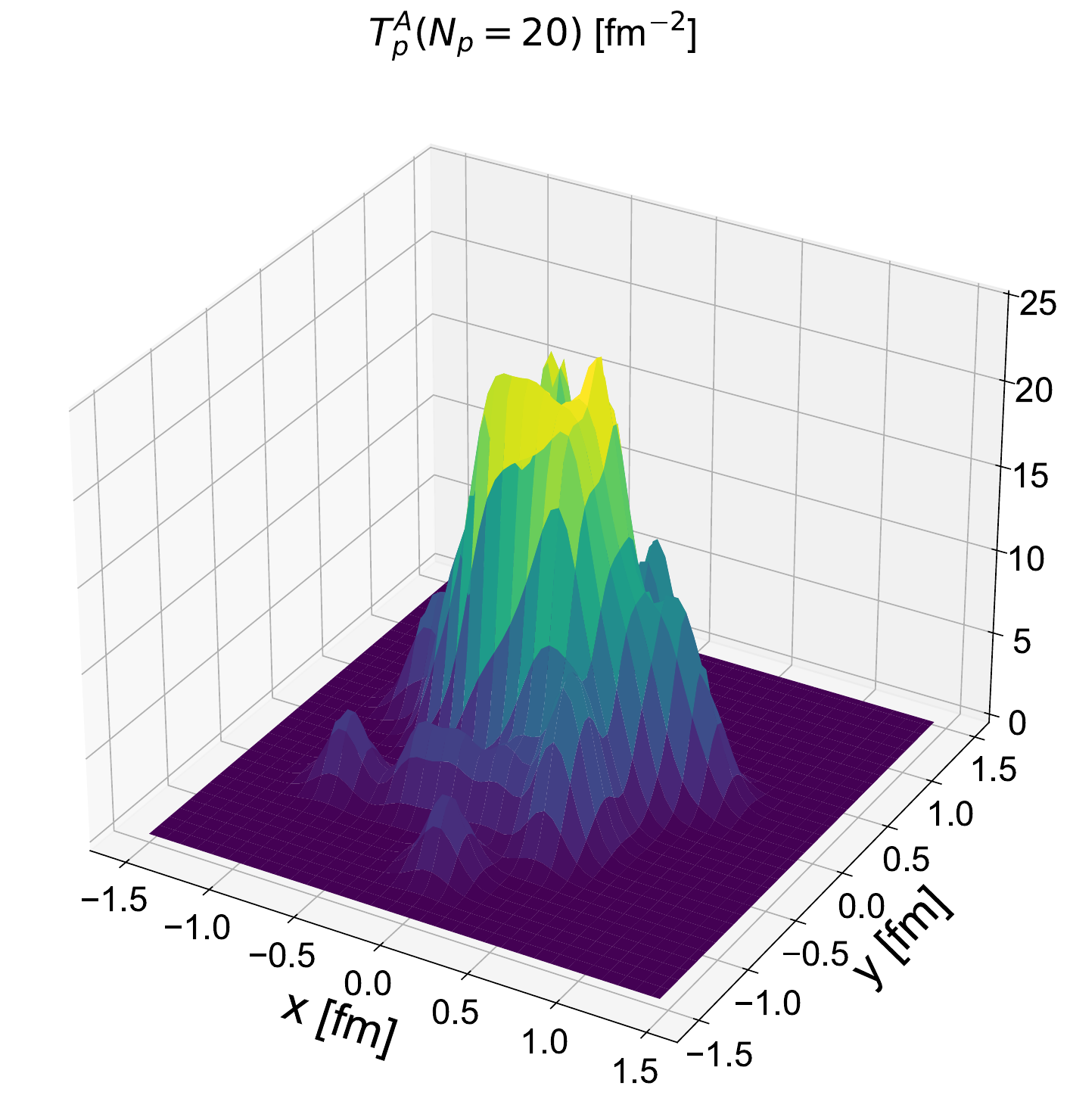}
    \caption{Thickness functions used for the generation of the color
    charges. In the \textbf{upper} panel we show $T_p$ for the proton
    (corresponding to $N_\mathrm{part}=1$),
    in the \textbf{middle} panel we show the case $N_\mathrm{part}=8$, corresponding to the average number of participants in 
    pA collisions at the LHC \cite{ALICE:2014xsp},
    and in the \textbf{lower} panel we show $T_p$ for the case $N_\mathrm{part}=20$.}
    \label{Fig:tp1e2}
\end{figure}

In Figure \ref{Fig:tp1e2} we show, for illustrative purposes, various charge profiles obtained with this model. Along with the profile of a proton (which is the same as in Fig. \eqref{fig:proton_distribution_Bq_Bcq}, top right), we also show the profiles we get for $N_\text{part}=8$ and $N_\text{part}=20$. The value $N_\text{part}=8$, in particular, corresponds to the average number of participants in pA collisions at LHC, cfr. \cite{ALICE:2014xsp}.

\begin{figure}[t!]
    \centering
\includegraphics[width=0.48\linewidth]{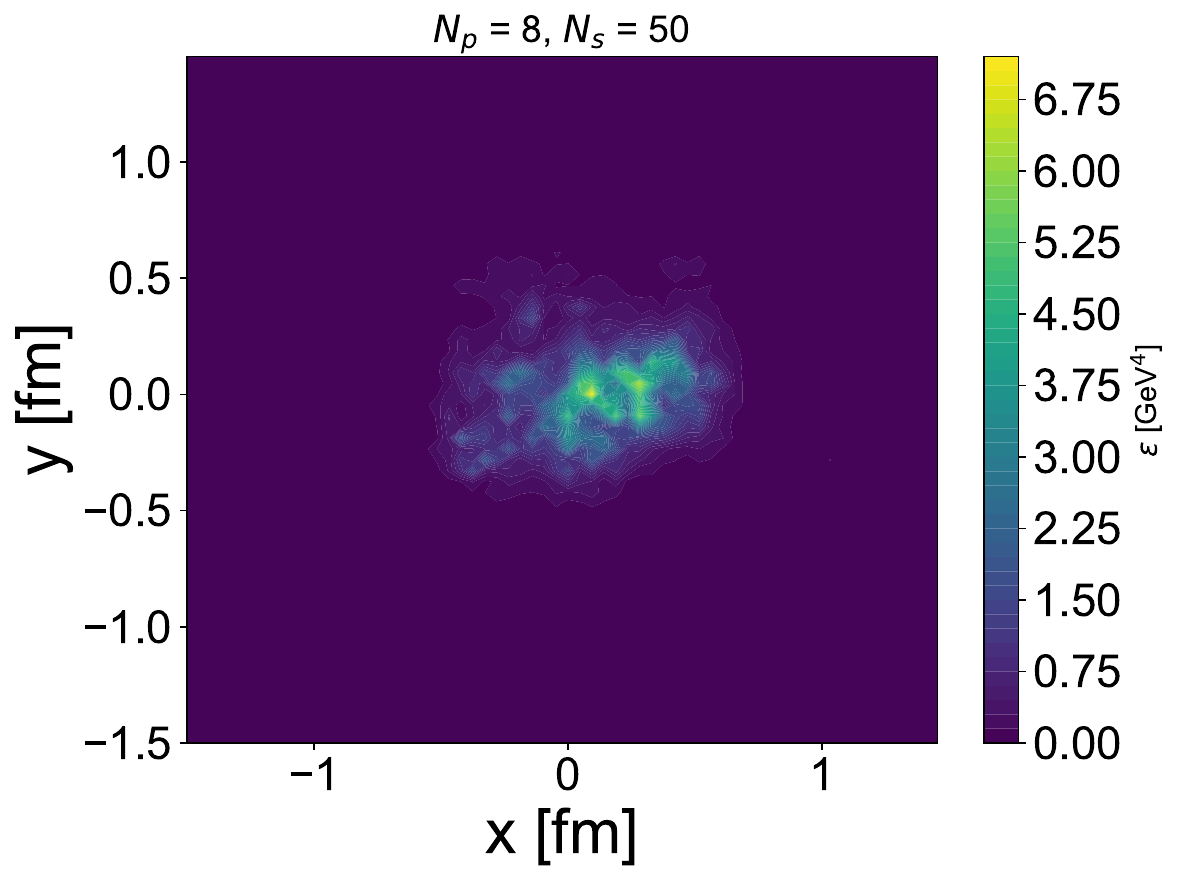}
\includegraphics[width=0.48\linewidth]{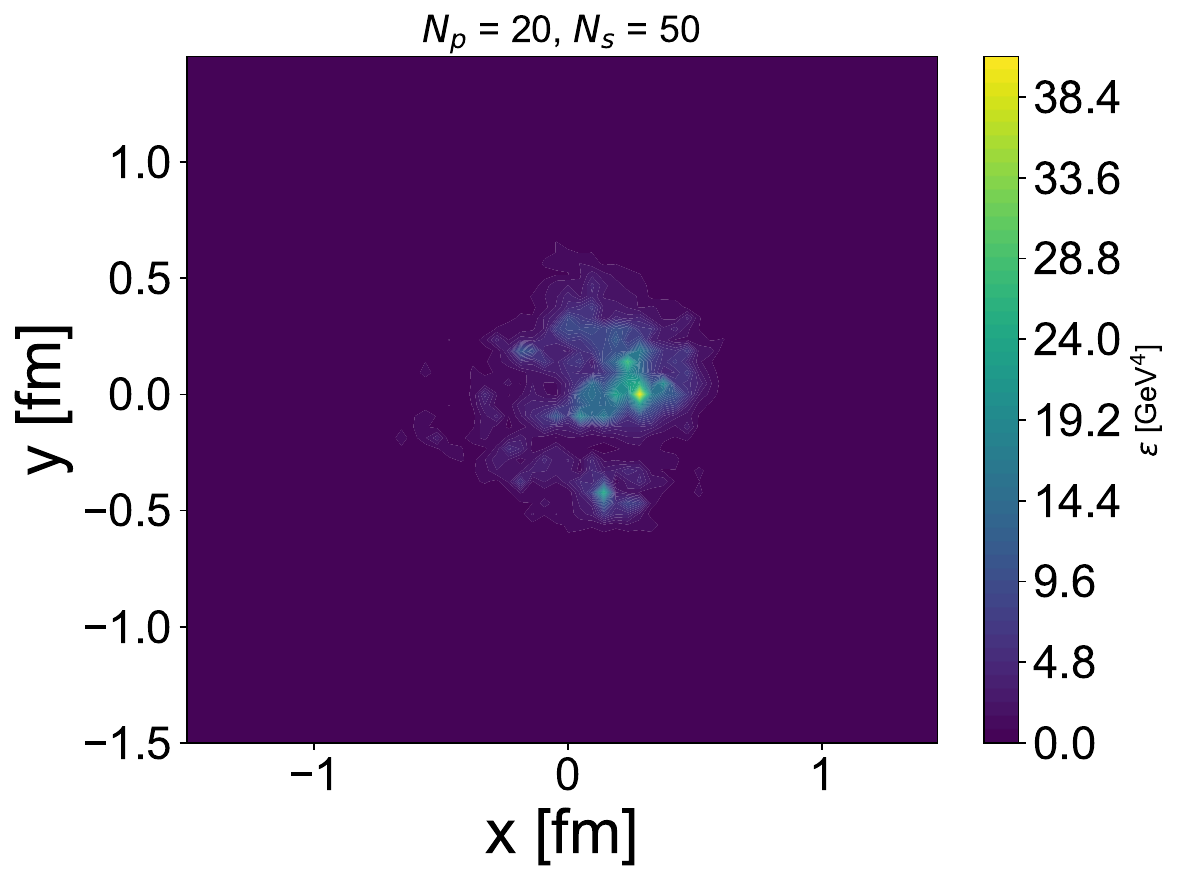}
    \caption{Initial energy density profiles for the glasma produced in pA, with $N_\mathrm{part}=8$
\textbf{(left)} and $N_\mathrm{part}=20$ \textbf{(right)}.}
    \label{Fig:epsilon}
\end{figure}

In Figure \ref{Fig:epsilon} we show the profile in energy density produced in pA collisions, using the aforementioned charge distributions. In the left panel we consider $N_\mathrm{part}=1$ for the proton and $N_\mathrm{part}=8$ for the nucleus, while in the right panel we consider $N_\mathrm{part}=1$ for the proton and $N_\mathrm{part}=20$ for the nucleus. In obtaining this plot, and for all the calculations in this Chapter, the transverse size has been fixed as $L_\perp=3$ fm, the number of lattice points is $N_\perp=64$ and we fix $N_s=50$ color sheets. What we observe here is that, as soon as we move on from the wall model for the heavy nucleus, the resulting energy density will be affected not only by the spatial inhomogeneities of the proton, but also by the ones of the heavy nucleus. This explains the visual difference between the profiles in Fig. \ref{Fig:epsilon} and that, for instance, in Fig. \ref{Fig:1} left: indeed, in the latter the energy density closely resembles the structure of the proton, since the charge distribution of the nucleus is homogeneous, while the profiles in Fig. \ref{Fig:epsilon} are much more complicated.

\section{\texorpdfstring{$R_{pA}$}{RpA} of heavy quarks}
\begin{figure}[t!]
\centering
\includegraphics[width=0.6\linewidth]{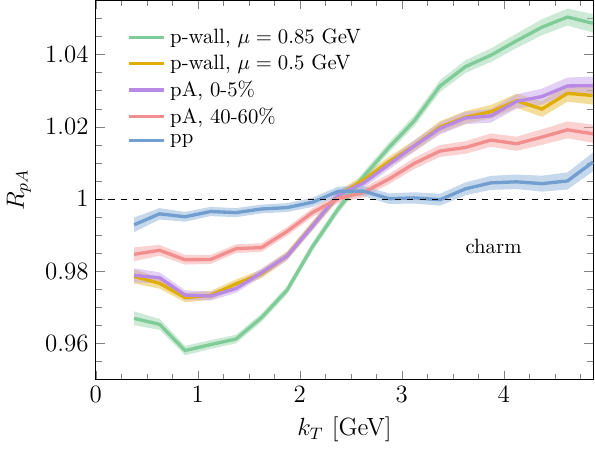}
\caption{Nuclear modification factor $R_{pA}$, defined in \eqref{eq:RpA_definition}, of charm quarks in glasma, computed at $\tau=0.4$ fm/c
for several collision systems.}
\label{Fig:RpA_charm_pWall_pA_pp}
\end{figure}

The first effect we want to highlight here is the modification of the HQ spectrum due to the action of the glasma fields. In particular, let us define the $R_{pA}$ as the ratio of the HQ spectrum at a time $\tau$ over the HQ spectrum at initial time:
\begin{equation}
    R_{pA}(k_T)=\frac{\left.dN/d^2\bm{k}_T(k_T)\right|_{\tau}}{\left.dN/d^2\bm{k}_T(k_T)\right|_{0}}.
    \label{eq:RpA_definition}
\end{equation}
We show our results for the $R_{pA}$ of charm quarks vs $k_T$ in Fig. \ref{Fig:RpA_charm_pWall_pA_pp}, for various collision systems, at $\tau=0.4$ fm/c. In particular, we show the $R_{pA}$ obtained for 'proton-wall' collisions (for two different values of the MV parameter, green $\mu=0.85$ GeV and yellow $\mu=0.5$ GeV), as well as for pA collisions at 0-5$\%$ and 40-60$\%$ centralities (purple and red curves), and pp collisions (blue curve). The aforementioned centralities are reproduced as p-$N_\mathrm{part}$ collisions with $N_\mathrm{part}=16$ and $N_\mathrm{part}=8$, respectively (check \cite{ALICE:2014xsp}). The evolution of the charm quarks is driven by the Wong equations \eqref{eq:wong1}, \eqref{eq:wong2} and \eqref{eq:wong3}, without any pair potential. Along with the ensemble average (obtained here over 1500 events, as well as for all the other plots which follow), we also show the error band, obtained as the standard deviation of the mean value.

Notice that in all cases we observe a value smaller than 1 for $R_{pA}$ for $k_T\lesssim 2.5$ GeV, while for $k_T\gtrsim 2.5$ GeV we observe that $R_{pA}$ is greater than 1. This goes to show that the action of the glasma on the spectrum is a net injection of energy, leading to a depletion of the spectrum at low momenta in favor of the higher momenta. A striking feature that we observe is that all the curves share a common value of transverse momentum at which the spectrum of the charm quarks is not modified by the action of the glasma fields, and we are able to locate such point around $k_T\sim 2.5$ GeV. We can clearly state that the value of $k_T$ at which this happens does not depend on either the system size or the values of the MV parameter, since these parameters vary from one curve to another, but rather only on the HQ mass, since this is the only dimensionful physical scale common to all curves.

Moving on to the differences among curves, we observe that the maximum deviation from the value $R_{pA}=1$ is achieved in the p-wall data at $\mu=0.85$ GeV (green curve), followed by $\mu=0.5$ GeV (yellow curve), which show a maximum displacement of $R_{pA}$ from 1 of around $5\%$ and $3\%$ in our $k_T$ range of reference, respectively. This suggests that the charm quarks experience a greater amount of spectrum modification in glasma systems with greater energy, or equivalently, with greater saturation scale $Q_s$. If we consider the sub-nucleonic structure, we observe that the smallest is the number of participants, the smallest is the modification in momentum spectrum.  This can be interpreted, once again, in terms of energy density, since a smaller number of participant nucleons implies a lower energy of the glasma, therefore a reduced modification of the initial HQ spectrum. The comparison among the p-wall results and the p-nucleons results show that, interestingly, the p-wall data at $\mu=0.5$ GeV (yellow) are compatible with the pA data at 0-5$\%$ centrality (purple), indicating that the energy densities of the color fields in the two systems are comparable. 

It is worth considering that, for all the systems considered, the spectrum modification is limited to a few percent, reaching at most a $5\%$ deviation with respect to the value 1. Such value is much smaller than the ones observed experimentally in proton-lead collisions at LHC, which achieve a maximum $R_{pPb}$ even of the order of 50$\%$ for $D$ mesons \cite{LHCb:2022dmh}. This goes to show that, in p-Pb collisions at LHC, most of the observed suppression of the transverse momentum spectrum is due to the later stages of the nuclear collision, while the initial stages have only a mild effect. In the following Section we will see that, on the other hand, the effect that the initial stages have on the observed momentum anisotropies of the charm quarks is quite substantial.

\section{Anisotropic flows}
\label{sec:anisotropic_flows}
After studying the modification of the spectrum, let us consider the anisotropic flows $v_n$. There are two main approaches which enable one to evaluate the anisotropic flows (although here we will show results for the elliptic flow $v_2$ only), that are known as event plane and 2-particle correlation methods, which we will now illustrate. After that, we will show the results obtained using these techniques.


\subsection{Event-Plane (EP) method}
\label{subsec:EP_method}

A first method for evaluating the $v_n$ is what is called the \emph{event-plane} (EP) method. It consists in evaluating the event plane angle $\Psi_n$ as a weighted average over the particle spectrum. In particular, we identify $\Psi_n$ as:
\begin{equation}
\Psi_n = \frac{1}{n}\arctan\frac{S_n}{C_n},
\label{eq:angoloPsin}
\end{equation}
with
\begin{align}
C_n &= \int \cos (n\phi) \frac{dN}{d^2\bm{k}_T} \, d^2\bm{k}_T, \\
S_n &= \int \sin (n\phi) \frac{dN}{d^2\bm{k}_T} \, d^2\bm{k}_T,
\end{align}
being $\phi$ the azimuthal angle of $\bm{k}_T$. From this, the anisotropic coefficient $v_n\{\mathrm{EP}\}$ at for a given momentum modulus $k_T$ is defined by integrating over all $\phi$ at fixed $k_T$:
\begin{equation}
v_n\{\mathrm{EP}\}(k_T) = \Bigg\langle\frac{\displaystyle{\int d\phi\, \frac{dN}{d^2\bm{k}_T}\, \cos[n(\phi - \Psi_n)]}}
{\displaystyle{\int d\phi\, \frac{dN}{d^2\bm{k}_T}}}\Bigg\rangle,
\label{eq:ep_vn}
\end{equation}
where the brackets denote the ensemble average.

\subsection{Two-Particle Correlation (2PC) method}
\label{subsec:2PC_method}

An alternative to the event-plane method for the extraction of anisotropic flow coefficients $v_{n}$ is provided by the so-called \emph{two-particle correlation} (2PC) method, which is based on experimental grounds. Differently from before, the basic idea here does not consist in explicitly calculating the event plane angle $\Psi_n$, here we rather indirectly infer its existence by studying the azimuthal correlation between pairs of particles. More specifically, for each $k_T$ we compare particles belonging to the bin centered in $k_T$ with particles in a broader ‘‘reference'' region. From this, we are able to derive the anisotropic flows.\\

Given the desired harmonic $v_n$, for a given event we define the quantity $Q_n$ as
\begin{equation}
Q_{n} = \sum_{j=i}^{M} w_{i}\, e^{i n \phi_{i}},
\end{equation}
where the sum runs over the $M$ particles, each with azimuthal angle $\phi_{j}$ and weight $w_{i}$. In our analysis the weights can be taken either from the particle spectrum $dN/d^2\bm{k}_T$ or, equivalently, set to $w_{j}=1$ if one wants to perform a pure particle count. Focusing on the latter case, for simplicity, let $N_{j}$ be the number of particles having momenta included in the bin centered at $k_{T}^{j}$, where $j$ runs over the number of bins. From these particles we can construct
\begin{equation}
    Q_{n}^{(j)} = \sum_{a=1}^{N_j} e^{i n \phi_{j,a}}.
\end{equation}
Now let us consider the so-called \emph{reference set}, defined by all particles with $k_{T}$ in a fixed interval $[k_{T}^{\mathrm{min}}, k_{T}^{\mathrm{max}}]$, with multiplicity $N_{\mathrm{ref}}$ and $Q_n$ defined as
\begin{equation}
    Q_{n}^{\mathrm{ref}} = \sum_{j\in \mathrm{ref}} Q_n^{(j)}=\sum_{j \in \mathrm{ref}} \sum_{a=1}^{N_j} e^{i n \phi_{j,a}}.
\end{equation}
The first sum is intended over the bins included in the reference interval.

The basic two-particle correlator between the bin of interest and the reference set is defined as
\begin{equation}
V_{n}^{\Delta}(k_{T}^{j}, \mathrm{ref}) = 
\frac{\Re \left[ Q_{n}^{(j)} \, Q_{n}^{\mathrm{ref}*} \right]}{N_{j}\,N_{\mathrm{ref}}} 
\label{eq:v22pc},
\end{equation}
where $\Re$ denotes the real part and $N_{j}\,N_{\mathrm{ref}}$ indicates the number of all possible particle pairs. Similarly, one can define the reference-reference correlator as
\begin{equation}
V_{n}^{\Delta}(\mathrm{ref}, \mathrm{ref}) = 
\frac{\Re \left[ Q_{n}^{\mathrm{ref}} \, Q_{n}^{\mathrm{ref}*} \right]}{N_{\mathrm{ref}}^{2}} 
= \frac{|Q_{n}^{\mathrm{ref}}|^{2}}{N_{\mathrm{ref}}^{2}}.
\end{equation}

Our purpose is to consider correlations among particle pairs. However, the definition above may include trivial ``self-correlation'' contributions when the bin of interest $k_{T}^{j}$ lies inside the reference interval. In such case, the same particle can appear both in $Q_{n}^{(j)}$ and in $Q_{n}^{\mathrm{ref}}$, producing additional spurious terms. To correct for this, one can explicitly subtract the self-correlation contributions by evaluating, for each particle in the bin $j$, the correlation with all reference particles except itself. This leads to the modified expression
\begin{equation}
V_{n}^{\Delta}(k_{T}^{j}, \mathrm{ref}) = \frac{ \Re \left[ Q_{n}^{(j)} Q_{n}^{\mathrm{ref}*} \right] - N_{j} }{ N_{j} (N_{\mathrm{ref}}-1)},
\end{equation}
valid if the bin $j$ lies within the reference interval. Here the subtraction of $N_{j}$ accounts for the removal of the $N_{j}$ diagonal terms, and the denominator uses $N_{\mathrm{ref}}-1$ since one particle is excluded from the reference set for each particle in the bin. On the other hand, if the bin of interest lies \emph{outside} the reference range, no particle is double-counted, and one should simply use the previously given definition \cref{eq:v22pc}. Also the reference-reference correlator can be corrected analogously by excluding the trivial self-pairs, yielding
\begin{equation}
V_{n}^{\Delta}(\mathrm{ref}, \mathrm{ref}) =
\frac{|Q_{n}^{\mathrm{ref}}|^{2} - N_{\mathrm{ref}}}{ N_{\mathrm{ref}}(N_{\mathrm{ref}}-1)}.
\end{equation}

Once the self-correlations have been subtracted, the two-particle estimate of the anisotropic flow coefficient for a given transverse momentum bin $k_T^j$ is given by
\begin{equation}
v_{n}\{\mathrm{2PC}\}(k_{T}^{j}) = 
\frac{ V_{n}^{\Delta}(k_{T}^{j}, \mathrm{ref}) }{ \sqrt{ V_{n}^{\Delta}(\mathrm{ref}, \mathrm{ref})}}.
\end{equation}
The final result is obtained via an average over the events.

\subsection{Results: glasma \texorpdfstring{$v_2$}{v2}}
\label{subsec:results_glasma}

Let us first deal with the elliptic flow generated by the glasma fields. Results for $v_2$ of glasma are already well known in the literature, see e.g. \cite{Schenke:2015aqa}. As already mentioned, the calculation of a spectrum for the glasma fields is not trivial: while for particles one is able to easily track their momentum and angular position, the matter formed in the initial stages of heavy ion collisions is made up of strong color fields, for which the procedure is not straightforward. In order to account for this issue, first of all we define a momentum spectrum of glasma fields: such definition will of course be constructed using the glasma gauge field and its conjugate momentum, i.e. the electric field, taken from Eqs. \eqref{12}, \eqref{eq:gauge_field_xy} and \eqref{eq:gauge_field_eta}. In this work we follow \cite{Lappi:2009xa} and take the gluon spectrum as
\begin{align}
\frac{\d N}{\d^2 \bm{k}_\perp \d y} = \frac{1}{(2 \pi)^2} \Tr \bigg\{ &
\frac{1}{\tau |\bm{k}_\perp|} E^i(\bm{k}_\perp)E^i(-\bm{k}_\perp) 
+ \tau |\bm{k}_\perp|  A_i(\bm{k}_\perp)A_i(-\bm{k}_\perp) 
\label{eq:first_line_spectrum_text}\\ &
+ \frac{1}{|\bm{k}_\perp|} \tau E^\eta(\bm{k}_\perp) E^\eta(-\bm{k}_\perp)
+ \frac{|\bm{k}_\perp|}{\tau} A_\eta(\bm{k}_\perp) A_\eta(-\bm{k}_\perp)
\label{eq:second_line_spectrum_text}\\ & \label{eq:multitrint_text}
+ i \Big[ E^i(\bm{k}_\perp) A_i(-\bm{k}_\perp) - A_i(\bm{k}_\perp)E^i(-\bm{k}_\perp) \Big]
\\ & \label{eq:multietaint_text}
 +i \Big[ E^\eta(\bm{k}_\perp) A_\eta(-\bm{k}_\perp) -  A_\eta(\bm{k}_\perp) E^\eta(-\bm{k}_\perp) \Big]
\bigg\},
\end{align}
where the fields are evaluated in the 2-dimensional Coulomb gauge. Notice that, while the first two lines \eqref{eq:first_line_spectrum_text} and \eqref{eq:second_line_spectrum_text} possess reflection symmetry, the interference terms \eqref{eq:multitrint_text} and \eqref{eq:multietaint_text} are odd under the 
transformation $\bm{k}_\perp \to -\bm{k}_\perp$. These do not contribute when the gluon spectrum is averaged over the angle of $\bm{k}_\perp$, integrated over $\bm{k}_\perp$ or averaged over configurations of the sources. Neglecting them, as for instance done in \cite{Lappi:2003bi}, would be justified if one was interested in evaluating the single inclusive gluon spectrum. In the case of two-gluon correlations they cannot, however, be neglected. Due to these interference terms, the symmetry $n(\bm{k}_\perp) = n(-\bm{k}_\perp)$ does not hold configuration by configuration, but only on average, thus the correlation function does not satisfy
$C(\bm{p}_\perp,\bm{q}_\perp) \neq C(\bm{p}_\perp,-\bm{q}_\perp)$. As pointed out in \cite{Lappi:2009xa}, there is a peak in the correlation at
$\bm{p}_\perp=\bm{q}_\perp$, which without the interference terms \eqref{eq:multitrint_text} and \eqref{eq:multietaint_text}
would, by symmetry, imply a similar peak at $\bm{p}_\perp=-\bm{q}_\perp$. The main numerical 
effect of including these terms is that the backward peak at $\bm{p}_\perp=-\bm{q}_\perp$ is significantly 
depleted. 
In any case, relations \eqref{eq:first_line_spectrum_text}--\eqref{eq:multietaint_text} we have just discussed act as a proxy for a particle distribution, and are not the only ones considered in the literature.  In Appendix \ref{appendix:spectrum_stuff} we will mention other approaches that are commonly employed.\\

\begin{figure}[t!]
\centering
\includegraphics[width=0.6\linewidth]{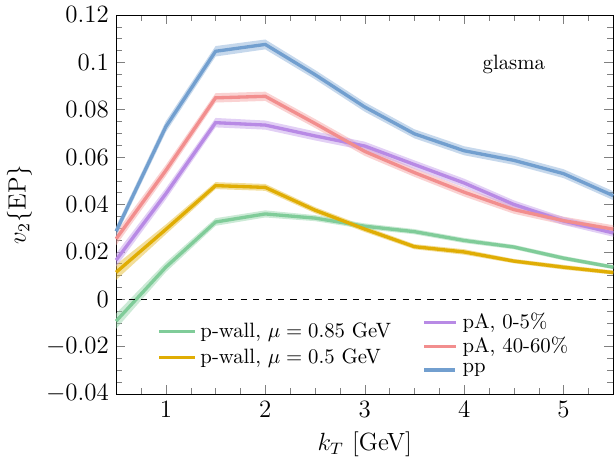}
\caption{Elliptic flow $v_2$ of the glasma fields versus $k_T$, computed at 
$\tau=0.4$ fm/c,
for several collision systems.}
\label{Fig:v2_glasma}
\end{figure}

In Fig. \ref{Fig:v2_glasma} we plot 
the glasma $v_{2}\{\mathrm{EP}\}$ versus $k_T$ for several collision systems
at $\tau=0.4$ fm/c. In particular, we show the elliptic flow obtained for p-wall collisions for two different values of the MV parameter $\mu=0.85,0.5$ GeV (green and yellow, respectively), as well as for pA collisions at 0-5$\%$ and 40-60$\%$ centralities (purple and red curves), and pp collisions (blue curve). As already mentioned, 0-5$\%$ and 40-60$\%$ centralities are reproduced as p-$N_\mathrm{part}$ collisions with $N_\mathrm{part}=16$ and $N_\mathrm{part}=8$, respectively (check \cite{ALICE:2014xsp}). We observe that all curves show similar qualitative behavior, reaching a maximum for a transverse momentum of around $k_T\sim 2$ GeV and then decreasing as $k_T$ increases. Let us now discuss the quantitative features. First of all, by focusing on the p-wall calculations, we observe that
the $v_2$ for $\mu=0.85$ GeV is smaller than the one for $\mu=0.5$ GeV: this is likely related to the smaller size of correlation domains in the former case, which implies an overall smaller anisotropy. Moving on to the curves obtained in the p-nucleons cases, we observe that for a smaller number of participants, the elliptic flow is higher. For the proton-proton case, in particular, the $v_2$ is quite significant, with a maximum of around $v_2\sim 0.11$ for $k_T\sim 2$ GeV. This ordering with the number of participants is straightforward to interpret, since a smaller number of participants implies a larger spatial anisotropy in the transverse plane (see e.g. Fig. \ref{Fig:tp1e2}), and this is in turn converted into a momentum space anisotropy. Finally, it is worth noting that our results for the $v_2$ obtained in pA collisions at 0-5$\%$ are compatible with the results from \cite{Schenke:2015aqa}.

\subsection{Results: charm quark \texorpdfstring{$v_2$}{v2}}

After dealing with the anisotropies of the glasma, we now move on to study the elliptic flow of charm quarks evolving in the glasma fields.

\begin{figure}[t!]
\centering
\includegraphics[width=0.6\linewidth]{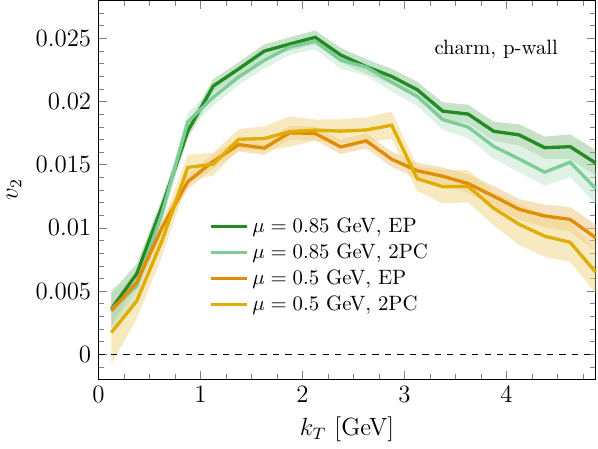}
\caption{Elliptic flow $v_2$ of charm quarks in p-wall collisions, computed at $\tau=0.4$ fm/c. We compare the results obtained for two different values of $\mu=0.5,0.85$ GeV, and with two different methods, Event-Plane (EP) and 2-Particle Correlations (2PC).}
\label{Fig:v2_charm_pWall_EPvs2PC}
\end{figure}

We first want to compare the two methods for the calculation of $v_2$ that we have outlined in Sections \ref{subsec:EP_method} and \ref{subsec:2PC_method}, namely EP and 2PC (including the removal of autocorrelations). In Figure \ref{Fig:v2_charm_pWall_EPvs2PC} we compare the charm quark $v_2$ calculated in p-wall collisions at $\tau=0.4$ fm/c. In particular, we compare the results obtained using EP and 2PC for two values of the MV parameter. What we see is that the extracted $v_{2}$ values from 2PC are in very good agreement with the EP results, each lying within the error bands of the other. This holds for both the $\mu=0.5$ GeV and the $\mu=0.85$ GeV results, showing that the observed azimuthal anisotropies are physical in origin and not artifacts of a particular analysis technique. Moving on to the physical features, we see that for both values of $\mu$ the $v_2$ reaches a maximum value for $k_T\sim 2$ GeV, after which we have a decreasing behavior. This closely resembles the generic trends seen in heavy-ion phenomenology, i.e. that elliptic flow is typically most pronounced at intermediate transverse momenta before flattening at higher values. The maximum value of elliptic flow is found to be around $v_2\sim$ 0.025 for $\mu=0.85$ GeV (green curves) and $v_2\sim$ 0.018 for $\mu=0.5$ GeV (orange curves), so at higher $\mu$ we observe higher values of $v_2$. This can be interpreted by noting that at higher $\mu$ the glasma fields have a greater energy density, therefore the color electric and magnetic fields in the initial stages are more efficient in transmitting their momentum anisotropy to the heavy quarks which evolve therein.

\begin{figure}[t!]
\centering
\includegraphics[width=0.6\linewidth]{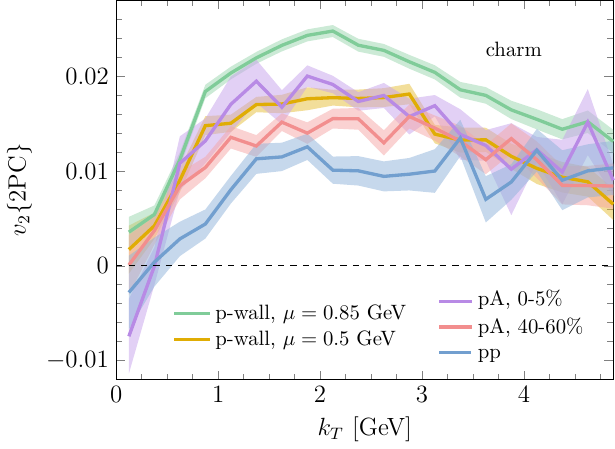}
\caption{Elliptic flow $v_2$ of charm quarks, computed at $\tau=0.4$ fm/c, for several collision systems.}
\label{Fig:v2_charm_pWall_pA_pp}
\end{figure}

The data we have just commented on have been obtained in p-wall collisions, but we can extend our results to include a sub-nucleonic structure within the heavy nucleus. In Figure \ref{Fig:v2_charm_pWall_pA_pp} we show the charm quark results of the elliptic flow $v_2$ vs $k_T$ for various collision systems using the 2PC method: along with the p-wall data (the same as in Fig. \ref{Fig:v2_charm_pWall_EPvs2PC}), we also show results for pA collisions at 0--5$\%$ and 40--60$\%$ centralities (purple and red curves), as well as pp collisions (blue curve). A striking feature is that all the curves follow the same qualitative behavior, reaching a maximum at around $k_T\sim 2$ GeV and then decreasing for higher $k_T$. Quantitatively, we observe that the smallest is the number of participants, the lowest is the obtained $v_2$ for charm quarks: once again, a smaller number of participant nucleons implies a lower energy of the glasma, hence the heavy quarks maintain a larger amount of their initial momentum isotropy. Notice that the ordering of the $v_2$ with respect to the number of participants is opposite to the one we have observed for the glasma $v_2$ in Fig. \ref{Fig:v2_glasma}. Interestingly, we also note that the results for pA collisions at 0--5$\%$ (purple curve) are comparable, both qualitatively and quantitatively, with the p-wall results at $\mu=0.5$ GeV (yellow), even though the first system is manifestly more anisotropic than the second (because of the presence of a sub-nucleonic structure). We can conclude that, in this comparison, the greater spatial anisotropy which is observed in pA collisions at 0-5$\%$ centrality is not transmitted to momentum as efficiently as it does in the p-wall case. This leads to a balancing effect which renders the yellow and the purple curves compatible one with the other, errors.

\begin{figure}[t]
\centering
\includegraphics[width=0.6\linewidth]{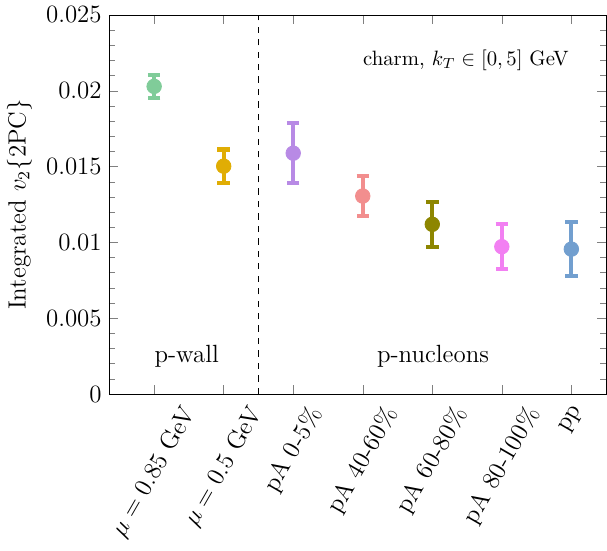}
\caption{Elliptic flow of charm quarks, computed at $\tau=0.4$ fm/c for several collision systems, integrated over the $k_T$ range $[0,5]$ GeV.}
\label{Fig:v2_charm_integrated}
\end{figure}

These conclusions are supported also in Fig. \ref{Fig:v2_charm_integrated}, in which we show the integrated elliptic flow of charm quarks at $\tau=0.4$ fm/c. Along with the same cases we have shown in the previous plot, here we also show results for the integrated $v_2$ obtained in pA collisions for 60--80$\%$ and 80--100$\%$ centrality classes, corresponding to p-$N_\mathrm{part}$ collisions with $N_\mathrm{part}=5$ and $N_\mathrm{part}=3$, respectively (check once again \cite{ALICE:2014xsp}). By ‘integrated' we refer to an averaging procedure of the $v_2$, weighted with the momentum spectrum, over the range $[0,5]$ GeV. We observe the same ordering that we have highlighted in Fig. \ref{Fig:v2_charm_pWall_pA_pp}: the smaller is the number of participants, or the value of $\mu$, the smaller is the $v_2$ acquired by the charm quarks. Once again, we note that the value of the integrated $v_2$ for pA collisions at 0-5$\%$ (purple) is compatible with the p-wall result at $\mu=0.5$ GeV (yellow). The former result has a larger statistical uncertainty, which is due to the random sampling of the positions of the constituent nuclei.

\begin{figure}[t]
\centering
\includegraphics[width=0.6\linewidth]{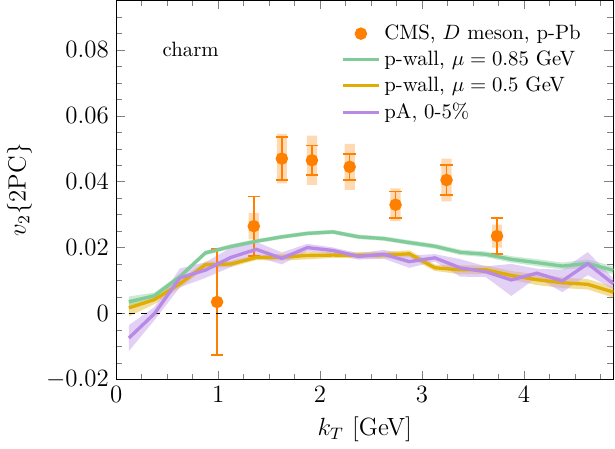}
\caption{Experimental data from CMS for the elliptic flow $v_2$ of $D^0$ mesons from \cite{CMS:2018loe}, compared with our calculations for the $v_2$ of charm quarks, computed at $\tau=0.4$ fm/c for several collision systems. In the CMS data the error bars correspond to statistical uncertainties, while the shaded areas denote the systematic uncertainties. The CMS data have been rescaled by a factor $2$ both in $k_T$ and in $v_2$, in order to account for quark coalescence.}
\label{Fig:v2_charm_ExpData}
\end{figure}

Our calculations for the $v_2$ give an insight on the impact that the initial stages of heavy ion collisions have on the experimental observables. In particular, the elliptic flow of charmed mesons has been measured in heavy-ion collisions at both RHIC and the LHC by several experiments. These measurements provide a direct probe of the degree to which quarkonium states couple to the medium formed in ultrarelativistic collisions, quantifying the momentum anisotropy that they inherit from it. In Fig. \ref{Fig:v2_charm_ExpData} we compare the elliptic flow $v_2$ obtained in our calculations with the experimental data from the CMS group at LHC (orange points), from \cite{CMS:2018loe}. The theoretical calculations that we show are the same appearing in Fig. \ref{Fig:v2_charm_pWall_pA_pp}, but considering only the results simulating central collisions, which are the p-wall simulations at $\mu=0.5,0.85$ GeV (yellow and green curves) and pA collisions in the 0-5$\%$ centrality class (purple curve). As far as the experimental data are concerned, these have been obtained by studying the $D^0$ meson production in proton-lead collisions at 8.16 TeV. In order to take into account for hadronization, so to obtain an estimate for the momentum anisotropy of the charm quark itself, in the experimental data we divided both the $k_T$ and the $v_2$ values by a factor 2, assuming pure coalescence \cite{Kolb:2004gi}. Along with the data, we also show the statistical and the systematic errors, as an error band and a shaded area respectively. First of all, we notice that our simulations reproduce the qualitative behavior of the experimental data, mimicking the observed rise of $v_{2}$ at intermediate transverse momenta. Most importantly, we observe that our simulations reach around $50$ to $80\%$ of the experimental $v_2$ at intermediate momenta ($k_T\sim 1.5$--$3.5$ GeV), depending on the theoretical model considered and on the specific value of transverse momentum. Even more remarkably, at the highest and lowest values of $k_T$ our calculations actually agree with the experimental data, within the uncertainties. This comparison suggests that a significant part of the charmed mesons anisotropy may indeed be traced back to the early-time dynamics of charm quarks in the glasma, which later feed into hadronization. The overall consistency between our theoretical trends and the experimental data supports the interpretation that the elliptic flow of quarkonia encodes valuable information on the collective behavior of heavy quarks in the evolving QCD medium.

	\chapter{Conclusions}\label{conclusion}
    
	In this Thesis we focused on the initial stages of heavy ion collisions, in particular on the effective model which describes it: the Color Glass Condensate. After having outlined the main features of this phase, we moved on to analyze the dynamics of heavy quarks in this medium.

	The first physical case we have discussed, in Chapter \ref{chap:Heavy quarks in anisotropically fluctuating glasma}, is the diffusion of heavy quarks
in the early stage of high-energy proton-nucleus
collisions. Our initial condition is based on 
the glasma picture,
and includes event-by-event
fluctuations of the color charges in the proton, 
as well as fluctuations that break the boost invariance.
We performed $3+1$D real-time statistical simulations 
of the SU(3) Yang-Mills fields produced immediately
after the collision.
Both the modeling of the proton and the
longitudinal fluctuations for an expanding geometry
in SU(3) have not been considered before
in the literature in the context of heavy quarks.
For simplicity, we limited ourselves to study the large-mass
limit of the heavy quarks: in this limit, 
the solution of the Wong equations for the momentum diffusion
amounts to computing the time-correlator of the
color-electric fields.
As a model of initial state field fluctuations,
we adopted a simple superposition of several harmonics as in \cite{Fukushima:2011nq}, with a free 
parameter $\Delta$ that measures the strength of the
fluctuating fields, see Eq.~\eqref{48_bis}.
The fact that the strength of the fluctuating
fields scales as $\Delta/|\mathcal{I}|$ in 
Eq.~\eqref{48_bis} allows us to keep the 
initial energy density carried by the fluctuations
unaffected by the inclusion of more harmonics.

We found that 
momentum broadening
both in longitudinal and in the transverse directions, $\langle\delta p_L^2\rangle$ and $\langle\delta p_T^2\rangle$ respectively, are unaffected (within error bars) 
by the inclusion of $\eta$-dependent fluctuations. 
Our interpretation of this result is that the fluctuations
do not alter in a substantial way the time  correlations of the color-electric fields, that are directly related
to momentum broadening.
Hence, our conclusion is
that the initial state fluctuations,
at least in the form introduced within our study,
might be not sufficient
to achieve isotropization 
of heavy quark momenta
within the early stage. This occurs despite the 
fact that for the (illustrative) case of intense fluctuations
considered in our work, the quantity $P_L/P_T$,
which is a measure of the amount of anisotropy
of the bulk gluon fields, may be orders of magnitude larger than one.

Additionally, we computed the anisotropy of the
angular momentum fluctuations 
that we quantified by the coefficient $\Delta_2$, first
introduced in~\cite{Pooja:2022ojj}.
In~\cite{Pooja:2022ojj} the $\Delta_2$ has been studied
for a static geometry, within the gauge group SU(2) and
without initial state fluctuations, hence the present work
improves these several aspects
with respect to \cite{Pooja:2022ojj}.
We found that $\Delta_2$ remains considerably large
during the whole early stage: 
in the time range where 
the picture based on
an evolving glasma is phenomenologically
relevant, that is 
for $\tau$ in the
range $[0.2,0.4]$ fm, $\Delta_2$ remains 
in the range $[0.4,0.6]$, signaling a substantial amount
of anisotropic angular momentum fluctuations. For larger times we can grasp a linear trend for $\Delta_2$, however such behavior not only may not be accessible phenomenologically, but it is also likely due to the expansion of the system, rather than a genuine feature of the glasma fields.\\

In Chapter \ref{chap:Melting of heavy quark pairs in glasma} we studied the melting of charm and beauty pairs in the 
pre-equilibrium stage of high-energy proton-nucleus collisions.
We modeled the early stage according to the color-glass-condensate
picture, choosing the glasma as the initial condition and solving 
the classical Yang-Mills equations to study its evolution.
The heavy quarks constituting the $c\bar c$ and $b\bar b$ pairs, 
which we assume to be produced in color-singlet states,
are then evolved on top of the classical gluon fields
via semi-classical equations of motion, known as the Wong equations.
Our main goal was the calculation of the percentage of pairs
that melt during their diffusion in the early stage.

Then we moved on to the evolution of the gauge-invariant correlator of the
color charges of the quark and its companion antiquark, $G(\tau)$.
At $\tau=\tau_\mathrm{form}$ we have $G(\tau_\mathrm{form})=1$
due to the color-singlet initialization of the pairs.
The interaction of the quarks with the dense
gluon environment leads to a degradation of the correlations.
Consequently, there are fluctuations to the color-octet state
at later times, and this process causes the melting of the pairs. 
The timescale of decorrelation,
$\tau_\mathrm{dec}$,
can be estimated, for example, 
by the requirement that $G(\tau_\mathrm{dec})=1/2$.
We found that $\tau_\mathrm{dec}-\tau_\mathrm{form}$ stays in the range
$[0.3,0.4]$ fm for both charm and beauty.
We also find that $\tau_\mathrm{dec}-\tau_\mathrm{form}$ 
for $b$ quarks is  larger than the similar quantity 
computed for the $c$ quarks: this is most likely due to the fact that
$b$ quarks are slower, so they spend more time within correlation domains
of the gluonic background. We find this result despite the fact that the beauty pairs are produced earlier ($\tau_\mathrm{form}\sim 1/2M$), so they actually experience a stronger diffusive effect glasma fields. We reformulated the decorrelation of color charges 
in HQ pairs in terms of color-singlet and color-octet projectors, 
$P_S$ and $P_O$, 
which appear in our scheme as the classical counterparts of the quantum projection operators onto the singlet and octet representations of SU(3). 
We identified $P_S$ and $P_O$ with the probabilities that the pair is in the singlet and octet states, respectively, 
thereby quantifying the singlet and octet components of the pair. Our findings indicate that the loss of color 
correlation leads to equilibration of color at large times,
similarly to the results of Ref.~\cite{Delorme:2024rdo}
for the QGP. Specifically, we found that at large times $\langle P_S\rangle\approx \langle P_O\rangle/8\approx 1/9$, which we interpret as the pair having equal 
probability of being in the singlet state or in one of the 
eight possible octet states.

The diffusion of the heavy quarks in the evolving glasma fields leads
to the gradual melting of the pairs. Within our framework, the survival probability of each pair depends
on the fluctuations of the color charges: they generate
a color-octet component for the pair and eventually break the pair itself.
These are related to the gauge field fluctuations
in the dense gluonic bath formed in the early stage. Within our work, we support the idea that even when the separation 
in coordinate space
of the quarks is small, one can refer to a
pair as ``separated'' if color-decorrelation is sufficiently large. In particular, in the early stage we find that the spatial separation is almost 
constant (see Fig. \ref{Fig:rreldifferentpotentials} and the left panel of Fig. \ref{fig:r_rel_c_cor_charm_vs_beauty}), as a combined effect of the attractive potential and the broadening due
to the gluons in the bath. We therefore find more effective to
define a separation in the color space rather than in the coordinate space. For each pair at a given time, we accept its survival with the
probability $\mathcal{P}_\mathrm{survival}$
defined in Eq.~\eqref{eq:conte_non_contento}.
This probability depends on the color charges of the two particles that form the pair: these quantities are directly obtained from the solution of the Wong equations.
$\mathcal{P}_\mathrm{survival}$ depends on one parameter, $\kappa$, 
that measures
how large the color fluctuations need to be in order to melt the pair.
We introduced a breaking time, $\tau_\mathrm{break}$, defined as the value
of the proper time at which half of the pairs are melted: as mentioned, it is color decorrelation that primarily 
drives the melting of the pair, so we relate the melting probability solely to the color fluctuations.
We found that for both 
$c\bar c$ and $b\bar b$ pairs, 
$\tau_\mathrm{break}-\tau_\mathrm{form}$ 
is approximately in the range 
$[0.4,0.5]$ fm.

This work represents many improvements 
in comparison to Ref.~\cite{Pooja:2024rnn}, since we go beyond the SU(2) glasma in a static box,
and more importantly, include for the first time a detailed study of
the color fluctuations in the pairs. Another improvement is the 
quark-antiquark potential for the pair,
with coupling proportional to $Q_a\bar Q_a$,
that evolves dynamically as a result of
the interaction of the pair with the background gluon field. 
This potential accounts for the fact that the pair does not remain in a pure singlet state during the evolution, allowing for the development of an octet component due to interactions with the background gluon field.
Finally,
we implemented the gauge-invariant formulation in terms of gauge links
and plaquettes, while in~\cite{Pooja:2024rnn} the authors used
the continuum formulation to solve the Yang-Mills equations.

The work presented in this Chapter can be continued along several directions.
Firstly, it is desirable to link the final stage of our
pre-equilibrium evolution to relativistic transport,
in order to analyze the dynamics in the quark-gluon plasma stage and allow a more direct comparison with experimental observables. 
It is also possible to adopt a different approach, based on
a master equation for the density matrix of the pairs, to analyze
the color decorrelation and the melting of the pairs within a 
quantum-mechanical approach. Moreover, it will be interesting
to analyze the same problem within different scenarios in which
boost invariance is broken by longitudinal fluctuations of the background gluon field, e.g. in a similar fashion as in Chapter \ref{chap:Heavy quarks in anisotropically fluctuating glasma}. Finally, in future developments also the interactions among all the quarks in the system, and not only within the quarks in each pair, may be taken into account. All these projects are currently under investigation and we plan to report on them soon.\\

Finally, in Chapter \ref{chap:Anisotropies of gluons and heavy quarks in glasma} we moved on to study momentum anisotropies of glasma and of heavy quarks evolving in such glasma.
The first result of the Chapter is the nuclear modification factor $R_{pA}$ of charm quarks evolving under the action of the glasma fields. We observe that the lower the number of participant nucleons and hence the energy density of the glasma, the lower is the discrepancy of $R_{pA}$ from 1, as expected since we are diminishing the ability of the glasma of modifying the momentum spectrum of HQs. More interestingly, we find a universal value of momentum for which all $R_{pA}$ are equal to 1, i.e. at which the charm momentum spectrum does not get modified by the glasma fields, regardless the collision energy or centrality.

Once we did so, we moved on to the elliptic flow $v_2$. First of all we evaluated the $v_2$ of the glasma fields, in order to check agreement with the existing literature. We find comparable results with \cite{Schenke:2015aqa} and, as expected, we see that the most spatially anisotropic systems (i.e. pA collisions at low centralities and pp collisions) are the ones carrying also the highest momentum anisotropy. The most compelling calculation is, however, the $v_2$ of charm quarks at the end of the glasma phase, i.e. at $\tau=0.4$ fm/c. We first show that the two main approaches followed in the literature, which are the event plane (EP) and the 2-particle correlation (2PC) methods, give compatible results with one another, showing that our results are physical in origin and do not suffer from non-physical artifacts. Moreover, we observed that the HQs are significantly affected by the anisotropy of the underlying glasma fields. Our calculations, which are the first results for the heavy quark $v_2$ in glasma, show that this quantity scales inversely with the number of participants, showing once again that the building up of $v_2$ requires energy from the glasma, and this is higher when we have more central collisions. Moreover, for the most central collisions, we see that the accumulated momentum anisotropy of the HQ throughout the glasma phase is quite high, reaching up to 50-80 $\%$ of the experimental data for $v_2$ of $D^0$ mesons at the CMS experiment, for intermediate $k_T$. This is striking evidence that the effect that the short-lived ($\tau\sim$ 0.3-0.4 fm/c) initial stage has on the final observables is far from irrelevant, when compared to the kinetic theory-driven later stages (which last for quite longer, even in small collision systems).

The results from this Chapter can be extended, for instance, by studying collisions among light ions, e.g. oxygen-oxygen and argon-argon collisions. Indeed, the forthcoming experimental runs at LHC using light ions aim to unravel the details of the nuclear structure and of the fluctuations in the nucleon positions, features which get instead washed out in heavy ion collisions. In particular, theoretical calculations in these kind of frameworks provide unique insight into the effect of the initial stages on the experimental observables. As already mentioned, in order to allow a more direct comparison with experimental observables, we also envisage a link of the initial glasma stage with a subsequent relativistic kinetic theory evolution, which aims to reproduce the dynamics of the quark-gluon plasma. This outlook is conceived as future work.\\

These have been the main research purposes of this Thesis. These topics are of great interest for a wide sector of the high-energy physics research, which aims to unravel the deepest features of heavy ion collisions and of the first-principles theory which describes them, that is Quantum ChromoDynamics. Despite the great interest, it is likely that these challenges will hold for many years to come.

\appendix
 \chapter{Notations and conventions}\label{app:conv}

In this Chapter of the appendix we summarize special notation and conventions that are used throughout the Thesis.

 \section{Natural units}
 
 Throughout this work we used the natural-units convention:
 \begin{equation}
     \hbar=1,\qquad c=1,\qquad k_B=1.
 \end{equation}
This convention allows us to express various physical quantities in terms of units of energy: momentum, mass, length, time, temperature. In particular, since the energies under consideration are very low compared to macroscopic scales, in particle physics one usually considers the {\sl
electron volt}, such that:
\begin{equation}
1 \text{ eV}=1.602\cdot 10^{-19}\text{ J}.
    \label{eq:1electronvolt}
\end{equation}
    In these units, for instance, the mass of the proton is 938 MeV=1.672$\cdot 10^{-27}$ Kg, its radius is of the order of (200 MeV)$^{-1}=10^{-15}$ m, while the temperatures reached at LHC are of the order of $1$ GeV $\sim 10^{13}$ K. A very useful relation allows us to pass from energy scales to length scales:
 \begin{equation}
      \hbar c= 0.19732\; \mathrm{GeV}\cdot\mathrm{fm}=1,
 \end{equation}
 being 1 fm=$10^{-15}$ m. In other terms, 1 fm=5.06791 GeV$^{-1}$.
 \section{SU(3) generators}
The generators of the Lie algebra of the SU(3) group in the principal representation represent, for obvious reason, the basis for the study of QCD. Those are closely related to the so-called Gell-Mann matrices, which are a set of eight linearlly independent traceless Hermitian matrices.

In particular, the generators of SU(3) are
\begin{equation}
T^a=\lambda^a/2,
    \label{eq:SU3_generators}
\end{equation}
where $\lambda^a$ with $a=1,\dots, 8$ are the Gell-Mann matrices:
\begin{gather*}
\lambda^1 = \matthree {0}{1}{0}{1}{0}{0}{0}{0}{0},\quad
\lambda^2 = \matthree {0}{-i}{0}{i}{0}{0}{0}{0}{0},\quad
\lambda^3 = \matthree {1}{0}{0}{0}{-1}{0}{0}{0}{0},\\[1ex]
\lambda^4 = \matthree {0}{0}{1}{0}{0}{0}{1}{0}{0},\quad
\lambda^5 = \matthree {0}{0}{-i}{0}{0}{0}{i}{0}{0},\quad
\lambda^6 = \matthree {0}{0}{0}{0}{0}{1}{0}{1}{0},\\[1ex]
\lambda^7 = \matthree {0}{0}{0}{0}{0}{-i}{0}{i}{0},\quad
\lambda^8 = \frac{1}{\sqrt{3}} \matthree {1}{0}{0}{0}{1}{0}{0}{0}{-2}
\end{gather*}

These obey the following relations

\begin{equation}
[T_a,T_b]=i f_{abc}T_c,~~~~~~\{T_a,T_b\}=\frac13 \delta_{ab}\mathbb{1}+d_{abc}T_c,
    \label{eq:comm_and_anticomm_relations}
\end{equation}
where $f_{abc}$ and $d_{abc}$ denote the {\sl antisymmetric} and {\sl symmetric structure constants}, respectively. Their values are reported in Table \ref{Table:antisymm_structure_constants} and \ref{Table:symm_structure_constants}.

For completeness, here we list the trace relations that the generators obey, which can be easily derived from \eqref{eq:comm_and_anticomm_relations}
\begin{equation}
    \text{Tr}[T^a]=0,
    \label{eq:Tr_Ta}
\end{equation}
\begin{equation}
    \text{Tr}[T^aT^b]=\frac12 \delta_{ab},
    \label{eq:Tr_TaTb}
\end{equation}
\begin{equation}
    \text{Tr}[T^aT^b T^c]=\frac{1}{4}(d_{abc}+if_{abc}).
    \label{eq:Tr_TaTbTc}
\end{equation}

\begin{table*}[t]
    \centering
    \fontsize{12pt}{25pt}\selectfont
    \begin{tabular}{||ccccccccc||}
        \hline\hline
 $f_{123}$ & $f_{147}$ & $f_{156}$ & $f_{246}$ & $f_{257}$ & $f_{345}$ & $f_{367}$ & $f_{458}$ & $f_{678}$ \\
\hline
           $1$ & $\tfrac{1}{2}$ & $-\tfrac{1}{2}$ & $\tfrac{1}{2}$ & $\tfrac{1}{2}$ & $\tfrac{1}{2}$ & $-\tfrac{1}{2}$ & $\tfrac{\sqrt{3}}{2}$ & $\tfrac{\sqrt{3}}{2}$ \\
\hline\hline
    \end{tabular}
    \caption{The antisymmetric structure constants for SU(3). All the others terms are either zero or obtained by cyclical permutation of indices.}
    \label{Table:antisymm_structure_constants}
\end{table*}

\begin{table*}[t]
    \centering
    \fontsize{12pt}{25pt}\selectfont
    \begin{tabular}{||cccccccc||}
        \hline \hline $d_{118}$ &  $d_{146}$ &$d_{157}$ &$d_{228}$ &$d_{247}$ &$d_{256}$ &$d_{338}$ &$d_{344}$\\\hline
        $\frac{1}{\sqrt{3}}$ & $\frac{1}{2}$ &
        $\frac{1}{2}$ &
        $\frac{1}{\sqrt{3}}$ &
        $-\frac{1}{2}$ &
        $\frac{1}{2}$ &
        $\frac{1}{\sqrt{3}}$ &
        $\frac{1}{2}$ \\\hline\hline
$d_{355}$ &$d_{366}$ &$d_{377}$ &$d_{448}$ &$d_{558}$ &$d_{668}$ &$d_{778}$ &$d_{888}$\\\hline
        $\frac{1}{2}$ &
        $-\frac{1}{2}$ &
        $-\frac{1}{2}$ &
        $-\frac{1}{2\sqrt{3}}$ &
        $-\frac{1}{2\sqrt{3}}$ &
        $-\frac{1}{2\sqrt{3}}$ &
        $-\frac{1}{2\sqrt{3}}$ &
        $-\frac{1}{\sqrt{3}}$ \\\hline\hline
    \end{tabular}
    \caption{The symmetrc structure constants for SU(3). All the others terms are either zero or obtained by cyclical permutation of indices.}
    \label{Table:symm_structure_constants}
\end{table*}

\section{Milne and Minkowski coordinates}
\label{sec:Milne and Minkowski coordinates}
In studying heavy-ion collisions, one can choose the most suitable system of coordinates, in order to exploit the maximum amount of symmetry. In particular, the coordinates which allow us to work with the minimum number of variables are the-so called {\sl Milne coordinates}: these identify the beam direction as the preferred one, and are particularly useful in the case of boost-invariance.

To pass from the familiar Minkowski coordinates $(t,x, y, z)$ to the Milne coordinates $(\tau, x, y,\eta)$, one employs the following transformations:
\begin{equation}
\tau = \sqrt{t^2 -z^2}, \qquad \eta = \tanh^{-1} \left(\frac{z}{t}\right)=\frac12 \log \frac{t+z}{t-z},
\label{eq:milne_coordinate_transformation}
\end{equation}
whose inverses are:
\begin{equation}
t=\tau \cosh\tau, \qquad z=\tau \sinh\tau.
\label{eq:milne_inverse_coordinate_transformation}
\end{equation}
The variable $\tau$ is the {\sl proper time}, i.e. the time as measured in the rest frame of the system, while $\eta$ is referred to as {\sl space-time rapidity}. Both these variables are properly defined only in the time-like region, where $t>z$, which is the physically interesting one in which particles can be produced and travel. Ideally, for massless particles we could have $t=z$ and therefore their trajectory lies in the light-like region: for these we would have $\tau=0$ and $\eta=+\infty$ always. In this case the local map fails, and we would have to resort to another system of coordinates (either move back to Minkowski or to a third different system), however this case is unphysical since these particles would be spectators of the collision and no detector would be able to detect particles with polar angle either $0$ or $\pi$.

The metric tensor of such a coordinate system is
\begin{equation*}
	g_{\mu\nu} = \begin{pmatrix}
		1 & 0 & 0 & 0\\
		0 & -1 & 0 & 0 \\
		0 & 0 & -1 & 0 \\
		0 & 0 & 0 & -\tau^2 
	\end{pmatrix}.
\end{equation*}
Notice that the metric is singular if $\tau=0$, reflecting what we have just said about the light-like limit.

\chapter{Gluon momentum spectrum in early stages}
\label{appendix:spectrum_stuff}

As mentioned in the main text, the approaches which are present in the literature to evaluate the gluon spectrum are various. In this Appendix we want to give a brief insight and show the rationale for each one of them.\\

In \cite{Lappi:2003bi, Jia:2022awu}, the gluon spectrum has been evaluated starting from the hamiltonian density of the glasma at mid-rapidity, which is
\begin{equation}
	\mathcal{H}={\rm Tr}\left[\frac{1}{\tau}E_{i}^2+\tau E_{\eta}^2+\frac{1}{\tau}B_{i}^2+\tau B_\eta^2\right].
    \label{eq2.2.15}
\end{equation}
Thus we can identify Hamiltonian by:
\begin{equation}
	H(\tau)=\int \d \bm{x}_{T}^2\,\mathcal{H}(\bm{x}_{T},\tau)=\int\frac{\d^2\bm{p}_{T}}{(2\pi)^2}w(\bm{p}_{T})n(\tau,\bm{p}_{T}),
\end{equation}
where $\omega_{p} = |\bm{p}_{T}|$ is the free dispersion relation of gluons at initial stage, and $n(\tau,\bm{p}_{T})$ is their occupation number. We use the equation above to estimate the occupation number in evolving glasma and the relation is as follows
\begin{equation}
	\begin{aligned}
		n(\bm{p}_{T},\tau)=\frac{1}{|\bm{p}_{T}|}\rm{Tr}\bigg[&\frac{1}{\tau}E_{i}(\bm{p}_{T},\tau)E_{i}(-\bm{p}_{T},\tau)+\tau E_{\eta}(\bm{p}_{T},\tau)E_{\eta}(-\bm{p}_{T},\tau)\\
		&\left.+\frac{1}{\tau}B_i(\bm{p}_{T},\tau)B_i(-\bm{p}_{T},\tau)+ \tau B_\eta(\bm{p}_{T},\tau)B_\eta(-\bm{p}_{T},\tau)\right].
        \label{eq17}
	\end{aligned}
\end{equation}
Taking advantage of the equipartition of energy in the classical 
theory, only the electric field 
parts of the numerical solution were used. The gluon multiplicity can therefore be taken to be
\begin{equation}\label{eq:defmultiee}
\frac{\d N}{\d^2 \bm{k}_\perp \d y} = \frac{1}{(2 \pi)^2}\frac{2}{|\bm{k}_\perp|} 
\text{Tr} \bigg[
\frac{1}{\tau} E^i(\bm{k}_\perp)E^i(-\bm{k}_\perp) 
+\tau E^{\eta}(\bm{k}_\perp) E^{\eta}(-\bm{k}_\perp)
\bigg],
\end{equation}

Let us move on to the expression of the spectrum which we have employed in the main text, namely Eqs. \eqref{eq:first_line_spectrum_text}--\eqref{eq:multietaint_text}. The starting point are high energy factorization theorems, which have been derived for 
inclusive multi-gluon production in a rapidity interval 
$\Delta y \lesssim 1/\alpha_s$~\cite{Gelis:2008rw,Gelis:2008ad} 
in A+A collisions. The result can be expressed very compactly 
as~\cite{Gelis:2008ad, Lappi:2009xa}
\begin{equation}
\left<\frac{\d^n  N_n}{\d^3\bm{p}_1\cdots \d^3\bm{p}_n}\right>_{_{\rm LLog}}
 =
  \int \big[D\rho_1\big]\big[D\rho_2\big]\,
  Z_{y}\big[\rho_1\big]\,
  Z_{y}\big[\rho_2\big]\,
  \left.\frac{\d N}{\d^3\bm{p}_1}\right|_{_{\rm LO}}\cdots\,
  \left.\,\frac{\d N}{\d^3\bm{p}_n}\right|_{_{\rm LO}} .
\label{eq:ngluon-LLog}
\end{equation}
The $Z$'s are gauge invariant weight functionals that describe the distribution of color sources at the rapidity of interest. They are obtained in full generality by evolving the 
JIMWLK equations from an initial rapidity close to the 
beam rapidity. In the large $N_c$ limit, the weight functionals $Z$ 
can instead be obtained from the simpler mean field Balitsky--Kovchegov 
(BK) equation \cite{Balitsky:1995ub,Balitsky:1998kc,Balitsky:1998ya,Kovchegov:1999yj,Kovchegov:1999ua,Balitsky:2001re}. in this case, they can be represented as non-local Gaussian distributions in 
the sources~\cite{Blaizot:2004wv}. By doing so for large nuclei, without significant small $x$ 
evolution, one recovers the local Gaussian distributions of the McLerran-Venugopalan 
(MV) model~\cite{McLerran:1993ni,McLerran:1993ka,McLerran:1994vd}. 
We emphasize that the validity of \eqref{eq:ngluon-LLog} is restricted to the kinematics where all the produced particles 
are measured within a rapidity interval $ \lesssim 1/\alpha_s$ from each other, so that we can evaluate
$Z$ at this same rapidity $y_1 \approx \cdots \approx y_n \approx y$. This is acceptable when one considers boost invariant glasma. The generalization to larger rapidity separations is non-trivial 
because one needs to account for the gluon radiation in the region between the tagged gluons. A formalism describing arbitrarily long range rapidity 
separations has been developed  e.g. in \cite{Gelis:2008sz,Lappi:2009fq}, but for simplicity we do not consider it here.

The leading order single particle distributions in 
\eqref{eq:ngluon-LLog} are given by 
\begin{multline}
\left.E_{\bm{p}} \frac{\d N}{\d^3\bm{p}}\right|_{_{\rm LO}} \Big[\rho_1,\rho_2\Big]=\frac{1}{16\pi^3}
\lim_{x_0\to+\infty}\int \d^3\bm{x} \, \d^3\bm{y}
\;e^{ip\cdot(x-y)}
\;(\partial_x^0-iE_{\bm{p}})(\partial_y^0+iE_{\bm{p}})
\\
\cdot\sum_{\lambda}
\epsilon_\lambda^\mu(\bm{p})\epsilon_\lambda^\nu(\bm{p})
A_\mu^{a, \rm cl.}[\rho_1,\rho_2](x)\;A_\nu^{a, \rm cl.}[\rho_1,\rho_2](y)\; .
\label{eq:AA}
\end{multline}
For each configuration of sources $\rho_1$ and $\rho_2$ of each of the nuclei, 
one can solve classical Yang--Mills equations to compute the gauge fields 
$A_\mu^{\rm cl.}[\rho_1,\rho_2]$ in the forward light cone~\cite{Krasnitz:1998ns,Krasnitz:1999wc,Krasnitz:2000gz,Krasnitz:2001qu,Krasnitz:2003jw,Lappi:2003bi,Lappi:2006hq,Krasnitz:2002ng,Krasnitz:2002mn,Lappi:2006xc}. 
Once the expression \eqref{eq:AA} for the corresponding single inclusive distribution is substituted 
in \eqref{eq:ngluon-LLog}, and the distributions are averaged over with the distributions $Z$, one has determined from first principles (to all orders in perturbation theory and to leading logarithmic accuracy\footnote{In \cite{Lappi:2009xa} the authors refer to ‘‘leading log of $x$'' as the resummation of the leading dependence in rapidity between the observed gluons and the projectiles, not between the different produced gluons.} 
in $x$), the $n$-gluon inclusive distribution in high energy A+A collisions at proper times 
$\tau\sim 1/Q_s$. As noted previously, \eqref{eq:ngluon-LLog} is valid only for 
$\Delta Y\lesssim 1/\alpha_s\sim$ 3--5 
units of rapidity in A+A collisions at RHIC and the LHC, respectively.

The formalism \eqref{eq:AA} is fundamentally based on the LSZ formalism, and it leads to a definition of the gluon multiplicity which slightly deviates from \eqref{eq:defmultiee}. In particular we get

\begin{align}\label{eq:defmulti}
\frac{\d N}{\d^2 \bm{k}_\perp \d y} = \frac{1}{(2 \pi)^2} \Tr \bigg\{ &
\frac{1}{\tau |\bm{k}_\perp|} E^i(\bm{k}_\perp)E^i(-\bm{k}_\perp) 
+ \tau |\bm{k}_\perp|  A_i(\bm{k}_\perp)A_i(-\bm{k}_\perp) 
\\ &
+ \frac{1}{|\bm{k}_\perp|} \tau E^\eta(\bm{k}_\perp) E^\eta(-\bm{k}_\perp)
+ \frac{|\bm{k}_\perp|}{\tau} A_\eta(\bm{k}_\perp) A_\eta(-\bm{k}_\perp)
\\ & \label{eq:multitrint}
+ i \Big[ E^i(\bm{k}_\perp) A_i(-\bm{k}_\perp) - A_i(\bm{k}_\perp)E^i(-\bm{k}_\perp) \Big]
\\ & \label{eq:multietaint}
 +i \Big[ E^\eta(\bm{k}_\perp) A_\eta(-\bm{k}_\perp) -  A_\eta(\bm{k}_\perp) E^\eta(-\bm{k}_\perp) \Big]
\bigg\},
\end{align}
where the fields are evaluated in the 2-dimensional Coulomb gauge. Notice the presence of the $\bm{k}_T$-odd terms \eqref{eq:multitrint} and \eqref{eq:multietaint}, on which we have already discussed in §\ref{subsec:results_glasma}.\\

In other works, e.g. \cite{Schenke:2015aqa, Berges:2013eia} the single particle distribution is evaluated as
\begin{equation}
\left. \frac{dN}{d^2 \bm{k}_\perp dy} \right|_\tau 
= \frac{1}{(2\pi)^2} \sum_{\lambda, a} 
\left| \tau\, g^{\mu\nu} \left( \xi^{\lambda, \bm{k}_\perp\, *}_\mu(\tau) \, \overset\leftrightarrow{\partial}_\tau A^a_\nu(\tau, \bm{k}_\perp) \right) \right|^2
\label{eq.c1}
\end{equation}
where the metric is $g^{\mu\nu} = (1, -1, -1, -\tau^{-2})$, $\lambda = 1, 2$ labels the two transverse polarizations, and $a = 1, \dots, N_c^2 - 1$ is the color index. The double arrow denotes $A \overset\leftrightarrow{\partial}_\tau B=A\, \partial_\tau B-B\, \partial_\tau A$. In Coulomb gauge ($A_\tau=0$ and $\partial_i A^i=0|_\tau$) the mode functions take the form
\begin{align}
\xi^{(1), \bm{k}_\perp}_\mu(\tau) &= \frac{\sqrt{\pi}}{2|\bm{k}_\perp|}
\begin{pmatrix}
- k_y \\
k_x \\
0
\end{pmatrix}
H_0^{(2)}(|\bm{k}_\perp| \tau),
\label{eq.c2} \\
\xi^{(2), \bm{k}_\perp}_\mu(\tau) &= \frac{\sqrt{\pi}}{2|\bm{k}_\perp|}
\begin{pmatrix}
0 \\
0 \\
k_\perp \tau
\end{pmatrix}
H_1^{(2)}(|\bm{k}_\perp| \tau),
\label{eq.c3}
\end{align}
where $\bm{k}_\perp = (k_x, k_y)$ and $H_\alpha^{(2)}$ denote the Hankel functions of the second type and order $\alpha$.

It is a useful to understand the connection between this expression and the others, when fully expanded in terms of the gauge field and of its conjugate momentum i.e. the electric field.
The full expansion of \eqref{eq.c1} in term of Hankel functions is:
\begin{align}
\nonumber
&\left.\frac{dN}{d^2 \bm{k}_\perp dy} \right|_\tau=\\\nonumber
&\frac{1}{8\pi}\text{Tr}\Big[|\bm{k}_\perp|^2\tau^2\left(H_1^{(2)*}H_1^{(2)}\right)(A_i(\bm{k}_\perp) A_i(-\bm{k}_\perp))
-\tau^2\left(H_1^{(2)*}H_1^{(2)}\right)(k_i A_i(\bm{k}_\perp))(k_j A_j(-\bm{k}_\perp))\\\nonumber
&+\left(H_0^{(2)*}H_0^{(2)}\right)E^i(\bm{k}_\perp) E^{i}(-\bm{k}_\perp)-\frac{1}{|\bm{k}_\perp|^2}\left(H_0^{(2)*}H_0^{(2)}\right)(k_i E^i(\bm{k}_\perp))(k_jE^{j}(-\bm{k}_\perp))\\\nonumber
&+|\bm{k}_\perp|\tau\left(H_1^{(2)*}H_0^{(2)}\right)A_i(\bm{k}_\perp)E^{i}(-\bm{k}_\perp)-\frac{\tau}{|\bm{k}_\perp|}\left(H_1^{(2)*}H_0^{(2)}\right)(k_iA_i(\bm{k}_\perp))(k_jE^{j}(-\bm{k}_\perp))\\\nonumber
&+|\bm{k}_\perp|\tau\left(H_1^{(2)}H_0^{(2)*}\right)A_i(-\bm{k}_\perp)E^{i}(\bm{k}_\perp)-\frac{\tau}{|\bm{k}_\perp|}\left(H_1^{(2)}H_0^{(2)*}\right)(k_iA_i(-\bm{k}_\perp))(k_jE^{j}(\bm{k}_\perp))\\\nonumber
&+A_\eta(\bm{k}_\perp) A_\eta(-\bm{k}_\perp)(M^*M)|\bm{k}_\perp|^2+\tau^2\left(H_1^{(2)}H_1^{(2)*}\right)E^\eta(\bm{k}_\perp) E^{\eta}(-\bm{k}_\perp)\\
&-|\bm{k}_\perp|\tau M^* H_1^{(2)}A_\eta(\bm{k}_\perp) E^{\eta}(-\bm{k}_\perp)-|\bm{k}_\perp|\tau M H_1^{(2)*}A_\eta(-\bm{k}_\perp) E^{\eta}(\bm{k}_\perp)\Big]
    \label{eq.c4}
\end{align}
where
\begin{equation}
M\equiv \frac{1}{|\bm{k}_\perp|\tau}H_1^{(2)}+\frac12 \left(H_0^{(2)}-H_2^{(2)}\right).
    \label{eq.c5}
\end{equation}
In the transverse Coulomb gauge, $k_i A^i = k_i E^i=0$, 
this result simplifies to
\begin{align}
\nonumber
\left.\frac{dN}{d^2 \bm{k}_\perp dy} \right|_\tau=&\frac{1}{8\pi}\text{Tr}\Big[|\bm{k}_\perp|^2\tau^2\left(H_1^{(2)*}H_1^{(2)}\right)(A_i(\bm{k}_\perp) A_i(-\bm{k}_\perp))
 \\\nonumber
&+\left(H_0^{(2)*}H_0^{(2)}\right)E^i(\bm{k}_\perp) E^{i}(-\bm{k}_\perp) \\\nonumber
&+|\bm{k}_\perp|\tau\left(H_1^{(2)*}H_0^{(2)}\right)A_i(\bm{k}_\perp)E^{i}(-\bm{k}_\perp) \\\nonumber
&+|\bm{k}_\perp|\tau\left(H_1^{(2)}H_0^{(2)*}\right)A_i(-\bm{k}_\perp)E^{i}(\bm{k}_\perp) \\\nonumber
&+A_\eta(\bm{k}_\perp) A_\eta(-\bm{k}_\perp)(M^*M)|\bm{k}_\perp|^2+\tau^2\left(H_1^{(2)}H_1^{(2)*}\right)E^\eta(\bm{k}_\perp) E^{\eta}(-\bm{k}_\perp)\\
&-|\bm{k}_\perp|\tau M^* H_1^{(2)}A_\eta(\bm{k}_\perp) E^{\eta}(-\bm{k}_\perp)-|\bm{k}_\perp|\tau M H_1^{(2)*}A_\eta(-\bm{k}_\perp) E^{\eta}(\bm{k}_\perp)\Big]
    \label{eq.c4_simple}
\end{align}
This formula can be expanded in the $|\bm{k}_\perp|\tau\to +\infty$, by exploiting the asymptotic behavior of the Hankel functions. By doing so, we get:
\begin{align}
\nonumber
\left.\frac{dN}{d^2 \bm{k}_\perp dy} \right|_\tau=\frac{1}{(2\pi)^2}\text{Tr}&\Big\{|\bm{k}_\perp|\tau A_i(\bm{k}_\perp) A_i(-\bm{k}_\perp)+\frac{1}{|\bm{k}_\perp|\tau}E^i(\bm{k}_\perp) E^{i}(-\bm{k}_\perp)\\\nonumber
&+\frac{|\bm{k}_\perp|}{\tau}A_\eta(\bm{k}_\perp) A_\eta(-\bm{k}_\perp)+\frac{\tau}{|\bm{k}_\perp|}E^\eta(\bm{k}_\perp) E^{\eta}(-\bm{k}_\perp)\\\nonumber
&+i[E^\eta(\bm{k}_\perp)A_\eta(-\bm{k}_\perp)-A_{\eta}(\bm{k}_\perp)E^{\eta}(-\bm{k}_\perp)]\\\nonumber
&+i[E^i(\bm{k}_\perp)A_i(-\bm{k}_\perp)-A_i(\bm{k}_\perp)E^{i}(-\bm{k}_\perp)]\\\nonumber
&+\frac{1}{2}\frac{1}{|\bm{k}_\perp| \tau}(E^i(\bm{k}_\perp)A_i(-\bm{k}_\perp)+A_i(\bm{k}_\perp) E^{i}(-\bm{k}_\perp))\\
&-\frac{1}{2}\frac{1}{|\bm{k}_\perp|\tau}(E^\eta(\bm{k}_\perp) A_\eta(-\bm{k}_\perp)+A_\eta(\bm{k}_\perp) E^{\eta}(-\bm{k}_\perp))\Big\}
\label{eq.c6}
\end{align}
We can note that the terms in the first four lines of \eqref{eq.c6} coincide with the spectrum in \cite{Lappi:2009xa}, 
which is the one we used in the main text (Eqs. \eqref{eq:first_line_spectrum_text}--\eqref{eq:multietaint_text}).

Similarly we can evaluate the spectrum from \eqref{eq.c1} in the limit of small times. Let us move on with the calculations:
\begin{equation}
    \xi_\mu^{\lambda,\bm{k}_\perp *}(\tau)\stackrel{\leftrightarrow}{\partial_\tau} A_\nu^a(\tau, \bm{k}_\perp)=\xi_\mu^{\lambda,\bm{k}_\perp *}(\tau)\left[\partial_\tau A_\nu^a(\tau, \bm{k}_\perp)\right]-\left[\partial_\tau\xi_\mu^{\lambda,\bm{k}_\perp *}(\tau)\right] A_\nu^a(\tau, \bm{k}_\perp).
    \label{eq:spectrum_small_times_1}
\end{equation}
As far as the first term is concerned we have:
\begin{align}
\tau &g^{\mu\nu}\xi_\mu^{\lambda,\bm{k}_\perp *}(\tau)\left[\partial_\tau A_\nu^a(\tau, \bm{k}_\perp)\right]\nonumber\\
&=-\tau \xi_1^{\lambda,\bm{k}_\perp *}(\tau)\left[\partial_\tau A_1^a(\tau, \bm{k}_\perp)\right]-\tau \xi_2^{\lambda,\bm{k}_\perp *}(\tau)\left[\partial_\tau A_2^a(\tau, \bm{k}_\perp)\right]-\tau \cdot \frac{1}{\tau^2}\xi_\eta^{\lambda,\bm{k}_\perp *}(\tau)\left[\partial_\tau A_\eta^a(\tau, \bm{k}_\perp)\right].
\label{eq:spectrum_small_times_2}
\end{align}
For each polarization we have:
\begin{itemize}
    \item[$\lambda=1:$]
    \begin{equation}
    -\tau \cdot \frac{\sqrt{\pi}}{2|\bm{k}_\perp|}(-k_y)H_0^{(2)*}(|\bm{k}_\perp|\tau)\partial_\tau A_1^a-\tau \cdot \frac{\sqrt{\pi}}{2|\bm{k}_\perp|}k_x H_0^{(2)*}(|\bm{k}_\perp|\tau)\partial_\tau A_2^a+0
    \label{eq:spectrum_small_times_3}
\end{equation}
    \item[$\lambda=2:$]
    \begin{equation}-\tau \cdot \frac{1}{\tau^2}\cdot \frac{\sqrt{\pi}}{2|\bm{k}_\perp|}(|\bm{k}_\perp| \tau)H_1^{(2)*}(|\bm{k}_\perp|\tau)\partial_\tau A_\eta^a=-\frac{\sqrt{\pi}}{2}H_1^{(2)*}(|\bm{k}_\perp|\tau)\partial_\tau A_\eta^a
    \label{eq:spectrum_small_times_4}
\end{equation}
\end{itemize}
For the second term we have instead:
\begin{align}
-\tau &g^{\mu\nu}\left[\partial_\tau\xi_\mu^{\lambda,\bm{k}_\perp *}(\tau)\right] A_\nu^a(\tau, \bm{k}_\perp)\nonumber\\
&=\tau \left[\partial_\tau\xi_1^{\lambda,\bm{k}_\perp *}(\tau)\right] A_1^a(\tau, \bm{k}_\perp)+\tau \left[\partial_\tau\xi_2^{\lambda,\bm{k}_\perp *}(\tau)\right] A_2^a(\tau, \bm{k}_\perp)+\tau \cdot \frac{1}{\tau^2}\left[\partial_\tau\xi_3^{\lambda,\bm{k}_\perp *}(\tau)\right] A_\eta^a(\tau, \bm{k}_\perp).
\label{eq:spectrum_small_times_5}
\end{align}
For each polarization we have:
\begin{itemize}
    \item[$\lambda=1:$]
    \begin{align}\tau &\left[\partial_\tau\xi_1^{(1),\bm{k}_\perp *}(\tau)\right] A_1^a(\tau, \bm{k}_\perp)+\tau \left[\partial_\tau\xi_2^{(2),\bm{k}_\perp *}(\tau)\right] A_2^a(\tau, \bm{k}_\perp)\nonumber \\
    &=\tau \frac{\sqrt{\pi}}{2 |\bm{k}_\perp|}(-k_y)\frac{\partial}{\partial y}H_0^{(2)*}(y)|\bm{k}_\perp|A_1^a+\tau \frac{\sqrt{\pi}}{2 |\bm{k}_\perp|}k_x\frac{\partial}{\partial y}H_0^{(2)*}(y)|\bm{k}_\perp|A_2^a
    \label{eq:spectrum_small_times_6}
\end{align}
    \item[$\lambda=2:$]
    \begin{align}\tau\cdot \frac{1}{\tau^2}\frac{\sqrt{\pi}}{2|\bm{k}_\perp|}|\bm{k}_\perp|H_1^{(2)*}(|\bm{k}_\perp|\tau)A_\eta^a+\tau\cdot \frac{1}{\tau^2}\frac{\sqrt{\pi}}{2 |\bm{k}_\perp|}(|\bm{k}_\perp|\tau) \frac{\partial}{\partial y}H_1^{(2)*}(y)\cdot |\bm{k}_\perp|A_\eta
    \label{eq:spectrum_small_times_7}
\end{align}

\end{itemize}
Putting everything together we therefore get:
\begin{align}
    &\left.\frac{dN}{d^2 \bm{k}_\perp dy}\right|_{\tau=0}=\nonumber\\
    &\frac{1}{(2\pi)^2}\left\{\sum_a\left|\tau \frac{\sqrt{\pi}}{2|\bm{k}_\perp|}H_0^{(2)*}(|\bm{k}_\perp|\tau)[k_y(\partial_\tau A_1^a)-k_x(\partial_\tau A_2^a)]-\tau \frac{\sqrt{\pi}}{2}\frac{\partial}{\partial y} H_0^{(2)*}(y)[k_yA_1^a-k_xA_2^a]\right|^2\right.\nonumber\\
    &+\left.\sum_a \left|-\frac{\sqrt{\pi}}{2}H_1^{(2)*}(|\bm{k}_\perp|\tau) \partial_\tau A_\eta^a+A_\eta^a\frac{\sqrt{\pi}}{2}\left[\frac{1}{\tau} H_1^{(2)*}(|\bm{k}_\perp|\tau)+|\bm{k}_\perp| \frac{\partial}{\partial y}H_1^{(2)*}(y)\right] \right|^2
\right\}\nonumber\\
&=\frac{1}{(2\pi)^2}\left\{\sum_a\left|\tau \frac{\sqrt{\pi}}{2|\bm{k}_\perp|}H_0^{(2)*}(|\bm{k}_\perp|\tau)[k_y(\partial_\tau A_1^a)-k_x(\partial_\tau A_2^a)]+\tau \frac{\sqrt{\pi}}{2} H_1^{(2)*}(|\bm{k}_\perp|\tau)[k_yA_1^a-k_xA_2^a]\right|^2+\right.\nonumber\\
&+\left.\sum_a \left|-\frac{\sqrt{\pi}}{2}H_1^{(2)*}(|\bm{k}_\perp|\tau) \partial_\tau A_\eta^a+A_\eta^a\frac{\sqrt{\pi}}{2}\left[\frac{1}{\tau} H_1^{(2)*}(|\bm{k}_\perp|\tau)+|\bm{k}_\perp| \frac{1}{2}[H_0^{(2)*}(|\bm{k}_\perp|\tau)-H_2^{(2)*}(|\bm{k}_\perp|\tau)]\right] \right|^2
\right\},
\label{eq:spectrum_small_times_8}
\end{align}
where we used the following properties of the Hankel functions:
\begin{equation}
    \frac{\partial}{\partial y} H_0^{(2)}(y)=- H_1^{(2)}(y),~~~~\frac{\partial}{\partial y} H_1^{(2)}(y)=\frac{1}{2}[H_0^{(2)}(y)-H_2^{(2)}(y)].
    \label{eq:spectrum_small_times_9}
\end{equation}
Let us now perform the $\tau\to 0$ limit. In this limit the Hankel functions behave as:
\begin{align}
    H_0^{(2)*}(|\bm{k}_\perp|\tau)&\to 1+\frac{i}{\pi}[2\gamma-2\log 2+2\log (|\bm{k}_\perp|\tau)],\nonumber\\
    H_1^{(2)*}(|\bm{k}_\perp|\tau)&\to -\frac{2i}{\pi(|\bm{k}_\perp|\tau)},\nonumber\\
    H_2^{(2)*}(|\bm{k}_\perp|\tau)&\to -\frac{4i}{\pi(|\bm{k}_\perp|\tau)^2}-\frac{i}{\pi}.
\label{eq:spectrum_small_times_10}
\end{align}
We obtain:
\begin{align}
\left.\frac{dN}{d^2 \bm{k}_\perp dy}\right|_{\tau=0}=&\frac{1}{(2\pi)^2}\left\{\sum_a\left|\tau \frac{\sqrt{\pi}}{2}\frac{2i}{\pi(|\bm{k}_\perp|\tau)}[k_yA_1^a-k_xA_2^a]\right|^2\right.\nonumber\\
&+\sum_a \left|\frac{\sqrt{\pi}}{2}\frac{2i}{\pi(|\bm{k}_\perp|\tau)} \partial_\tau A_\eta^a+A_\eta^a\frac{\sqrt{\pi}}{2}\left[-\frac{1}{\tau} \frac{2i}{\pi(|\bm{k}_\perp|\tau)}\right.\right.\nonumber\\
&\left.\left.\left.+\frac{1}{2}|\bm{k}_\perp| \left(1+\frac{i}{\pi}\left(2\gamma-2\log 2+2\log (|\bm{k}_\perp|\tau)+\frac{4i}{\pi(|\bm{k}_\perp|\tau)^2}+\frac{i}{\pi}\right)\right)\right] \right|^2
\right\}\nonumber\\
=&\frac{1}{(2\pi)^2}\left\{\sum_a\left|\frac{1}{\sqrt{\pi}|\bm{k}_\perp|}[k_yA_1^a-k_xA_2^a]\right|^2\right.\nonumber\\
&+\sum_a \left|+\frac{\sqrt{\pi}}{2}\frac{2i}{\pi(|\bm{k}_\perp|\tau)} (-2\tau A^{\eta,a}-\tau^2\partial_\tau A^{\eta,a})-\tau^2 A^{\eta,a}\frac{\sqrt{\pi}}{2}\left[-\frac{1}{\tau} \frac{2i}{\pi(|\bm{k}_\perp|\tau)}\right.\right.\nonumber\\
&\left.\left.\left.+\frac{1}{2}|\bm{k}_\perp| \left(1+\frac{i}{\pi}\left(2\gamma-2\log 2+2\log (|\bm{k}_\perp|\tau)+\frac{4i}{\pi(|\bm{k}_\perp|\tau)^2}+\frac{i}{\pi}\right)\right)\right] \right|^2
\right\}\to\nonumber\\
\to& \frac{1}{(2\pi)^2}\left\{\sum_a\left|\frac{1}{\sqrt{\pi}|\bm{k}_\perp|}[k_yA_1^a-k_xA_2^a]\right|^2\right.\nonumber\\
&\left.+\sum_a \left|-\frac{2i}{\sqrt{\pi}(|\bm{k}_\perp|)} A^{\eta,a}-A^{\eta,a}\left[- \frac{i}{\sqrt{\pi}(|\bm{k}_\perp|)}+\frac{i}{\sqrt{\pi}(|\bm{k}_\perp|)}\right] \right|^2
\right\}\nonumber\\
=&\frac{1}{(2\pi)^2}\left\{\sum_a\left|\frac{1}{\sqrt{\pi}|\bm{k}_\perp|}[k_yA_1^a-k_xA_2^a]\right|^2+\sum_a \left|-\frac{2i}{\sqrt{\pi}(|\bm{k}_\perp|)} A^{\eta,a}\right|^2
\right\}.
\label{eq:spectrum_small_times_11}
\end{align}

\addcontentsline{toc}{chapter}{Bibliography}
\bibliographystyle{mprsty}
\bibliography{biblio.bib}

\end{document}